\documentclass{jfm}
\usepackage{amsmath}
\usepackage{amssymb}
\usepackage{alphalph}
\usepackage{array}
\usepackage{graphicx}
\usepackage{xcolor}
\usepackage{comment}
\usepackage{float}
\usepackage{subcaption}
\usepackage{placeins}
\usepackage{nameref}
\usepackage[colorlinks=true,citecolor=blue,linkcolor=blue,filecolor=blue]{hyperref}
\usepackage[nameinlink,noabbrev]{cleveref}
\usepackage[normalem]{ulem}

\crefname{section}{§}{§§}
\Crefname{section}{§}{§§}
\crefformat{section}{§#2#1#3}
\crefformat{subfigure}{#2figure~(#1)#3}
\Crefformat{subfigure}{#2Figure~(#1)#3}

\definecolor{app-crimson}{rgb}{0.6350 0.0780 0.1840}
\definecolor{app-green}{rgb}{0.4660 0.6740 0.1880}
\definecolor{app-purple}{rgb}{0.4940 0.1840 0.55600}
\definecolor{du-peach}{rgb}{0.8 0.4 0.4}
\definecolor{C15-blue}{rgb}{0.0290 0.6940 1}
\definecolor{C18-yellow}{rgb}{0.9290 0.6940 0}

\title{Near-wall turbulence alteration with the Transpiration-Resistance Model}

\author{Seyed Morteza Habibi Khorasani\aff{1}\corresp{\email{smhk2@mech.kth.se}}, U\v{g}is L\={a}cis\aff{1,2}, Simon Pasche\aff{3}, Marco Edoardo Rosti\aff{4} \and Shervin Bagheri\aff{1}}

\shortauthor{S.M. Habibi Khorasani, U. L\={a}cis, S. Pasche, M.E. Rosti and S. Bagheri}

\affiliation{\aff{1}Department of Engineering Mechanics, FLOW Centre, KTH Royal Institute of Technology, SE-100 44 Stockholm, Sweden
\aff{2}Institute of Atomic Physics and Spectroscopy, University of Latvia, LV-1586 Riga, Latvia
\aff{3}Leclanché SA, Avenue des Sports 42, CH-1400 Yverdon-les-Bains, Switzerland
\aff{4}Okinawa Institute of Science and Technology Graduate University, 1919-1 Tancha, Onna-son, Okinawa 904-0495, Japan}

\begin{document}

\maketitle

\begin{abstract}
A set of boundary conditions called the Transpiration-Resistance Model (TRM) are investigated in altering near-wall turbulence. The TRM has been previously proposed by \citet{Lacis2020} as a means of representing the net effect of surface micro-textures on their overlying bulk flows. It encompasses conventional Navier-slip boundary conditions relating the streamwise and spanwise velocities to their respective shears through the slip lengths $\ell_x$ and $\ell_z$. In addition, it features a transpiration condition accounting for the changes induced in the wall-normal velocity by expressing it in terms of variations of the wall-parallel velocity shears through the transpiration lengths $m_x$ and $m_z$. Greater levels of drag increase occur when more transpiration takes place at the boundary plane, with turbulent transpiration being predominately coupled to the spanwise shear component for canonical near-wall turbulence. The TRM can reproduce the effect of a homogeneous and structured roughness up to ${k^+}\,{\approx18}$. In this transitionally rough flow regime, the transpiration lengths of the TRM must be empirically determined. The \emph{transpiration factor} is defined as the product between the slip and transpiration lengths, i.e. $(m\ell)_{x,z}$. This factor contains the compound effect of the wall-parallel velocity occurring at the boundary plane and increased permeability, both of which lead to the transport of momentum in the wall-normal direction. A linear relation between the transpiration factor and the roughness function is observed for regularly textured surfaces in the transitionally rough regime of turbulence.
The relations obtained between the transpiration factor and the roughness function show that such effective flow quantities can be suitable measures for characterizing rough surfaces in this flow regime.
\end{abstract}

\section{Introduction} \label{sec:intro}
Flows over surfaces are of engineering significance owing to their extensive technological applications. This has led to a vast body of work dedicated to this subject, with particular attention given to turbulent flow over rough surfaces \citep{Schlichting,Panton,jimenez_rough,chung_2021}. Within this large field, surface roughness may be categorized into regular and random (irregular) types, in which the former category is relevant to this work. Many direct numerical simulation (DNS) studies have been dedicated to understanding the changes caused in a turbulent flow due to the presence of surface roughness  \citep{leonardi_orlandi_smalley_djenidi_antonia_2003,orlandi_leonardi_antonia_2006,Forooghi_rough_2018,Abderrahaman2019}. Such studies have involved geometrical representations of the surface roughness, a necessity for capturing all of the flow physics down to the scale of the surface texture elements.

In the context of drag-reducing surfaces, such as riblets, a different line of inquiry pursued was characterizing the effect of textured surfaces in terms of physically meaningful flow parameters and eschewing the minute details of the flow within the region of the texture. \citet{bechert_bartenwerfer_1989} studied the drag-reducing effect of riblets experimentally and established that their influence on the longitudinal flow was as if it perceived a plain wall at a depth below the riblet tips and called this distance ``protrusion height". \citet{luchini_manzo_pozzi_1991} demonstrated that the protrusion height of the cross-flow, $h_{\perp}$, was less than that of the longitudinal flow, $h_{\parallel}$, for riblets. In other words, the former perceives a shallower plane wall (virtual origin) than the latter. They then demonstrated that the only physically pertinent parameter for characterising drag change becomes their difference, $\Delta{h}$. Analogous to protrusion heights, slip lengths have been used in combination with Navier slip velocity boundary conditions to imitate the behaviour of (idealized) superhydrophobic surfaces (SHS) on turbulent flows \citep{Min_2004,Fukagata_2006,busse_2012,luchini_2015,fairhall_abderrahaman-elena_garcia-mayoral_2019}.

The concepts of slip lengths and protrusion heights were later established as being equivalent \citep{luchini_2015,Garcia_Mayoral_2019}. Both are based on the principle that the viscous sublayer of a turbulent boundary layer is similar to a Couette flow. So long as the surface texture elements are small enough to reside within this layer, their effect can be well represented using slip lengths. Inspired by the success of these \textit{effective representations} for drag-reducing surfaces, this paper investigates if a similar approach can be used to model rough surface textures, such as a regular array of posts, that increase turbulent drag. While the idea of using velocity boundary conditions has seen usage for turbulence wall modeling in large eddy simulations \citep{bose_2018}, it has yet to see wider adoption in CFD codes as models for non-smooth surfaces. The reason is mainly that an appropriate form of such boundary conditions is not clearly established. For roughness, homogeneous slip boundary conditions have proven to be inadequate at capturing the interaction between the surface and the overlying turbulent flow \citep{zampogna_2019,bottaro_2019}, as they do not account for the transport of momentum in the wall-normal direction, i.e. the transpiration \citep{gomez_2018}. Hence, the objective of this work is to investigate the turbulence alteration induced by a particular set of effective boundary conditions called the \emph{Transpiration-Resistance Model} (TRM) that accounts for transpiration.

The TRM encapsulates the effect of surface micro-textures and was proposed by \citet{Lacis2020}. It falls under the category of homogenization approaches \citep{bottaro_2019} and originates from conditions that are rigorously derived from an asymptotic analysis under creeping-flow assumption \citep{sudhakar2020}. It is comprised of Robin boundary conditions, with Navier-slip type conditions for the wall-parallel velocities and a transpiration condition coupling the wall-normal velocity to the changes in shear of the other two velocity components. In addition, the boundary conditions contain several coefficients, such as slip and transpiration lengths, which need to be determined for each particular roughness geometry. For viscous-dominated flows, it is already established that the TRM can represent the effect of real roughness \citep{Lacis2020,sudhakar2020} and that the associated coefficients can be obtained from carrying out analyses on unit-cells that contain one periodic sample of the surface texture \citep{Lacis2020}. For turbulent flows however, the validity of the TRM as a surrogate model of real roughness is not established. It is also much more difficult to determine the associated TRM coefficients due to the inertial and unsteady flow around the surface texture. Further insight into these aspects will be provided by the present investigation. The extent of the applicability of a model which is comprised of boundary conditions for all three velocity components has not, to the best of the authors' knowledge, been explored for turbulent flows over rough surfaces in a manner similar to what has been done for SHS \citep{fairhall_abderrahaman-elena_garcia-mayoral_2019} using slip-only boundary conditions.

In the fully rough regime of turbulence, form-induced (pressure) drag is dominant and the cycle of canonical near-wall turbulence becomes entirely disrupted \citep{jimenez_rough}. Such effects cannot be captured by homogenized planar boundary conditions and must be supplemented \citep{bottaro_2019}, such as with the addition of volumetric forcing terms in the Navier-Stokes equations \citep{Forooghi_2018}. Likewise, in the upper transitionally rough regime ($20 \lesssim k_s^+ \lesssim 50$, where $k_s^+$ is the equivalent sand-grain roughness) the texture-coherent flow directly interacts with the overlying turbulence, limiting the use of homogenized conditions as the near-wall dynamics will no longer be ``smooth-wall-like'' \citep{Abderrahaman2019}.
Therefore, the focus of this work is on the low to intermediate range of transitional roughness ($5 \lesssim  k_s^+ \lesssim 20$), where the drag modification due to roughness is still dominated by viscous drag. It will be shown that textured surfaces can be effectively represented with the TRM in this regime.

The lower transitionally rough regime is important for turbulent applications operating at low and intermediate Reynolds numbers. By using the principle of mass conservation and the virtual-origin framework of \citet{ibrahim2020smoothwalllike}, in addition to assessing the effects of the TRM, the role of rough-wall-induced transpiration in the departure from the regime where near-wall turbulence remains smooth-wall-like will be highlighted.

\section{Governing equations and numerical methods} \label{sec:governing}

\subsection{The TRM boundary conditions} \label{TRMboundarycondition}

The flow considered is governed by the incompressible Navier-Stokes equations
\begin{gather}
\cfrac{\partial \boldsymbol{u}}{ \partial t} + \boldsymbol{u}\cdot\nabla{\boldsymbol{u}} = -\nabla{p} + \cfrac{1}{Re}{\nabla}^2{\boldsymbol{u}}, \label{NS}\\
\nabla\cdot{\boldsymbol{u}} = 0,\label{cont}
\end{gather}
where $\boldsymbol{u}=(u, v, w)$ is the fluid velocity with $u$, $v$ and $w$ being the velocity components along the streamwise ($x$), wall-normal ($y$) and spanwise ($z$) coordinates, $p$ is the pressure and \emph{Re} is the bulk Reynolds number. The effect of a rough surface texture on a turbulent channel flow is emulated by using the TRM.

The TRM wall boundary conditions for the three velocity components are comprised of the following Robin boundary conditions
\begin{gather}
u = \ell_x \cfrac{\partial u}{ \partial y}\Big\rvert_{y=0}, \label{eq:TRMu}\\
w = \ell_z \cfrac{\partial w}{ \partial y}\Big\rvert_{y=0}, \label{eq:TRMw}\\
v = -m_x \cfrac{\partial u}{\partial x}\Big\rvert_{y=0} - m_z \cfrac{\partial w}{\partial z}\Big\rvert_{y=0}. \label{eq:TRMv}
\end{gather}
The conditions for the streamwise and spanwise velocities  (\ref{eq:TRMu}-\ref{eq:TRMw}) are the familiar Navier-slip condition. What is distinctive, is the boundary condition for the wall-normal velocity, i.e. the transpiration boundary condition \eqref{eq:TRMv}. 
The coefficients $m_x$ and $m_z$, in analogy to the slip lengths $\ell_x$ and $\ell_z$, are called the transpiration lengths. In hydrodynamic terms, the transpiration length represents approximately the distance below the interface to which wall-normal fluid motion can penetrate, while the slip length is the distance where the velocity profile decays to zero when linearly extrapolated.
From here on, the superscript ``$+$'' indicates scaling in ``inner units"; i.e. normalized using the friction velocity based on the fluid stress at the wall, $u_{\tau} = \sqrt{\tau_w/\rho}$, and the kinematic viscosity $\nu$.

%%%%%%%%%%%%%%%%%%%%%%%%%%%%%%%%%%%%%%%%%%%%%%%%%%%%%%%%%%%%%%%
\begin{figure}
    \begin{center}
    \begin{subfigure}[tbp]{0.45\textwidth}
        \includegraphics[width=1\linewidth]{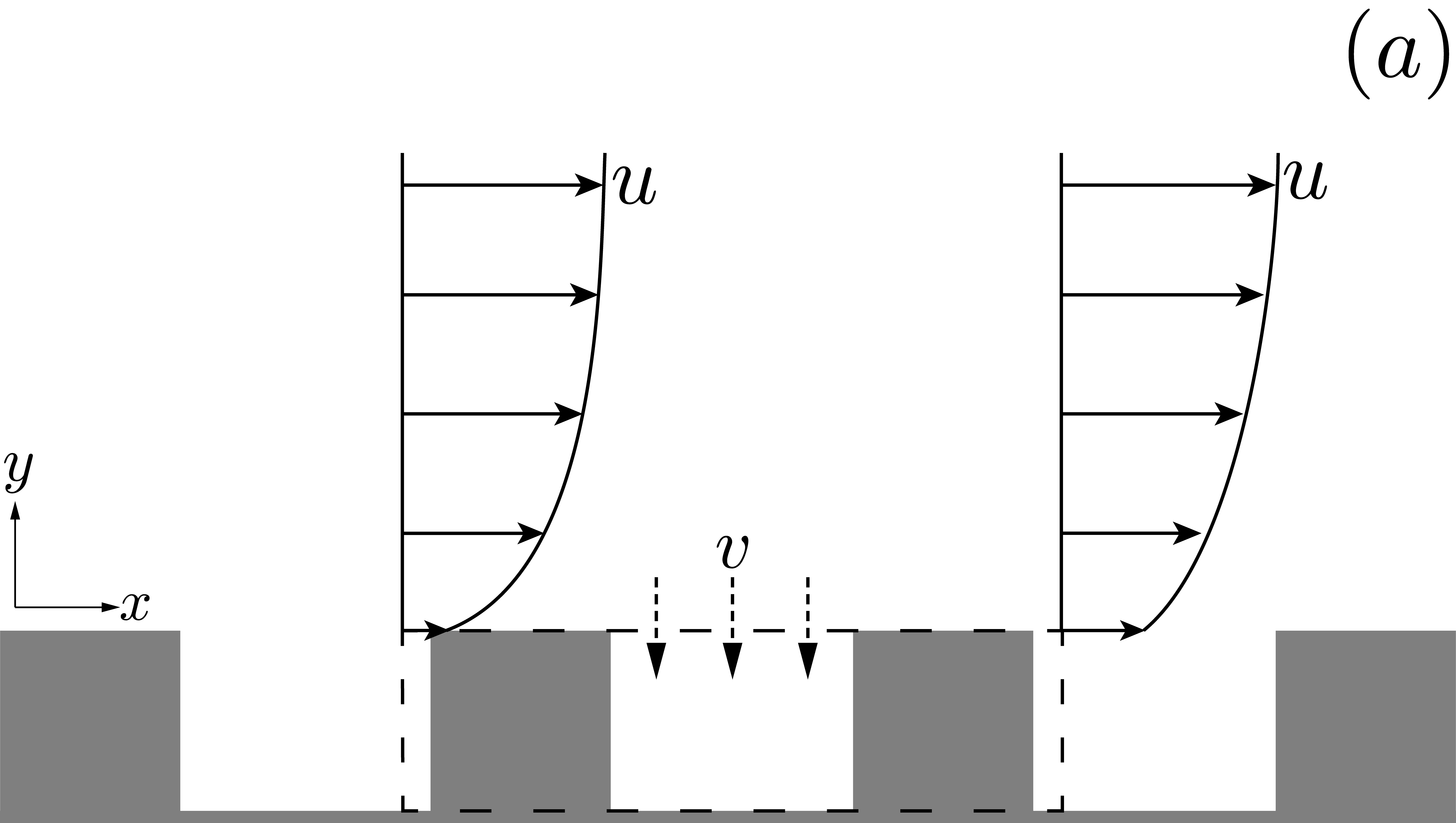}
        {\phantomcaption\label{fig:TRM:concept}}
    \end{subfigure}%
    \hspace*{5mm}
    \begin{subfigure}[tbp]{0.45\textwidth}
        \includegraphics[width=1\linewidth]{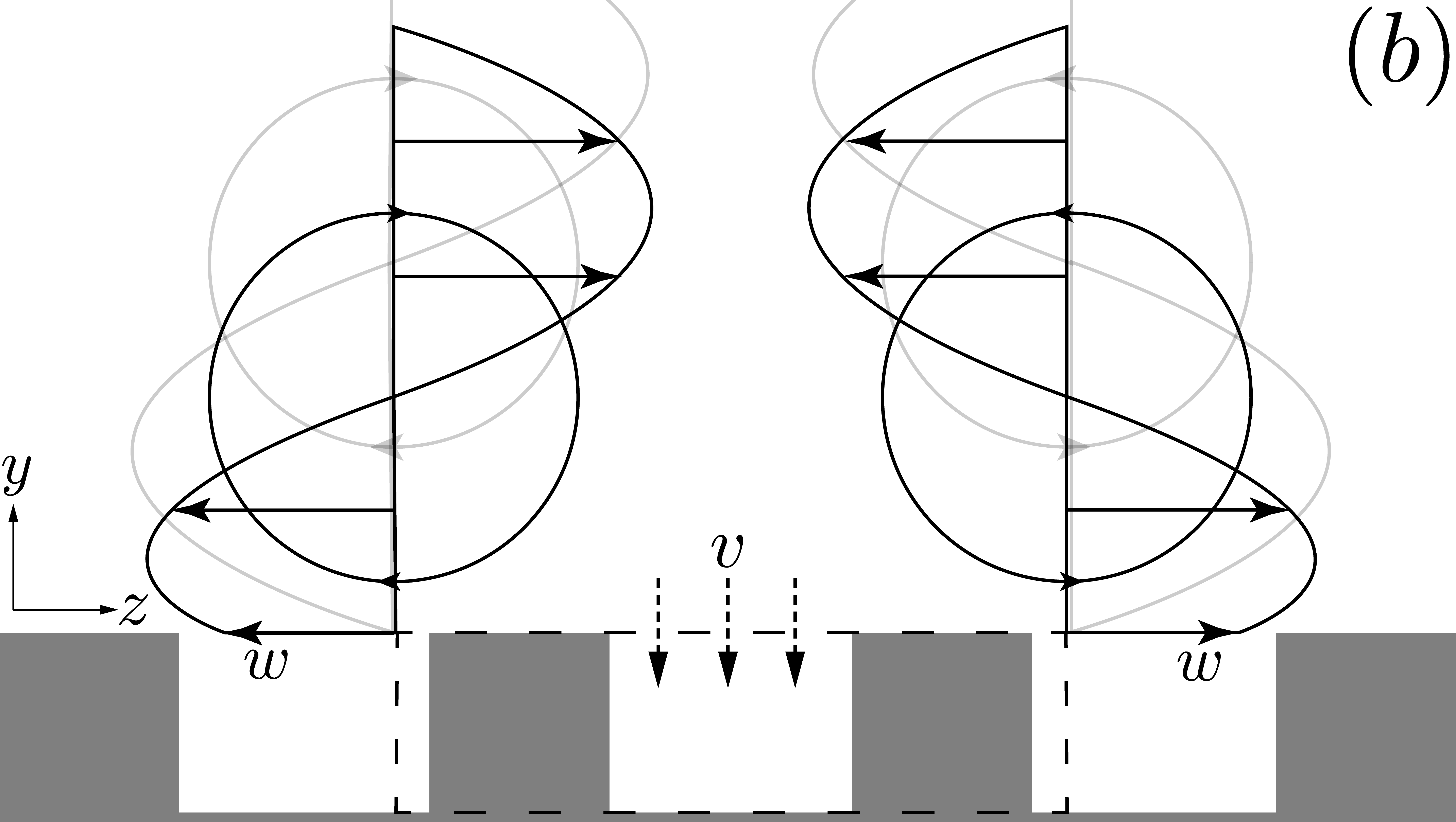}
        {\phantomcaption\label{fig:TRM:displacement}}
    \end{subfigure}
    \vspace*{-4mm} 
    \caption{Conceptual illustration of the mass conservation argument for transpiration generation. In ($a$), the streamwise slip velocity at the crest plane of the textured surface varies in the streamwise direction. Consequently, the in-going and out-going streamwise fluxes through the vertical faces of the control volume (dashed lines) will differ, which due to mass conservation induces a transpiration. ($b$) shows the displacement of quasi-streamwise vortices towards the textured surface due to the relaxation of both spanwise no-slip and wall-normal impermeability. The varying spanwise slip velocity at the crest plane results in a net outgoing spanwise flux in the control volume, which has to be compensated by transpiration.} 
    \end{center}
    \label{TRM}
\end{figure}
%%%%%%%%%%%%%%%%%%%%%%%%%%%%%%%%%%%%%%%%%%%%%%%%%%%%%%%%%%%%%%%

\citet{sudhakar2020} performed an asymptotic expansion for Stokes flow ($Re \ll 1$) over a surface with small textures and showed that the transpiration part of the TRM appears as a $\mathcal{O}(\epsilon^2)$ term, whereas the tangential slip components are $\mathcal{O}(\epsilon^1)$ terms. Here, $\epsilon\ll 1$ is the ratio between the characteristic length scale of the surface  and the system (e.g. channel height). However, the transpiration term in the TRM can also be derived from a mass conservation argument, without any use of asymptotic expansions \citep{Lacis2020}. The physical picture depicting this argument is shown in \cref{fig:TRM:concept}, where a control volume is indicated by the dashed line. Due to mass conservation in this control volume, the varying slip velocity experienced by the flow as it moves over the textured surface must be balanced by a proportional transpiration. A similar reasoning was employed by \citet{garcia-mayoral_2011} in deriving the transpiration velocity for flow over riblets. A picture more relevant for near-wall turbulence is sketched in \cref{fig:TRM:displacement}, where two counter-rotating quasi-streamwise vortices and a control-volume (dashed line) are shown in the cross-plane. The spanwise velocity slip due the presence of the vortices results in a flux through the vertical faces of the control volume; the difference between these fluxes must then be compensated by a flux through the horizontal face of the control volume, leading to the occurrence of transpiration. 

To the best of the authors' knowledge, transpiration boundary conditions have not been thoroughly investigated for the aim of modeling roughness. This is despite the fact that studies of turbulent flow over regular and random roughness \citep{orlandi_2_3, orlandi_leonardi_antonia_2006, Forooghi_rough_2018} have established that greater levels of drag are driven by a pronounced presence of wall-normal velocity fluctuations within the roughness region. It should be mentioned that in the context of manipulating near-wall turbulence,
\citet{gomez_2018}, \citet{GOMEZDESEGURA2020} and \citet{ibrahim2020smoothwalllike} investigated Robin boundary conditions of Navier-slip form for all three velocity components,
\begin{gather}
u = \ell_x \cfrac{\partial u}{ \partial y}\Big\rvert_{y=0}\:,\qquad w = \ell_z \cfrac{\partial w}{ \partial y}\Big\rvert_{y=0}\:,\qquad v = \ell_y \cfrac{\partial v}{ \partial y}\Big\rvert_{y=0}.\quad \label{eq:robin_bdry_cond}
\end{gather}
Note that for isotropic transpiration, i.e. $m_x = m_z$, the TRM transpiration condition \eqref{eq:TRMv} becomes the same as the transpiration condition of \eqref{eq:robin_bdry_cond}.

The boundary condition for the wall-normal velocity \eqref{eq:TRMv} can be rewritten by substituting the wall-parallel velocities with their respective slip boundary conditions (\ref{eq:TRMu}-\ref{eq:TRMw}), giving
\begin{equation} \label{eq:TRMv2}
v = -m_x\ell_x \cfrac{\partial^2 u}{ \partial x \partial y}\Big\rvert_{y=0} - m_z\ell_z \cfrac{\partial^2 w}{ \partial z \partial y}\Big\rvert_{y=0}.
\end{equation}
\\
In this form, the transpiration velocity's definition as being due to the variation of the shear-rates of the wall-parallel velocities becomes explicit. The terms $m_x\ell_x=(m\ell)_x$ and $m_z\ell_z=(m\ell)_z$ are the streamwise and spanwise \emph{transpiration factors}, respectively. These factors contain the compound effect of slip and transpiration lengths and effectively measure the momentum exchange across the crest plane of the roughness. 
For example, the spanwise transpiration factor can be increased either through a larger spanwise slip length, $\ell_z$, or through a larger spanwise transpiration length, $m_z$.
The latter allows a deeper penetration of wall-normal momentum into the texture, while the former indicates a deeper distance below the crest plane before the spanwise velocity component diminishes. Both of these effects increase momentum transport into the roughness region. Finally, it should be emphasized that the TRM models the surface homogeneously and is applied at all locations on the wall using the same coefficients (i.e. it is applied uniformly).

\subsection{Numerical method}
Equations \eqref{NS}-\eqref{cont} are discretized using second-order central finite differences on a staggered Cartesian grid. A standard fractional-step pressure-correction method \citep{Chorin1968,kim1985} updates the solution at each time step of the simulation by first calculating an intermediate velocity field which does not satisfy continuity. A correction pressure is then calculated by solving a Poisson equation using a computationally efficient FFT-based solver \citep{costa2018}. The correction pressure then projects the intermediate velocity onto a divergence free space, thus giving the final velocity field which is divergence free. Temporal integration of the solution utilizes a fully explicit third-order triple sub-step low-storage Runge-Kutta method \citep{SPALART1991,Wesseling2009}. The time-step for the temporal integration is determined using the following stability criteria given in \citet{Wesseling2009},
\begin{gather}
\cfrac{\Delta{t}}{\mathrm{CFL}} = \,\mathrm{min}\left\{{ \cfrac{1.65}{4\nu}\left({{\cfrac{1}{{\Delta{x}}^2}+{\cfrac{1}{\Delta{y}^2}}+\cfrac{1}{{\Delta{z}}^2}}}\right)^{-1},\cfrac{\sqrt{3}}{\vert{u}\vert{\Delta{x}}^{-1} + \vert{v}\vert{\Delta{y}}^{-1} + \vert{w}\vert{\Delta{z}}^{-1}} }\right\},
\end{gather}
with a $\mathrm{CFL} = 0.5$ being used. The computational domain has conventional dimensions of $L_x = 2\pi\delta$, $L_y = 2\delta$ and $L_z = \pi\delta$, where $\delta$ is the channel half-height. The number of grid points are $N_x = 192$, $N_y = 144$ and $N_z = 160$; evenly spaced along $x$, $z$ and unevenly along $y$. The spatial resolution in the horizontal directions are $\Delta{x}^+ = 5.9$ and $\Delta{z}^+ = 3.5$, while the wall-normal resolution varies from $\Delta{y}^+ = 0.6$ at the boundaries to $\Delta{y}^+ = 4.3$ at the channel mid-plane. Parallelization of the computational domain is achieved through a two-dimensional pencil-like decomposition using the 2DECOMP\&FTT library \citep{2DECOMPFFTA}, which uses MPI. Simulations were conducted at a fixed friction Reynolds number of $Re_{\tau}={u_{\tau}\delta}/{\nu}=180$, with the flow driven by an imposed constant mean pressure gradient. Two simulations at $Re_{\tau}=550$ were also conducted, maintaining the previously mentioned spatial resolutions but requiring $N_x = 448$, $N_y = 448$ and $N_z = 448$ grid points. Statistically converged smooth-wall solutions were used as the initial conditions and the simulations were advanced for $110\;\delta/u_{\tau}$, with statistics gathered over at least $50\;\delta/u_{\tau}$.

The boundary conditions along $x$ and $z$ are periodic while the TRM boundary conditions are imposed on both domain boundaries at $y=[0,2\delta]$. The coupling between the slip lengths and the wall-normal velocity gradients of $u$ and $w$ (\ref{eq:TRMu}-\ref{eq:TRMw}) are explicit within the solver, as are the transpiration lengths and double derivatives \eqref{eq:TRMv2}. Due to the explicit imposition of the boundary conditions and the coupling between the velocity and pressure, numerical instabilities can arise when advancing the solution. To avoid this, an iterative relaxation sub-loop gradually updates $v$ at the boundaries using the solution from the previous step as its starting value until convergence is achieved. The updated transpiration velocity then in conjunction with the Neumann condition of the pressure field (zero-gradient) satisfies continuity at the boundaries. Only the pressure fluctuations at the wall-normal boundaries are used to ensure zero net mass flux at them, as was similarly done by \citet{jimenez_2001}.

The numerical solver and the implementation of the boundary conditions were validated and the results are reported in \cref{app:validation}, along with an assessment of solution grid independence.

\section{TRM coefficients and virtual origins} \label{sec:TRM}

Given appropriate TRM coefficients ($\ell_x$, $\ell_z$, $m_x$, $m_z$), one may impose the TRM boundary conditions (\ref{eq:TRMu}-\ref{eq:TRMv}) to induce the macroscopic effects of a textured surface on the overlying flow. \cref{sec:eval} discusses two approaches to determine the TRM coefficients of a given textured surface; \textit{a priori} through a so-called unit cell approach or \textit{a posteriori} by matching DNS data. \cref{sec:virtual} describes the approach of characterizing the modification of turbulence in terms of the virtual-origins of different flow quantities \citep{ibrahim2020smoothwalllike}.  If turbulence remains overall similar to that of canonical wall-bounded flow (smooth-wall-like), the roughness function (a measure of drag change) is directly quantifiable in terms of the virtual origins of the mean flow and the turbulence.

\subsection{Obtaining slip and transpiration lengths for non-smooth surfaces}

\label{sec:eval}
%%%%%%%%%%%%%%%%%%%%%%%%%%%%%%%%%%%%%%%%%%%%%%%%%%%%%%%%%%%%%%%
\begin{figure}
    \begin{center}
    \hspace*{-5mm}
    \begin{subfigure}[tbp]{.32\textwidth}
        \includegraphics[width=1\linewidth]{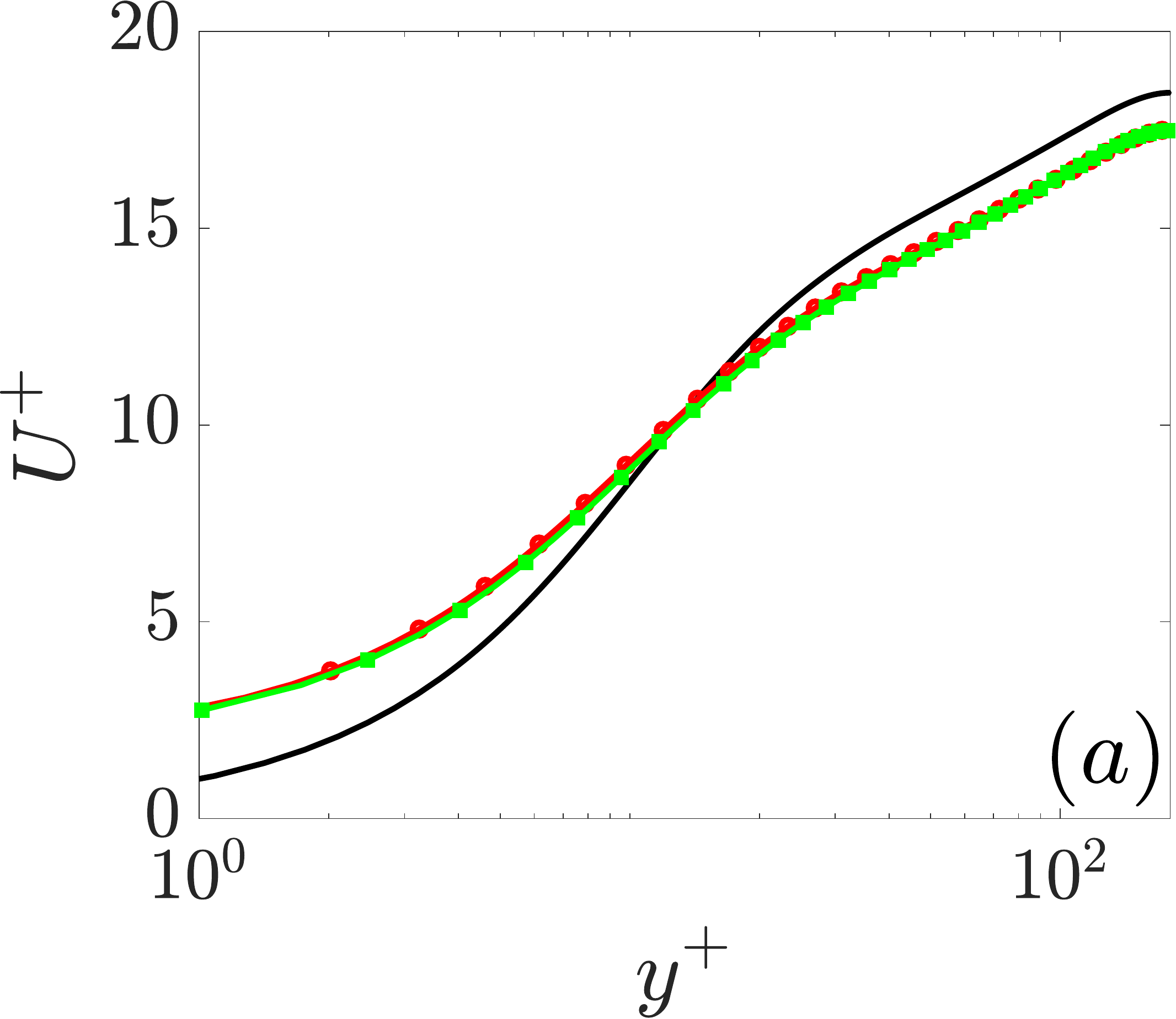}
        {\phantomcaption\label{fig:TRMvsGeoRes:mean}}
        \vspace*{-2mm}
    \end{subfigure}%
    \hspace*{5mm}
    \begin{subfigure}[tbp]{.33\textwidth}
        \includegraphics[width=1\linewidth]{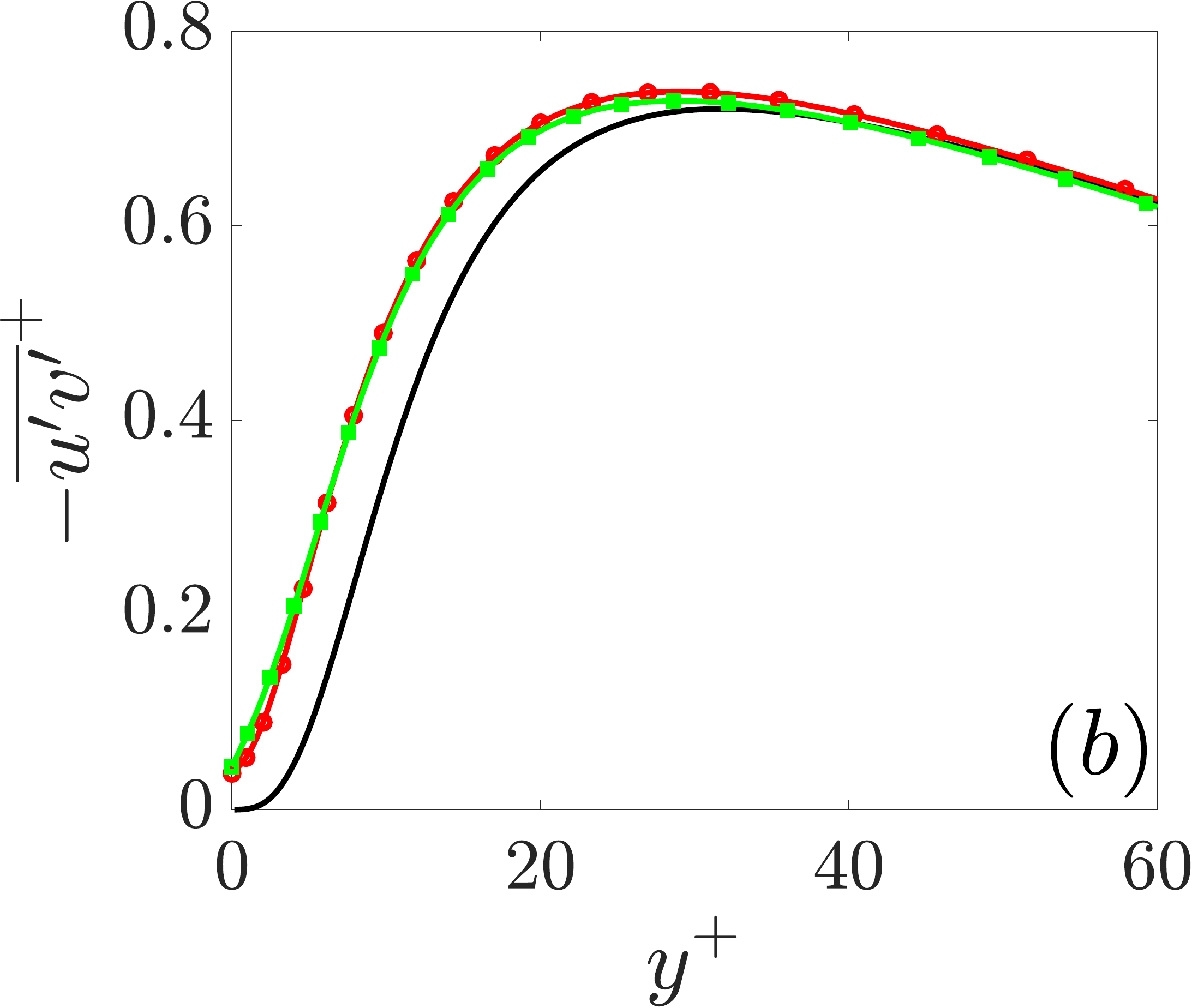}
        {\phantomcaption\label{fig:TRMvsGeoRes:reynolds_stress}}
        \vspace*{-2mm}
    \end{subfigure}
    \hspace*{-4mm}
    \begin{subfigure}[tbp]{.32\textwidth}
        \includegraphics[width=1\linewidth]{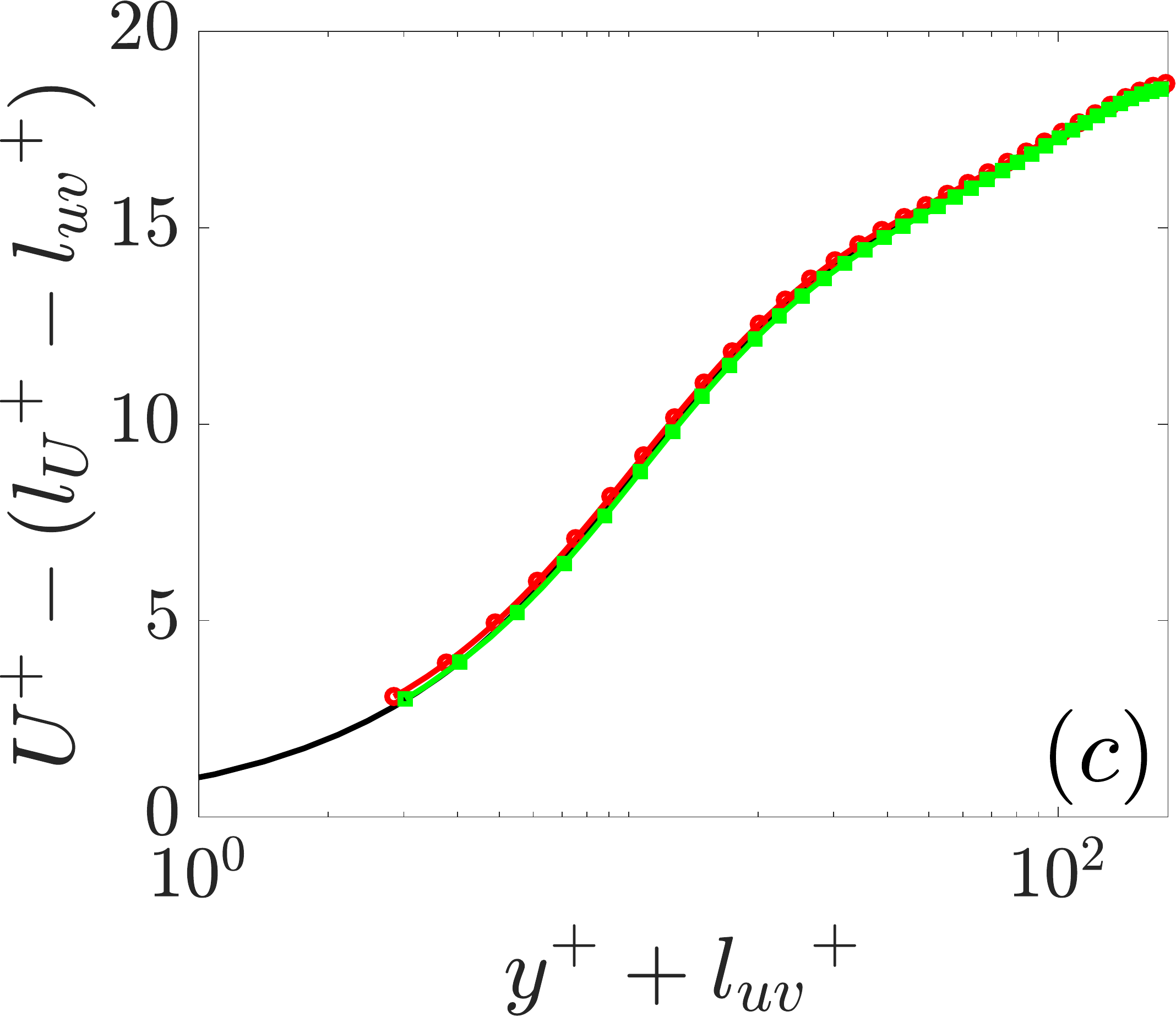}
        {\phantomcaption\label{fig:TRMvsGeoRes:mean_shifted}}
    \end{subfigure}%
    \hspace*{5mm}
    \begin{subfigure}[tbp]{.33\textwidth}
        \includegraphics[width=1\linewidth]{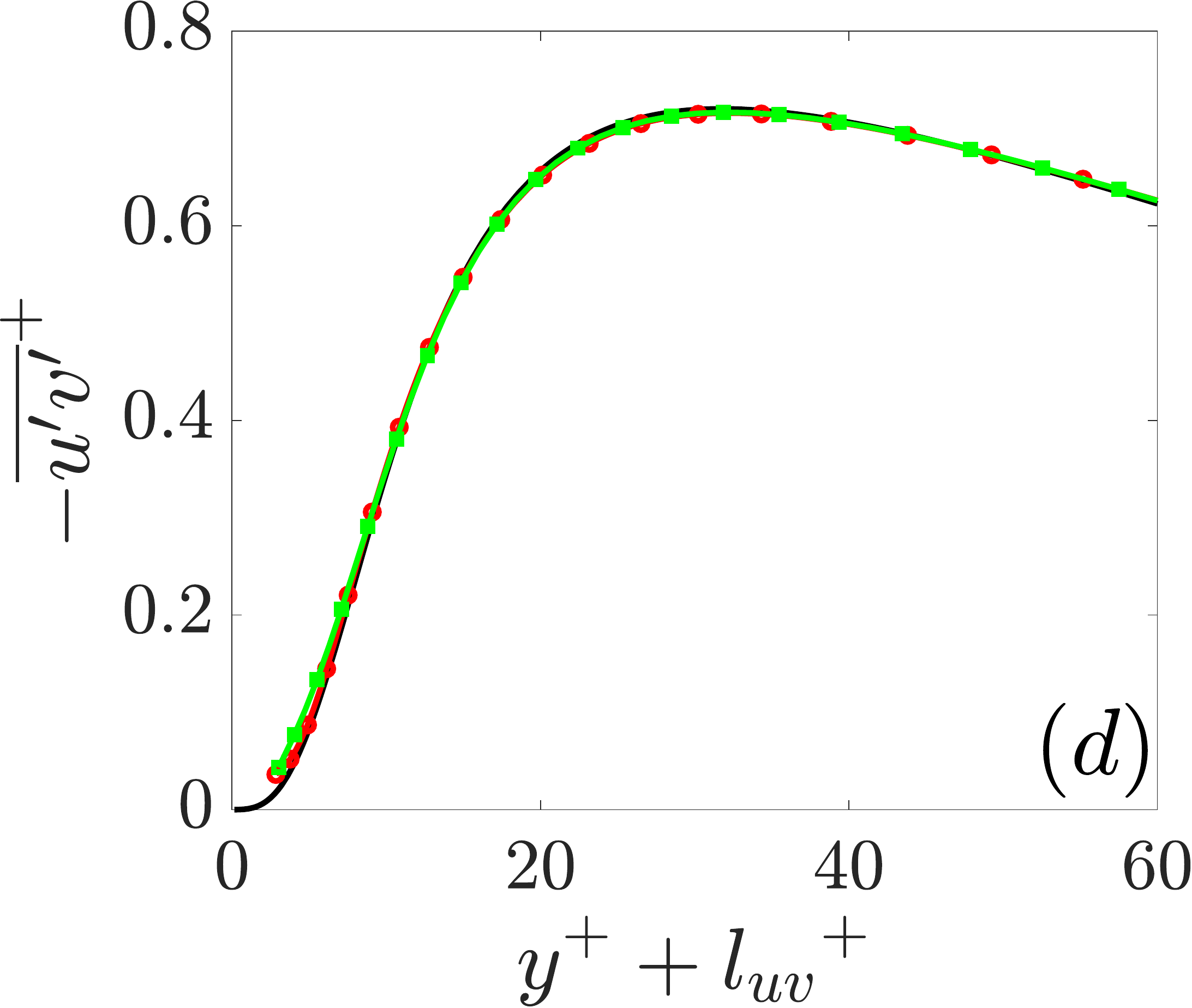}
        {\phantomcaption\label{fig:TRMvsGeoRes:reynolds_stress_shifted}}
    \end{subfigure}
    \vspace*{-4mm}
    \caption{Top row shows the mean velocity ($a$) and Reynolds shear stress ($b$) profiles for $-\!-$, smooth-wall; $\textcolor{red}{-}\!\textcolor{red}{\bullet}\!\textcolor{red}{-}$, geometry-resolving DNS of \citet{Lacis2020}; {\small$\textcolor{green}{-}$\!{$\scriptstyle\textcolor{green}{\blacksquare}$}\!$\textcolor{green}{-}$} DNS with the TRM coefficients ($\ell^+=2$, $m^+=6$). For the same data, the bottom row shows the shifted mean velocity ($c$) and Reynolds shear stress ($d$) profiles for the origin set to $y^+ = {-\ell_{uv}}^+$ and scaled with its associated friction velocity using \eqref{eq:u_tau_turb}.}
    \label{fig:TRMvsGeoRes}
    \end{center}
    \vspace*{-2mm}
\end{figure}

Assuming that the local flow in the region of a textured surface has $Re\ll 1$, the slip and transpiration lengths of the TRM boundary conditions can be obtained by carrying out a Stokes flow analysis \citep{bottaro_2019,Lacis2020}. The procedure requires numerical simulations of one or a few texture elements, and thus a relatively small computational box, i.e. a representative element volume (REV) or unit cell. By an averaging of the flow quantities in the unit cell, homogenized slip and transpiration lengths associated with that surface can be determined. When the Reynolds number is small, the TRM coefficients are a property of the surface alone, and independent of the dynamics of the overlying flow.

An attempt to understand the accuracy of the Stokes-based unit-cell approach for modeling roughness in turbulent flows was made by \citet{Lacis2020}. The surface was comprised of collocated cuboids of height ${k^+}\!\!\approx\!7$ \citep[more specifications of the surface are found in figure 11 and table 5 of][]{Lacis2020}. A Stokes analysis in a REV containing a single cuboid provided a slip length of $\ell=\ell_x^+=\ell_z^+=2.0$ and transpiration length of $m=m_x^+=m_z^+=2.9$. Note that due to the isotropic distribution of the collocated cuboids, the spanwise and streamwise TRM coefficients otained from the REV analysis became equal.
\citet{Lacis2020} conducted a channel flow DNS using the TRM with these calculated lengths which, while demonstrating the same effect, did not achieve quantitative agreement with a geometry-resolving DNS featuring the rough surface.
\Cref{fig:TRMvsGeoRes:mean,fig:TRMvsGeoRes:reynolds_stress} show, respectively, the mean velocity and Reynolds shear stress for a TRM simulation which achieves good quantitative agreement with the geometry-resolving data of \citet{Lacis2020}.
The coefficients used in this TRM simulation ($\ell^+=2$, $m^+=6$) were determined \emph{a posteriori} by comparing TRM simulations with different transpiration lengths to the geometry-resolving DNS data. The results show that the transpiration length obtained using the Stokes-based REV analysis is underestimated. A probable explanation for the underestimated transpiration length is the omission of the potentially small yet significant advective effects in the REV analysis. The larger transpiration length $m^+$ for surfaces exposed to turbulent flow also means a larger transpiration factor, ${{({m}{\ell})}^+}$. The latter depends on the volume-averaged flow within the surface texture region \citep[equation 2.14]{Lacis2020}. In turbulent conditions, it can be expected that the flow in-between textures is on average larger in magnitude compared to a laminar flow due to enhanced momentum transfer into the rough surface incurred by turbulent mixing. 

In contrast to the transpiration, the Stokes-based estimation of the slip length $\ell^+=2.0$ does not differ from that belonging to the TRM simulation which matches the geometry-resolving DNS. Investigations of \citet{fairhall_abderrahaman-elena_garcia-mayoral_2019} using only slip boundary conditions to model super-hydrophobic surfaces have shown that a Stokes flow analysis provides accurate estimates for textures with characteristic lengths ${L^+}\!\!\lesssim\!5$, which corresponds to slip lengths ${{\ell}^+}\!\!\lesssim\!2$. For larger textures, advective effects start to become significant. However, performing laminar flow simulations in an REV extends the validity of the estimates up to ${L^+}\!\!\lesssim\!15$. \citet{Abderrahaman2019} demonstrated a similar extent of validity for slip lengths calculated for roughness. Assessing whether laminar flow analysis resolves the discrepancy of the Stokes-estimated transpiration length has yet to be carried out and is not part of this work's objectives.

In the reminder of this study, the transpiration length is determined by matching the results of TRM simulations to those of geometry-resolving DNS, as was demonstrated in \Cref{fig:TRMvsGeoRes:mean,fig:TRMvsGeoRes:reynolds_stress}. In general, this is not a practical approach to obtain the TRM coefficients for a given surface since it requires performing several simulations. However, it is appropriate for addressing the purpose of the present work, namely, to demonstrate the applicability and extent of validity of the TRM for modeling physically realizable drag-increasing surfaces. 

\subsection{Drag change in terms of virtual origins} \label{sec:virtual}

The virtual-origin framework of \citet{ibrahim2020smoothwalllike} provides a method of quantifying drag change based on how small surface textures affect the near-wall turbulence structures. The measure of drag change adopted here is the roughness function $\Delta{U}^+$.
As is conventionally understood, the presence of surface textures causes a vertical shift, $\Delta{U}^+$, of the logarithmic region of the mean velocity profile \citep{CLAUSER19561},
\begin{gather}
    {U}^+ = \cfrac{1}{\kappa}\, \mathrm{log}\, \left ({y}^+\right) + B + \Delta{U}^+. \label{eq:lawofwalldelta}
\end{gather}
Here, $\kappa\approx 0.4$ is the von K{\'a}rm{\'a}n constant and $B\approx 5.0$ is the offset of the logarithmic region from the wall origin. A negative $\Delta{U}^+$ indicates a downward shift of the mean velocity profile which is observed for drag-increasing surfaces. This shift is considered to be the appropriate quantity in measuring the drag change caused by textures small enough for their effect to remain limited to the near-wall region. As long as the characteristic texture size remains the same when scaled in inner units, the corresponding value of $\Delta{U}^+$ will also remain fixed regardless of Reynolds number \citep{Spalart_2011,Garcia_Mayoral_2019}.

The virtual origin framework quantifies the roughness function in terms of two so-called \textit{virtual origins},
\begin{gather}
\Delta{U}^+\!={\ell_U}^+-{\ell_{T}}^+. \label{eq:virtual-origin-0}
\end{gather}
Here, ${\ell_U}^+$ is the virtual origin of the mean flow and ${\ell_T}^+$ the virtual origin of near-wall turbulence. The concept of separate virtual origins for the mean and turbulent components of the flow was originally established by \citet{luchini_1996}, with ${\ell_U}^+$ describing an imaginary impermeable smooth-wall perceived by the mean flow at the position ${y^+=-\ell_{U}}^+$ and ${\ell_T}^+$ describing the same for the near-wall turbulence.
The physical reasoning behind this, is that the turbulence dynamics in this region are driven by the quasi-streamwise vortices (\hyperref[fig:TRM:displacement]{figure \ref*{fig:TRM:displacement}}), which undergo a displacement due to a weakening of both cross-flow shear and impermeability owing to the presence of small surface textures. This is due to the first-order effect of the vortices at the boundary plane being the generation of cross-flow shear while transpiration is their second-order effect \citep{GOMEZDESEGURA2020,ibrahim2020smoothwalllike}. Thus, the turbulence undergoes a rigid translation by a distance ${\ell_T}^+$ but otherwise remains smooth-wall-like. This is analogous to it perceiving a smooth-wall at $y=-{\ell_T}^+$. 

As demonstrated by \citet{GOMEZDESEGURA2020},  once the origin for the wall-normal coordinate is set at $y^+={-\ell_T}^+$ and the friction velocity calculated at this origin,
\pagebreak
\begin{gather}
{u_{\tau}}\rvert_{y={-\ell_{T}}} = {u_{\tau}}\rvert_{y=0}\sqrt{ \cfrac{\delta + \ell_{T}}{\delta}}, \label{eq:u_tau_turb}
\end{gather}
is used for rescaling the flow quantities, the resulting mean velocity profile mirrors that of a smooth-wall turbulent flow and is only offset from it by $\Delta{U}^+\!={\ell_U}^+-{\ell_T}^+$.

Virtual origins can be defined for the streamwise (${\ell_u}^+$), spanwise (${\ell_w}^+$) and wall-normal (${\ell_v}^+$) velocities as well as for the Reynolds shear stress (${\ell_{uv}}^+$). \citet{ibrahim2020smoothwalllike} established that the appropriate choice for $\ell_T^+$ is the virtual origin of the Reynolds shear stress, i.e.~${\ell_T}^+={\ell_{uv}}^+$, ultimately giving
\begin{gather}
\Delta{U}^+\!={\ell_U}^+-{\ell_{uv}}^+ \label{eq:virtual-origin}.
\end{gather}
\Cref{fig:TRMvsGeoRes:mean_shifted} demonstrates this for the collocated cuboids of \citet{Lacis2020} and thus confirms that the turbulence modification caused by the surface texture in this case merely amounts to a displacement of the near-wall vortices. The choice of turbulence origin being ${\ell_T}^+={\ell_{uv}}^+$ is also commensurate with the fact that for the mean flow, the stress terms which appear in the mean momentum equation are those of viscous and Reynolds shear, with the latter being non-existent at the wall. This associates the wall for canonical smooth-wall turbulence with the condition that $\overline{u^{\prime}v^{\prime}}^+=0$. It follows from this line of reasoning that if the mean and turbulent components of the flow undergo displacements but remain otherwise smooth-wall-like, then the proper choice of wall-normal origin will be $y^+={-\ell_{uv}}^+$, the plane where $-\overline{u^{\prime}v^{\prime}}^+$ perceives an imaginary smooth-wall.

The relation between the roughness function and virtual-origins \eqref{eq:virtual-origin} can be derived from the mean momentum or RANS equation. The procedure is detailed in \citet{gomez_2019} and may be refereed to by the interested reader.
In this work, the virtual origin framework has been leveraged when analyzing the results of turbulent channel flow DNS where the TRM boundary conditions have been used. The virtual origins are obtained \emph{a posteriori}. Following \citet{ibrahim2020smoothwalllike}, ${\ell_U}^+$ corresponds to the slip velocity, ${U_{slip}}^+$, at $y^+=0$, while ${\ell_{uv}}^+$ represents the shift of $-\overline{u^{\prime}v^{\prime}}^+$ relative to that of a smooth-wall solution giving the best fit in the region of $10 < y^+ < 25$. The virtual origins of the velocity fluctuations, ${\ell_u}^+$, ${\ell_w}^+$ and $\ell^+_v$, also calculated \emph{a posteriori}, were obtained via extrapolation of their r.m.s. profiles. The curvature of the profiles were taken into account and hence linear extrapolation was not used. This is particularly important for the ${v^{\prime}}^+$ profile as it is strongly quadratic very close to the boundary.

\section{DNS of turbulent channel flow with the TRM} \label{Results_TRM}

%%%%%%%%%%%%%%%%%%%%%%%%%%%%%%%%%%%%%%%%%%%%%%%%%%%%%%%%%%%%%%%%%%%%%%%%%
%%%%%%%%%%%%%%%%%%%%%%%%%%%%%%%%%%%%%%%%%%%%%%%%%%%%%%%%%%%%%%%%%%%%%%%%%
\begin{table}
  \centering
  \begin{tabular}[t]{ m{1.5cm}m{0.8cm}m{0.9cm}m{0.8cm}m{0.8cm}m{0.8cm}m{0.6cm}|m{0.76cm}m{0.75cm}m{0.75cm}m{0.75cm}m{0.75cm}m{0.75cm} }
       Case & $Re_{\tau}$ & $Re_{\tau}^{\delta'}$ & ${\ell_x}^+$ & ${\ell_z}^+$ & ${m_x}^+$ & ${m_z}^+$ & ${\ell_u}^+$  & ${\ell_w}^+$ & ${\ell_v}^+$ & ${\ell_U}^+$ & ${\ell_{uv}}^+$ & \,\,\,${{\Delta}{U}}^+$\\
       &&&&&&&&&&&\\
       L2M0     & $180$ & $182$ & $2.0$ & $2.0$ & $0.0$ & $0.0$ & $2.0$ & $1.7$ & $0.0$ & $2.0$ & $1.3$ & $+0.7$\\
       L2M2     & $180$ & $183$ & $2.0$ & $2.0$ & $2.0$ & $2.0$ & $1.9$ & $1.7$ & $2.4$ & $2.0$ & $2.0$ & $\,\,\,\;0.0$\\
        L2M5    & $180$ & $184$ & $2.0$ & $2.0$ & $5.0$ & $5.0$ & $1.8$ & $1.7$ & $5.8$ & $1.9$ & $2.9$ & $-1.0$\\
        L5M0    & $180$ & $183$ & $5.0$ & $5.0$ & $0.0$ & $0.0$ & $4.6$ & $3.7$ & $0.0$ & $4.9$ & $2.2$ & $+2.7$\\
    L5M5    & $180$ & $187$ & $5.0$ & $5.0$ & $5.0$ & $5.0$ & $3.5$ & $3.5$ & $6.6$ & $4.4$ & $4.7$ & $-0.3$\\
     L5M10    & $180$ & $190$ & $5.0$ & $5.0$ & $10.0$ & $10.0$ & $3.3$ & $3.4$ & $11.6$ & $3.7$ & $6.6$ & $-2.9$\\
     L10M10   & $180$ & $191$ & $10.0$ & $10.0$ & $10.0$ & $10.0$ & $5.1$ & $5.4$ & $10.6$ & $6.1$ & $7.7$ & $-1.6$\\
       L2M2HR     & $550$ & $553$ & $2.0$ & $2.0$ & $2.0$ & $2.0$ & $1.9$ & $1.7$ & $2.0$ & $1.9$ & $1.9$ & $\,\,\,\;0.0$\\
        L5M5HR    & $550$ & $557$ & $5.0$ & $5.0$ & $5.0$ & $5.0$ & $3.3$ & $3.3$ & $4.7$ & $4.3$ & $4.6$ & $-0.3$\\
          &&&&&&&&&&&\\
       L2MX2    & $180$ & $182$ & $2.0$ & $2.0$ & $2.0$ & $0.0$ & $2.0$ & $1.7$ & $1.0$ & $2.0$ & $1.3$ & $+0.7$\\
       L2MX5   & $180$ & $182$ & $2.0$ & $2.0$ & $5.0$ & $0.0$ & $2.0$ & $1.7$ & $1.6$ & $2.0$ & $1.2$ & $+0.8$\\
       L5MX5   & $180$ & $183$ & $5.0$ & $5.0$ & $5.0$ & $0.0$ & $4.9$ & $3.7$ & $2.2$ & $4.9$ & $2.0$ & $+2.9$\\
       L5MX10    & $180$ & $183$ & $5.0$ & $5.0$ & $10.0$ & $0.0$ & $5.0$ & $3.8$ & $3.5$ & $4.9$ & $2.0$ & $+2.9$\\
       &&&&&&&&&&&\\
    L2MZ2    & $180$ & $183$ & $2.0$ & $2.0$ & $0.0$ & $2.0$ & $1.8$ & $1.7$ & $2.7$ & $2.0$ & $2.0$ & $\,\,\,\;0.0$\\
    L2MZ5   & $180$ & $185$ & $2.0$ & $2.0$ & $0.0$ & $5.0$ & $1.7$ & $1.7$ & $7.3$ & $1.9$ & $3.6$ & $-1.7$\\
     L5MZ5   & $180$ & $189$ & $5.0$ & $5.0$ & $0.0$ & $5.0$ & $3.3$ & $3.5$ & $9.8$ & $4.1$ & $5.8$ & $-1.7$\\
       L5MZ10   & $180$ & $193$ & $5.0$ & $5.0$ & $0.0$ & $10.0$ & $3.1$ & $3.5$ & $17.8$ & $3.2$ & $8.7$ & $-5.5$\\
  \end{tabular}
  \captionof{table}{Summary of the TRM simulations performed with their slip, ${\ell_{x}}^+$ and ${\ell_{z}}^+$, and transpiration lengths, ${m_{x}}^+$ and ${m_{z}}^+$. Here, $Re_{\tau}$ is based on ${u_{\tau}}\rvert_{y=0}$ and $Re_{\tau}^{\delta'}$ on ${u_{\tau}}\rvert_{y={-\ell_{uv}}}$ with $\delta' = \delta + {\ell_{uv}}^+$. The virtual origins of the velocity fluctuations, the mean flow and the Reynolds shear stress (${\ell_u}^+$, ${\ell_w}^+$, ${\ell_v}^+$, ${\ell_U}^+$, ${\ell_{uv}}^+$) are calculated \emph{a posteriori} as established in \cref{sec:virtual}. The reported roughness function  ${\Delta{U}}^+$ is computed from the virtual origins ${\ell_U}^+$ and ${\ell_{uv}}^+$ using \eqref{eq:virtual-origin}. The first group of 9 simulations use the TRM with isotropic transpiration lengths at $Re_\tau=180$ and $Re_\tau=550$. The second group of 5 simulations use the TRM with only streamwise transpiration imposed while the last group use the TRM with only spanwise transpiration imposed.}
  \label{tab:dns}
\end{table}
%%%%%%%%%%%%%%%%%%%%%%%%%%%%%%%%%%%%%%%%%%%%%%%%%%%%%%%%%%%%%%%%%%%%%%%%%
%%%%%%%%%%%%%%%%%%%%%%%%%%%%%%%%%%%%%%%%%%%%%%%%%%%%%%%%%%%%%%%%%%%%%%%%%

In \cref{sec:TRM_isotropic}, the results from a number of TRM-based DNS conducted by imposing different conditions ($\ell_x$, $\ell_z$, $m_x$, $m_z$) are reported and analyzed. This is to understand how turbulence is gradually modified as the TRM coefficients are varied, and if the observed modifications are consistent with those of textured surfaces in the transitionally rough regime. The actual connection of the TRM coefficients to physical textured surfaces is addressed in \cref{sec:TRM_vs_rough_DNS}.

As explained in \cref{sec:virtual}, the concept of virtual origins feature heavily in the works of \citet{gomez_2018,GOMEZDESEGURA2020} and the more recent work of \citet{ibrahim2020smoothwalllike} which formalized it into a framework. This will naturally invite comparisons to be made with the results herein and therefore a detailed cross-analysis has been gathered in \cref{app:comp_to_virtual_origin_framework} for the interested reader, but omitted here for the sake of brevity. The TRM boundary conditions' implementation has also been validated against matching cases from \citet{GOMEZDESEGURA2020} and \citet{ibrahim2020smoothwalllike}. The results are gathered in \cref{app:validation}.

The cases through which the TRM is examined are listed in \cref{tab:dns}. Each case is denoted with L$<\!\!\cdot\!\!>$M$<\!\!\cdot\cdot\!\!>$, where the digit following L refers to the slip lengths, ${\ell_x}^+$ and ${\ell_z}^+$, while the letter and digit following M denotes the transpiration length(s) imposed and their values; X for ${m_x}^+\neq0$, ${m_z}^+=0$; Z for ${m_x}^+=0$, ${m_z}^+\neq0$ and no letter for $m^+={m_x}^+={m_z}^+$. Note that for all simulations considered, the streamwise and spanwise slip lengths are equal ($\ell^+={\ell_x}^+={\ell_z}^+$) and anisotropy is investigated only for the transpiration lengths.
\FloatBarrier

\subsection{TRM with isotropic transpiration lengths} \label{sec:TRM_isotropic}

%%%%%%%%%%%%%%%%%%%%%%%%%%%%%%%%%%%%%%%%%%%%%%%%%%%%%%%%%%%%%%%%%%%%%%%%%
%%%%%%%%%%%%%%%%%%%%%%%%%%%%%%%%%%%%%%%%%%%%%%%%%%%%%%%%%%%%%%%%%%%%%%%%%
\begin{figure}
    \begin{center}
    \hspace*{-3mm}
    \begin{subfigure}[tbp]{.4\textwidth}
        \includegraphics[width=1\linewidth]{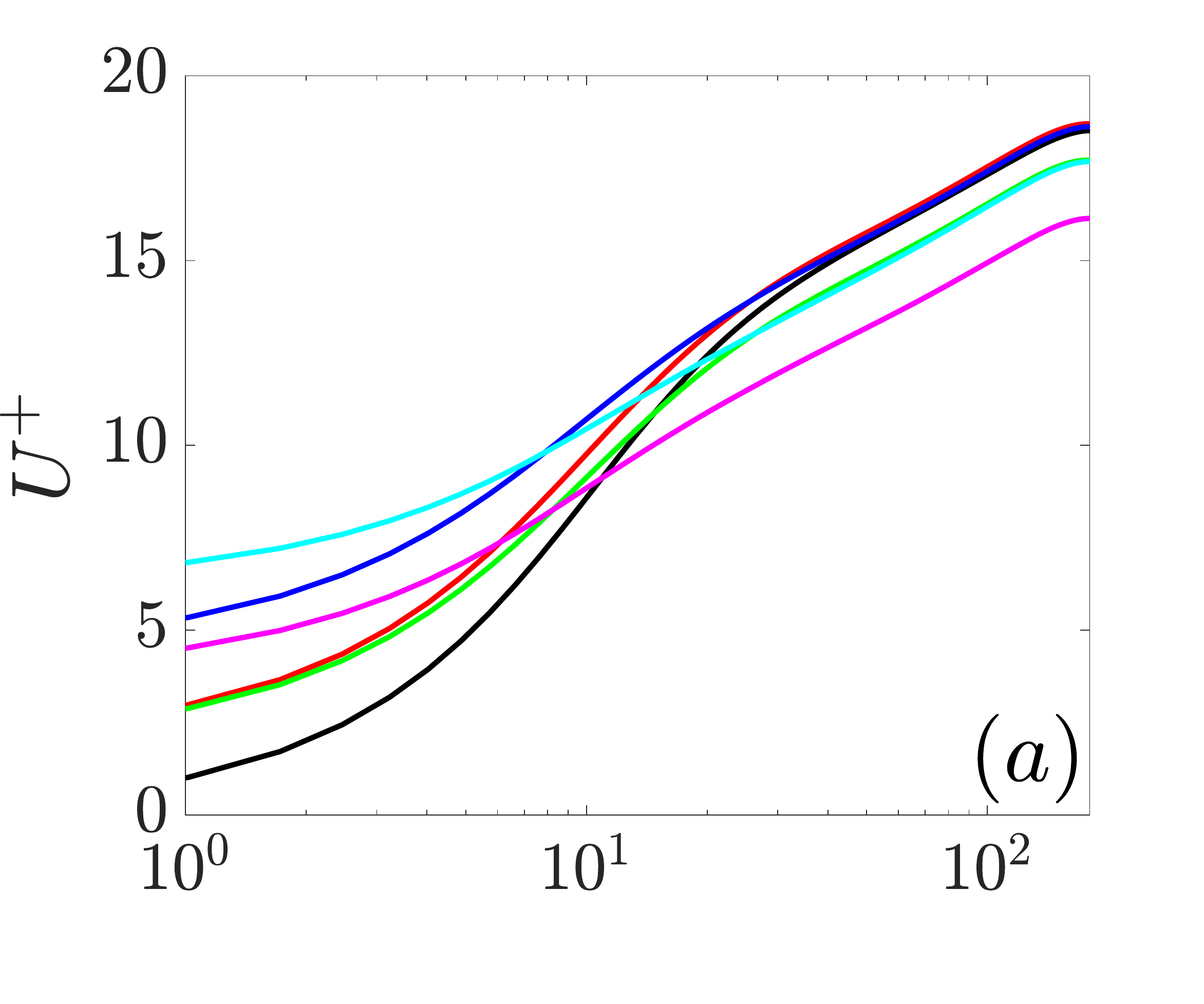}
        {\phantomcaption\label{fig:group1:mean}}
        \vspace*{-8mm}
    \end{subfigure}%
    \hspace*{-1mm}
    \begin{subfigure}[tbp]{.4\textwidth}
        \includegraphics[width=1\linewidth]{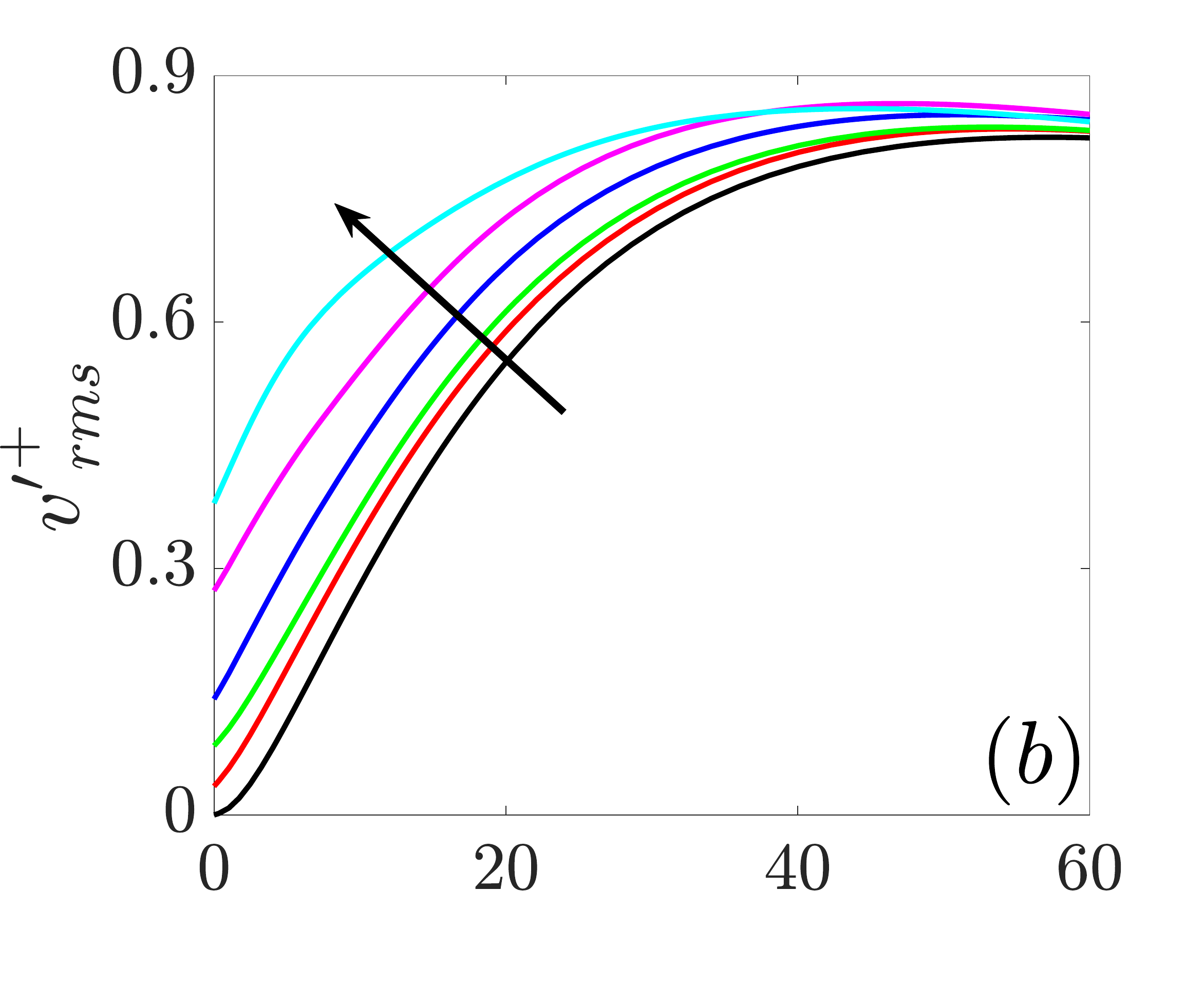}
        {\phantomcaption\label{fig:group1:v_rms}}
        \vspace*{-8mm}
    \end{subfigure}
    \begin{subfigure}[tbp]{.4\textwidth}
        \includegraphics[width=1\linewidth]{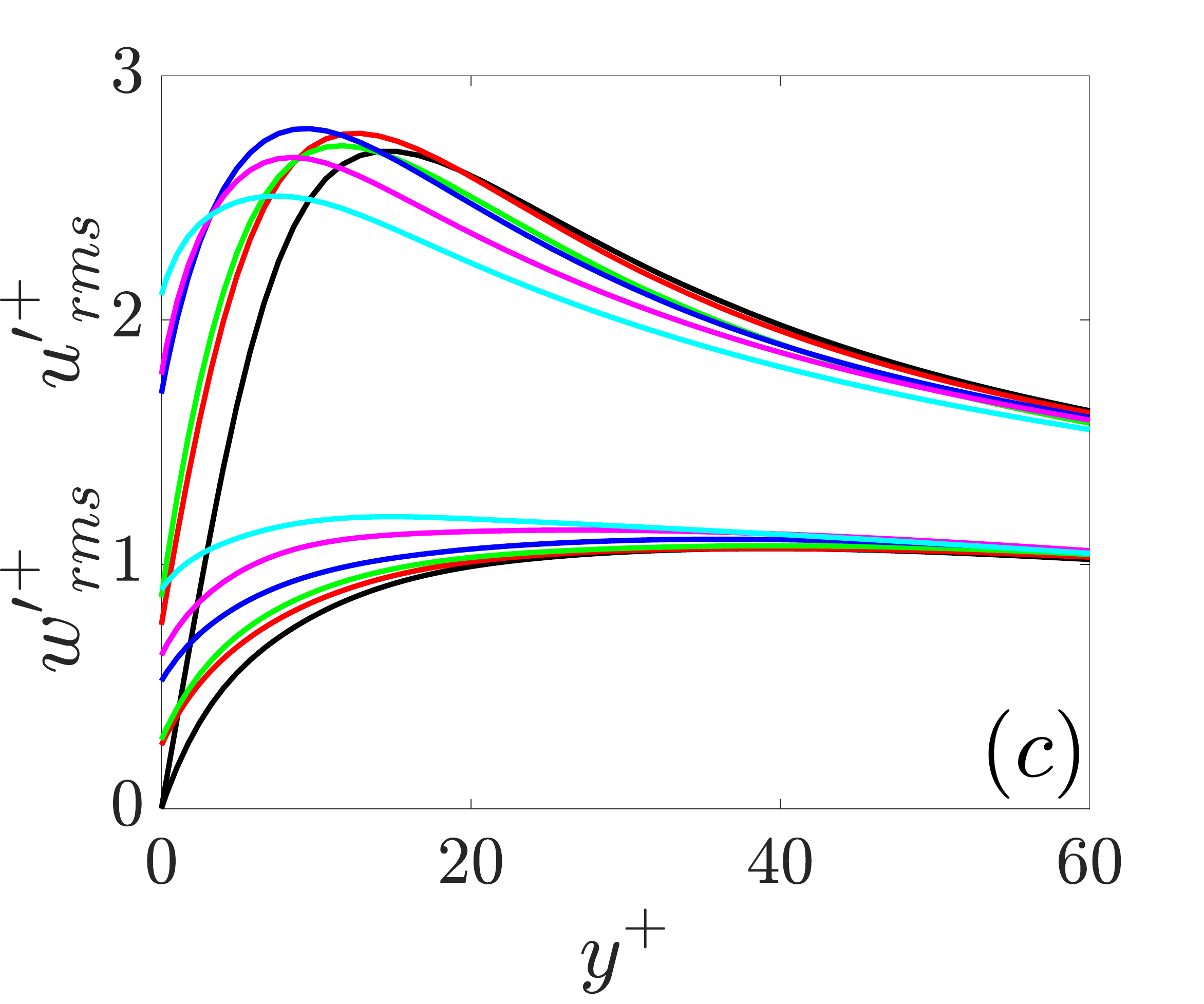}
        {\phantomcaption\label{fig:group1:u_w_rms}}
    \end{subfigure}%
    \begin{subfigure}[tbp]{.4\textwidth}
        \includegraphics[width=1\linewidth]{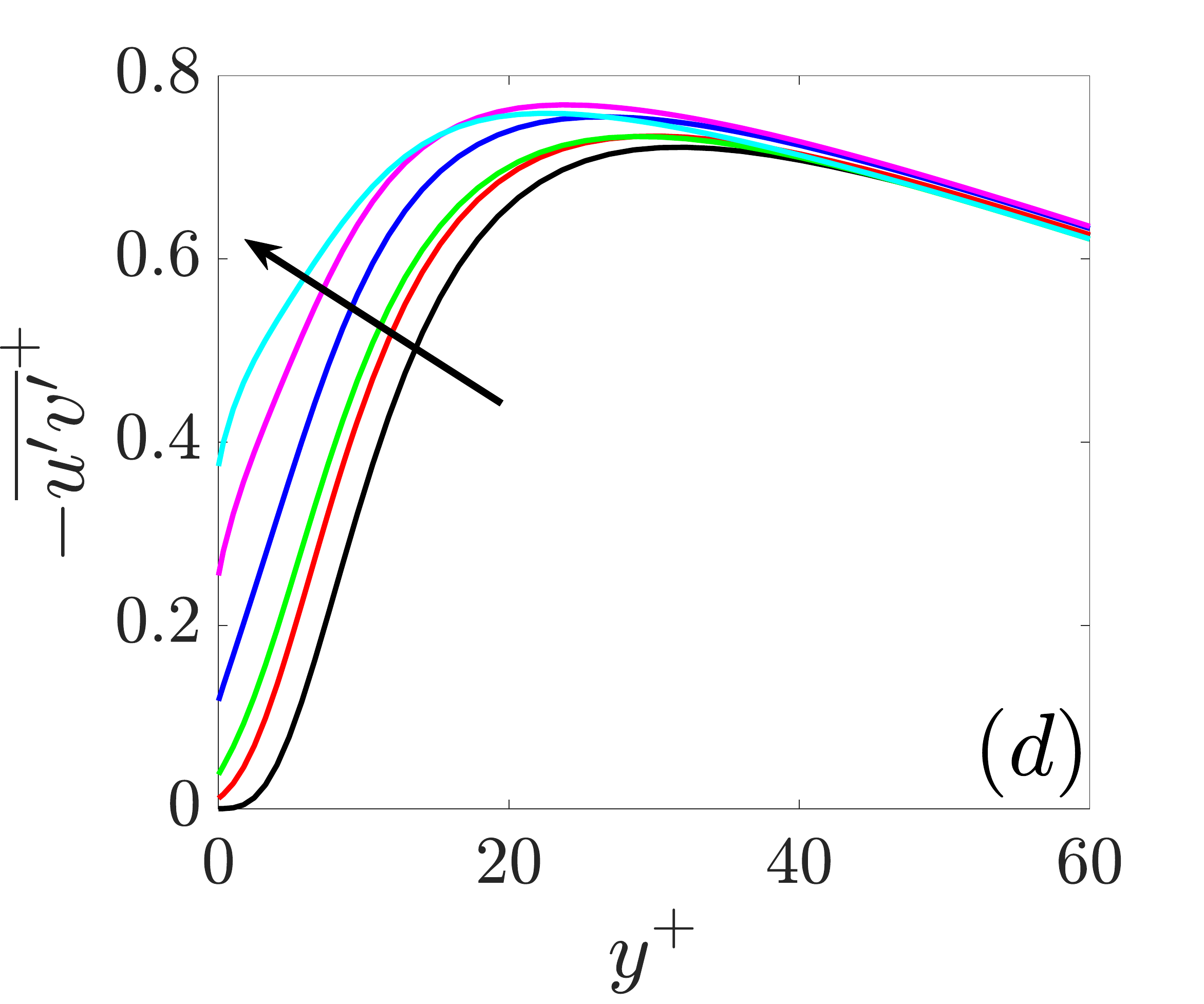}
        {\phantomcaption\label{fig:group1:reynolds_stress}}
    \end{subfigure}
    \vspace*{-5mm}
    \caption{Mean velocity ($a$), r.m.s. velocity fluctuation ($b$-$c$) and Reynolds shear stress ($d$) profiles of the L$<\!\!\cdot\!\!>$M$<\!\!\cdot\!\!>$ cases. \textcolor{red}{$-\!-$}, L2M2; \textcolor{green}{$-\!-$}, L2M5; \textcolor{blue}{$-\!-$}, L5M5; \textcolor{magenta}{$-\!-$}, L5M10; \textcolor{cyan}{$-\!-$}, L10M10; $-\!-$, smooth-wall data. Arrows indicate increasing $m^+$.}
    \label{fig:group1}
    \end{center}
    \vspace*{-1mm}
\end{figure}
%%%%%%%%%%%%%%%%%%%%%%%%%%%%%%%%%%%%%%%%%%%%%%%%%%%%%%%%%%%%%%%%%%%%%%%%%
%%%%%%%%%%%%%%%%%%%%%%%%%%%%%%%%%%%%%%%%%%%%%%%%%%%%%%%%%%%%%%%%%%%%%%%%%

Cases with equal transpiration lengths are considered first (the 9 initial rows in \cref{tab:dns}). This implies that the contributions made to the overall transpiration by the streamwise and spanwise flows are of equal proportion, with no preferentiality given to either.

\subsubsection{Smooth-wall-like regime }
The zero-transpiration case of L2M0 is similar to the many slip-only simulations found throughout the literature \citep{Min_2004,Fukagata_2006,busse_2012,GOMEZDESEGURA2020}. It causes a slight reduction in drag as evident by the excess momentum it has relative to the smooth-wall solution (${{{\Delta}{U}}^+}=0.7$). Indeed, the virtual origin of the mean flow (${\ell_U}^+=2$) is larger than the virtual origin of the Reynolds shear stress (${\ell_{uv}}^+=1.3$), which according to \eqref{eq:virtual-origin} will result in a decrease of drag. Note that while the impermeability condition is maintained for this case, and hence $\overline{u^{\prime}v^{\prime}}^+=0$ at the boundary, it's near-wall distribution undergoes a change, which results in ${\ell_{uv}}^+=1.3$. The drag-increasing effect here is due to the imposed spanwise slip, which, as argued in \cref{sec:virtual}, causes the quasi-streamwise vortices to become displaced and generate wall-normal mixing in the region closer to the boundary plane.

\Cref{fig:group1:mean} shows the mean velocity profiles of simulations with different slip and transpiration lengths. Increasing the streamwise slip length (${\ell_{x}}^+\in[2, 5, 10]$) causes the mean flow virtual origin (${\ell_U}^+$) to become deeper. Further away from the wall the mean velocity either conforms to the smooth-wall profile (drag neutral) or becomes shifted downwards (drag increase) depending on the amount of imposed transpiration length (${m^+}\in [2,5,10]$).
A non-zero transpiration length in case L2M2 results in both ${v^{\prime}}^+$ and $-\overline{u^{\prime}v^{\prime}}^+$ having finite values at the boundary plane ( \textcolor{red}{$-\!-$} in \hyperref[fig:group1:v_rms]{figures \ref*{fig:group1:v_rms}} and \hyperref[fig:group1:reynolds_stress]{\ref*{fig:group1:reynolds_stress}}).
Increased transpiration leads to greater Reynolds shear stress closer to the boundary and a decrease of the drag reduction observed in case L2M0, neutralizing it almost entirely (${\ell_U}^+={\ell_{uv}}^+=2$ and $\Delta U^+=0$ for L2M2).
Increasing the transpiration lengths further in case L2M5 (${\ell}^+=2, {m}^+=5$), such that they now exceed their corresponding slip lengths, amounts to a downward shift of the velocity profile ($\Delta U^+=-1$) and drag increase. As can be observed in \cref{fig:group1:mean,fig:group1:reynolds_stress} (\textcolor{green}{$-\!-$}), case L2M5 results in greater amounts of Reynolds shear stress in the near-boundary region (${\ell_{uv}}^+=2.9$), whereas the mean velocity slip (${\ell_U}^+=1.9$) has essentially the same value of the lower transpiration case L2M2. Therefore, the increase in Reynolds shear stress overcomes the beneficial effect of the mean velocity slip and leads to a momentum deficit of the flow (${{{\Delta}{U}}^+}<0$).

%%%%%%%%%%%%%%%%%%%%%%%%%%%%%%%%%%%%%%%%%%%%%%%%%%%%%%%%%%%%%%%%%%%%%%%%%
%%%%%%%%%%%%%%%%%%%%%%%%%%%%%%%%%%%%%%%%%%%%%%%%%%%%%%%%%%%%%%%%%%%%%%%%%
\begin{figure}
    \begin{center}
    \hspace*{-2mm}
    \begin{subfigure}[tbp]{.498\textwidth}
        \includegraphics[width=1\linewidth]{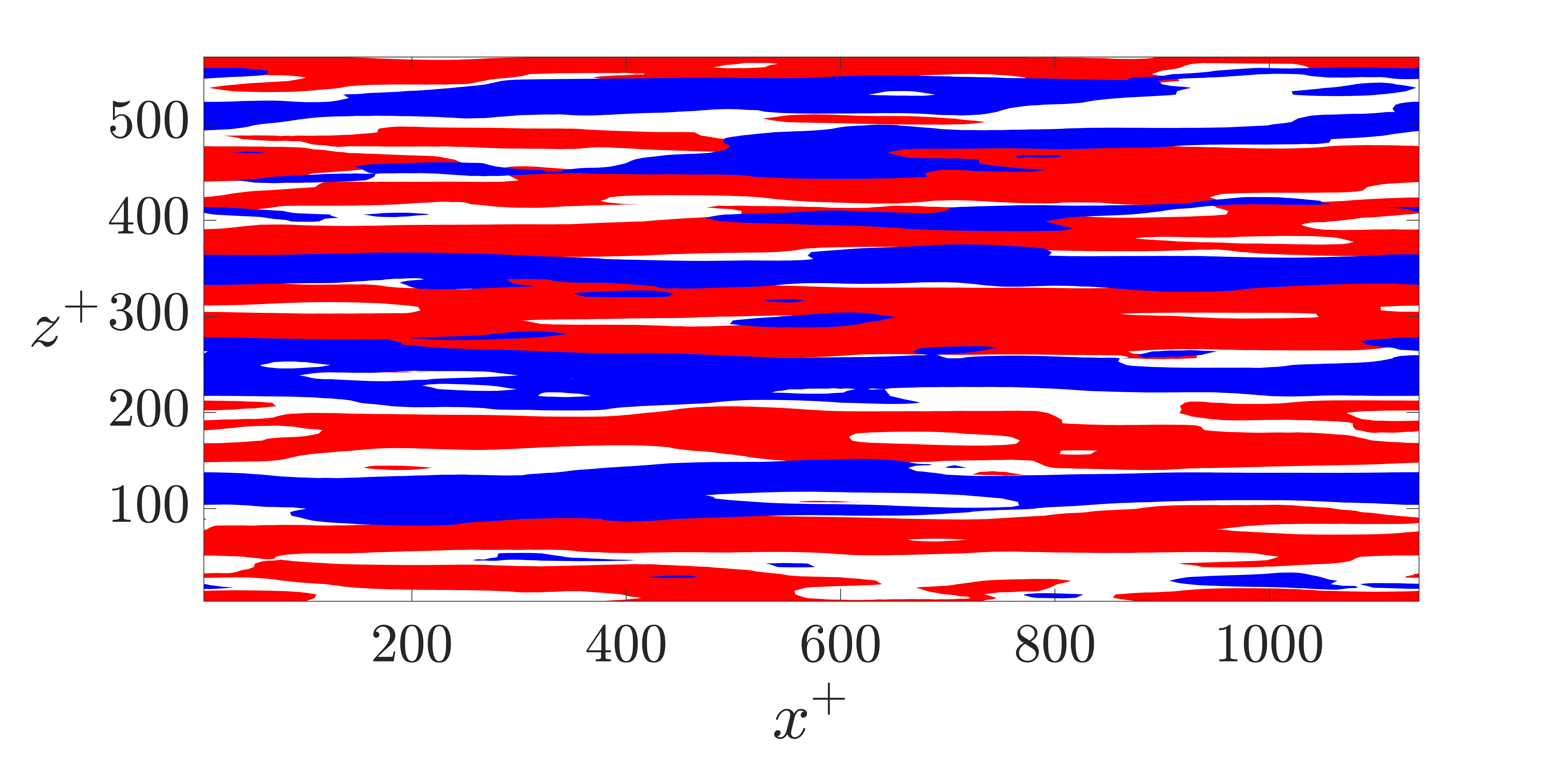}
        \vspace*{-11mm}
        {\captionsetup{position=top, labelfont=it,textfont=normalfont,singlelinecheck=off,justification=raggedleft,labelformat=parens}
        \caption{}\label{fig:boundary_velocity_fluctuations:v}}
    \end{subfigure}%
    \begin{subfigure}[tbp]{.498\textwidth}
         \includegraphics[width=1\linewidth]{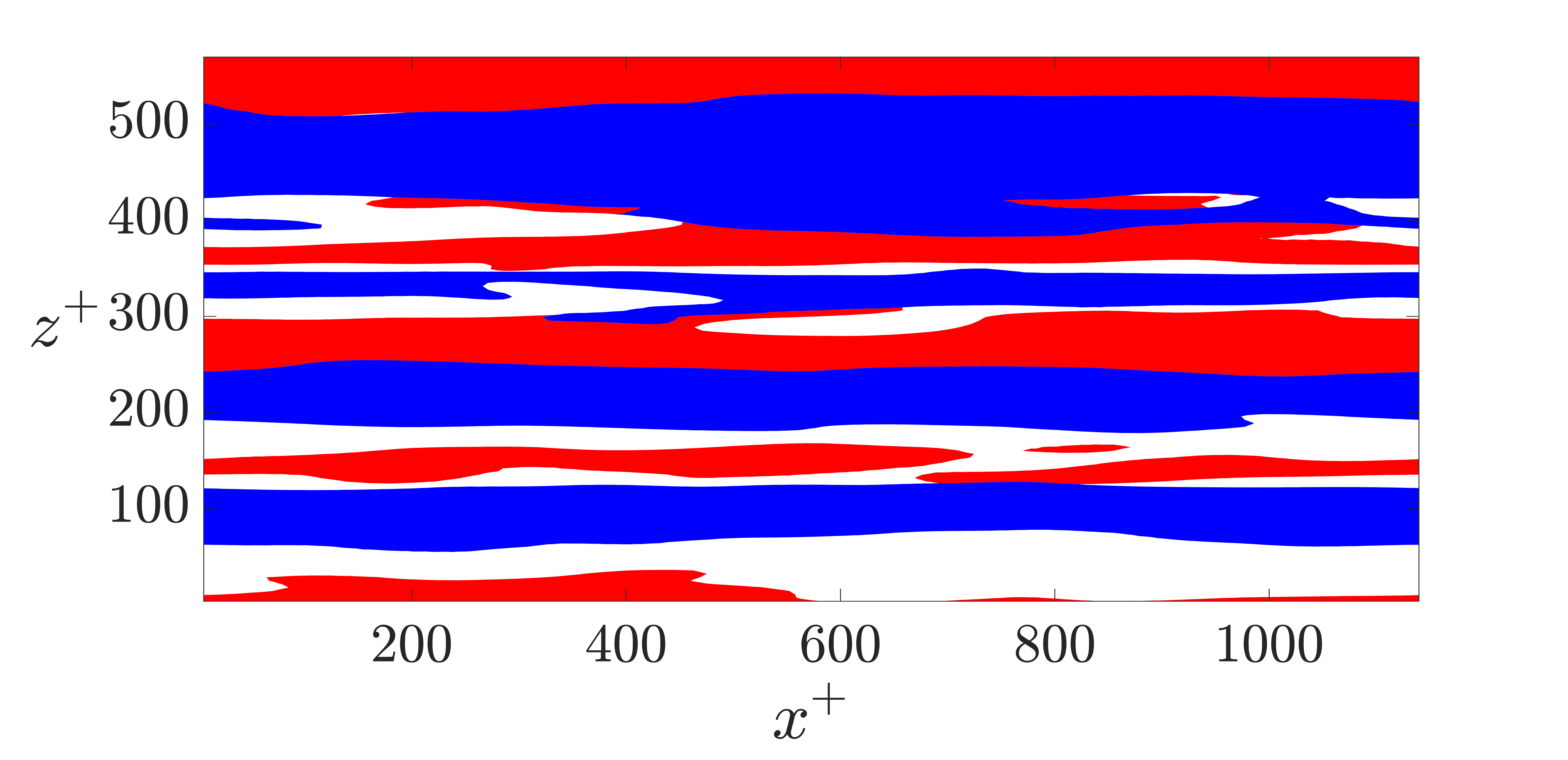}
         \vspace*{-11mm}
         {\captionsetup{position=top, labelfont=it,textfont=normalfont,singlelinecheck=off,justification=raggedleft,labelformat=parens}
         \caption{}\label{fig:boundary_velocity_fluctuations:w}}
    \end{subfigure}
    \vspace*{-2mm}
    \caption{Flow-field of case L2M5 at ${y^+}=0$, averaged over time and conditioned for positive (red) and negative (blue) velocity fluctuations of maximum magnitude: Wall-normal ($a$) and spanwise ($b$) velocity fluctuations.} 
    \label{fig:boundary_velocity_fluctuations}
    \end{center}
    \vspace*{-1mm}
\end{figure}
%%%%%%%%%%%%%%%%%%%%%%%%%%%%%%%%%%%%%%%%%%%%%%%%%%%%%%%%%%%%%%%%%%%%%%%%%
%%%%%%%%%%%%%%%%%%%%%%%%%%%%%%%%%%%%%%%%%%%%%%%%%%%%%%%%%%%%%%%%%%%%%%%%%

For case L2M5, \cref{fig:boundary_velocity_fluctuations:v} shows the wall-normal fluctuations at the boundary plane (${y^+}=0$), averaged over time and conditioned for ${v^{\prime}}^+>0$ (red) and ${v^{\prime}}^+<0$ (blue) of maximum intensity. There are alternating patches of ${v^{\prime}}^+$ along the spanwise direction. The distance between two patches of similar sign is ${\Delta{z^+}}\approx100$, consistent with the spacing of the near-wall streaks \citep{kline_1967}. The spacing between consecutive regions of opposite sign is ${50}\lesssim{{\lambda_{z}}^+}\lesssim{60}$, consistent with twice the diameter of a quasi-streamwise vortice \citep{kim_moin_moser_1987}. \Cref{fig:boundary_velocity_fluctuations:w} shows the same time-averaged flow-field but conditioned for ${w^{\prime}}^+>0$ (red) and ${w^{\prime}}^+<0$ (blue). Comparing both figures, one observes a change from ${{v^{\prime}}^+}>0$ to ${{v^{\prime}}^+}<0$ being coincident with a patch of ${{w^{\prime}}^+}<0$ and vice-versa. These observations are in accord with the depiction of the transpiration mechanism in \cref{fig:TRM:displacement}.

The level of transpiration generated due to the TRM (\ref{eq:TRMu}-\ref{eq:TRMv}) is influenced by slip lengths imposed on the tangential velocity components. To demonstrate this, for case L5M5 the transpiration lengths of L2M5 have been kept but the slip lengths increased by a factor of $2.5$. 
Since the transpiration factors in \eqref{eq:TRMv2} contain slip lengths, an increase in the latter modifies the intensity of wall-normal fluctuations at the boundary plane (\textcolor{blue}{$-\!-$} in \hyperref[fig:group1:v_rms]{figure \ref*{fig:group1:v_rms}}).
Greater levels of wall-normal fluctuations in turn result in greater levels of Reynolds-shear stress (\hyperref[fig:group1:reynolds_stress]{figure \ref*{fig:group1:reynolds_stress}}). This is also reflected in the calculated virtual origins; with ${\ell_v}^+=6.6$ and ${\ell_{uv}}^+=4.7$ for case L5M5, while being ${\ell_v}^+=5.8$ and $\ell^+_{uv}=2.9$ for case L2M5.

\subsubsection{Deviations from smooth-wall turbulence}
A pertinent question here is the extent of the TRM's applicability as a model for surfaces in the transitionally rough regime which exceed the thickness of the viscous sublayer. The transpiration length was therefore increased to $m^+=10$ for different slip lengths to assess the resulting modification of near-wall turbulence. The high-transpiration cases in \cref{tab:dns} are L5M10 (\textcolor{magenta}{$-\!-$}) and L10M10 (\textcolor{cyan}{$-\!-$}), shown in \cref{fig:group1}.

%%%%%%%%%%%%%%%%%%%%%%%%%%%%%%%%%%%%%%%%%%%%%%%%%%%%%%%%%%%%%%%%%%%%%%%%%
%%%%%%%%%%%%%%%%%%%%%%%%%%%%%%%%%%%%%%%%%%%%%%%%%%%%%%%%%%%%%%%%%%%%%%%%%
\begin{figure}
    \begin{center}
    \begin{subfigure}[tbp]{.35\textwidth}
        \includegraphics[width=1\linewidth]{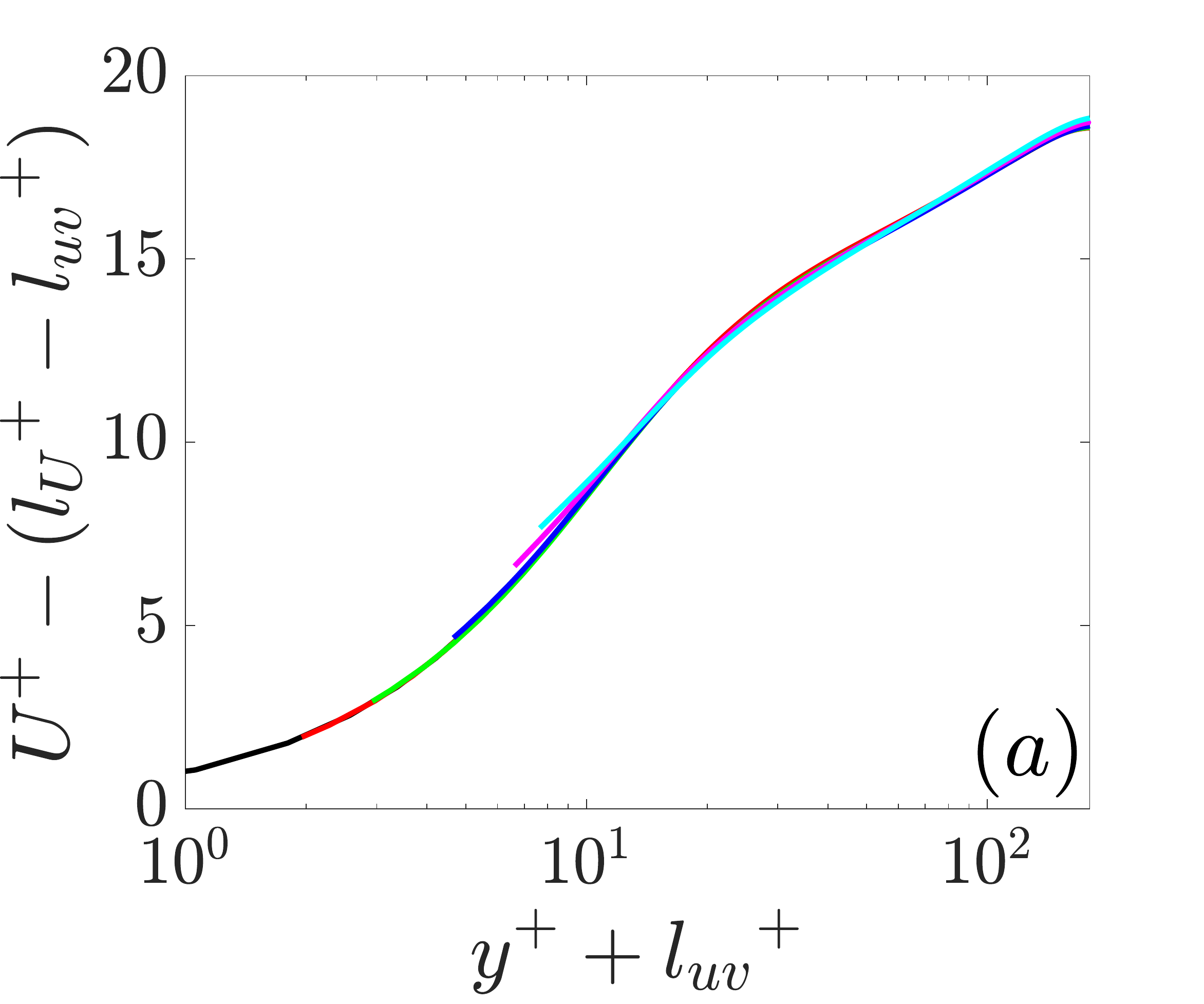}
        {\phantomcaption\label{fig:group1:shifted:sub1}}
        \vspace*{-1mm}
    \end{subfigure}%
    \hspace*{-4mm}
    \begin{subfigure}[tbp]{.35\textwidth}
        \includegraphics[width=1\linewidth]{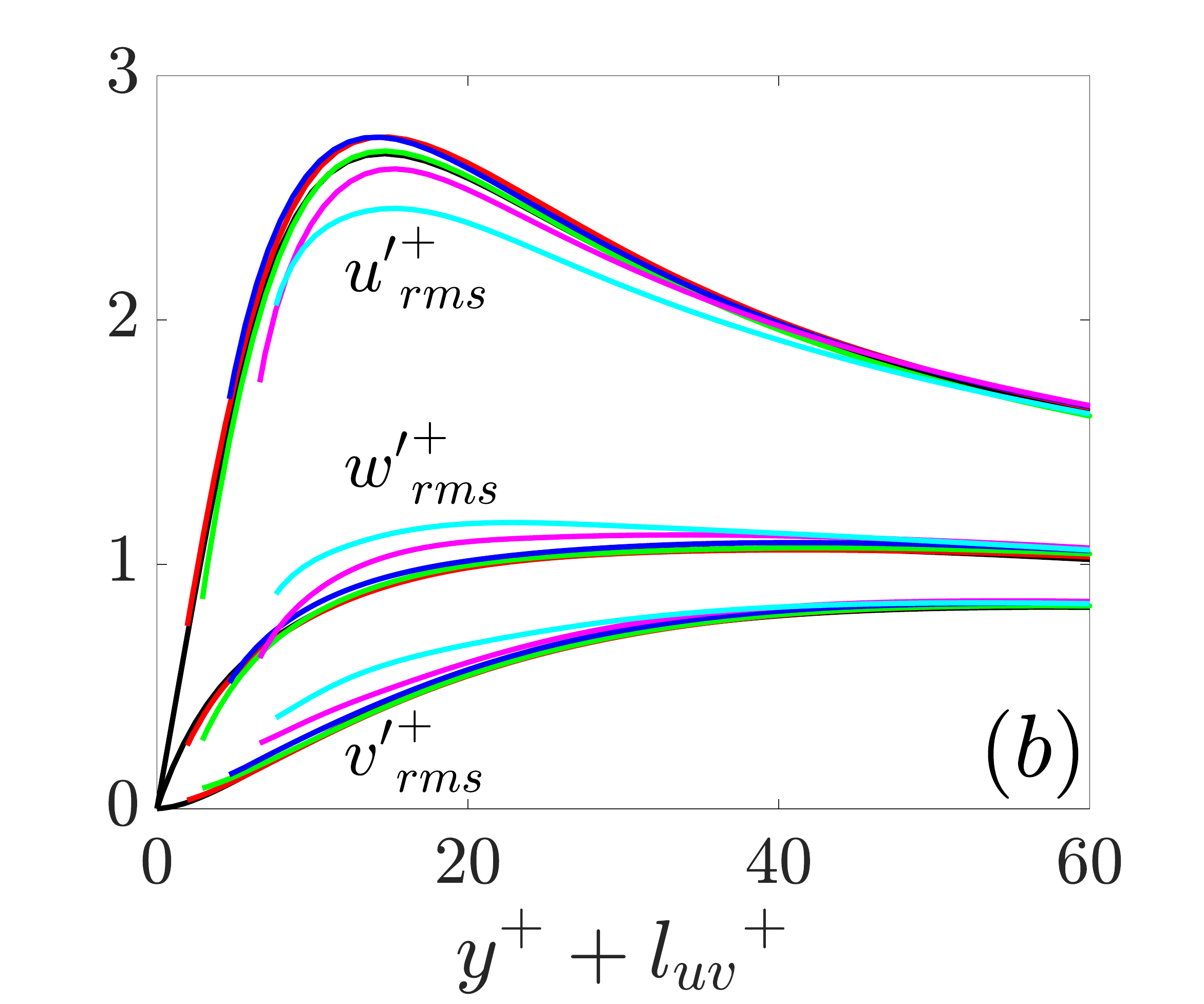}
        {\phantomcaption\label{fig:group1:shifted:sub2}}
        \vspace*{-1mm}
    \end{subfigure}%
    \hspace*{-2mm}
    \begin{subfigure}[tbp]{.35\textwidth}
        \includegraphics[width=1\linewidth]{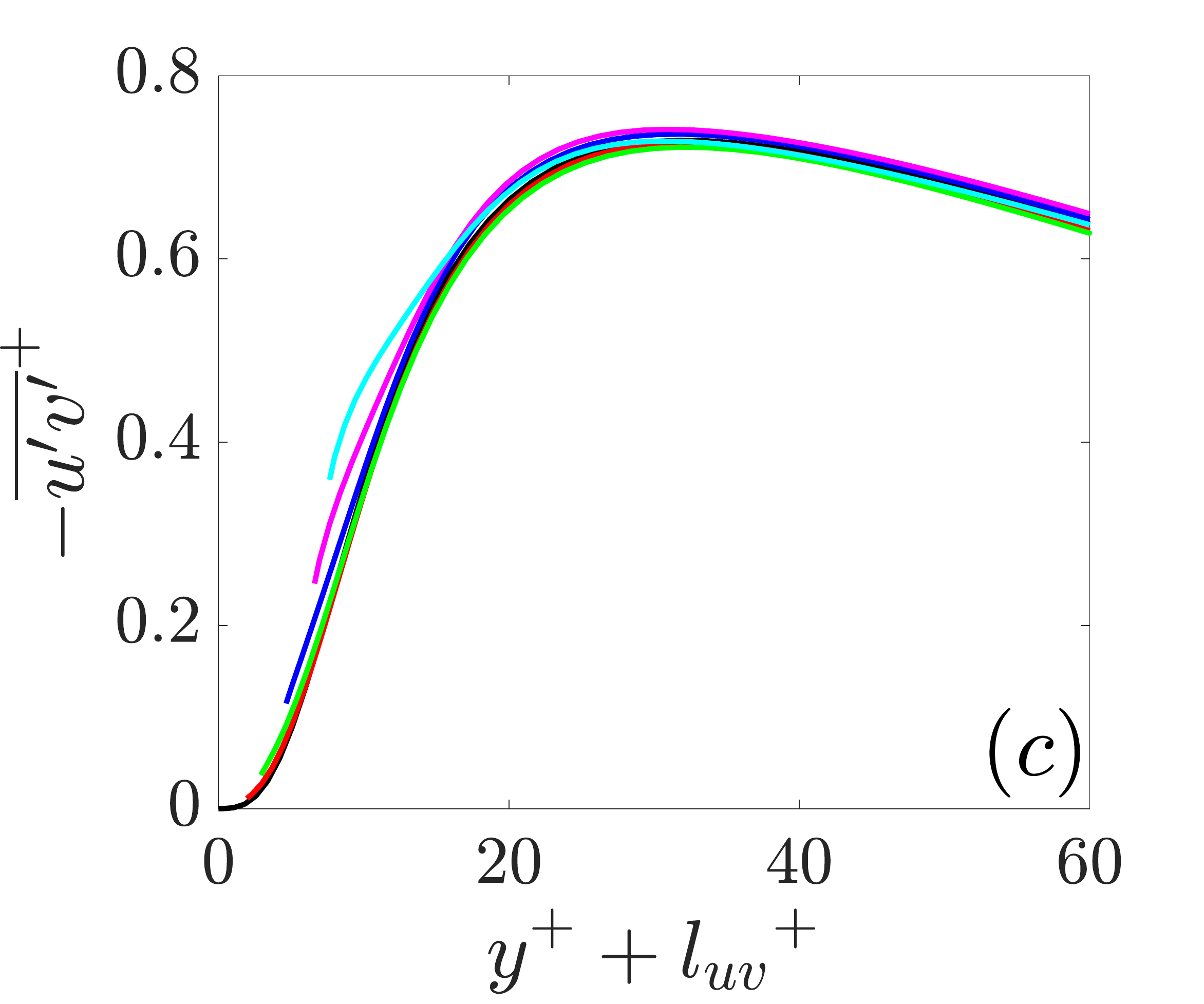}
        {\phantomcaption\label{fig:group1:shifted:sub3}}
        \vspace*{-1mm}
    \end{subfigure}
    \vspace*{-4mm}
    \caption{Mean velocity ($a$), r.m.s. velocity fluctuation ($b$) and Reynolds shear stress ($c$) profiles with the origin at ${y}^+ = -{\ell_{uv}}^+$ and rescaled with the ${u_{\tau}}$ at that plane. \textcolor{red}{$-\!-$}, L2M2; \textcolor{green}{$-\!-$}, L2M5; \textcolor{blue}{$-\!-$}, L5M5; \textcolor{magenta}{$-\!-$}, L5M10; \textcolor{cyan}{$-\!-$}, L10M10; $-\!-$, smooth-wall data.} 
    \label{fig:group1:shifted}
    \end{center}
    \vspace*{-2mm}
\end{figure}
%%%%%%%%%%%%%%%%%%%%%%%%%%%%%%%%%%%%%%%%%%%%%%%%%%%%%%%%%%%%%%%%%%%%%%%%%
%%%%%%%%%%%%%%%%%%%%%%%%%%%%%%%%%%%%%%%%%%%%%%%%%%%%%%%%%%%%%%%%%%%%%%%%%

Case L5M10 results in ${{{\Delta}{U}}^+}=-2.9$, which is nearly half way into the transitionally rough regime (assuming the fully rough regime to correspond to ${{{\Delta}{U}}^+}\approx6$, as shown by \citealt{jimenez_rough}). The resulting virtual origins of the mean flow and Reynolds shear stress are ${\ell_U}^+=3.7$ and ${\ell_{uv}}^+=6.6$. The latter indicates a significant modification of the near-wall turbulence. Compared to the cases with ${m^+}<10$, the intensities of ${v^{\prime}}^+$ and ${w^{\prime}}^+$ have increased while that of ${u^{\prime}}^+$ has decreased, demonstrating a move towards turbulence isotropization (\hyperref[fig:group1:v_rms]{figures \ref*{fig:group1:v_rms}} and \hyperref[fig:group1:u_w_rms]{\ref*{fig:group1:u_w_rms}}). The near-wall distribution of $-\overline{u^{\prime}v^{\prime}}^+$ also undergoes a noticeable outward rise compared to its smooth-wall counterpart (\hyperref[fig:group1:reynolds_stress]{figure \ref*{fig:group1:reynolds_stress}}). An even stronger modification of the turbulence is observed for case L10M10 ($\ell^+_{uv}=7.7$), although the drag increase (${{{\Delta}{U}}^+}=-1.6$) is smaller owing to the large mean flow slip ($\ell_U^+=6.1$).

\Cref{fig:group1:shifted} shows smooth-wall statistics are recoverable for the majority of the cases considered thus far after accounting for the virtual origin effect. However, differences from smooth-wall turbulence are noticeably present in cases L5M10 and L10M10 (\hyperref[fig:group1:shifted:sub2]{figure \ref*{fig:group1:shifted:sub2}}). 

\begin{figure}
    \begin{center}
    \hspace*{-1mm}
    \begin{subfigure}[tbp]{.2705\textwidth}
        \includegraphics[width=1\linewidth]{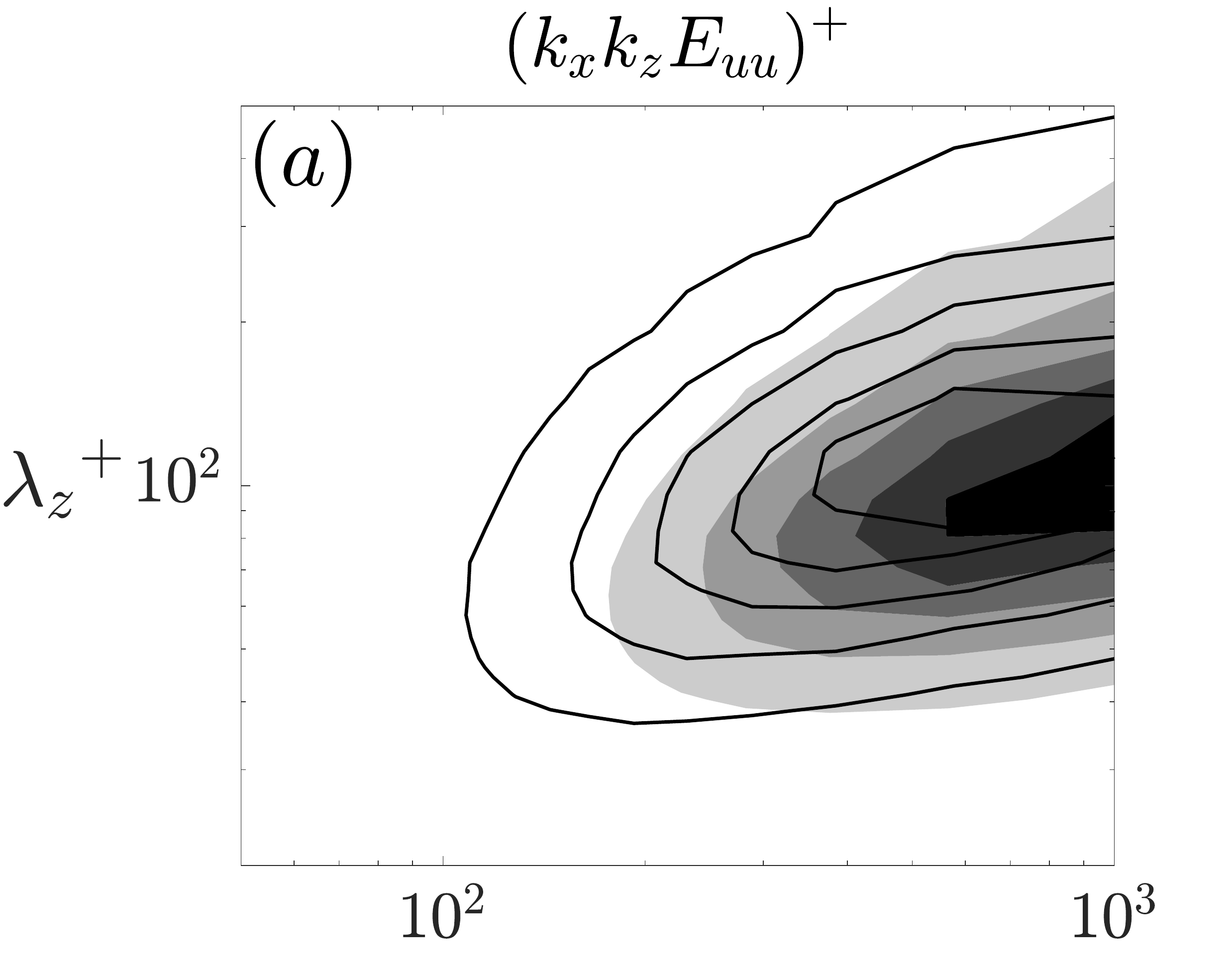}
        {\phantomcaption\label{spectra_L10M10:u}}
        \vspace*{-4.95mm}
    \end{subfigure}%
    \hspace*{-3mm}
    \begin{subfigure}[tbp]{.257\textwidth}
        \includegraphics[width=1\linewidth]{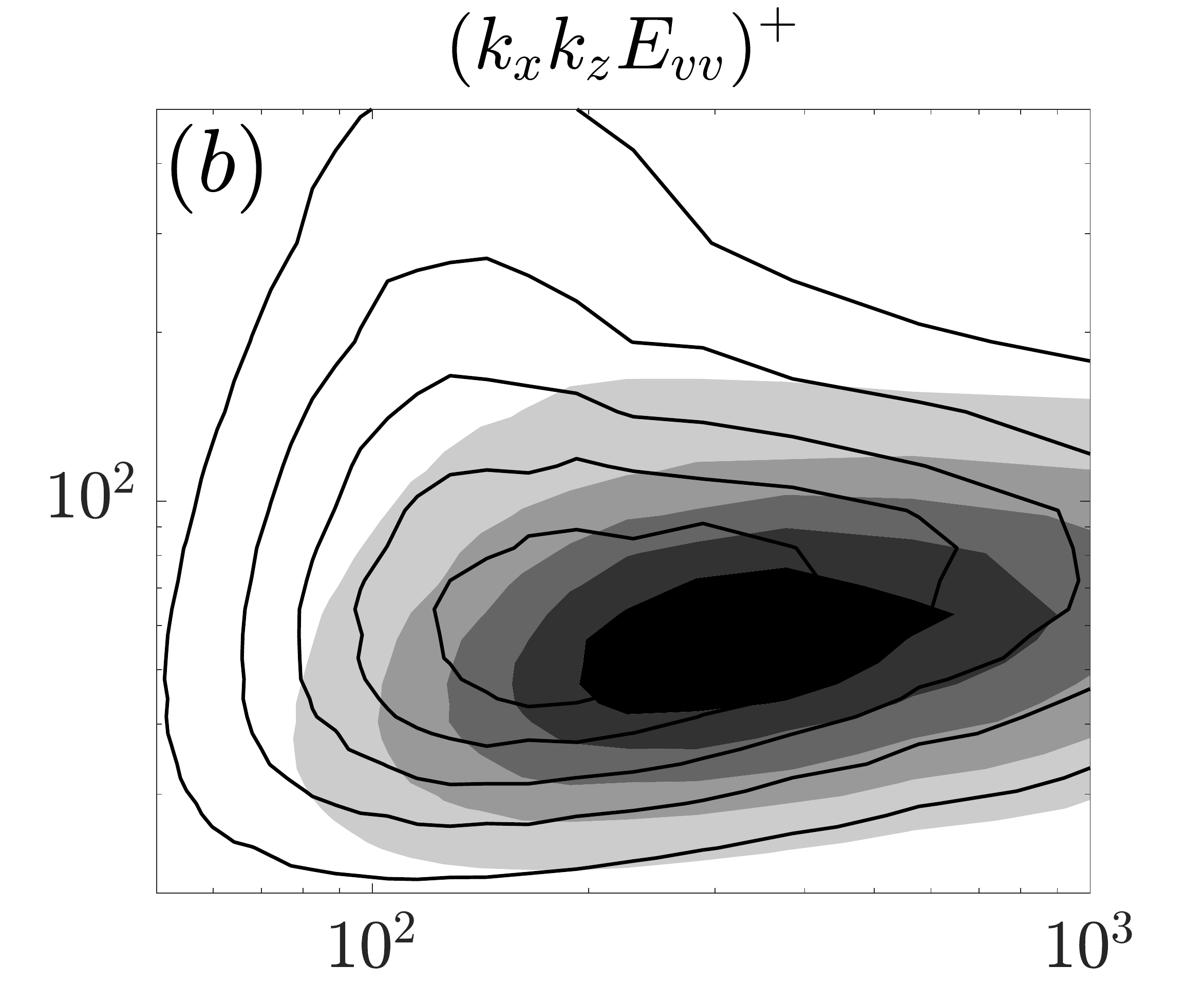}
        {\phantomcaption\label{spectra_L10M10:v}}
        \vspace*{-4.95mm}
    \end{subfigure}%
    \hspace*{-3mm}
    \begin{subfigure}[tbp]{.257\textwidth}
        \includegraphics[width=1\linewidth]{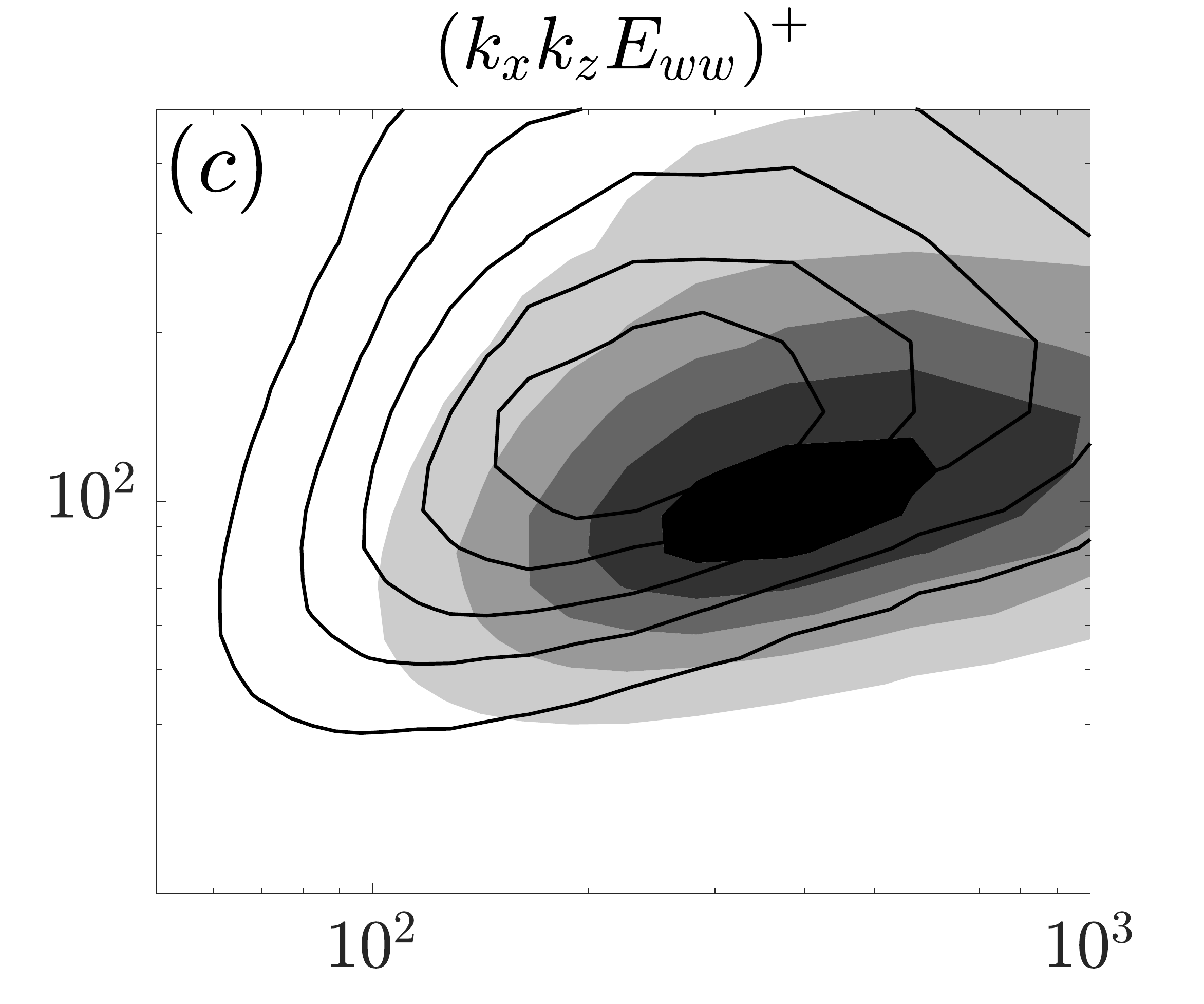}
        {\phantomcaption\label{spectra_L10M10:w}}
        \vspace*{-4.95mm}
    \end{subfigure}%
    \hspace*{-3mm}
    \begin{subfigure}[tbp]{.257\textwidth}
        \includegraphics[width=1\linewidth]{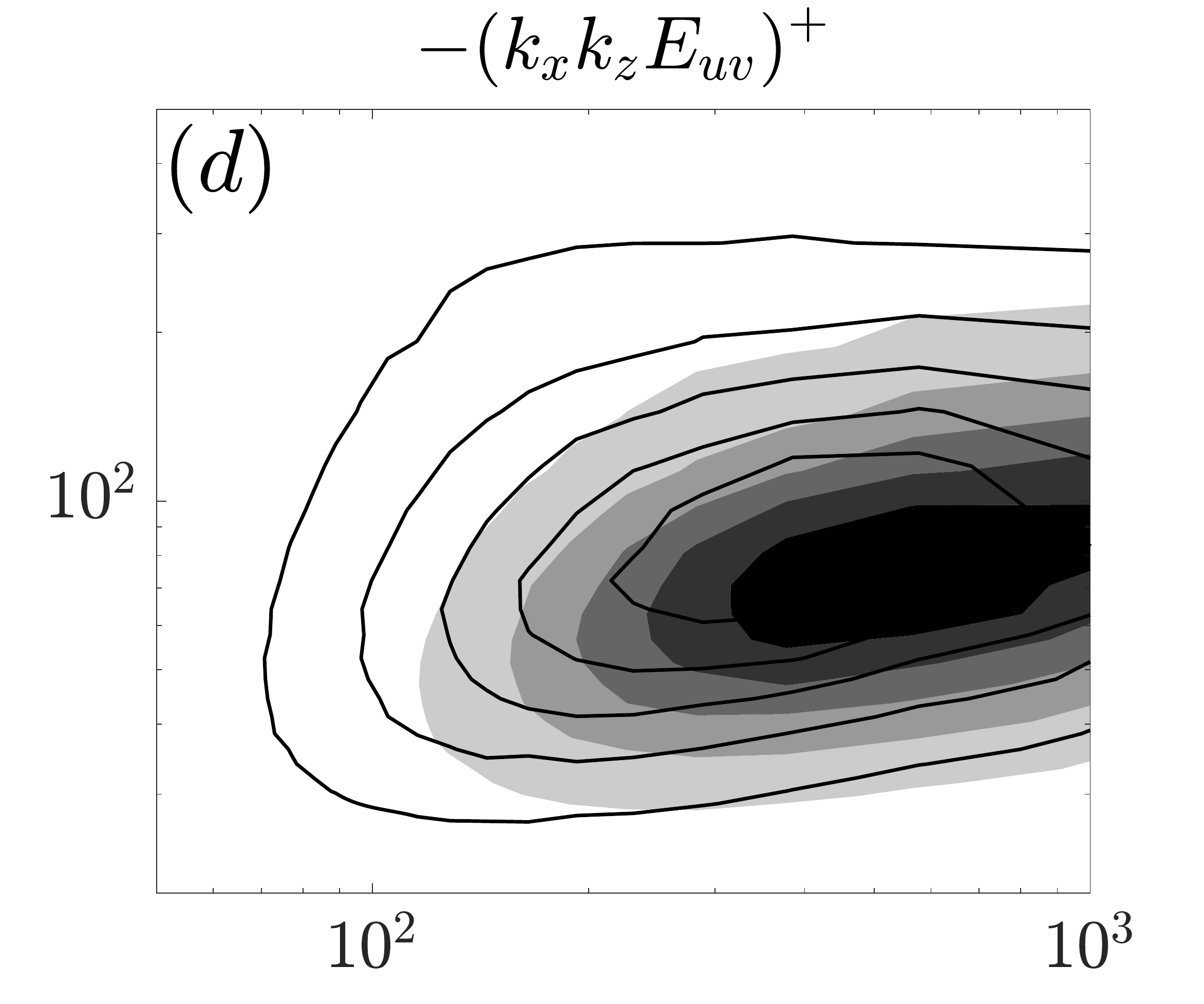}
        {\phantomcaption\label{spectra_L10M10:uv}}
        \vspace*{-4.95mm}
    \end{subfigure}
    \hspace*{-1mm}
    \begin{subfigure}[tbp]{.2705\textwidth}
        \includegraphics[width=1\linewidth]{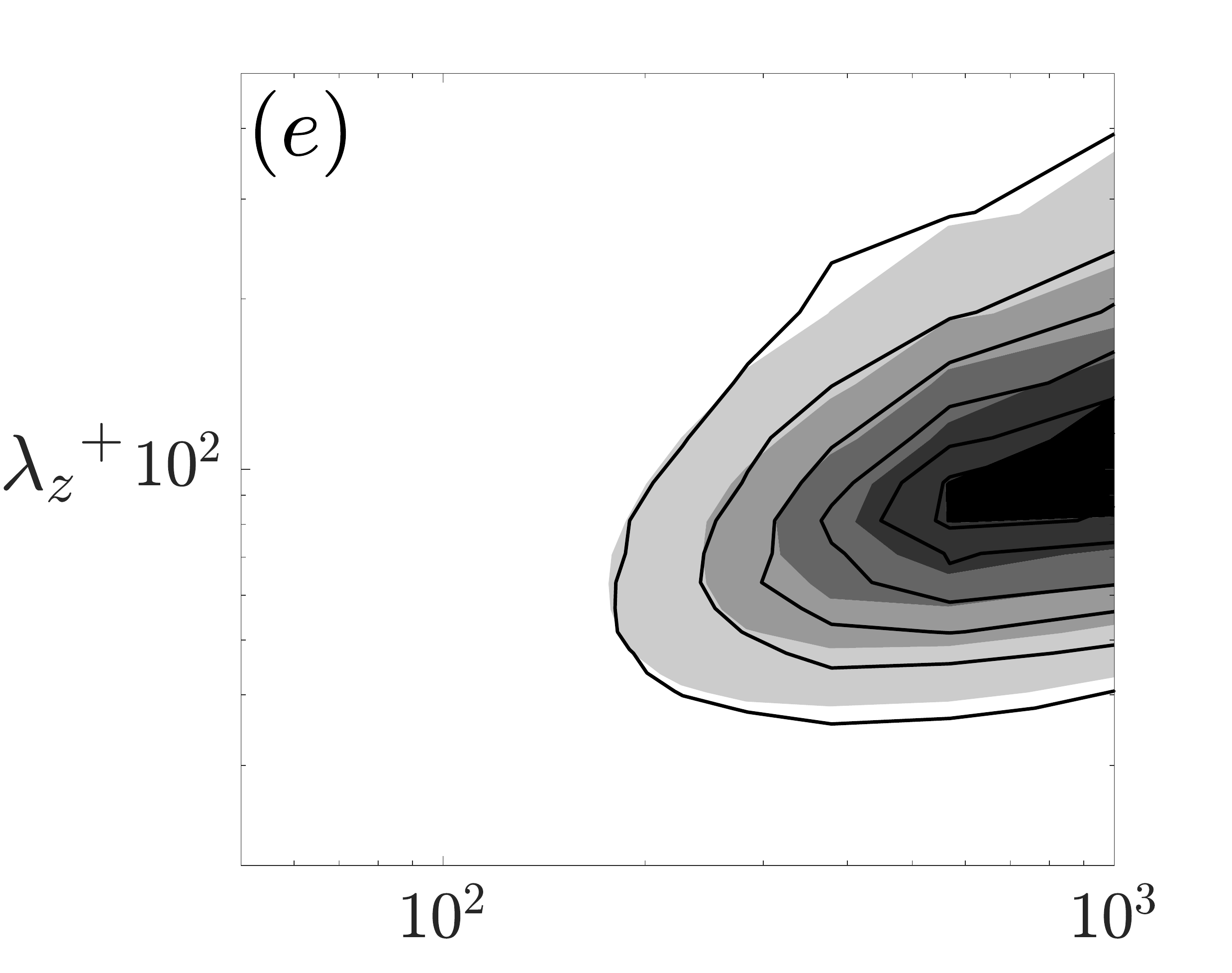}
        {\phantomcaption\label{spectra_L5M5:u}}
        \vspace*{-4.95mm}
    \end{subfigure}%
    \hspace*{-3mm}
    \begin{subfigure}[tbp]{.257\textwidth}
        \includegraphics[width=1\linewidth]{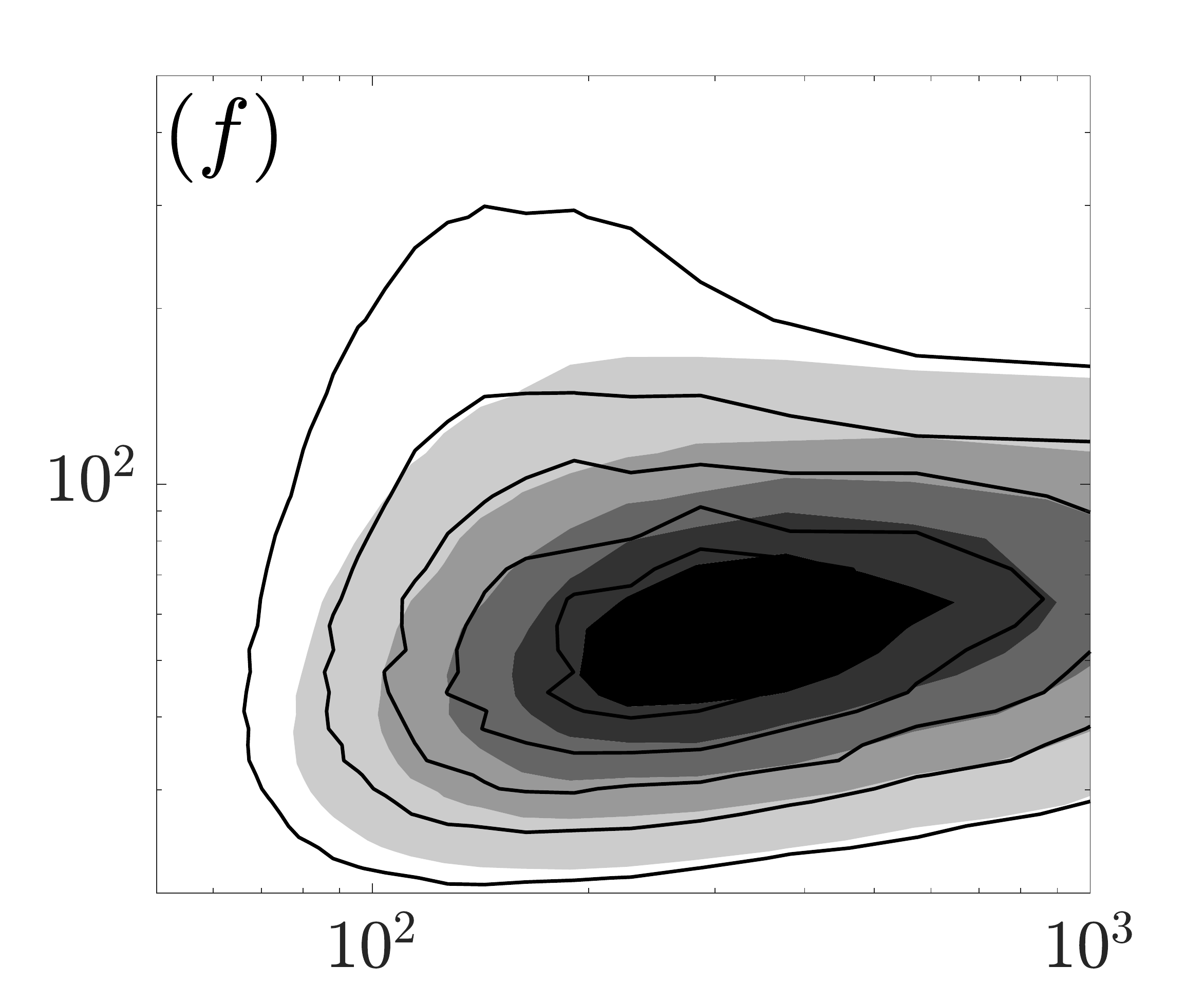}
        {\phantomcaption\label{spectra_L5M5:v}}
        \vspace*{-4.95mm}
    \end{subfigure}%
    \hspace*{-3mm}
    \begin{subfigure}[tbp]{.257\textwidth}
        \includegraphics[width=1\linewidth]{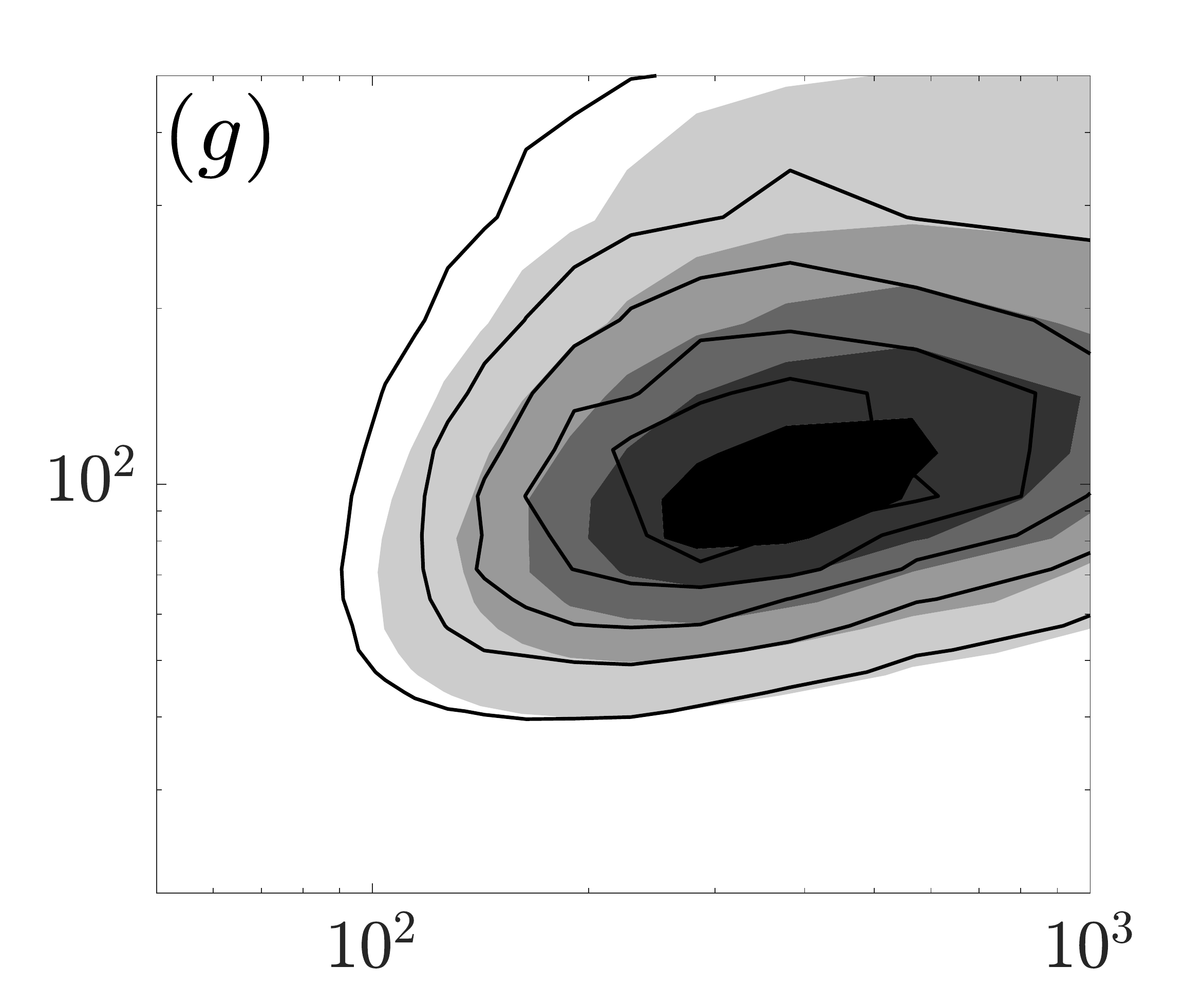}
        {\phantomcaption\label{spectra_L5M5:w}}
        \vspace*{-4.95mm}
    \end{subfigure}%
    \hspace*{-3mm}
    \begin{subfigure}[tbp]{.257\textwidth}
        \includegraphics[width=1\linewidth]{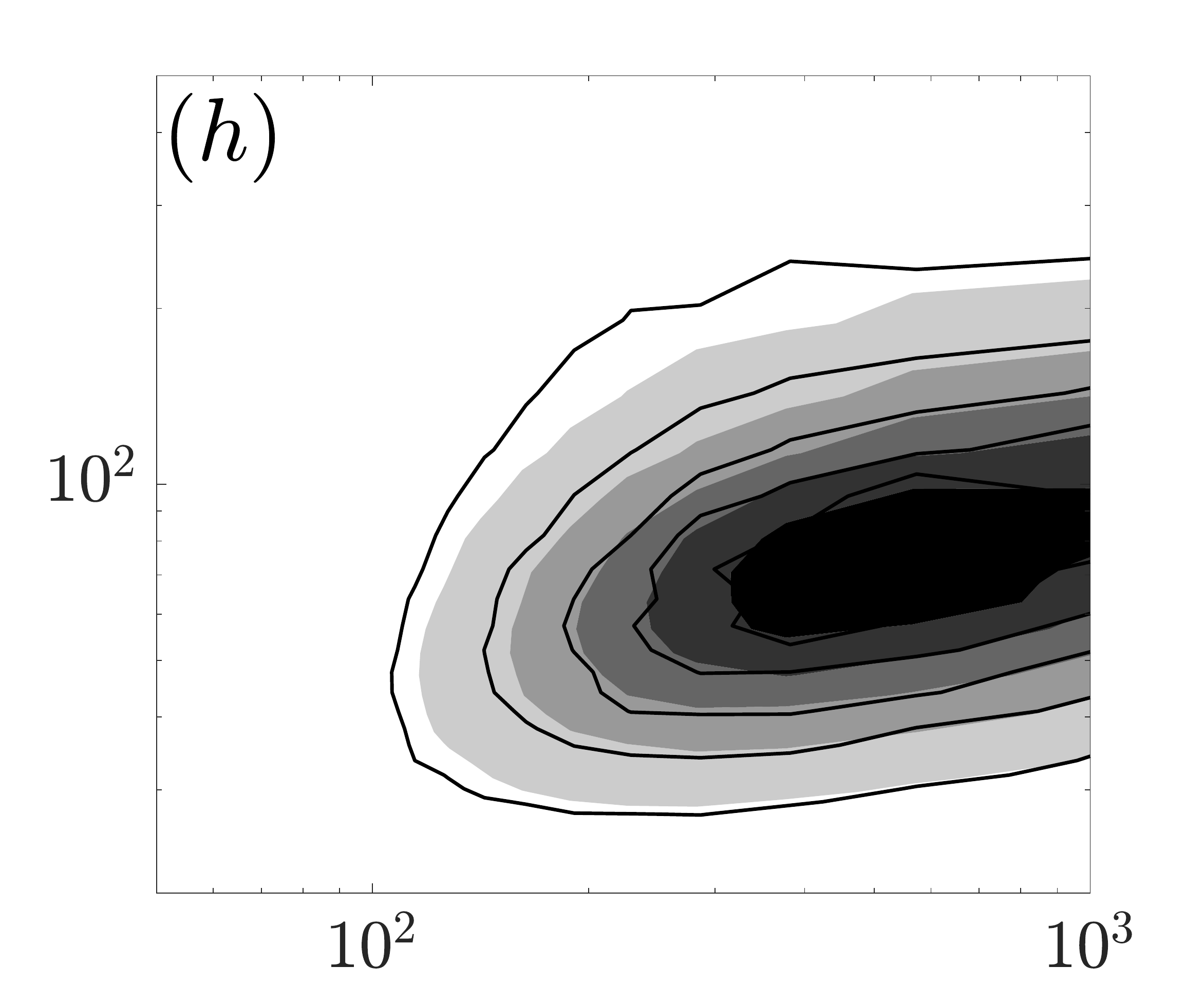}
        {\phantomcaption\label{spectra_L5M5:uv}}
        \vspace*{-4.95mm}
    \end{subfigure}
    \hspace*{-1mm}
    \begin{subfigure}[tbp]{.2705\textwidth}
        \includegraphics[width=1\linewidth]{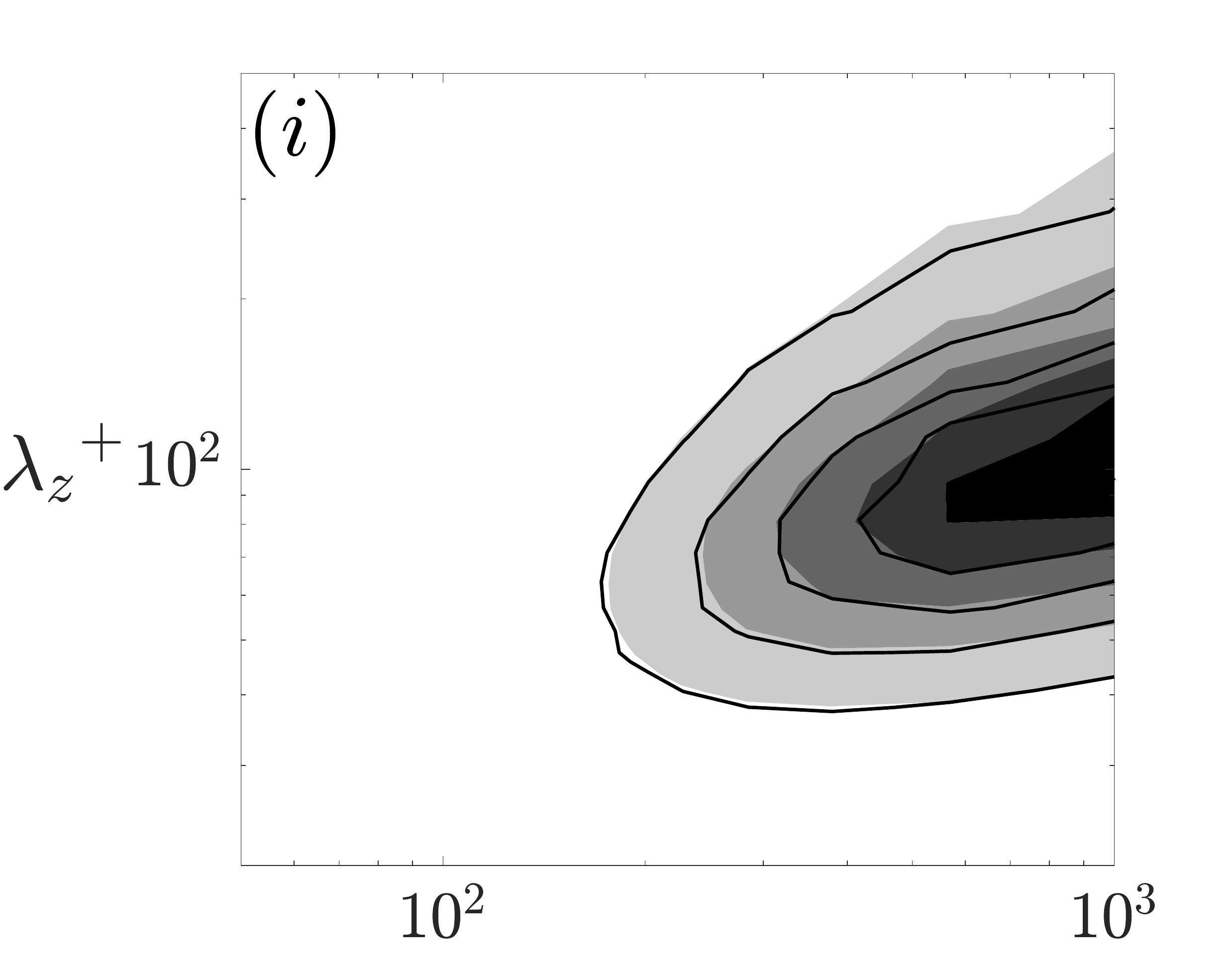}
        {\phantomcaption\label{spectra_L2M5:u}}
        \vspace*{-4.95mm}
    \end{subfigure}%
    \hspace*{-3mm}
    \begin{subfigure}[tbp]{.257\textwidth}
        \includegraphics[width=1\linewidth]{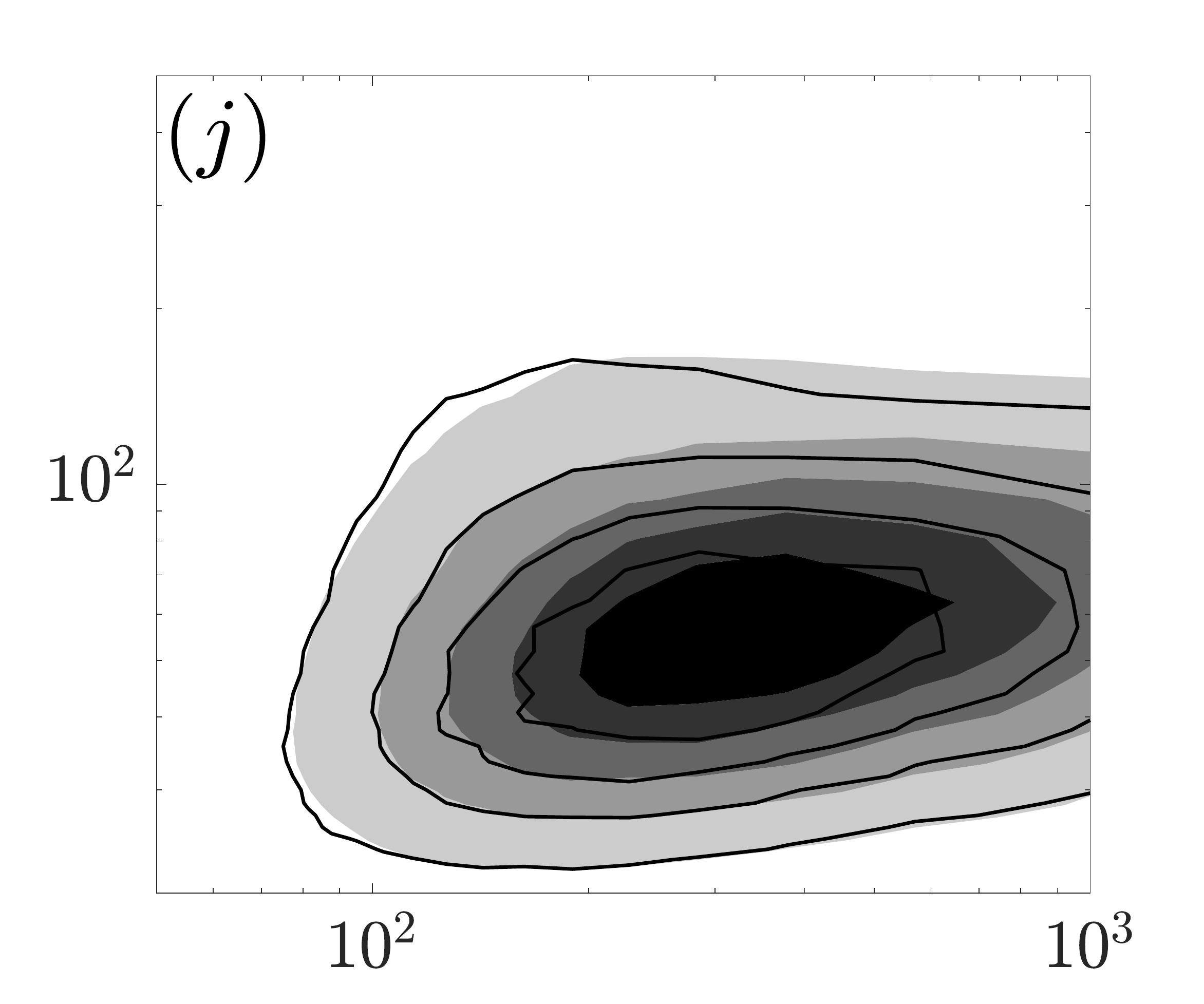}
        {\phantomcaption\label{spectra_L2M5:v}}
        \vspace*{-4.95mm}
    \end{subfigure}%
    \hspace*{-3mm}
    \begin{subfigure}[tbp]{.257\textwidth}
        \includegraphics[width=1\linewidth]{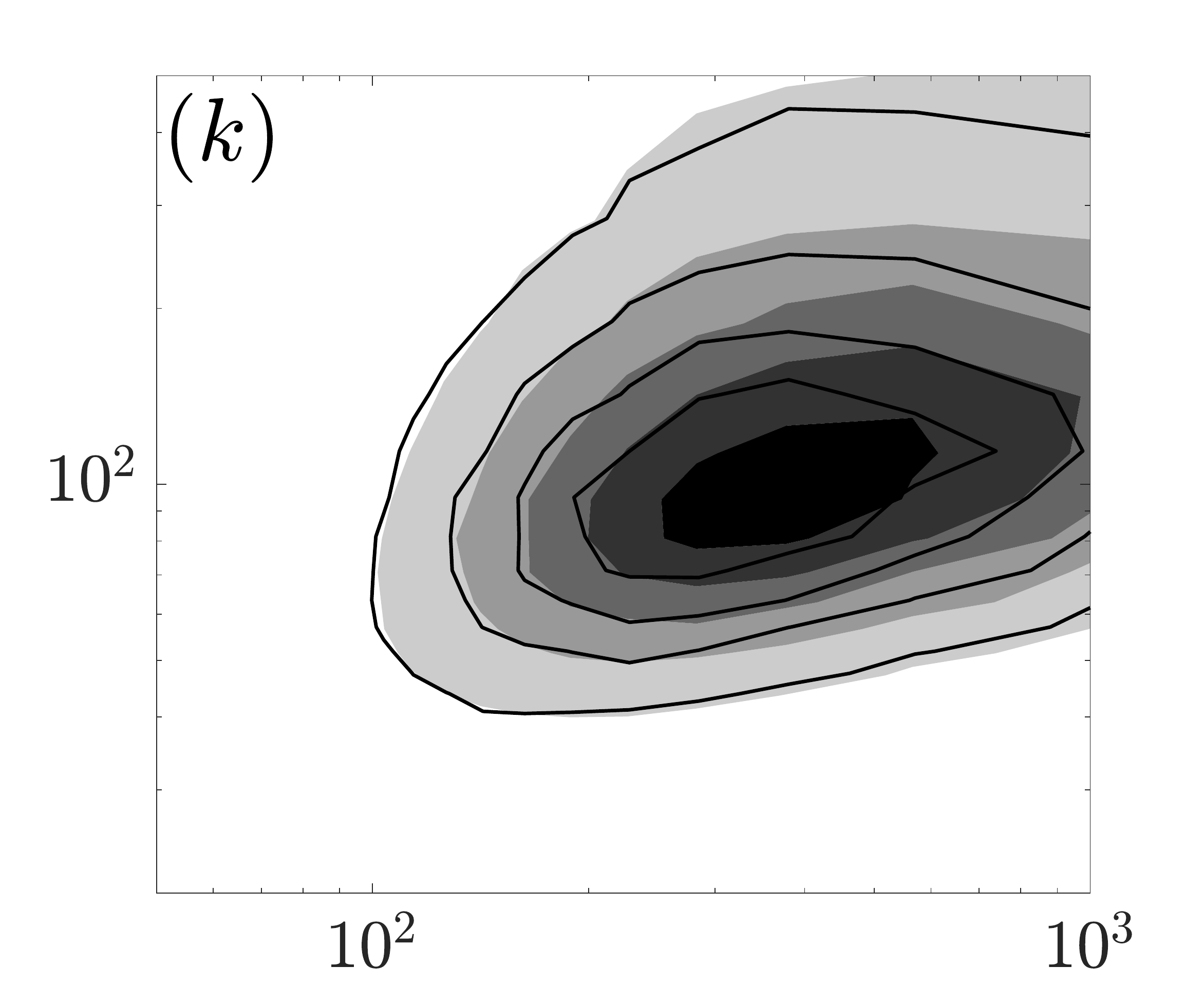}
        {\phantomcaption\label{spectra_L2M5:w}}
        \vspace*{-4.95mm}
    \end{subfigure}%
    \hspace*{-3mm}
    \begin{subfigure}[tbp]{.257\textwidth}
        \includegraphics[width=1\linewidth]{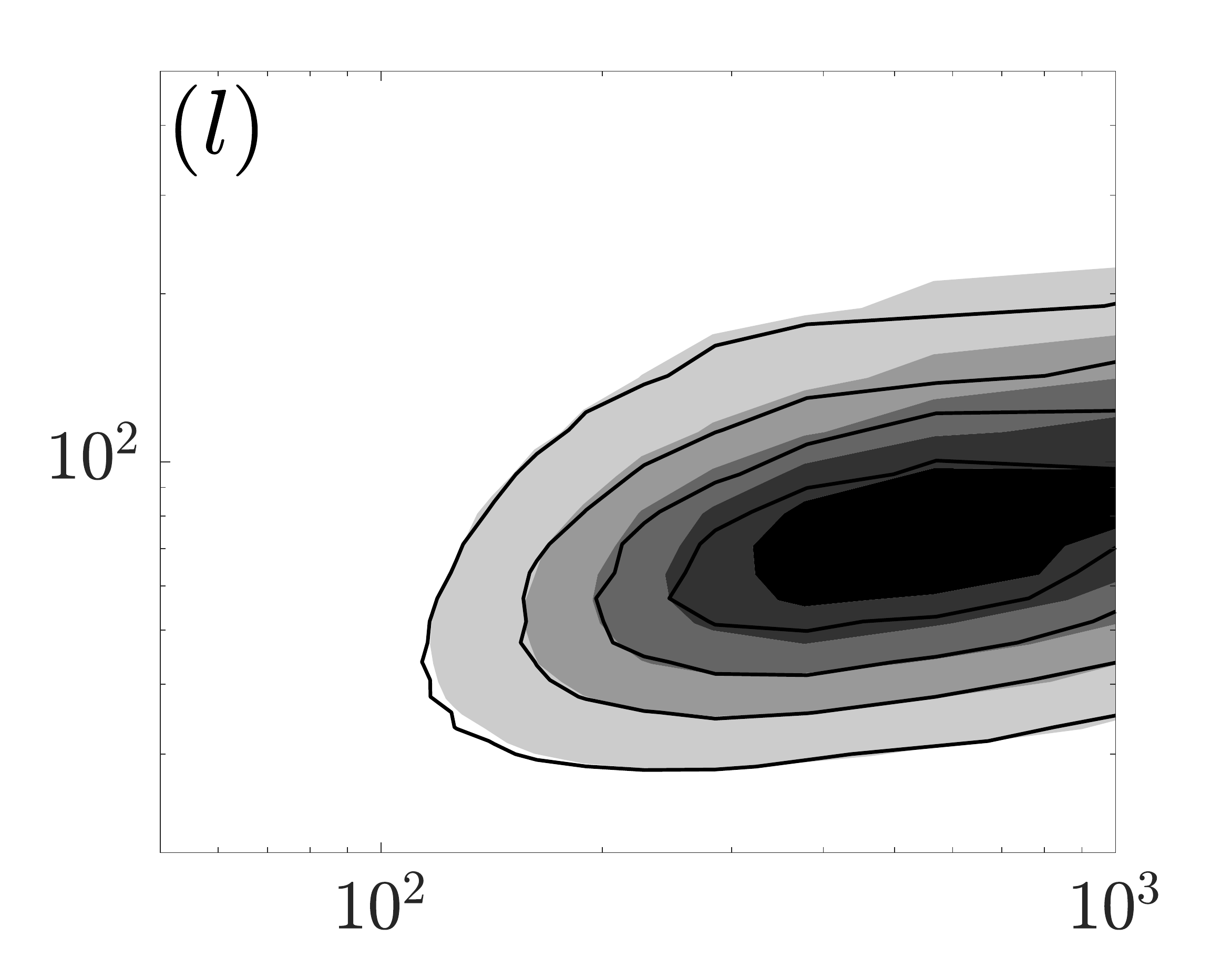}
        {\phantomcaption\label{spectra_L2M5:uv}}
        \vspace*{-4.95mm}
    \end{subfigure}
    \hspace*{-1mm}
    \begin{subfigure}[tbp]{.2705\textwidth}
        \includegraphics[width=1\linewidth]{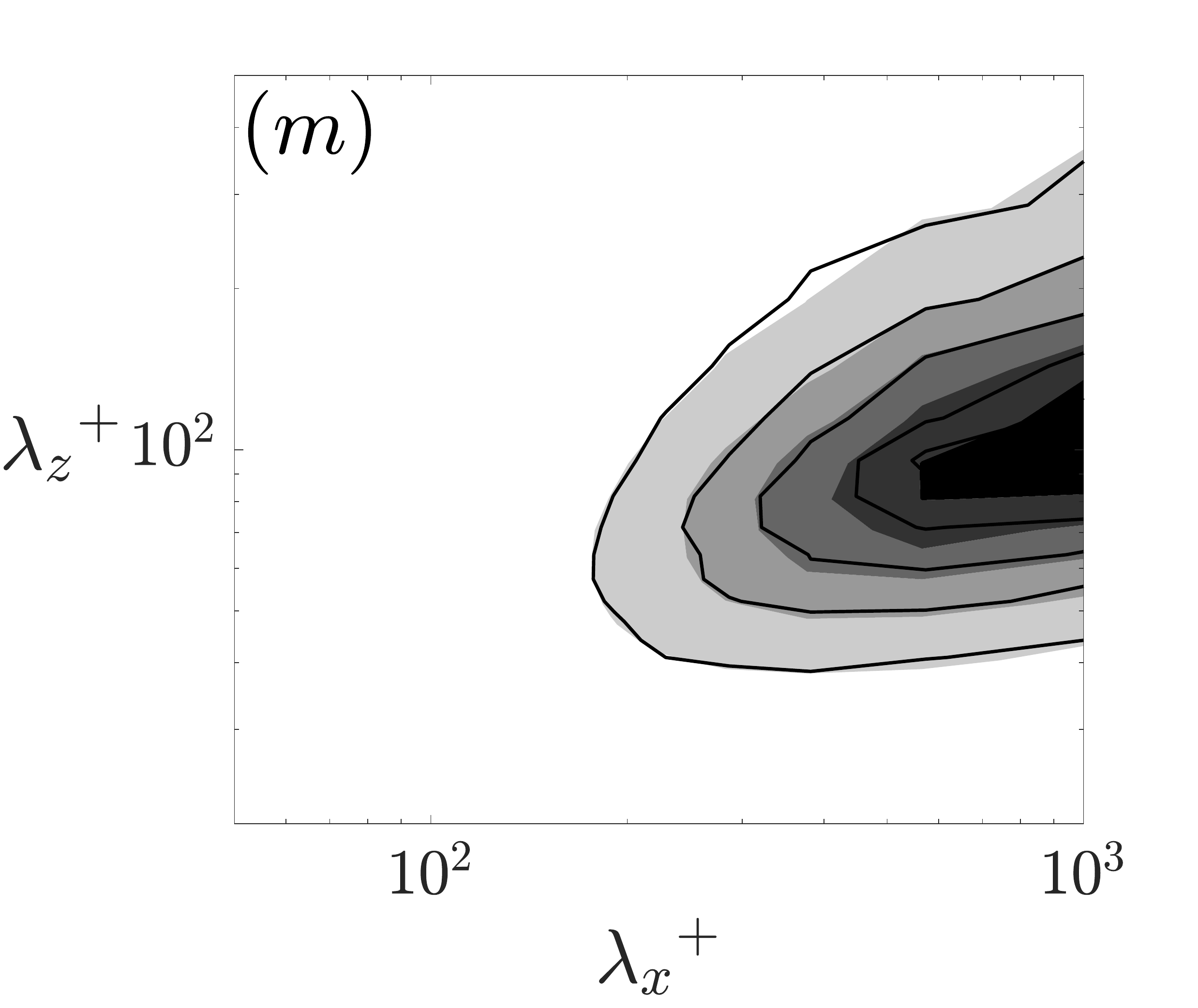}
        {\phantomcaption\label{spectra_L2M2:u}}
    \end{subfigure}%
    \hspace*{-3mm}
    \begin{subfigure}[tbp]{.257\textwidth}
        \includegraphics[width=1\linewidth]{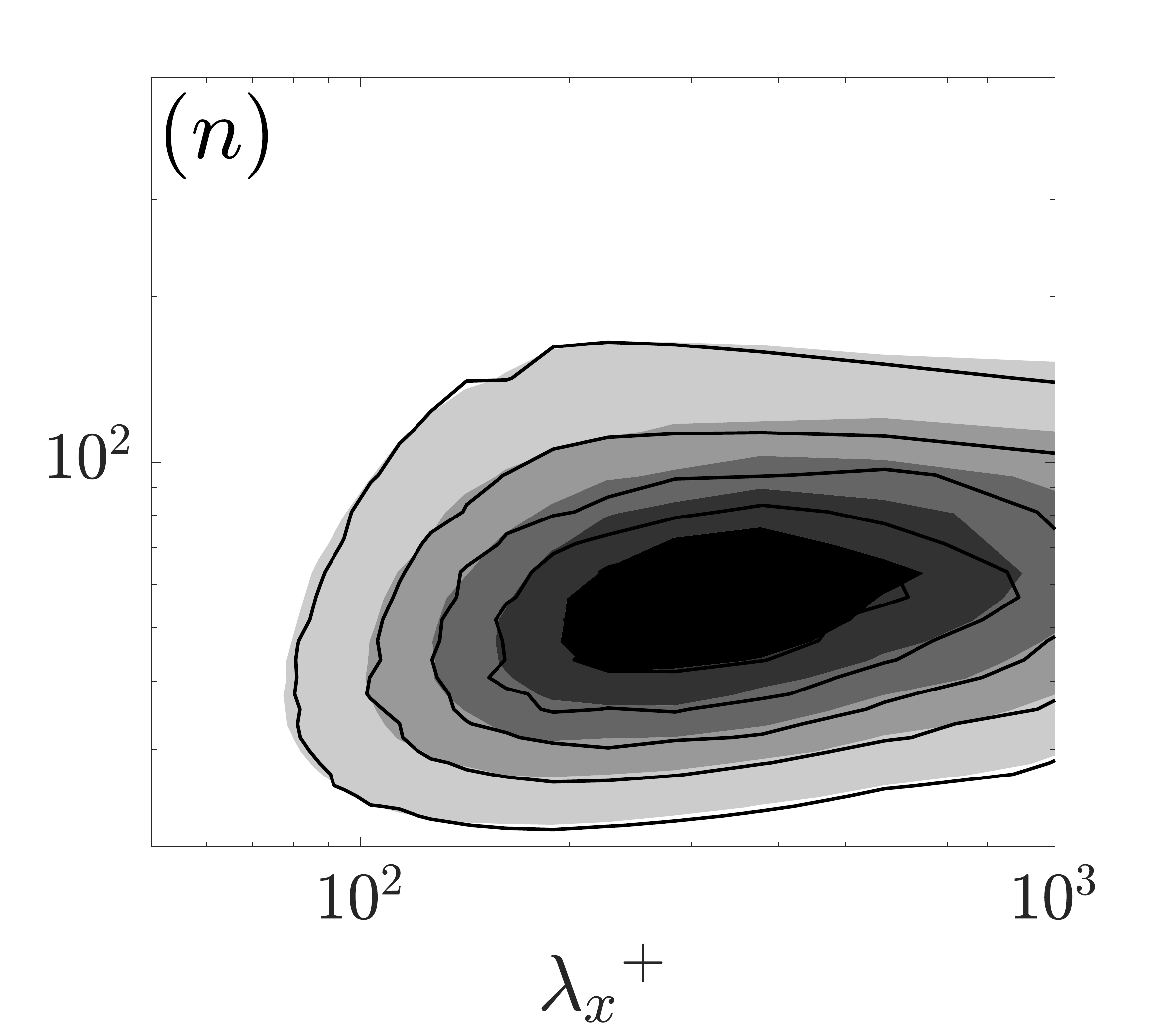}
        {\phantomcaption\label{spectra_L2M2:v}}
    \end{subfigure}%
    \hspace*{-3mm}
    \begin{subfigure}[tbp]{.257\textwidth}
        \includegraphics[width=1\linewidth]{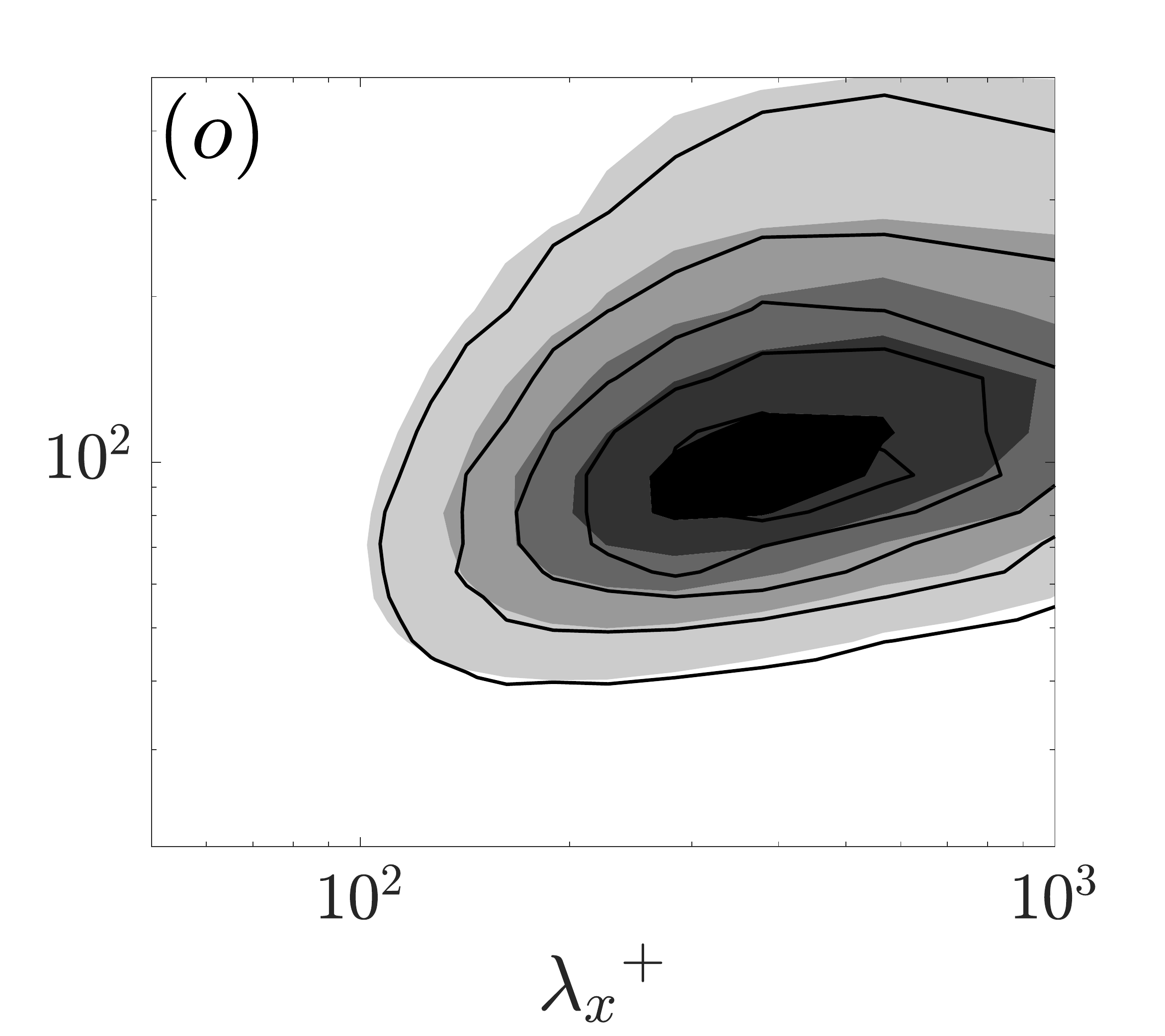}
        {\phantomcaption\label{spectra_L2M2:w}}
    \end{subfigure}%
    \hspace*{-3mm}
    \begin{subfigure}[tbp]{.257\textwidth}
        \includegraphics[width=1\linewidth]{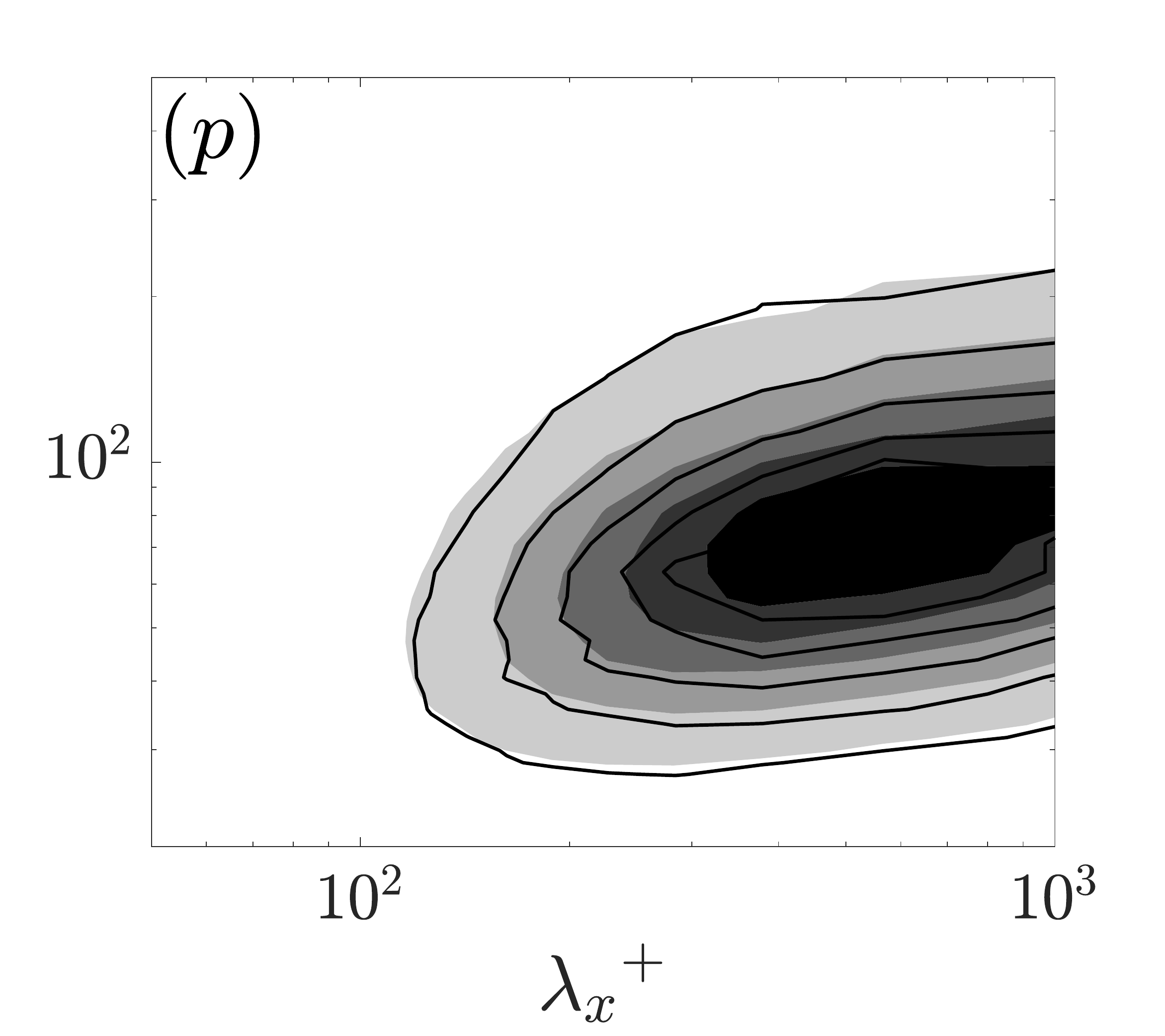}
        {\phantomcaption\label{spectra_L2M2:uv}}
    \end{subfigure}
    \vspace*{-7mm}
    \caption{Pre-multiplied two dimensional spectral densities of ${u}^2$, ${v}^2$, ${w}^2$ and ${uv}$. L10M10 ($a$\textbf{--}$d$); L5M5 ($e$\textbf{--}$h$); L2M5 ($i$\textbf{--}$l$); L2M2 ($m$\textbf{--}$p$); Shaded regions are the smooth-wall solution at $y^+\!\approx15$ and solid lines are the TRM cases at $y^+\!+{\ell_{uv}}^+\!\approx15$ scaled using the $u_{\tau}$ at $y^+\!=-{\ell_{uv}}^+$.}
    \label{fig:spectra1}
    \end{center}
    \vspace*{-3mm}
\end{figure}

To better determine the extent of such differences the pre-multiplied energy spectra may be examined. \Cref{fig:spectra1} shows (from top to bottom) the energy spectra for L10M10, L5M5, L2M5 and L2M2. The spectra of L2M2 and L2M5 conforms to the smooth wall spectra closely, reaffirming that the turbulence is smooth-wall-like in these cases. In contrast, the spectra of case L5M5 exhibits differences for ${v}^2$ and ${w}^2$ (\hyperref[spectra_L5M5:v]{figures \ref*{spectra_L5M5:v}} and \hyperref[spectra_L5M5:w]{\ref*{spectra_L5M5:w}}), while the co-spectra of its Reynolds shear stress (\hyperref[spectra_L5M5:uv]{figure \ref*{spectra_L5M5:uv}}) remains largely smooth-wall-like. Case L10M10 shows large differences across all spectra, consistent with the deviations observed in \cref{fig:group1:shifted}. The deviations observed in the co-spectra of the wall-normal fluctuations are typical for surfaces that induce a Kelvin-Helmholtz type of instability. This is further examined in \cref{changesinturbulence} where it is also observed for cases with anisotropic transpiration lengths.

%%%%%%%%%%%%%%%%%%%%%%%%%%%%%%%%%%%%%%%%%%%%%%%%%%%%%%%%%%%%%%%%%%%%%%%%%
%%%%%%%%%%%%%%%%%%%%%%%%%%%%%%%%%%%%%%%%%%%%%%%%%%%%%%%%%%%%%%%%%%%%%%%%%
%%%%%%%%%%%%%%%%%%%%%%%%%%%%%%%%%%%%%%%%%%%%%%%%%%%%%%%%%%%%%%%%%%%%%%%%%
%%%%%%%%%%%%%%%%%%%%%%%%%%%%%%%%%%%%%%%%%%%%%%%%%%%%%%%%%%%%%%%%%%%%%%%%%

\subsubsection{Reynolds number scaling} \label{sec:Reynolds_scaling}
%%%%%%%%%%%%%%%%%%%%%%%%%%%%%%%%%%%%%%%%%%%%%%%%%%%%%%%%%%%%%%%%%%%%%%%%%
%%%%%%%%%%%%%%%%%%%%%%%%%%%%%%%%%%%%%%%%%%%%%%%%%%%%%%%%%%%%%%%%%%%%%%%%%
\begin{figure}
    \begin{center}
    \hspace*{-2mm}
    \begin{subfigure}[tbp]{.268\textwidth}
        \includegraphics[width=1\linewidth]{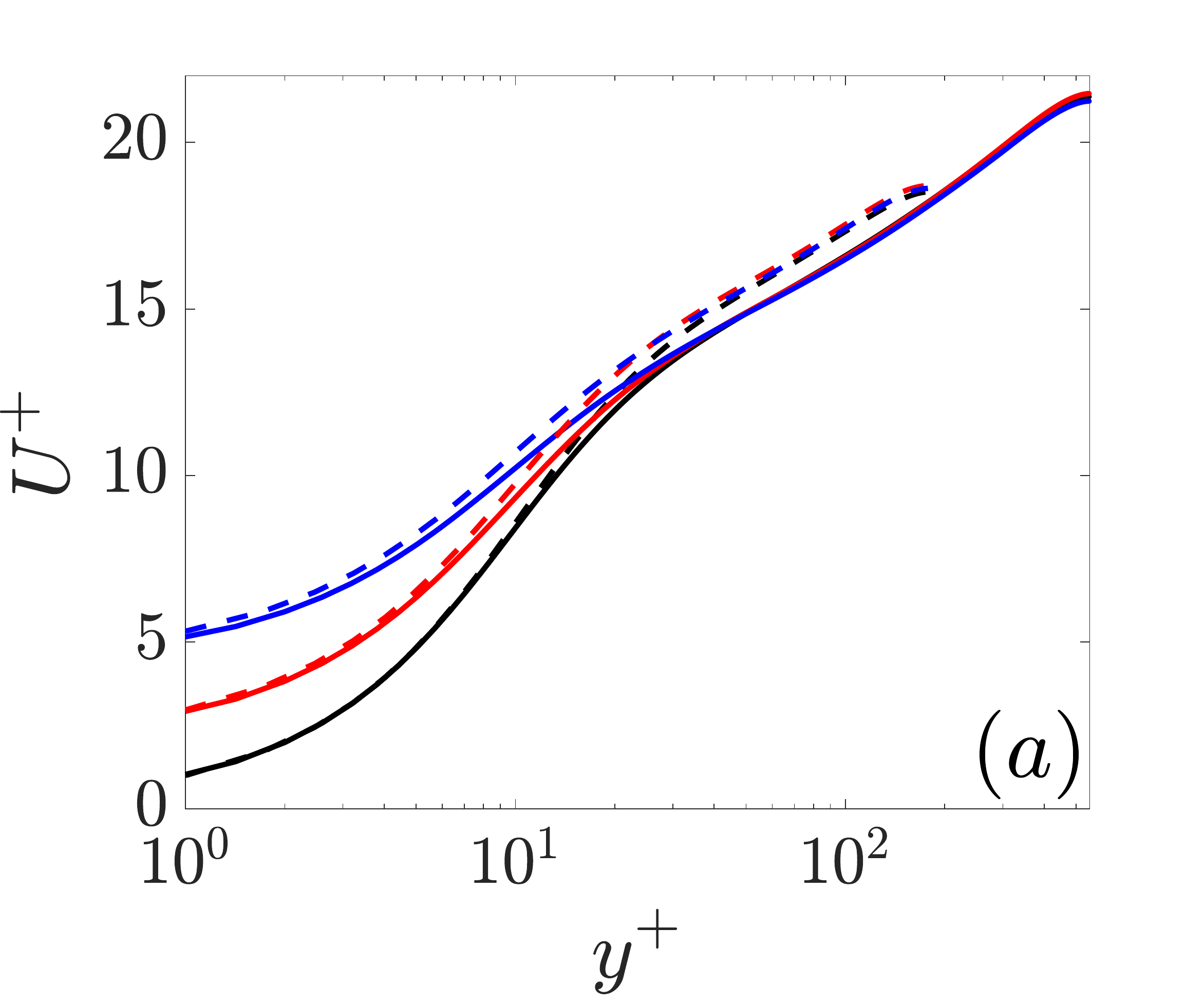}
        \vspace*{-4.2mm}
        {\phantomcaption\label{fig:Reynolds_scaling:sub1}}
    \end{subfigure}%
    \hspace*{-4mm}
    \begin{subfigure}[tbp]{.268\textwidth}
        \includegraphics[width=1\linewidth]{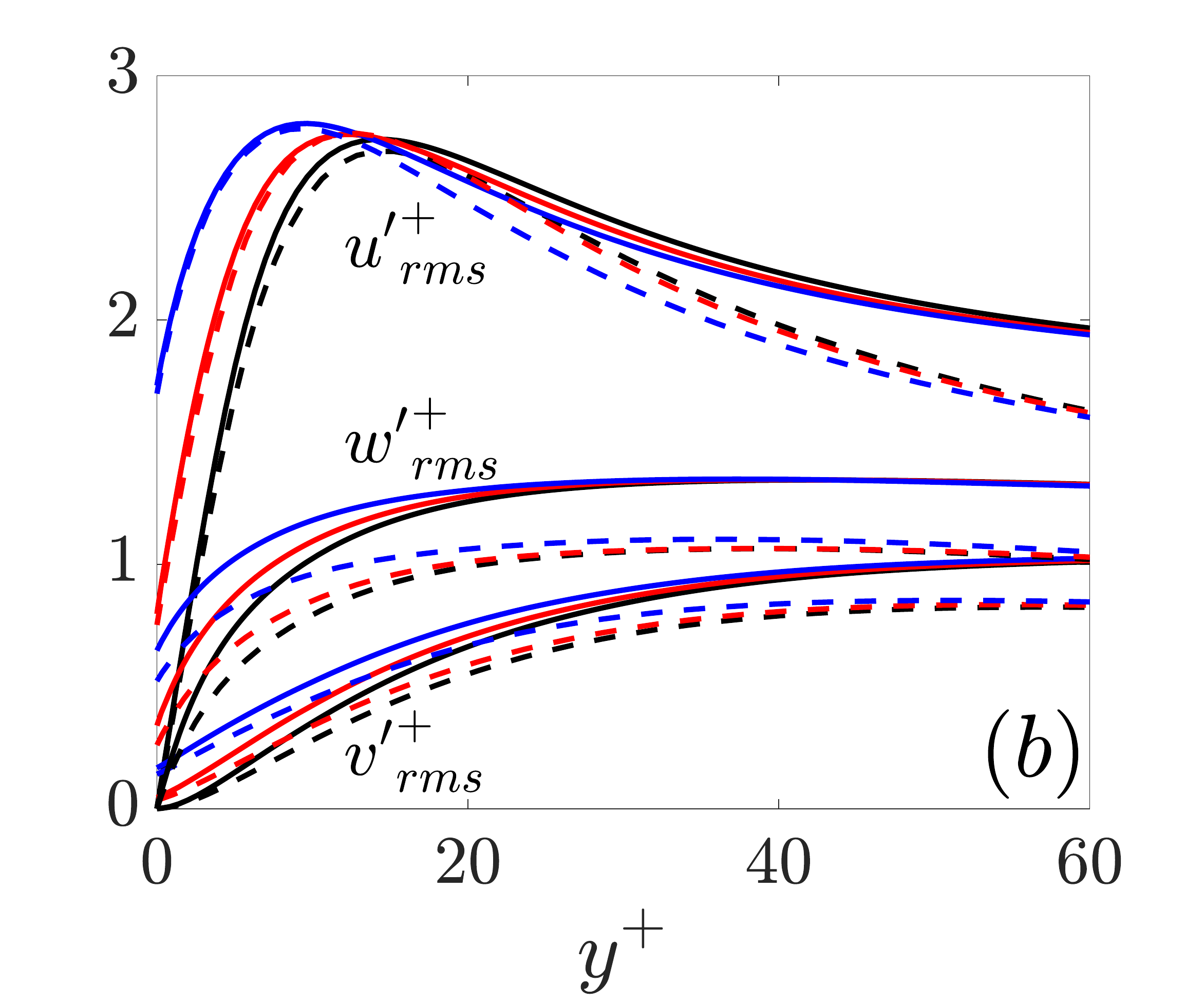}
        \vspace*{-4.2mm}
        {\phantomcaption\label{fig:Reynolds_scaling:sub2}}
    \end{subfigure}%
    \hspace*{-2mm}
    \begin{subfigure}[tbp]{.268\textwidth}
        \includegraphics[width=1\linewidth]{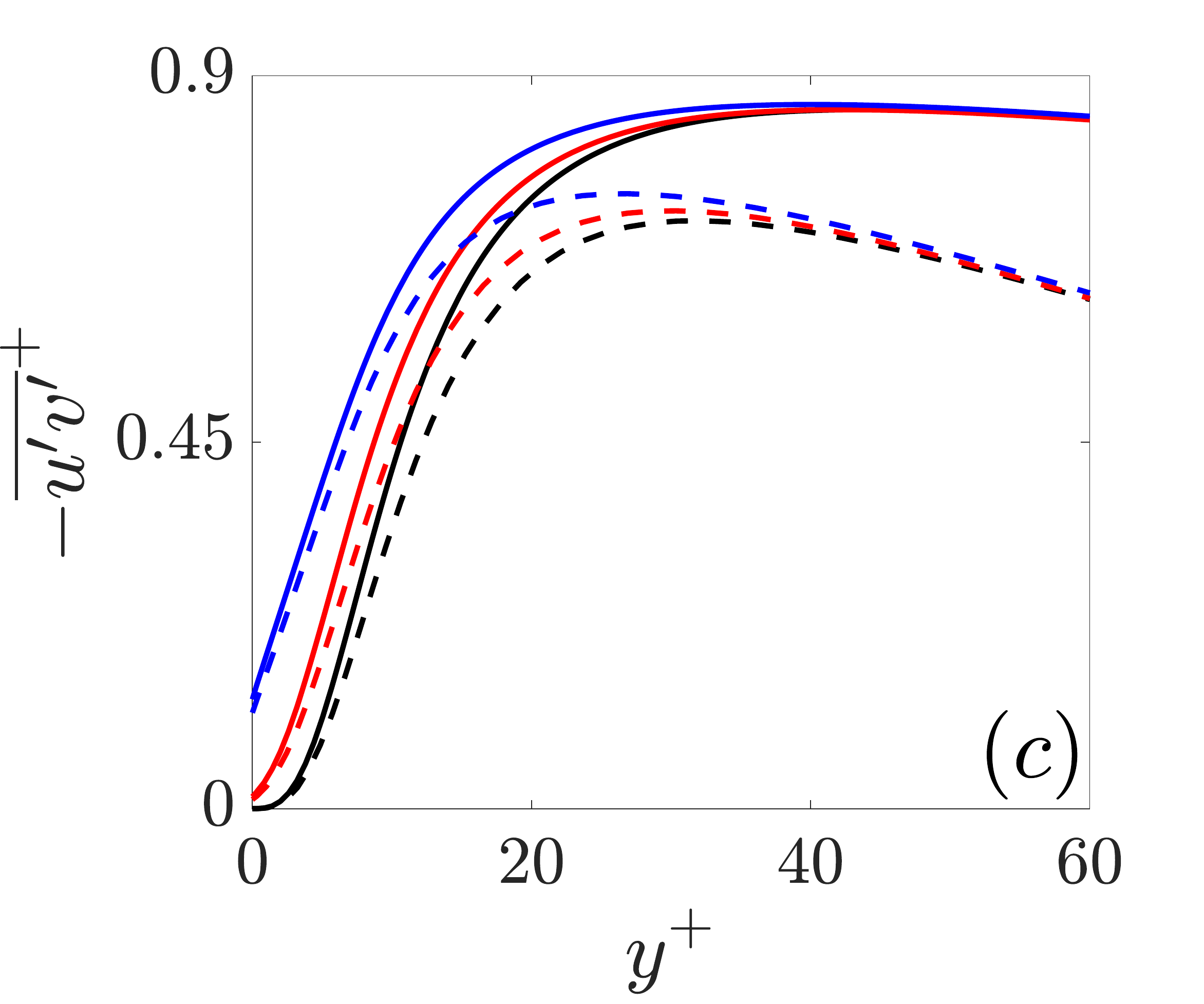}
        \vspace*{-4.2mm}
        {\phantomcaption\label{fig:Reynolds_scaling:sub3}}
    \end{subfigure}
    \hspace*{-1mm}
    \begin{subfigure}[tbp]{.268\textwidth}
        \includegraphics[width=1\linewidth]{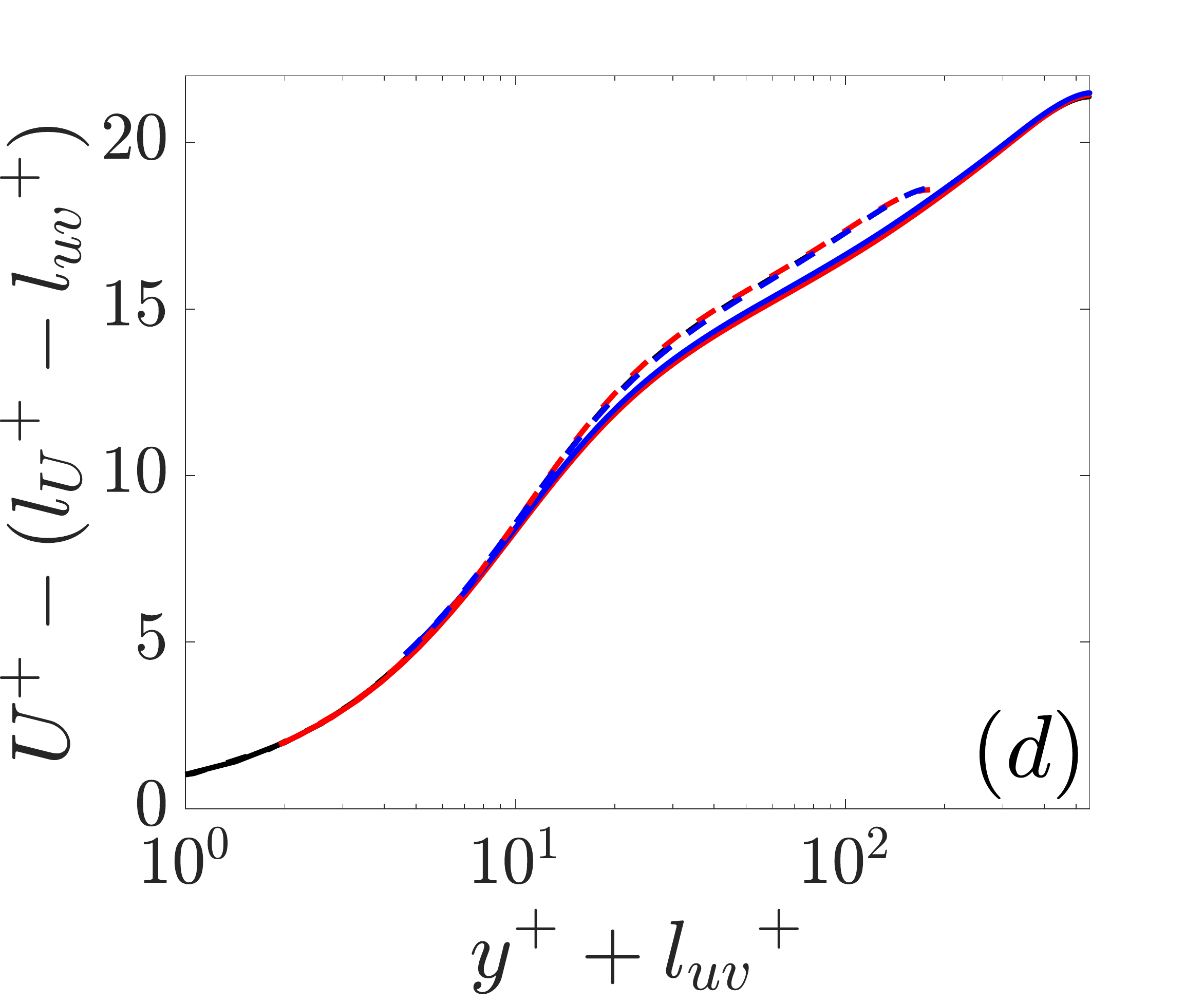}
        {\phantomcaption\label{fig:Reynolds_scaling:sub4}}
    \end{subfigure}%
    \hspace*{-4mm}
    \begin{subfigure}[tbp]{.268\textwidth}
        \includegraphics[width=1\linewidth]{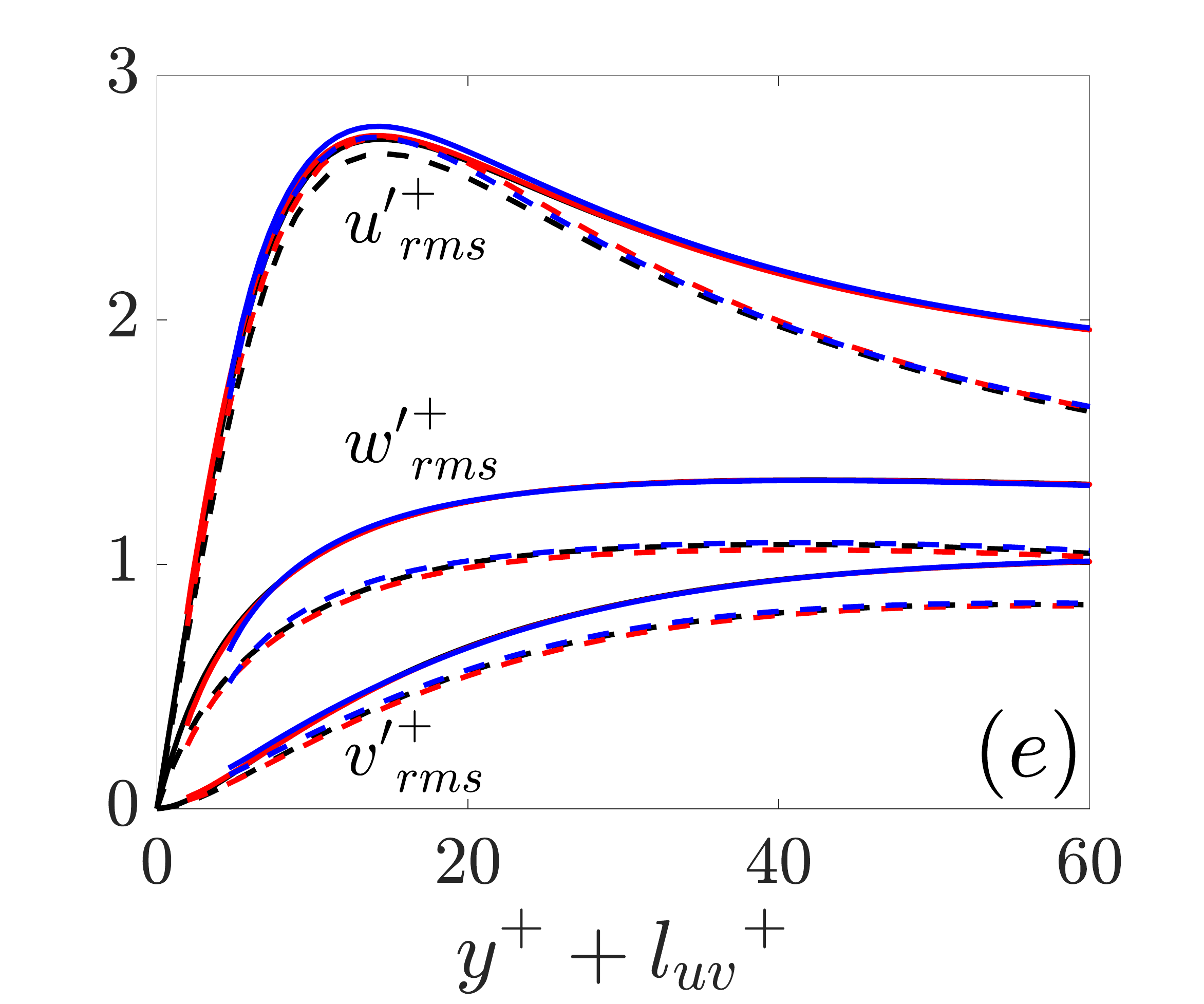}
        {\phantomcaption\label{fig:Reynolds_scaling:sub5}}
    \end{subfigure}%
    \hspace*{-2mm}
    \begin{subfigure}[tbp]{.268\textwidth}
        \includegraphics[width=1\linewidth]{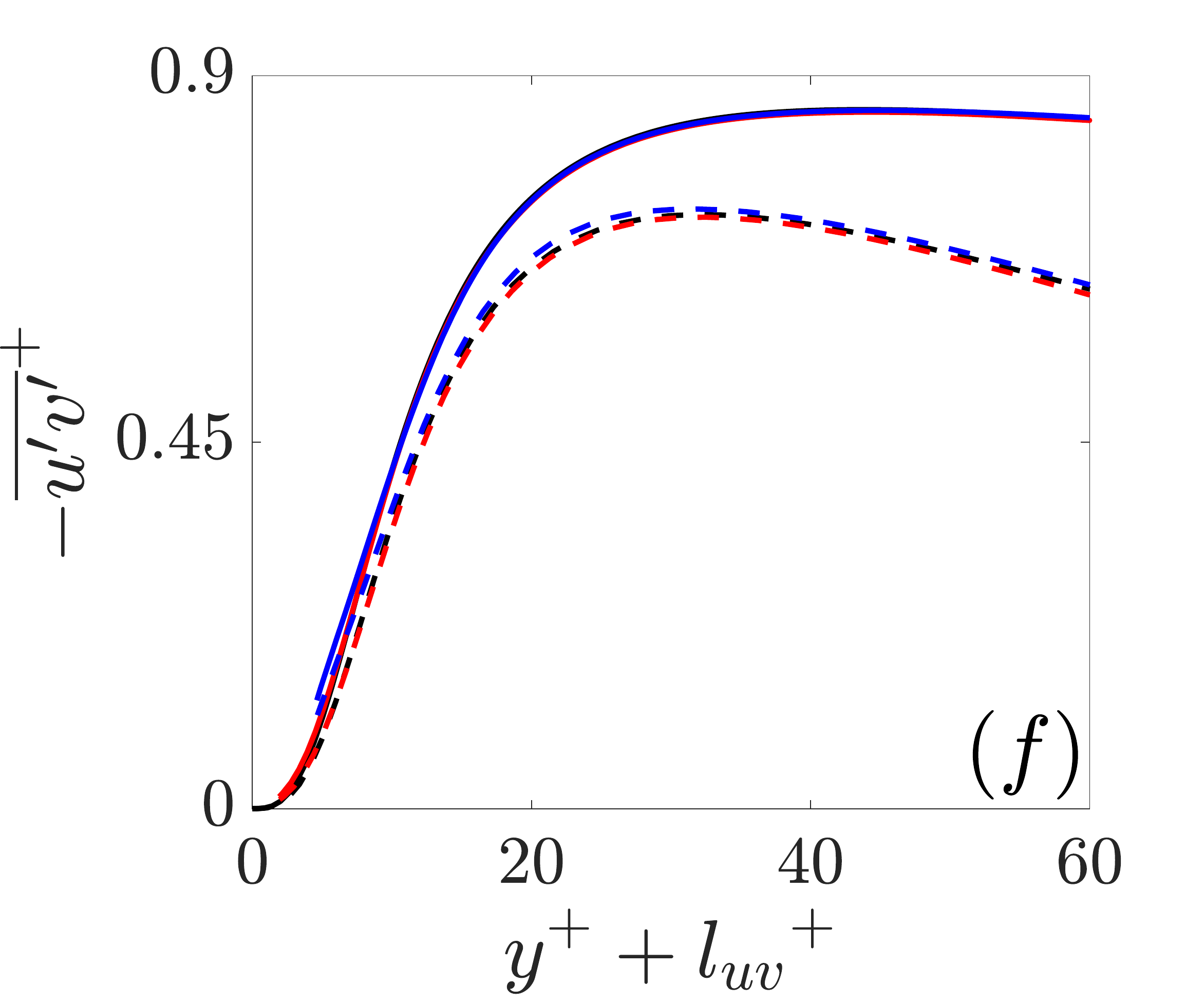}
        {\phantomcaption\label{fig:Reynolds_scaling:sub6}}
    \end{subfigure}
    \vspace*{-6mm}
    \caption{Mean velocity, r.m.s. velocity fluctuation and Reynolds shear stress profiles of the L$<\!\!\cdot\!\!>$M$<\!\!\cdot\!\!>$HR cases. Origin at ${y}^+ = 0$ ($a$-$c$), origin at ${y}^+ = -{\ell_{uv}}^+$ ($d$-$f$). \textcolor{red}{$-\!-$}, L2M2HR; \textcolor{blue}{$-\!-$}, L5M5HR; \textcolor{red}{$-\;-$}, L2M2; \textcolor{blue}{$-\;-$}, L5M5; \textcolor{black}{$-\!-$}, smooth-wall data at $Re_{\tau} = 550$; \textcolor{black}{$-\;-$}, smooth-wall data at $Re_{\tau} = 180$.} 
    \label{fig:Reynolds_scaling}
    \end{center}
\end{figure}
%%%%%%%%%%%%%%%%%%%%%%%%%%%%%%%%%%%%%%%%%%%%%%%%%%%%%%%%%%%%%%%%%%%%%%%%%
%%%%%%%%%%%%%%%%%%%%%%%%%%%%%%%%%%%%%%%%%%%%%%%%%%%%%%%%%%%%%%%%%%%%%%%%%

As described in \cref{sec:virtual}, $\Delta{U}^+$ should be independent of Reynolds number so long as the characteristic size of a given surface texture remains constant in inner units. Hence, for Robin boundary conditions which impose lengths of different magnitude on the velocity components, the effect should scale in inner units with Reynolds number.
This is investigated for cases L2M2 and L5M5 by simulating them at a higher Reynolds number but keeping their slip and transpiration lengths fixed in inner units. These higher Reynolds number cases are designated L2M2HR and L5M5HR in \cref{tab:dns}. Both were simulated at $Re_{\tau} = 550$ and the data are shown in \cref{fig:Reynolds_scaling}. The mean velocity  and Reynolds shear stress achieve a similar level of conformity to their respective smooth-wall solutions once the origin is shifted to $y^+ = -{{\ell}_{uv}}^+$ and the flow quantities are rescaled with the friction velocity at that position. A similar agreement is observed for the r.m.s. velocity fluctuations. The mean velocity shift, $\Delta{U}^+ = {{\ell}_U}^+ - {{\ell}_{uv}}^+$, for the two sets of slip and transpiration lengths remain the same at both Reynolds numbers, demonstrating that their effect is independent of Reynolds number.

\section{Applicability to geometrical roughness} \label{sec:TRM_vs_rough_DNS}

It has thus far been demonstrated that the TRM can cause changes in near-wall turbulence which lead to changes in drag. In this section, the applicability of the TRM at reproducing the effect of actual rough surfaces in the transitionally rough turbulence regime is assessed. The potential of this had been demonstrated by \citet{Lacis2020} as discussed in \cref{sec:eval}. Here, a more in-depth analysis is provided by considering a number of different rough surfaces.

\subsection{Matching TRM with geometry-resolving data}

%%%%%%%%%%%%%%%%%%%%%%%%%%%%%%%%%%%%%%%%%%%%%%%%%%%%%%%%%%%%%%%%%%%%%%%%%
%%%%%%%%%%%%%%%%%%%%%%%%%%%%%%%%%%%%%%%%%%%%%%%%%%%%%%%%%%%%%%%%%%%%%%%%%
\begin{figure}
    \begin{center}
    \hspace*{-5mm}
    \begin{subfigure}[tbp]{.35\textwidth}
        \includegraphics[width=1\linewidth]{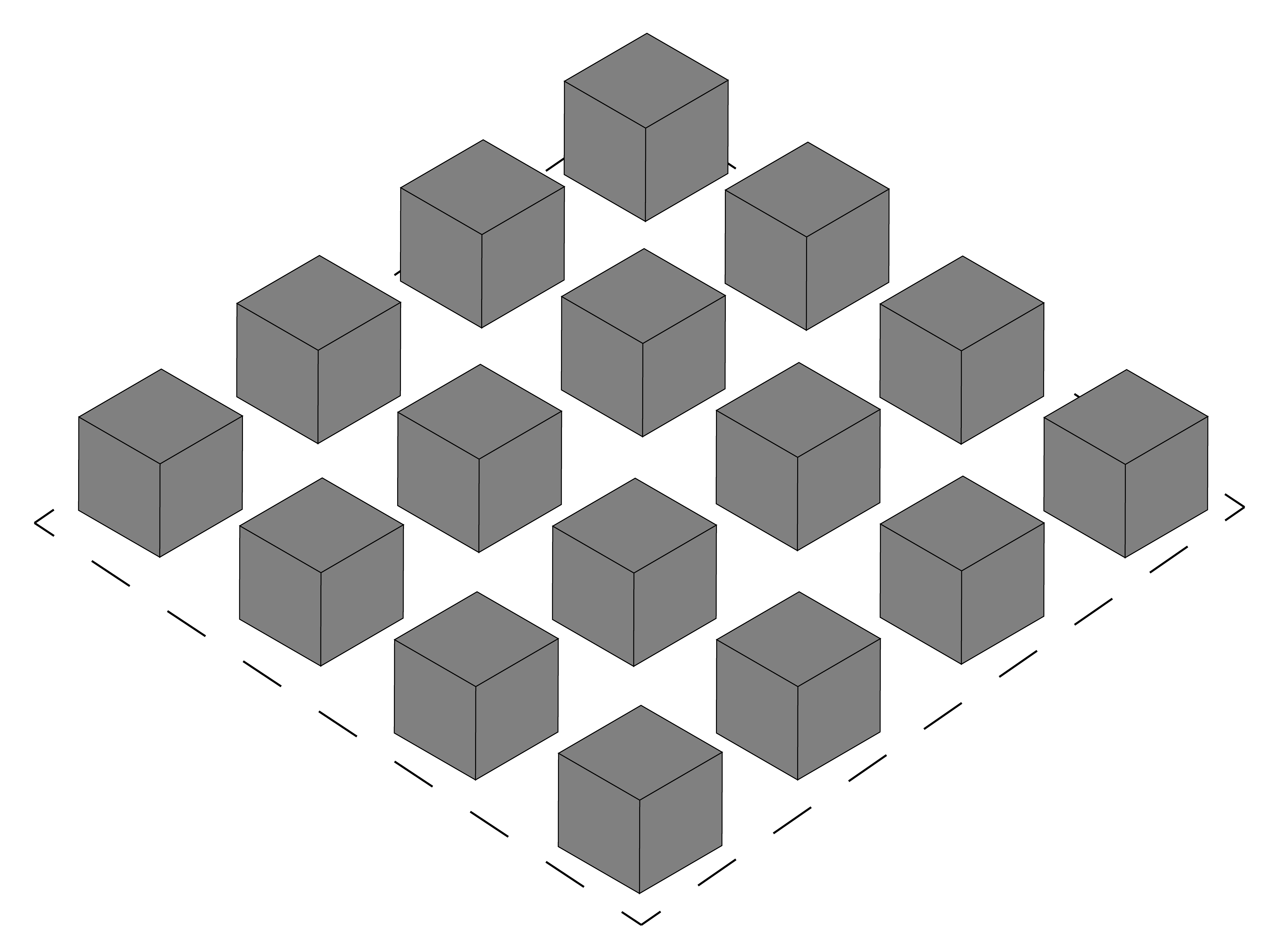}
        {\phantomcaption}
    \end{subfigure}%
    \hspace*{5mm}
    \begin{subfigure}[tbp]{.4\textwidth}
        \includegraphics[width=1\linewidth]{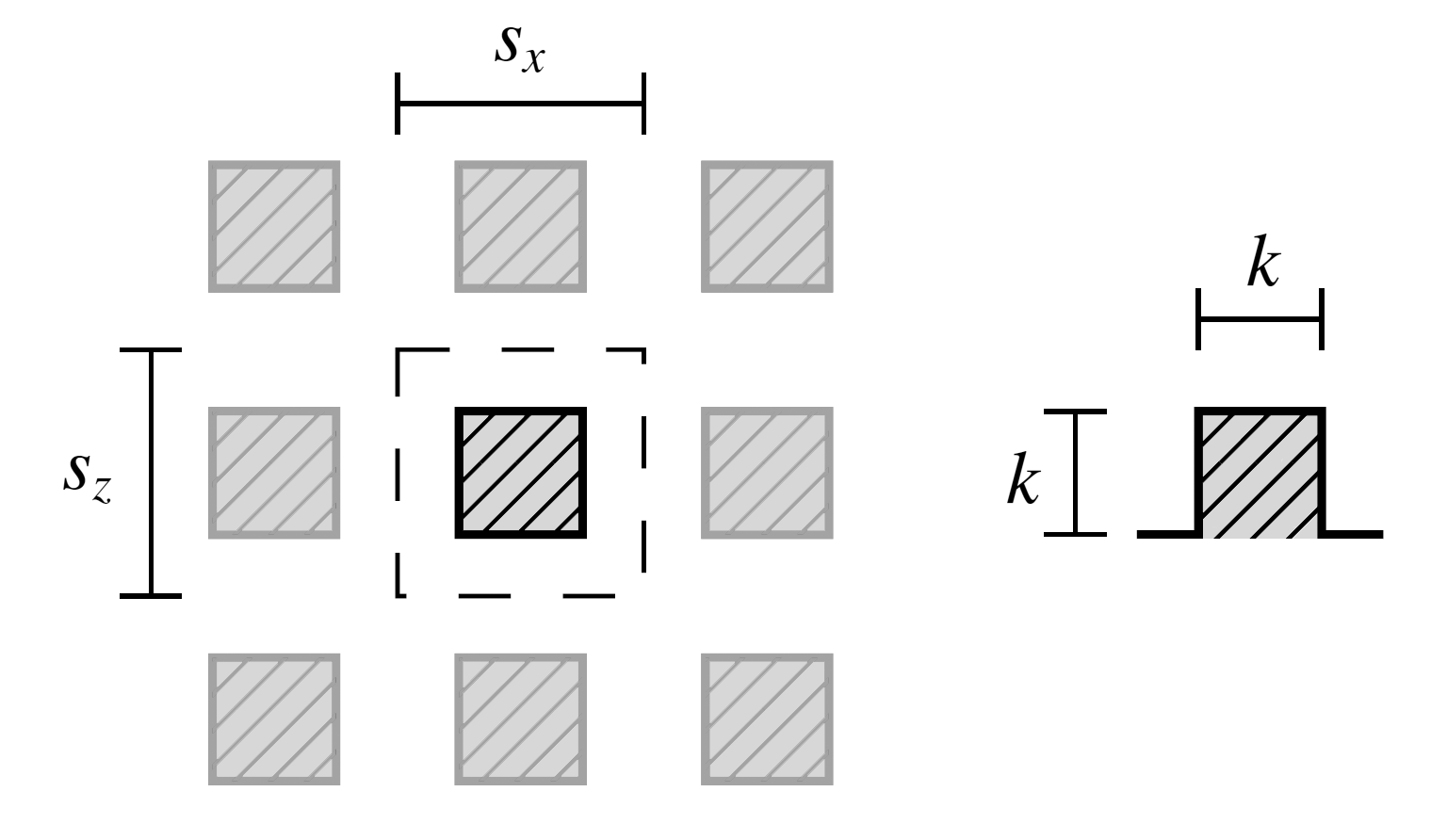}
        {\phantomcaption}
    \end{subfigure}
    \vspace*{-3mm}
    \caption{Roughness pattern of the DNS cases of \citet{Abderrahaman2019} consisting of collocated posts.}
    \label{fig:RoughGeometry}
    \end{center}
\end{figure}
%%%%%%%%%%%%%%%%%%%%%%%%%%%%%%%%%%%%%%%%%%%%%%%%%%%%%%%%%%%%%%%%%%%%%%%%%
%%%%%%%%%%%%%%%%%%%%%%%%%%%%%%%%%%%%%%%%%%%%%%%%%%%%%%%%%%%%%%%%%%%%%%%%%

Several geometries have been chosen from \citet{Abderrahaman2019}, who performed DNS of turbulent channel flow over textured surfaces. The texture geometry of the main case chosen for evaluation is depicted schematically in \cref{fig:RoughGeometry}. The surface consists of a homogeneous distribution of collocated posts of height $k$ with equal streamwise and spanwise pitch lengths ($s_x=s_z$). The data of \citet{Abderrahaman2019} also has a triple decomposition applied to it such that the footprint of the roughness region is removed from the turbulence statistics. This makes it more suitable for comparison as the TRM simulations likewise do not have a region occupied by any discrete geometry. It should be emphasized however that this does not modify the statistics above the roughness region or the macro-scale effect of the roughness on the bulk flow. The characteristics of the geometries considered (C06-C12) are listed in \cref{tab:rough_cases_abdrhmn}. 

%%%%%%%%%%%%%%%%%%%%%%%%%%%%%%%%%%%%%%%%%%%%%%%%%%%%%%%%%%%%%%%%%%%%%%%%%
%%%%%%%%%%%%%%%%%%%%%%%%%%%%%%%%%%%%%%%%%%%%%%%%%%%%%%%%%%%%%%%%%%%%%%%%%
\begin{table}
  \centering
  \begin{tabular}{ m{2.0cm}m{1.5cm}m{1.5cm}m{1.0cm}|m{0.5cm}m{1.4cm}m{1.4cm}m{1.4cm}m{0.7cm} }
       Geo-Res & ${s_x/k}$ & ${s_z/k}$ & \,\,${k}^+$ && ${{\ell_{U}}^+}$ & ${\ell_{x}}^+$ & ${\ell_{uv}}^+$ & \,\,\,${{{\Delta}{U}}^+}$\\
       &&&&&&&\\
       \,\,\,\,\,C06 & \,\,$2.0$ & \,\,$2.0$ & $6.0$  && $0.5$ & $0.5$ & $1.2$ & $-0.5$\\
       \,\,\,\,\,C09 & \,\,$2.0$ & \,\,$2.0$ & $8.8$  && $0.7$ & $0.7$ & $1.5$ & $-0.7$\\
       \,\,\,\,\,C12 & \,\,$2.0$ & \,\,$2.0$ & $11.7$ && $1.1$ & $1.2$ & $3.2$ & $-1.5$\\
       \,\,\,\,\,C15 & \,\,$2.0$ & \,\,$2.0$ & $14.4$ && $1.3$ & $1.5$ & $4.5$ & $-2.4$\\
       \,\,\,\,\,C18 & \,\,$2.0$ & \,\,$2.0$ & $17.4$ && $1.5$ & $1.9$ & $6.3$ & $-3.5$\\
  \end{tabular}
  \captionof{table}{Characteristics of the geometry resolving (Geo-Res) DNS cases of \citet{Abderrahaman2019} for the roughness pattern depicted in \cref{fig:RoughGeometry}. The definitions of the various terms are as described in \cref{tab:dns} with the exception of ${{\Delta}{U}}^+$, which here, is simply the mean velocity shift calculated at the channel center-line without any coordinate shift or rescaling having been applied. Data adapted from \citet{Abderrahaman2019}.}
  \label{tab:rough_cases_abdrhmn}
\end{table}
%%%%%%%%%%%%%%%%%%%%%%%%%%%%%%%%%%%%%%%%%%%%%%%%%%%%%%%%%%%%%%%%%%%%%%%%%
%%%%%%%%%%%%%%%%%%%%%%%%%%%%%%%%%%%%%%%%%%%%%%%%%%%%%%%%%%%%%%%%%%%%%%%%%

Owing to the homogeneous pattern of the roughness, both the streamwise and spanwise slip lengths used in the TRM boundary conditions are set equal to that calculated by \citet{Abderrahaman2019} for the mean streamwise flow ($\ell^+=\ell^+_x=\ell^+_z$). The values of $\ell_x^+$ are reported in \cref{tab:rough_cases_abdrhmn}. The transpiration lengths, $m^+={m_{x}}^+={m_{z}}^+$, are incrementally increased until the displacement of the Reynolds shear stress, i.e. the turbulence virtual origin ${\ell_{uv}}^+$, closely matches those reported by \citet{Abderrahaman2019}.

\begin{figure}
    \begin{center}
    \hspace*{-1mm}
    \begin{subfigure}[tbp]{.26\textwidth}
        \includegraphics[width=1\linewidth]{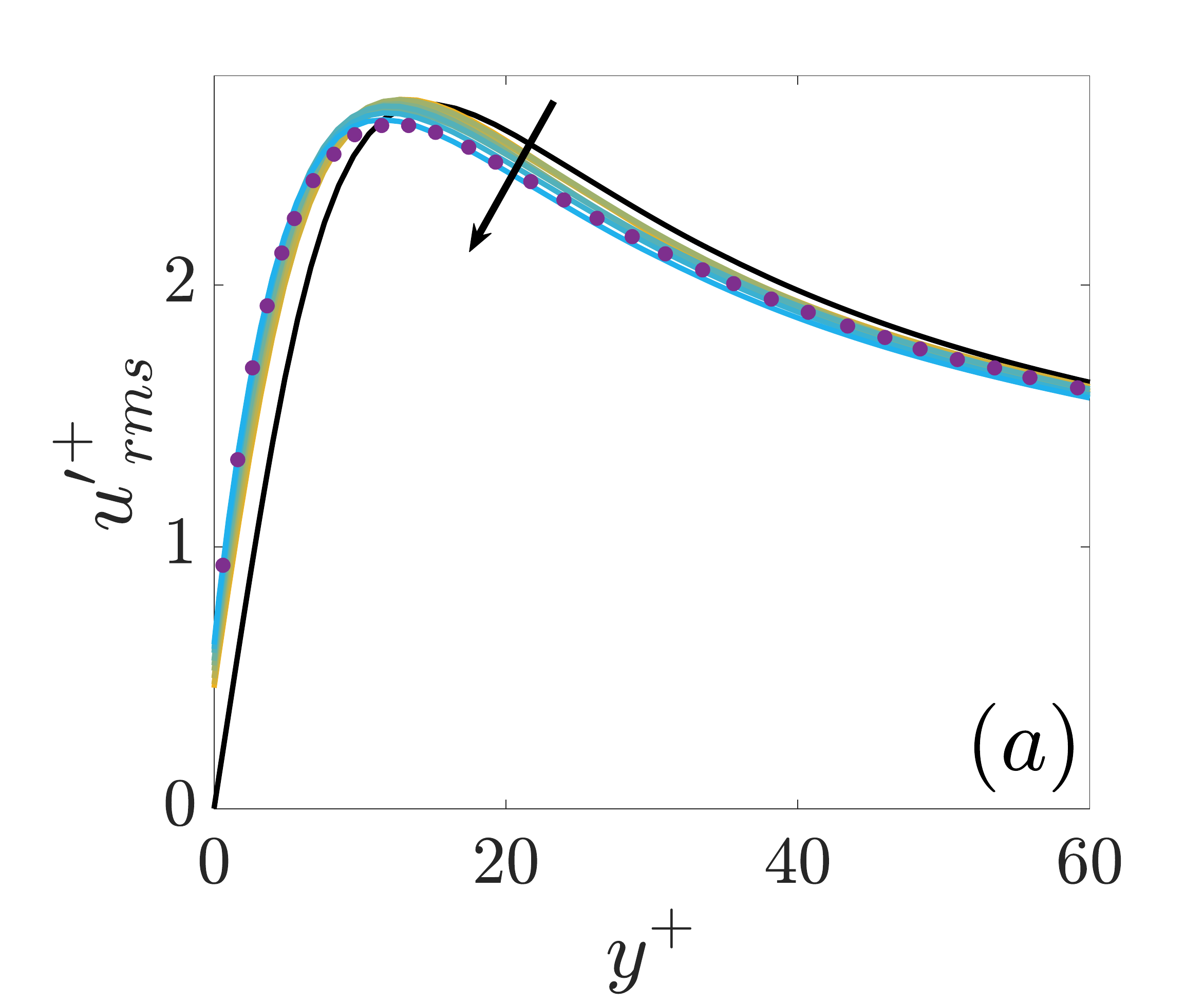}
        {\phantomcaption\label{CompToDNS:sub1}}
        \vspace*{-3mm}
    \end{subfigure}%
    \hspace*{-3mm}
    \begin{subfigure}[tbp]{.26\textwidth}
        \includegraphics[width=1\linewidth]{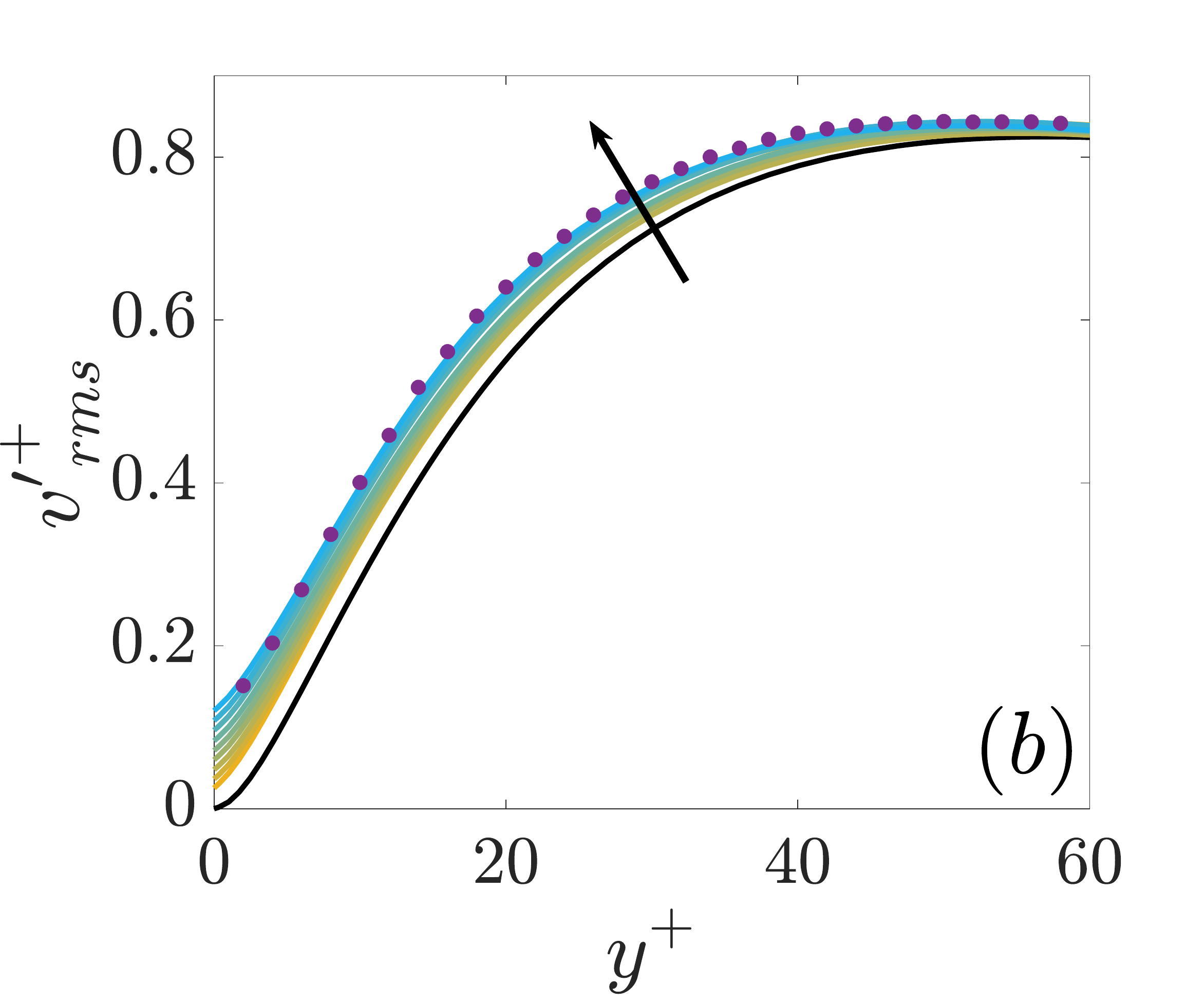}
        {\phantomcaption\label{CompToDNS:sub2}}
        \vspace*{-3mm}
    \end{subfigure}%
    \hspace*{-3mm}
    \begin{subfigure}[tbp]{.26\textwidth}
        \includegraphics[width=1\linewidth]{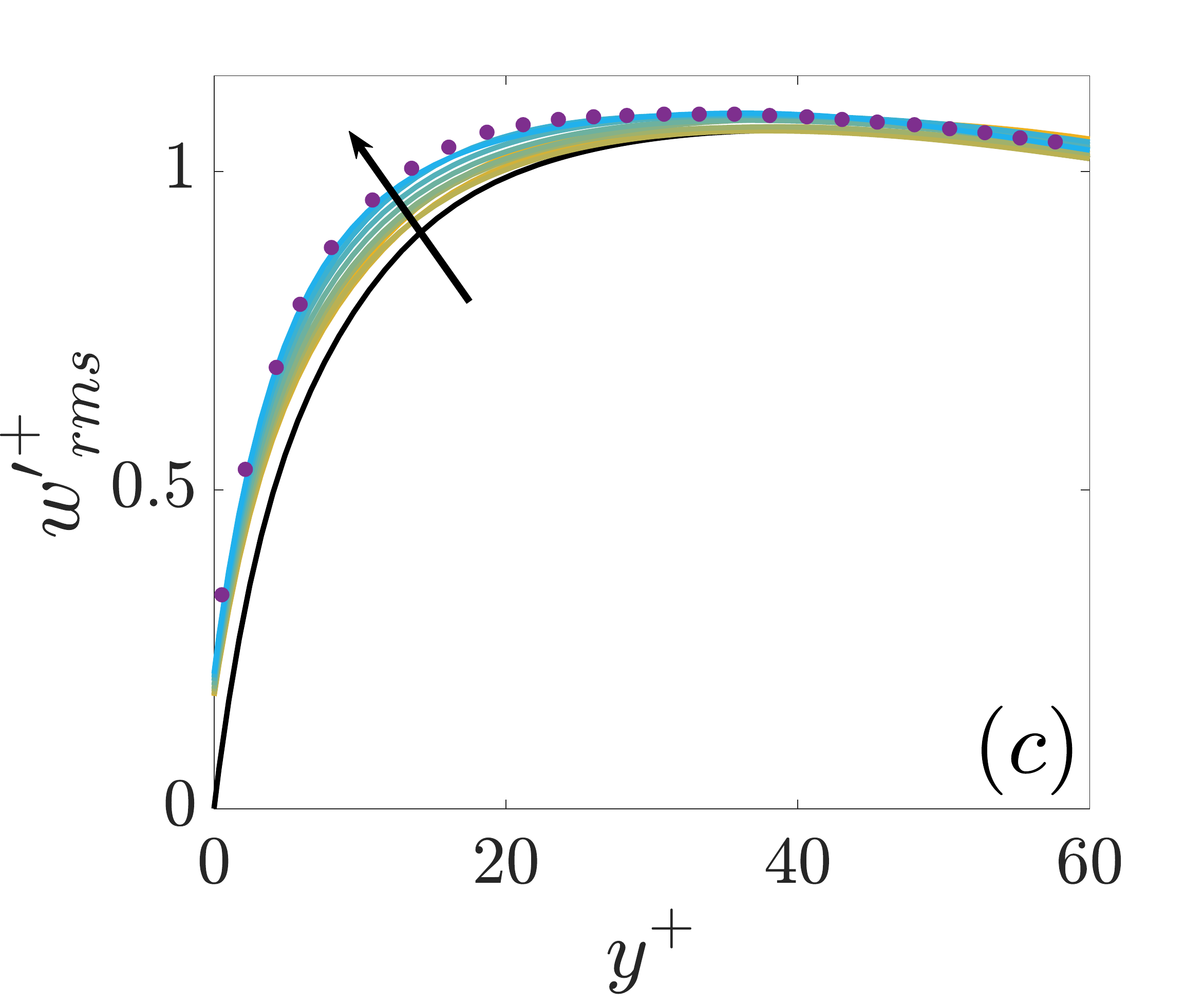}
        {\phantomcaption\label{CompToDNS:sub3}}
        \vspace*{-3mm}
    \end{subfigure}%
    \hspace*{-3mm}
    \begin{subfigure}[tbp]{.26\textwidth}
        \includegraphics[width=1\linewidth]{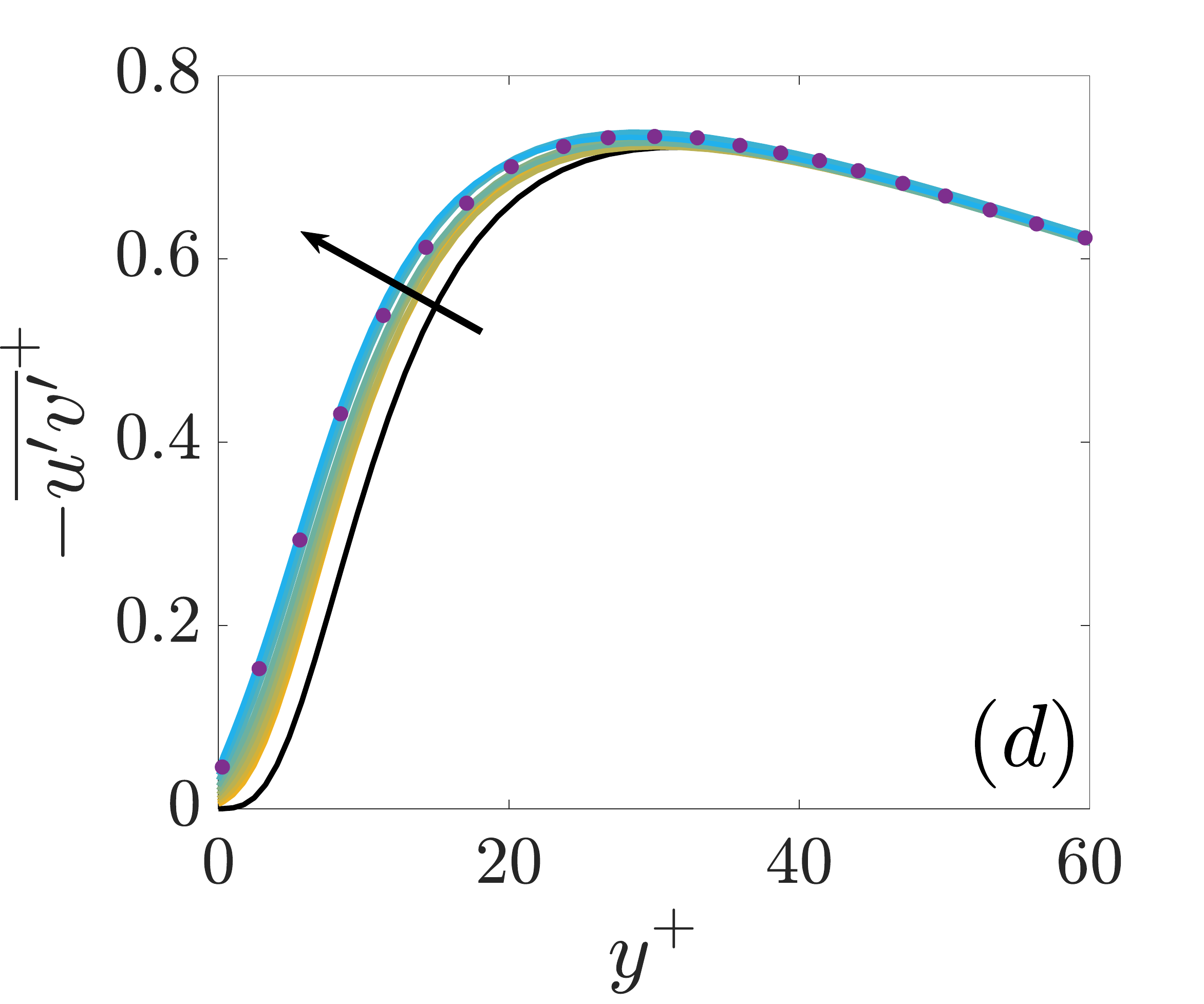}
        {\phantomcaption\label{CompToDNS:sub4}}
        \vspace*{-3mm}
    \end{subfigure}
    \hspace*{-1mm}
    \begin{subfigure}[tbp]{.26\textwidth}
        \includegraphics[width=1\linewidth]{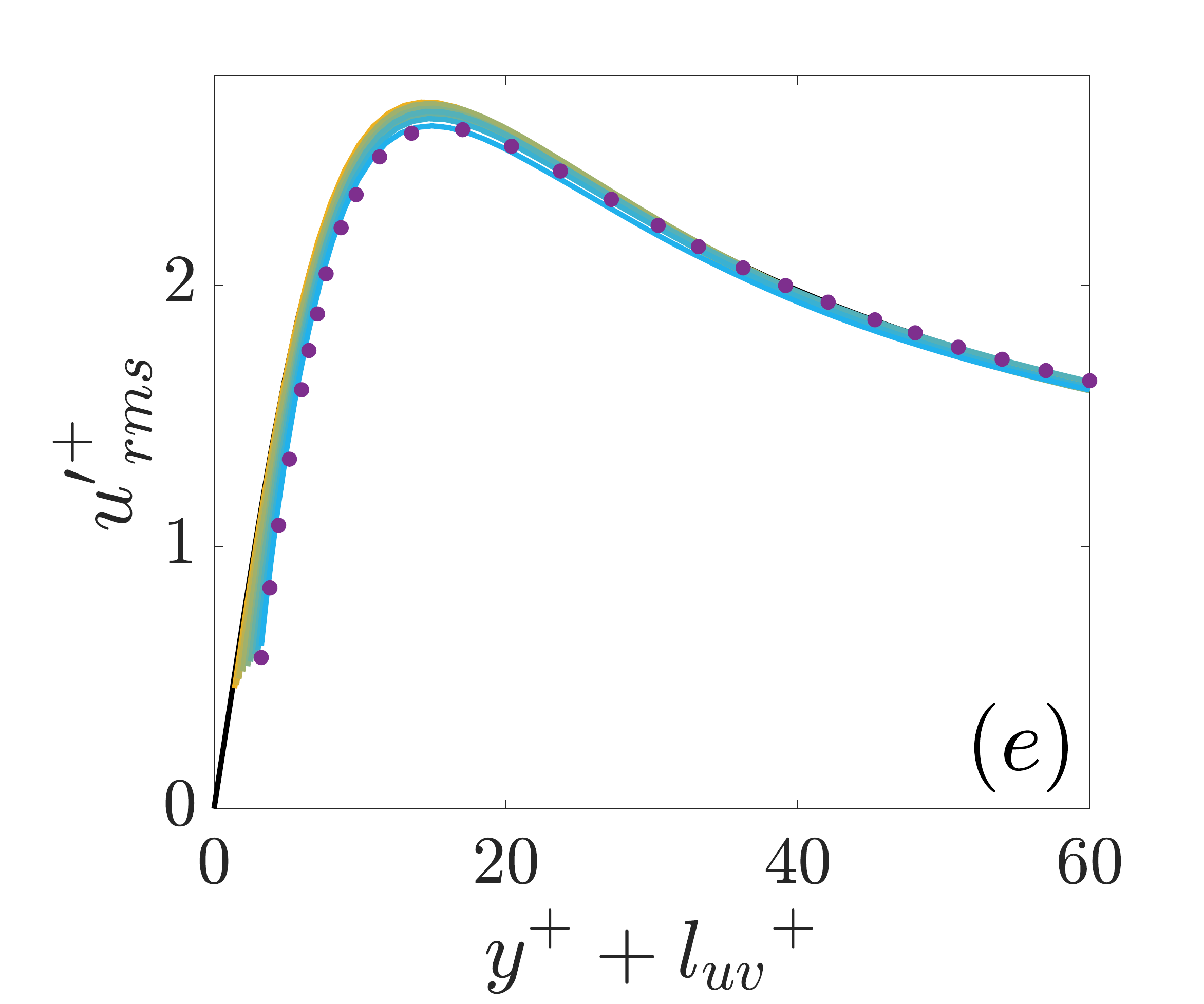}
        {\phantomcaption\label{CompToDNS:sub5}}
    \end{subfigure}%
    \hspace*{-3mm}
    \begin{subfigure}[tbp]{.26\textwidth}
        \includegraphics[width=1\linewidth]{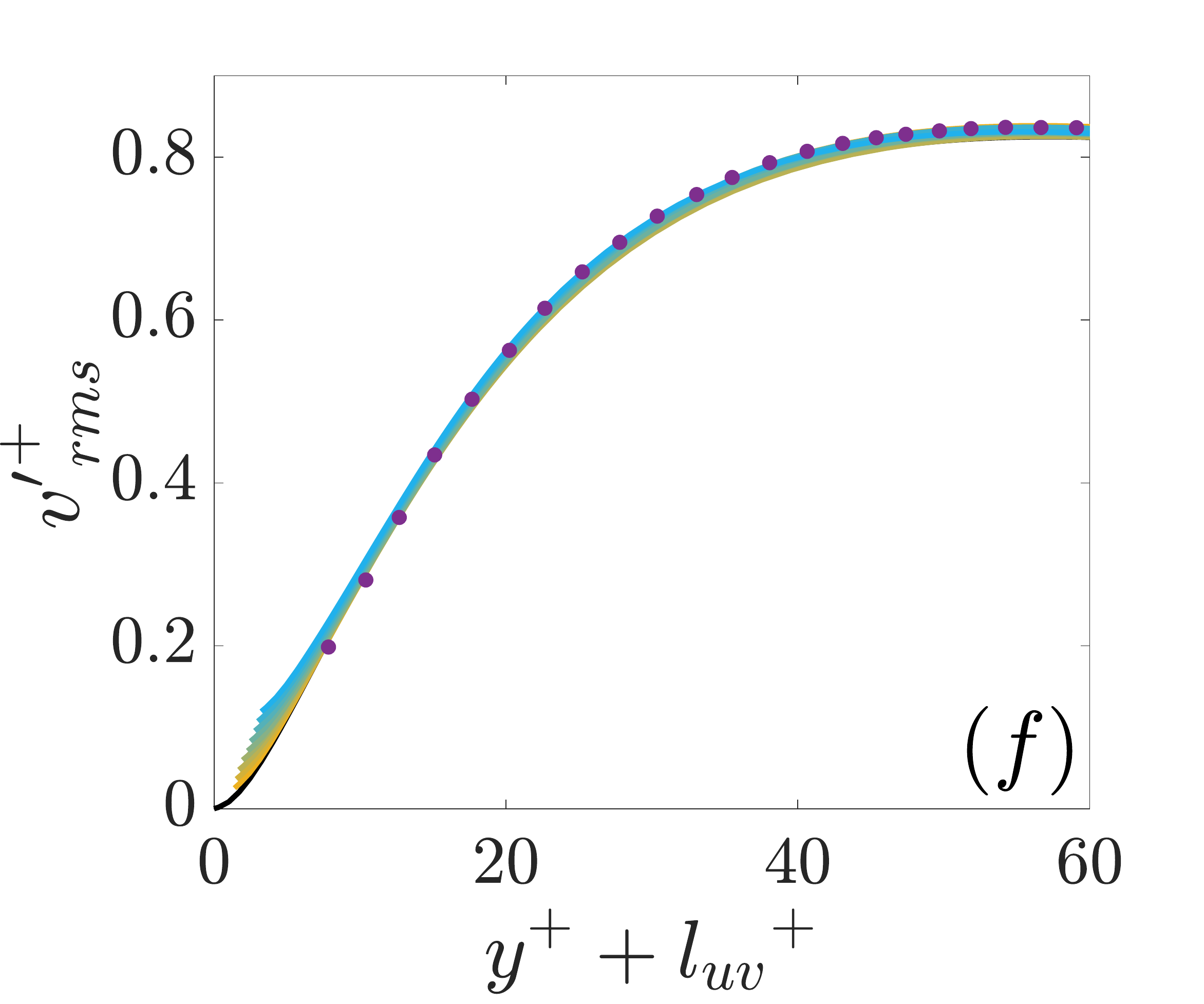}
        {\phantomcaption\label{CompToDNS:sub6}}
    \end{subfigure}%
    \hspace*{-3mm}
    \begin{subfigure}[tbp]{.26\textwidth}
        \includegraphics[width=1\linewidth]{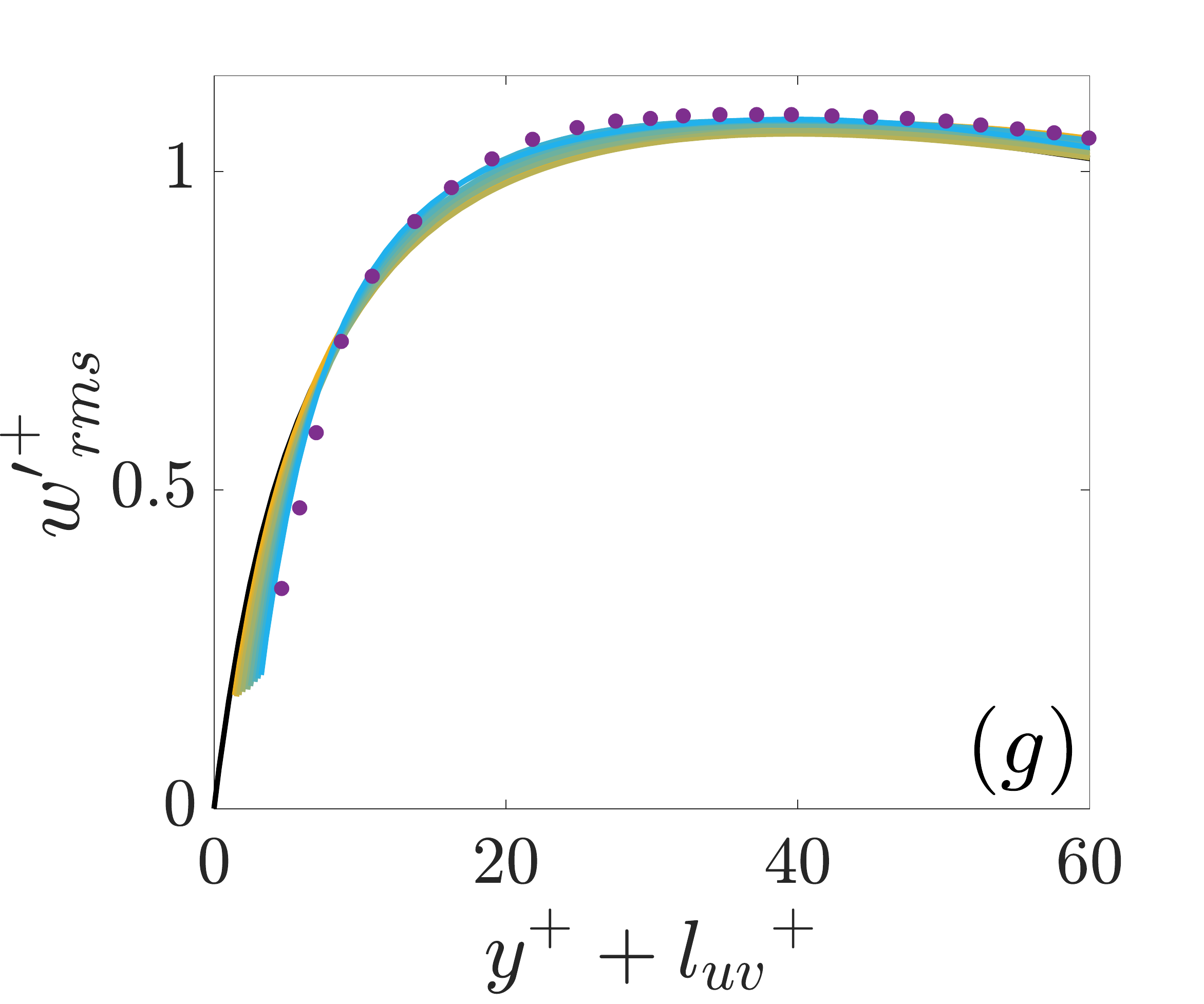}
        {\phantomcaption\label{CompToDNS:sub7}}
    \end{subfigure}%
    \hspace*{-3mm}
    \begin{subfigure}[tbp]{.26\textwidth}
        \includegraphics[width=1\linewidth]{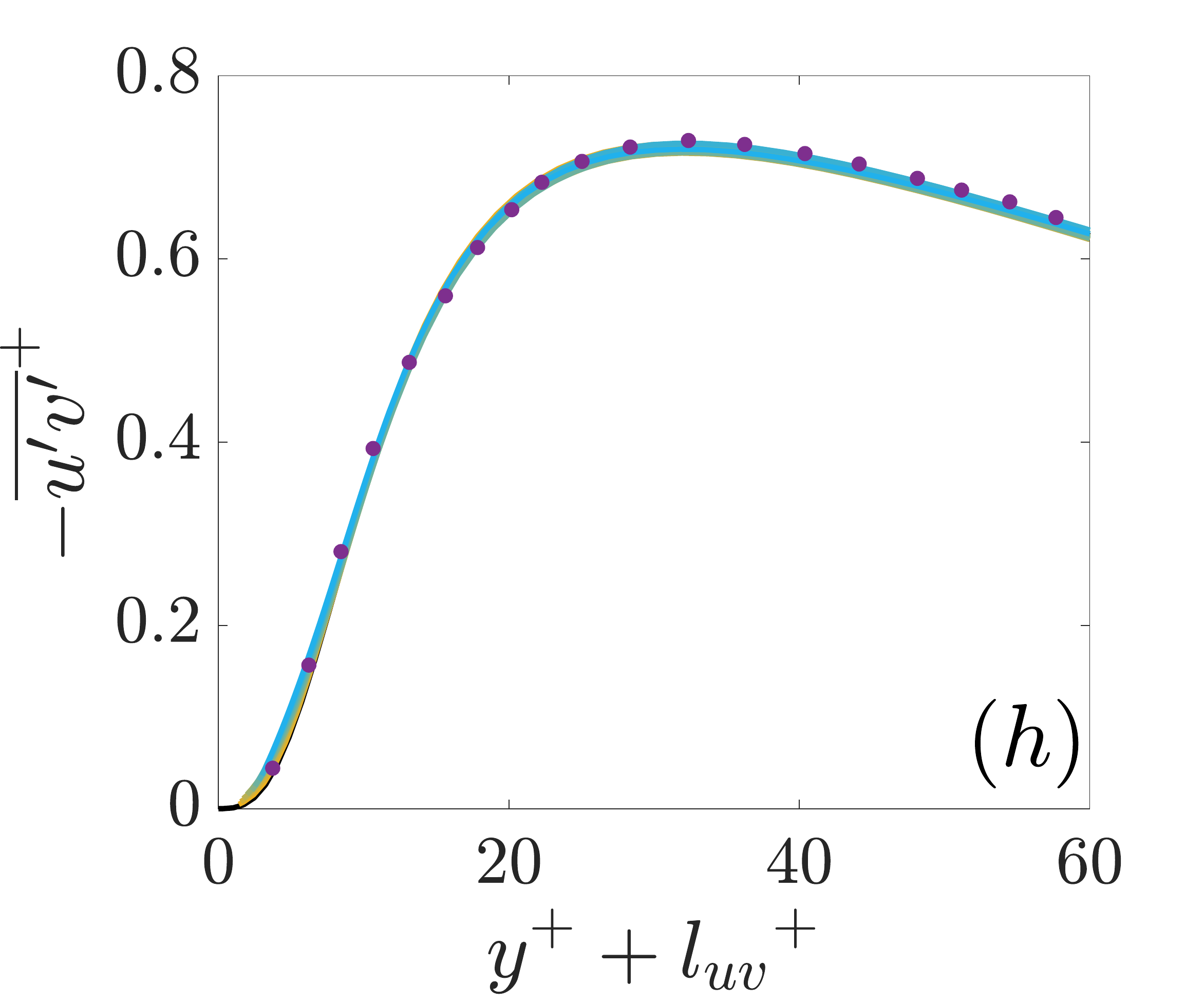}
        {\phantomcaption\label{CompToDNS:sub8}}
    \end{subfigure}
    \vspace*{-4mm}
    \caption{The r.m.s. velocity fluctuation and Reynolds shear stress profiles of cases L$1.2$M$<\!\!2-10\!\!>$. Origin at ${y}^+ = 0$ ($a$, $c$, $d$, $e$), origin at ${y}^+ = -{\ell_{uv}}^+$ and rescaled with the corresponding ${u_{\tau}}$ ($b$, $f$, $g$, $h$). Symbols represent geometry-resolving DNS case C12 from \citet{Abderrahaman2019}, arrows indicate increasing $m^+$.}
    \label{CompToDNS}
    \end{center}
\end{figure}

\begin{figure}
    \begin{center}
    \begin{subfigure}[tbp]{.35\textwidth}
        \includegraphics[width=1\linewidth]{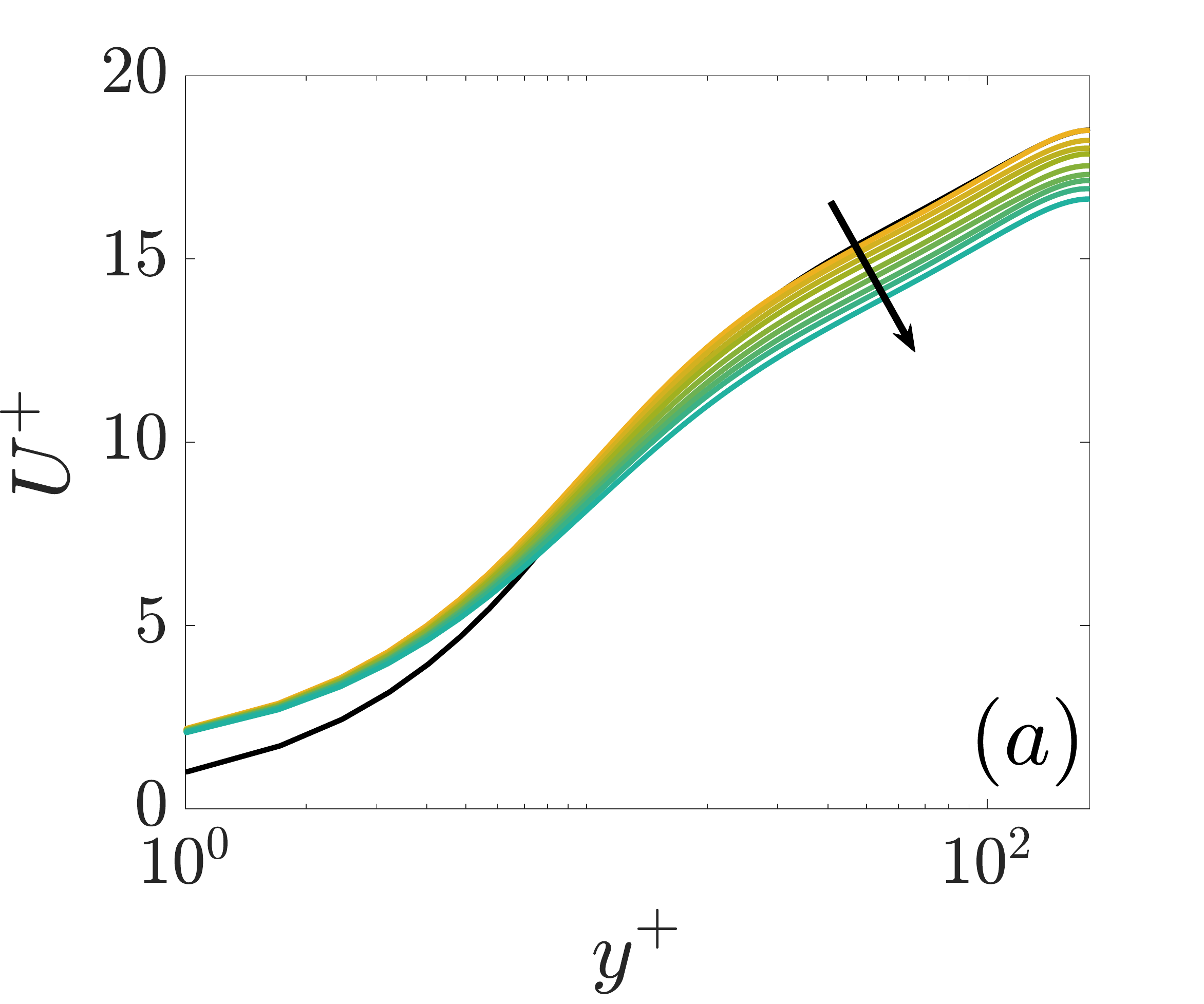}
        {\phantomcaption\label{CompToDNSMean:sub1}}
    \end{subfigure}%
    \hspace*{5mm}
    \begin{subfigure}[tbp]{.35\textwidth}
        \includegraphics[width=1\linewidth]{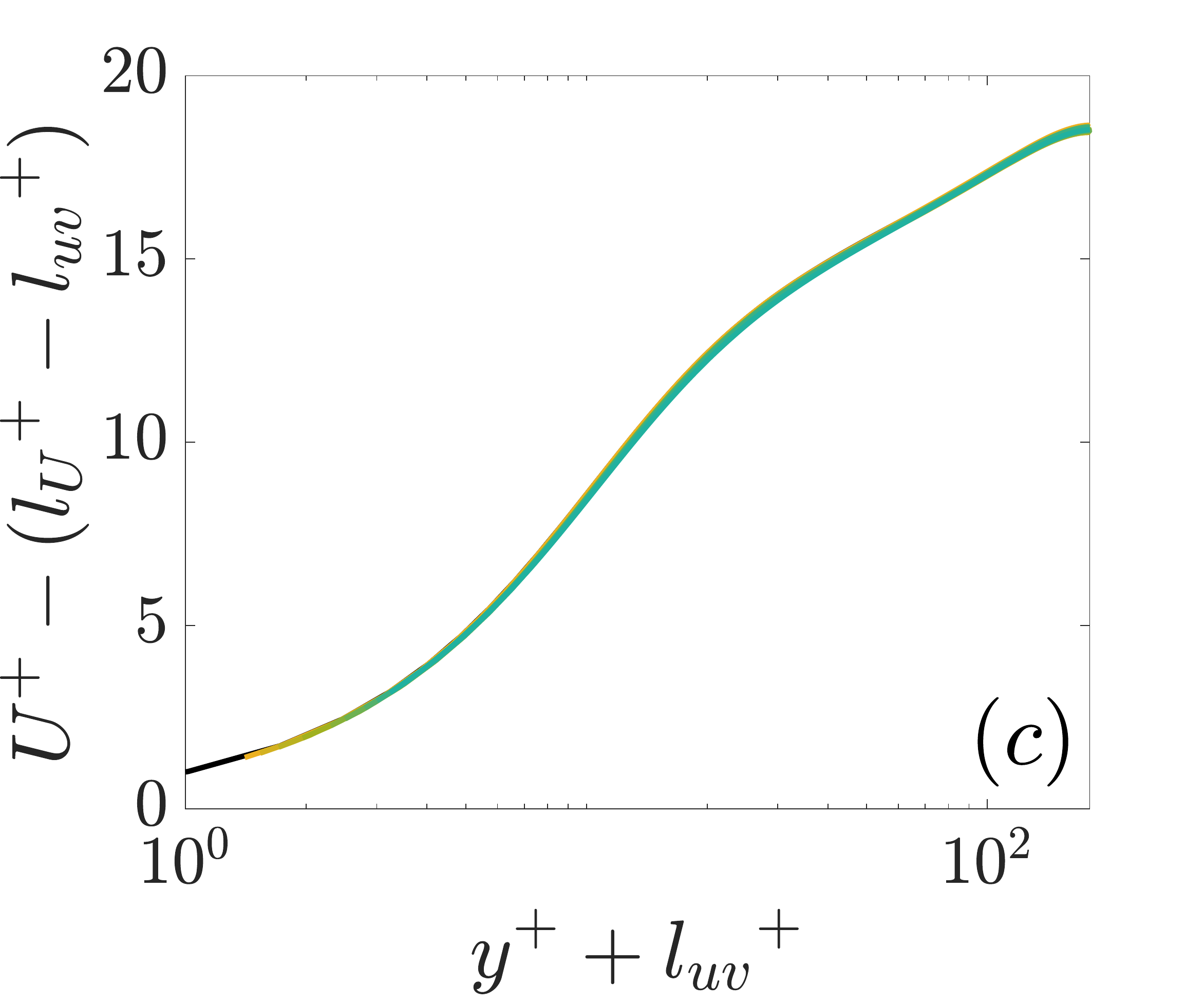}
        {\phantomcaption\label{CompToDNSMean:sub3}}
    \end{subfigure}
    \vspace*{-4mm}
    \caption{Mean velocity profiles of cases L$1.2$M$<\!\!2-10\!\!>$. Origin at ${y}^+ = 0$ ($a$); origin at ${y}^+ = -{\ell_{uv}}^+$, rescaled with the corresponding ${u_{\tau}}$ and with $\Delta{U}^+={\ell_{U}}^+-\,{\ell_{uv}}^+$ subtracted ($b$). Arrows indicate increasing $m^+$.}
    \label{CompToDNSMean}
    \end{center}
\end{figure}

\Cref{CompToDNS} demonstrates the process for case C12 by comparing the data of the TRM simulations to the geometry-resolving ones. The transpiration lengths used ranged from $m^+=2$ to $m^+=10$. The turbulence statistics produced using $m^+=10$ (case L1.2M10 in \cref{tab:trm_rough}) agree well with the geometry-resolving DNS data for C12, demonstrating that the TRM is capable of reproducing such transitionally rough effects. All of the TRM cases here fall within the regime of smooth-wall-like turbulence, as evident by the collapse of the Reynolds shear stress profiles onto that of a smooth-wall's when adjusted for ${\ell_{uv}}^+$ in \cref{CompToDNS:sub8}. It follows that the mean velocity profiles (\hyperref[CompToDNSMean:sub3]{figure \ref*{CompToDNSMean:sub3}}) also collapse onto the smooth-wall one (\hyperref[CompToDNSMean:sub3]{figure \ref*{CompToDNSMean:sub3}}). Aside from differences in the region closest to the boundary due to each velocity component decaying to a different virtual origin \citep{Abderrahaman2019}, the fluctuations also collapse onto their smooth-wall counter-parts.

%%%%%%%%%%%%%%%%%%%%%%%%%%%%%%%%%%%%%%%%%%%%%%%%%%%%%%%%%%%%%%%%%%%%%%%%%
%%%%%%%%%%%%%%%%%%%%%%%%%%%%%%%%%%%%%%%%%%%%%%%%%%%%%%%%%%%%%%%%%%%%%%%%%
%%%%%%%%%%%%%%%%%%%%%%%%%%%%%%%%%%%%%%%%%%%%%%%%%%%%%%%%%%%%%%%%%%%%%%%%%
%%%%%%%%%%%%%%%%%%%%%%%%%%%%%%%%%%%%%%%%%%%%%%%%%%%%%%%%%%%%%%%%%%%%%%%%%
%%%%%%%%%%%%%%%%%%%%%%%%%%%%%%%%%%%%%%%%%%%%%%%%%%%%%%%%%%%%%%%%%%%%%%%%%
%%%%%%%%%%%%%%%%%%%%%%%%%%%%%%%%%%%%%%%%%%%%%%%%%%%%%%%%%%%%%%%%%%%%%%%%%
%%%%%%%%%%%%%%%%%%%%%%%%%%%%%%%%%%%%%%%%%%%%%%%%%%%%%%%%%%%%%%%%%%%%%%%%%
%%%%%%%%%%%%%%%%%%%%%%%%%%%%%%%%%%%%%%%%%%%%%%%%%%%%%%%%%%%%%%%%%%%%%%%%%
%%%%%%%%%%%%%%%%%%%%%%%%%%%%%%%%%%%%%%%%%%%%%%%%%%%%%%%%%%%%%%%%%%%%%%%%%
%%%%%%%%%%%%%%%%%%%%%%%%%%%%%%%%%%%%%%%%%%%%%%%%%%%%%%%%%%%%%%%%%%%%%%%%%
\begin{table}
  \centering
  \begin{tabular}{ m{1.5cm}m{2.0cm}m{1.4cm}m{1.25cm}m{1.2cm}|m{0.5cm}m{1.4cm}m{1.3cm}m{0.7cm} }
%   {llccc|ccc|c}
       Geo-Res & \,\,TRM & $Re_{\tau}$ & ${\ell}^+$ & ${m}^+$ && ${\ell_U}^+$ & ${\ell_{uv}}^+$ & \,\,\,${{\Delta}{U}}^+$\\
       &&&&&&&&\\
     \,\,\,\,\,C06 & L0.5M8  & $180$ & $0.5$ & $8.0$  && $0.5$ & $1.2$ & $-0.6$\\
     \,\,\,\,\,C09 & L0.7M7  & $180$ & $0.7$ & $7.0$  && $0.7$ & $1.5$ & $-0.7$\\
     \,\,\,\,\,C12 & L1.2M10 & $180$ & $1.2$ & $10.0$ && $1.1$ & $3.2$ & $-1.8$\\
     \,\,\,\,\,C15 & L1.5M12 & $180$ & $1.5$ & $12.0$ && $1.4$ & $4.4$ & $-2.6$\\
     \,\,\,\,\,C18 & L1.9M15 & $180$ & $1.9$ & $15.0$ && $1.6$ & $6.0$ & $-3.7$\\
  \end{tabular}
  \captionof{table}{Summary of TRM simulations conducted for comparison with the rough-wall DNS of \citet{Abderrahaman2019}. The definitions of the various terms are as described in \cref{tab:dns} with the exception of ${{\Delta}{U}}^+$ which follows the definition brought in \cref{tab:rough_cases_abdrhmn}.}
  \label{tab:trm_rough}
\end{table}
%%%%%%%%%%%%%%%%%%%%%%%%%%%%%%%%%%%%%%%%%%%%%%%%%%%%%%%%%%%%%%%%%%%%%%%%%
%%%%%%%%%%%%%%%%%%%%%%%%%%%%%%%%%%%%%%%%%%%%%%%%%%%%%%%%%%%%%%%%%%%%%%%%%
%%%%%%%%%%%%%%%%%%%%%%%%%%%%%%%%%%%%%%%%%%%%%%%%%%%%%%%%%%%%%%%%%%%%%%%%%
%%%%%%%%%%%%%%%%%%%%%%%%%%%%%%%%%%%%%%%%%%%%%%%%%%%%%%%%%%%%%%%%%%%%%%%%%
\begin{figure}
    \begin{center}
    \hspace*{-2mm}
    \begin{subfigure}[tbp]{.27\textwidth}
        \includegraphics[width=1\linewidth]{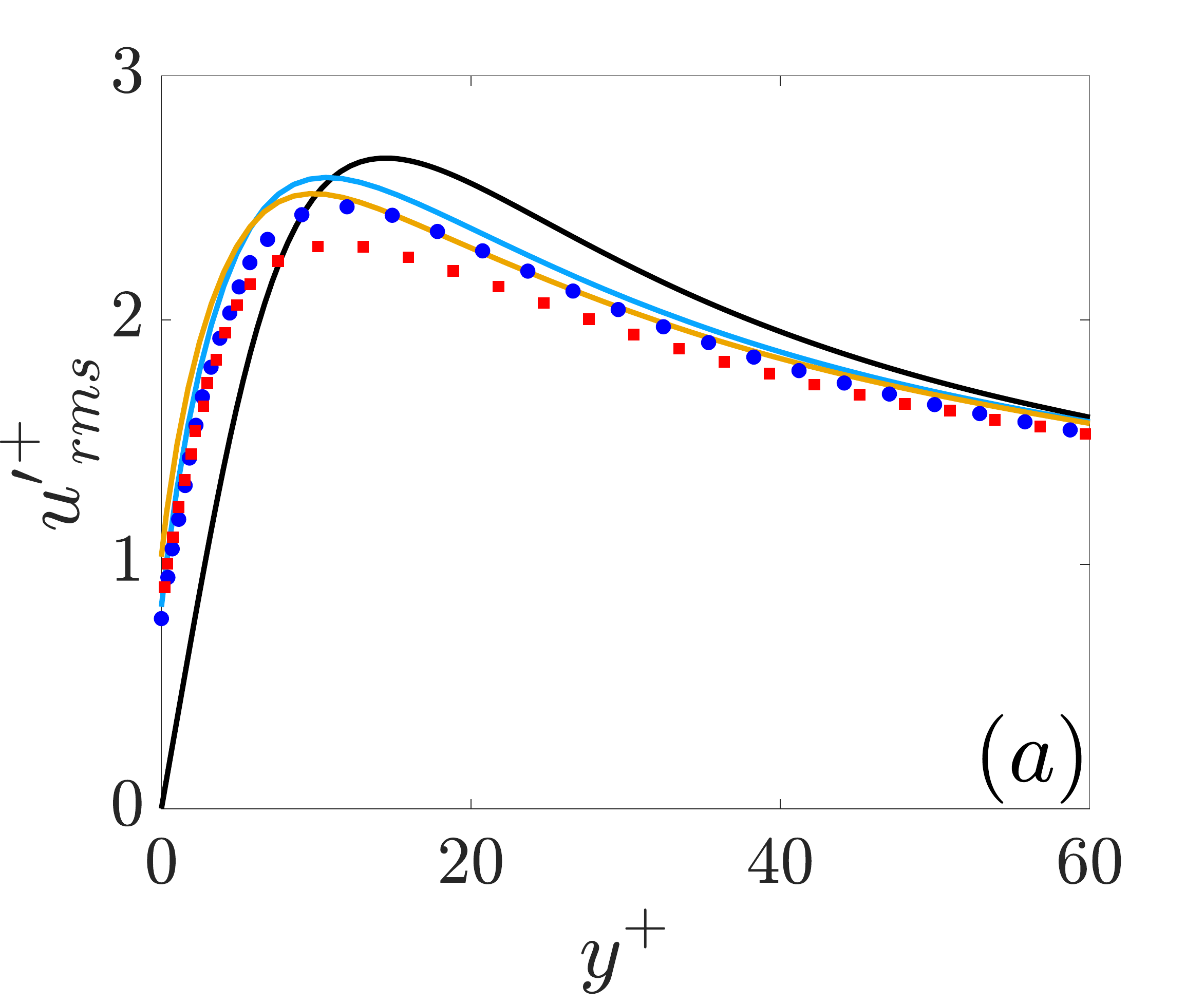}
        {\phantomcaption\label{fig:TRM_vs_C15_C18:u}}
    \end{subfigure}%
    \hspace*{-3mm}
    \begin{subfigure}[tbp]{.27\textwidth}
        \includegraphics[width=1\linewidth]{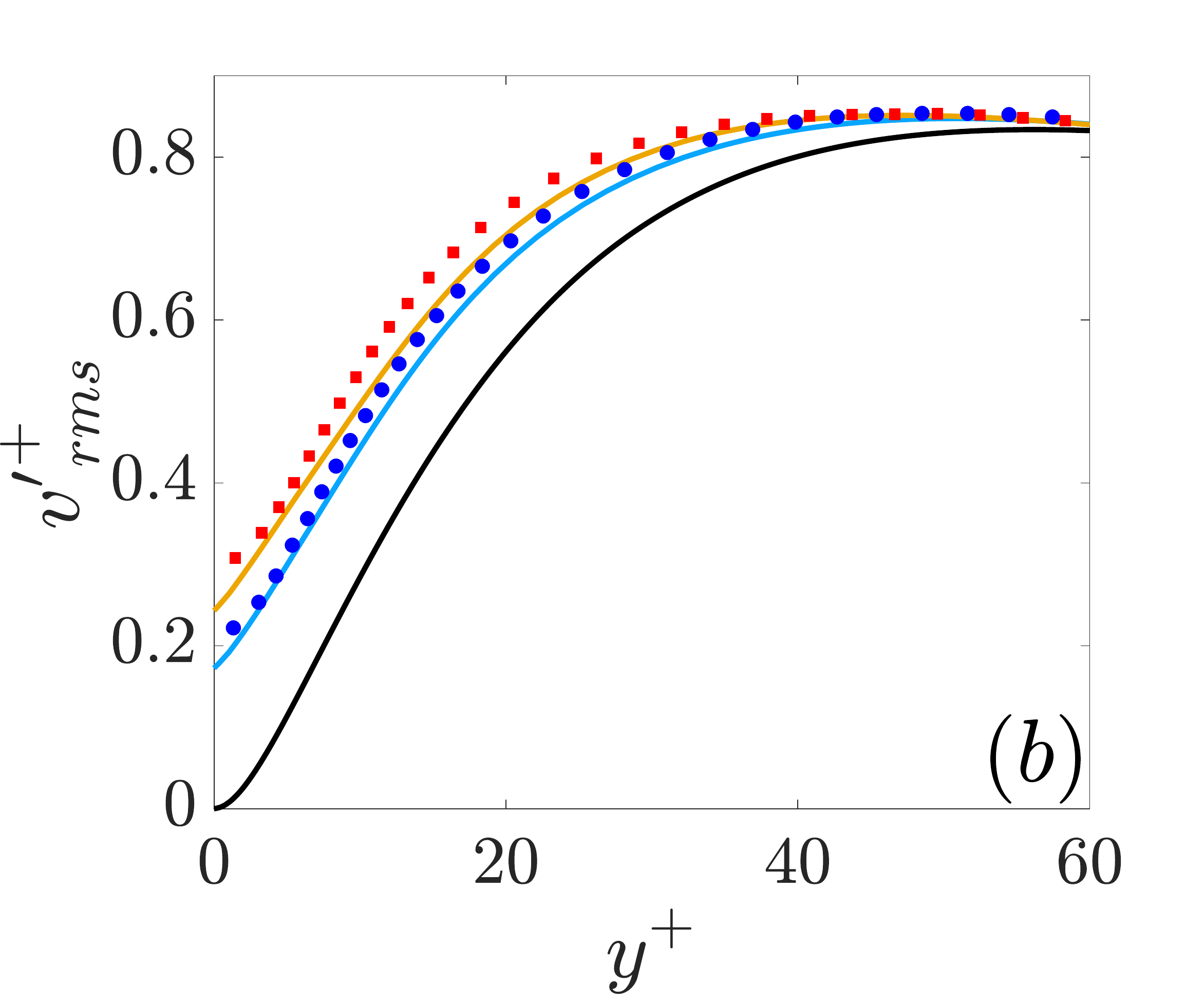}
        {\phantomcaption\label{fig:TRM_vs_C15_C18:v}}
    \end{subfigure}%
    \hspace*{-3mm}
    \begin{subfigure}[tbp]{.27\textwidth}
        \includegraphics[width=1\linewidth]{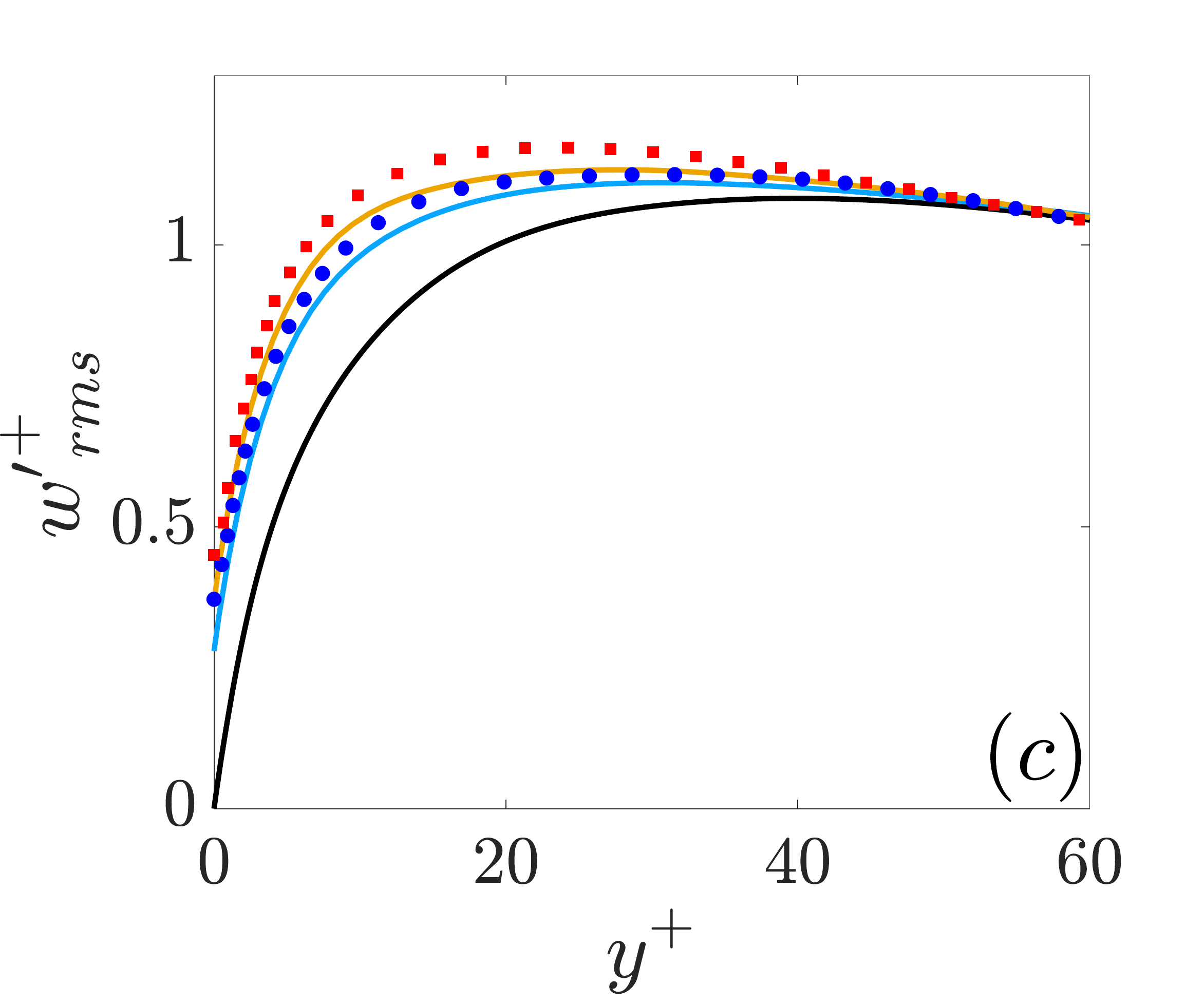}
        {\phantomcaption\label{fig:TRM_vs_C15_C18:w}}
    \end{subfigure}%
    \hspace*{-3mm}
    \begin{subfigure}[tbp]{.27\textwidth}
        \includegraphics[width=1\linewidth]{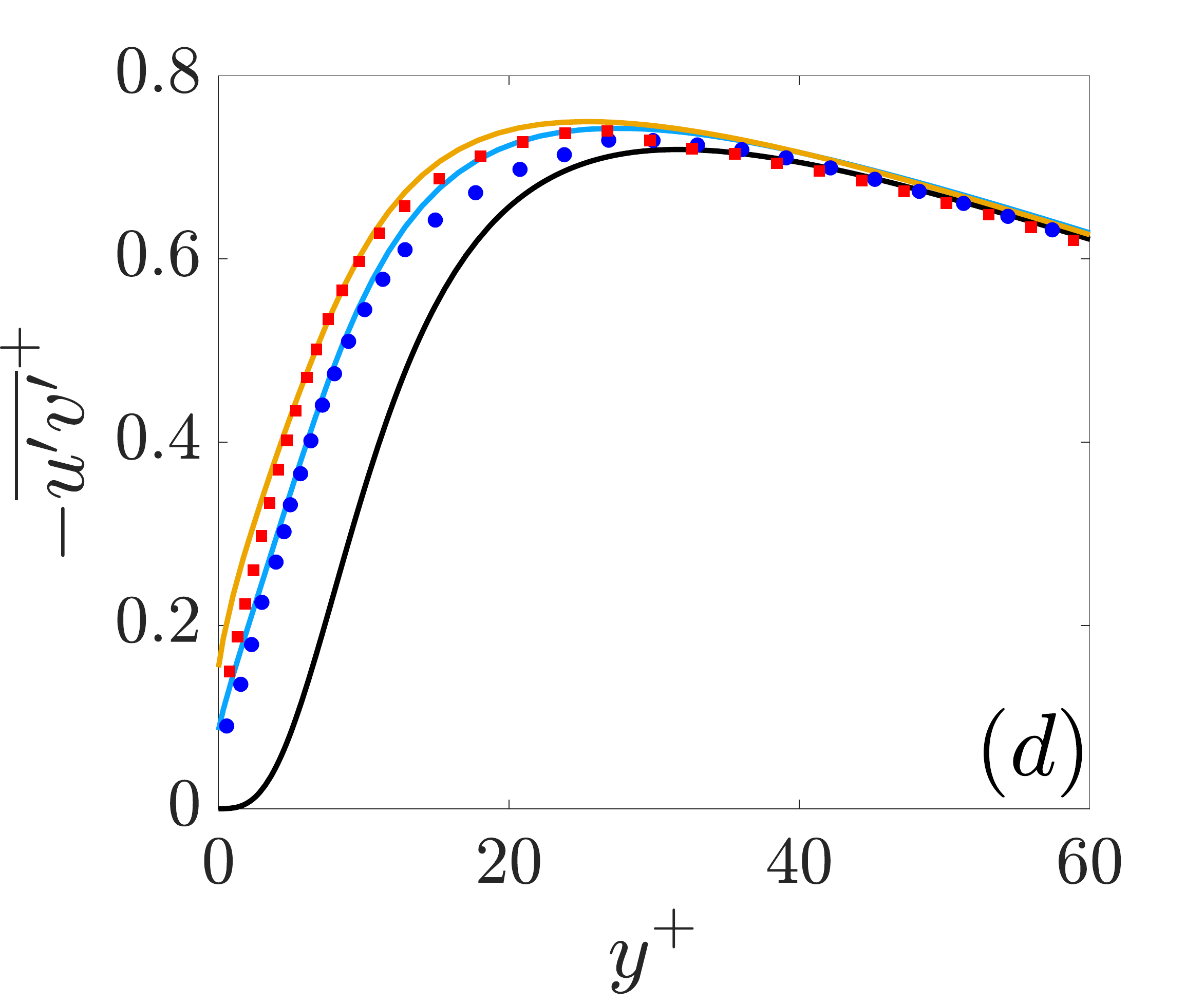}
        {\phantomcaption\label{fig:TRM_vs_C15_C18:uv}}
    \end{subfigure}
    \vspace*{-5mm}
    \caption{Comparison between the TRM DNS and the geometry-resolving rough wall DNS of \citet{Abderrahaman2019}. r.m.s. velocity fluctuation ($a$-$c$) and Reynolds shear stress ($d$) profiles. {$\textcolor{blue}{\bullet}$}, C15; {$\scriptstyle\textcolor{red}{\blacksquare}$}, C18; \textcolor{C15-blue}{$-\!-$}, L1.5M12; \textcolor{C18-yellow}{$-\!-$}, L1.9M15; \textcolor{black}{$-\!-$}, smooth-wall data.}
    \label{fig:TRM_vs_C15_C18}
    \end{center}
\end{figure}
%%%%%%%%%%%%%%%%%%%%%%%%%%%%%%%%%%%%%%%%%%%%%%%%%%%%%%%%%%%%%%%%%%%%%%%%%
%%%%%%%%%%%%%%%%%%%%%%%%%%%%%%%%%%%%%%%%%%%%%%%%%%%%%%%%%%%%%%%%%%%%%%%%%

The TRM simulations corresponding to the remaining geometries C06, C09, C15 and C18 are respectively L0.5M8, L0.7M7, L1.5M12 and L1.9M15 as listed in \cref{tab:trm_rough}. In other words, these cases produced similar levels of drag as their geometry-resolving counter-parts from \citet{Abderrahaman2019}. However, the geometry-resolving DNS cases C15 and C18 exhibit near-wall turbulence structures which are no longer smooth-wall-like \citep{Abderrahaman2019}. Therefore, it is important to assess if the corresponding TRM cases produce a similar type of turbulence and not only a matching change in drag. The comparisons are made in \cref{fig:TRM_vs_C15_C18}; some discrepancy exists, most notably for the r.m.s. velocity fluctuations of $u$ in the near-wall region (\hyperref[fig:TRM_vs_C15_C18:u]{figure \ref*{fig:TRM_vs_C15_C18:u}}) where the TRM results have peaks of greater magnitude. The distribution of the $v$ and $w$ r.m.s. velocities match quite well, as do those of the Reynolds shear stresses. Aside from the boundary conditions themselves being the principle source of these discrepancies, other potentially contributing factors may be the higher resolution of the geometry-resolving DNS simulations as well as the solver of those simulations using a pseudo-spectral spatial discretization.

\subsection{The role of the transpiration factor} \label{role_of_transpriation_factor}
%%%%%%%%%%%%%%%%%%%%%%%%%%%%%%%%%%%%%%%%%%%%%%%%%%%%%%%%%%%%%%%%%%%%%%%%%
%%%%%%%%%%%%%%%%%%%%%%%%%%%%%%%%%%%%%%%%%%%%%%%%%%%%%%%%%%%%%%%%%%%%%%%%%
\begin{figure}
    \begin{center}
    \begin{subfigure}{.33\textwidth}
     \includegraphics[width=1\linewidth]{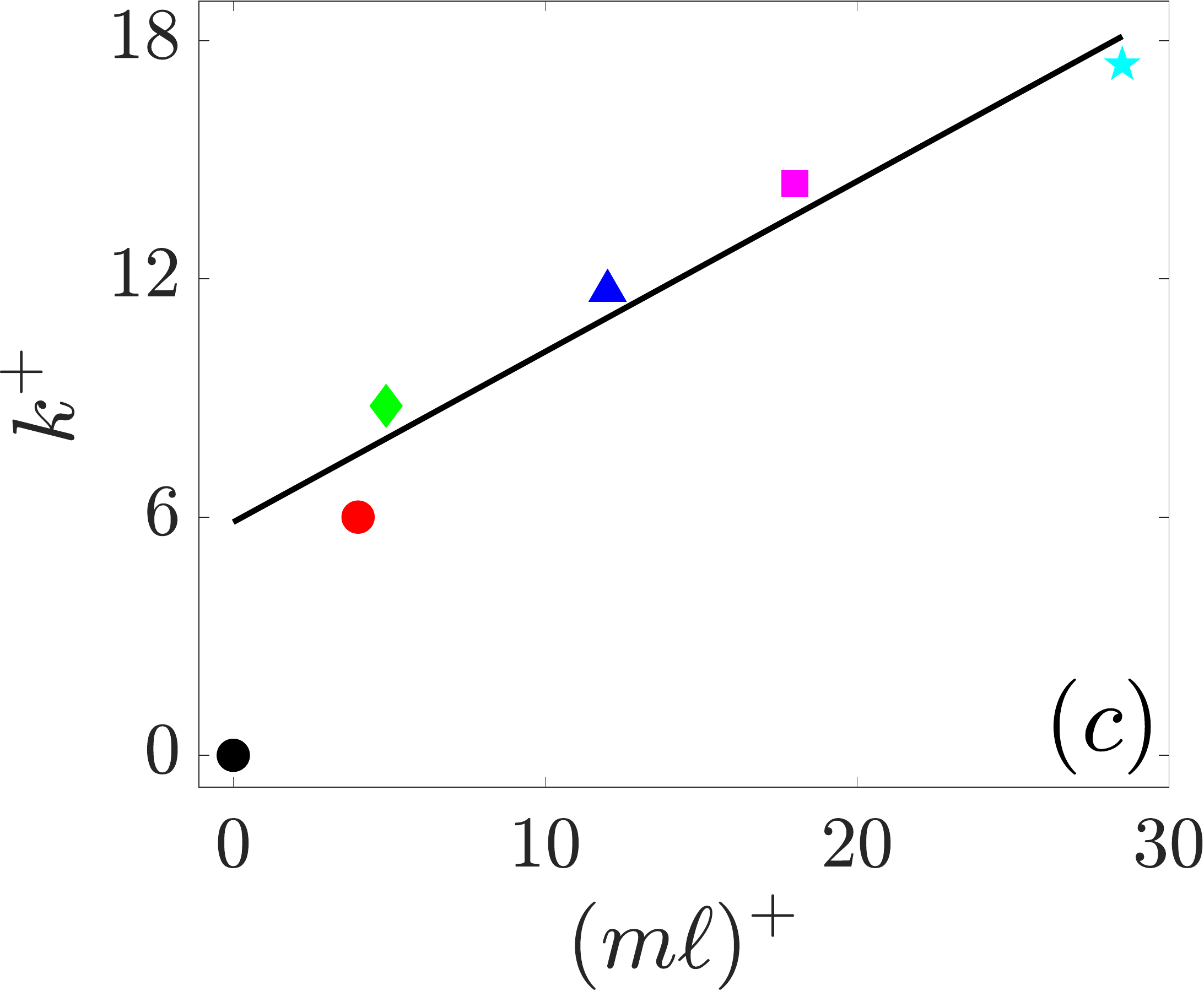}
     {\phantomcaption\label{fig:TranspirationScalingAbdrhmn:sub1}}
     \vspace*{-1mm}
    \end{subfigure}%
    \begin{subfigure}{.33\textwidth}
     \includegraphics[width=1\linewidth]{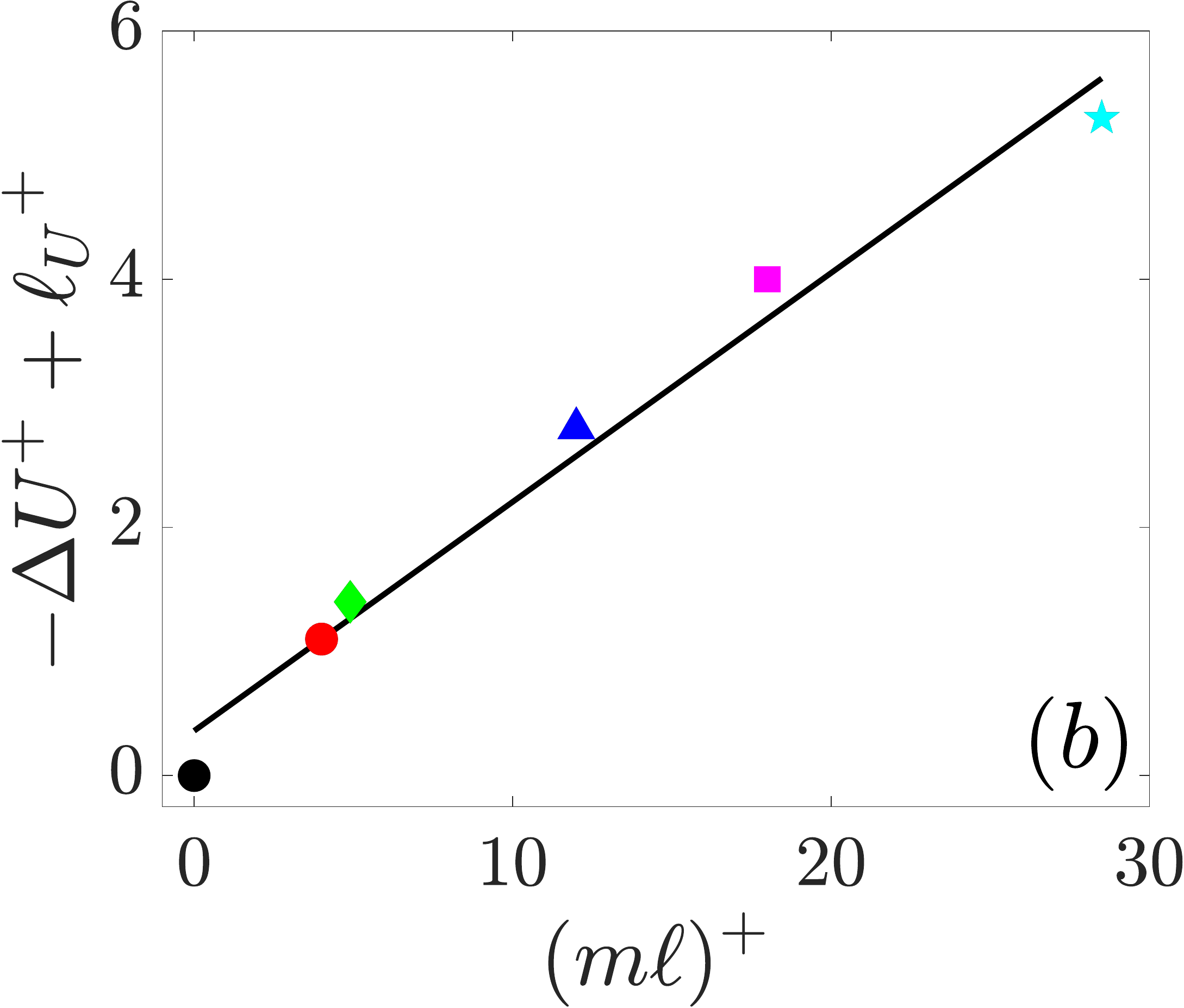}
     {\phantomcaption\label{fig:TranspirationScalingAbdrhmn:sub2}}
    \end{subfigure}%
    \begin{subfigure}{.33\textwidth}
     \includegraphics[width=1\linewidth]{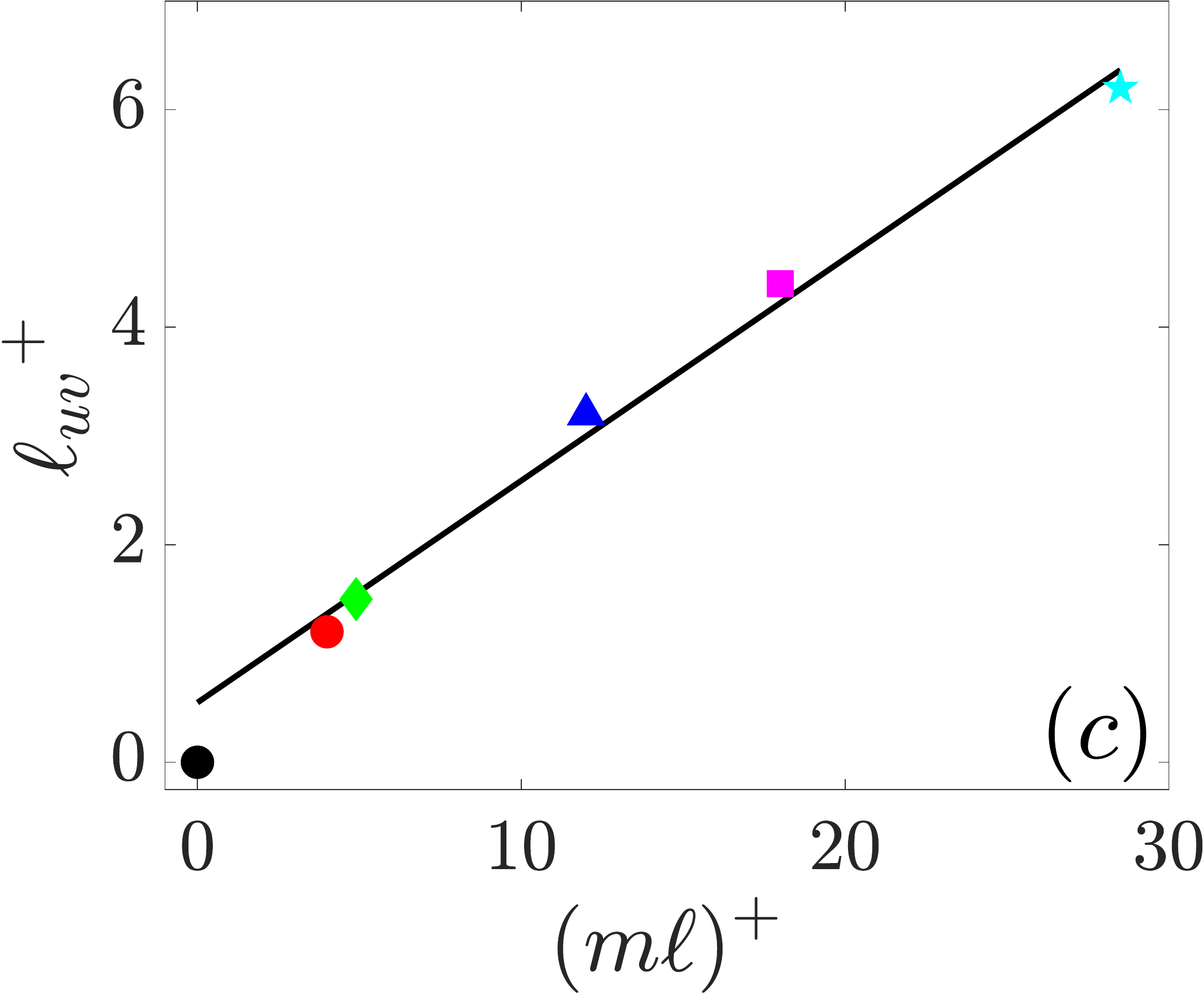}
     {\phantomcaption\label{fig:TranspirationScalingAbdrhmn:sub3}}
     \vspace*{-1mm}
    \end{subfigure}
    \vspace*{-4mm}
    \caption{Height of the collocated posts ($k^+$), roughness function relative to mean velocity slip (${-\Delta{U}^+}\!\!+{{\ell_{U}}^+}$) and the virtual origin of the Reynolds shear stress (${\ell_{uv}}^+$) as functions of the transpiration factor shown in ($a$), ($b$) and ($c$), respectively. {\Large\textcolor{black}{$\bullet$}}, smooth-wall; {\Large\textcolor{red}{$\bullet$}}, L0.5M8; \textcolor{green}{${\blacklozenge}$}, L0.7M7; \textcolor{blue}{${\blacktriangle}$}, L1.2M10; \textcolor{magenta}{${\blacksquare}$}, L1.5M12; \textcolor{cyan}{${\bigstar}$}, L1.9M15. \textcolor{black}{$-\!-$}, linear fits ${0.204}({m}{\ell})^+$; ${0.185}({m}{\ell})^+$ and ${0.429}({m}{\ell})^+$ in ($a$), ($b$) and ($c$) respectively.}
    \label{fig:TranspirationScalingAbdrhmn}
    \end{center}
\end{figure}
%%%%%%%%%%%%%%%%%%%%%%%%%%%%%%%%%%%%%%%%%%%%%%%%%%%%%%%%%%%%%%%%%%%%%%%%%
%%%%%%%%%%%%%%%%%%%%%%%%%%%%%%%%%%%%%%%%%%%%%%%%%%%%%%%%%%%%%%%%%%%%%%%%%

\Cref{fig:TranspirationScalingAbdrhmn:sub1} shows how the height of the posts (i.e. the roughness height), $k^+$, scales with respect to $(m\ell)^+$. A monotonic increase is observed along with a nearly linear proportionality. A larger height of the posts translates to deeper cavities in-between them, creating more room for the wall-normal velocity to penetrate below the crest plane. This is the effect quantified by the transpiration length, $m^+$. The spacing between the posts, ${s_x}^+$ and ${s_z}^+$, leads to the existence of a slip velocity at the roughness crests, which is the effect quantified by the slip lengths ${\ell_x}^+$ and ${\ell_z}^+$. Clearly, changes in the geometrical features of the surface are reflected in the TRM's coefficients ($\ell^+$, $m^+$) in a physically consistent way. In particular, the compound effect of the texture height and pitch, which determines the overall degree of transpiration that can take place at the crest plane, is captured by the transpiration factor ${(m\ell)}^+$.

Earlier studies on both regular and random roughness \citep{orlandi_2_3, Forooghi_rough_2018} have demonstrated a linear relation between the wall-normal velocity fluctuations at the crest plane and the roughness function. A similar relation between the transpiration factor ${(m\ell)}^+$ and the roughness function can be expected. As discussed in \cref{TRMboundarycondition}, ${(m\ell)}^+$ is an effective measure of the volume-averaged (using an appropriate REV) 
%wall-normal flux across the crest plane of a
flow within the textured surface. \Cref{fig:TranspirationScalingAbdrhmn:sub2} shows how the roughness function (reported here relative to the slip velocity, ${-\Delta{U}^+}\!\!+{{\ell_{U}}^+}$, following \citealt{orlandi_2_3} and \citealt{Abderrahaman2019}) scales with respect to the transpiration factor ${(m\ell)}^+$. An overall linear proportionality is observed for the TRM simulations that reproduce the collocated posts DNS cases C06-C18 in \cref{tab:rough_cases_abdrhmn}.

Since the roughness function demonstrates an even stronger linear relation with ${\ell_{uv}}^+$, a linear relation between ${(m\ell)}^+$ and ${\ell_{uv}}^+$ can also be expected, which is shown in \cref{fig:TranspirationScalingAbdrhmn:sub3}. As will be shown in the next section, the transpiration is predominately driven by changes in spanwise shear, as this leads to the displacement of quasi-streamwise vortices which generate wall-normal fluctuations for canonical turbulence (\hyperref[fig:TRM:displacement]{figure \ref*{fig:TRM:displacement}}). Since this displacement is quantified by ${\ell_{uv}}^+$, the virtual origin of turbulence and the transpiration factor are linearly proportional to each other. This proportionality also persists for the cases which fall outside of the smooth-wall-like regime.

\section{TRM with anisotropic transpiration lengths} \label{sec:TRM_anisotropic}

\begin{figure}
    \begin{center}
    \hspace*{-3mm}
    \begin{subfigure}[tbp]{.4\textwidth}
        \includegraphics[width=1\linewidth]{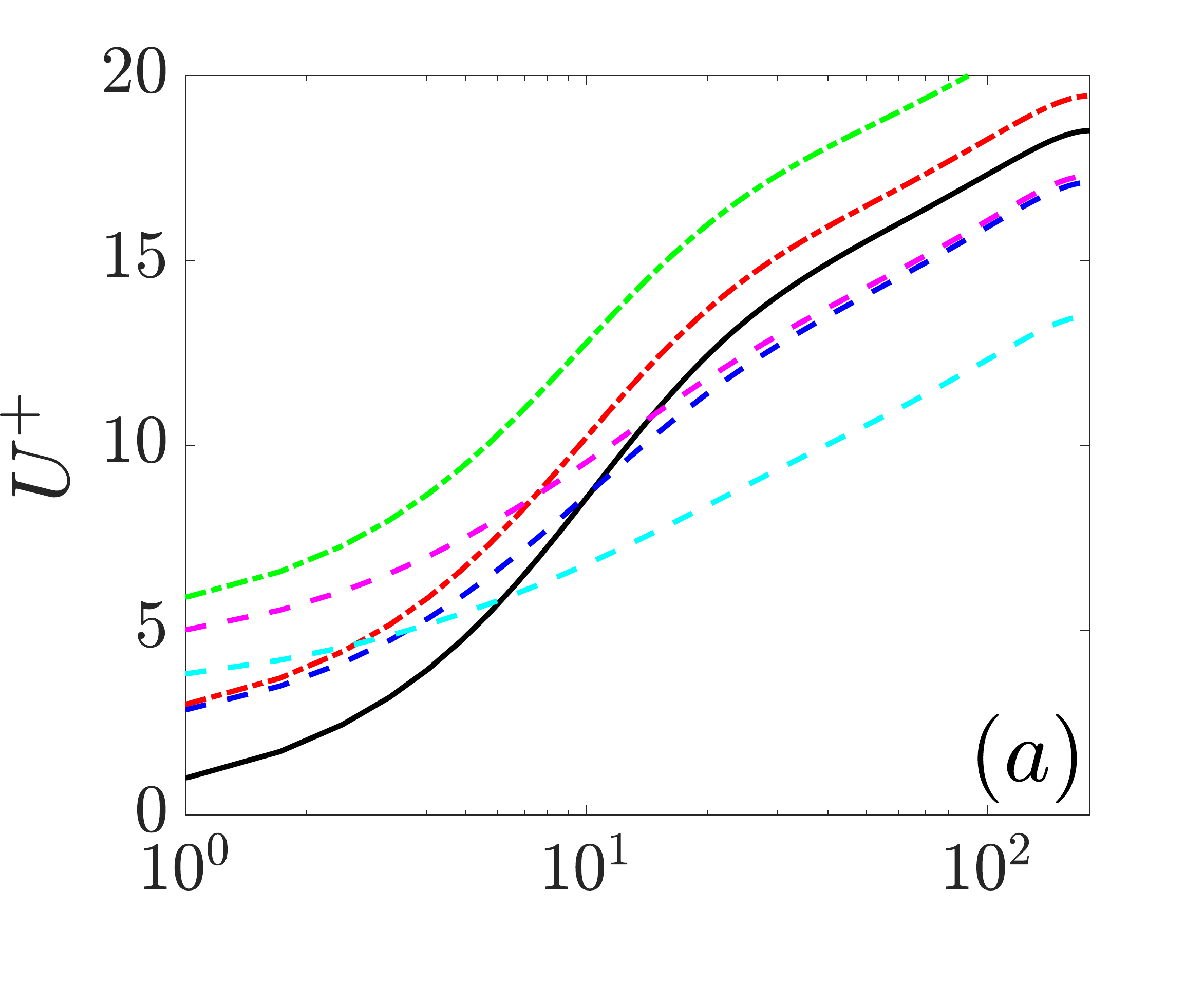}
        {\phantomcaption\label{fig:group2:mean_velocity}}
        \vspace*{-8mm}
    \end{subfigure}%
    \hspace*{-1mm}
    \begin{subfigure}[tbp]{.4\textwidth}
        \includegraphics[width=1\linewidth]{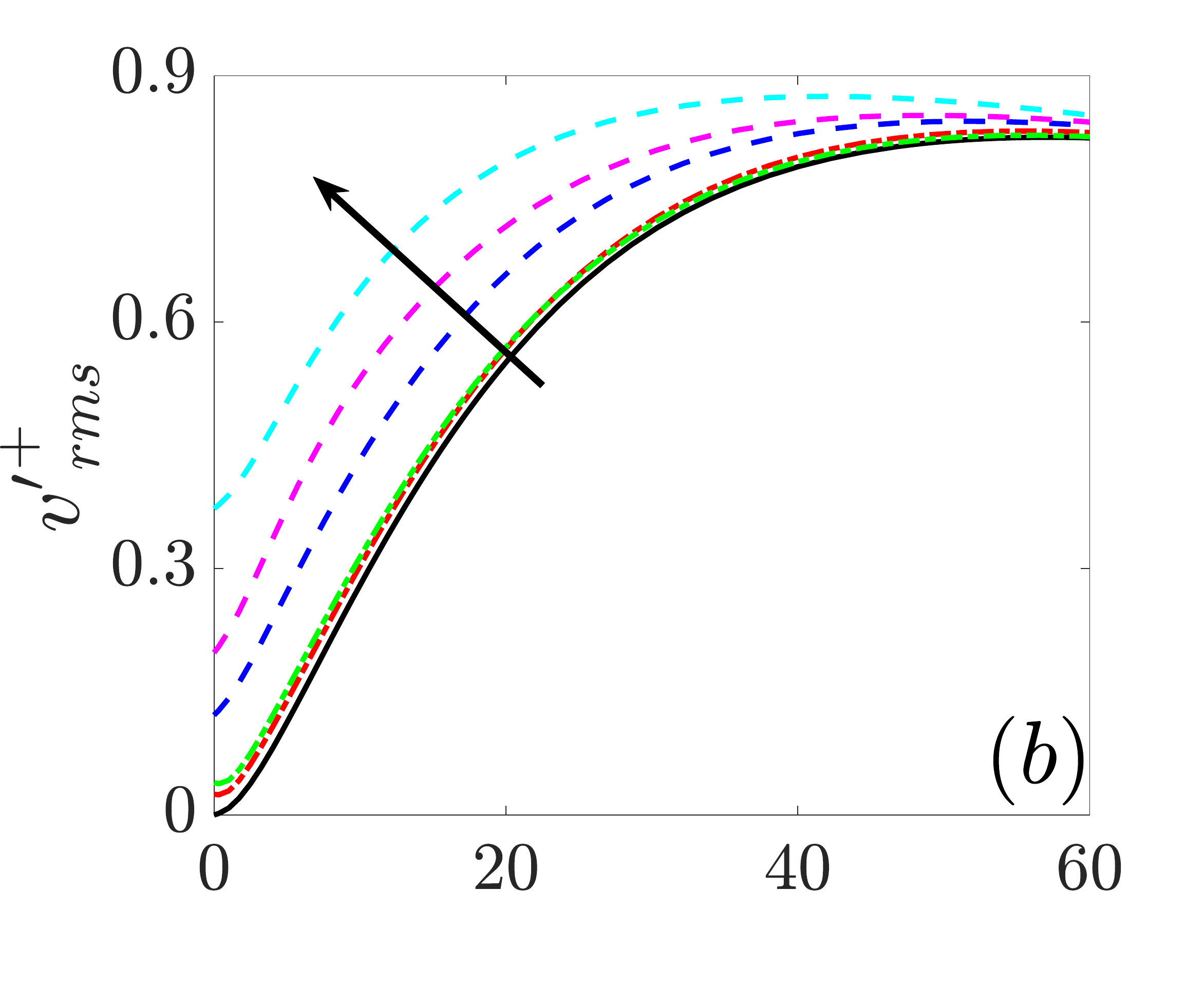}
        {\phantomcaption\label{fig:group2:v_rms}}
        \vspace*{-8mm}
    \end{subfigure}
    \begin{subfigure}[tbp]{.4\textwidth}
        \includegraphics[width=1\linewidth]{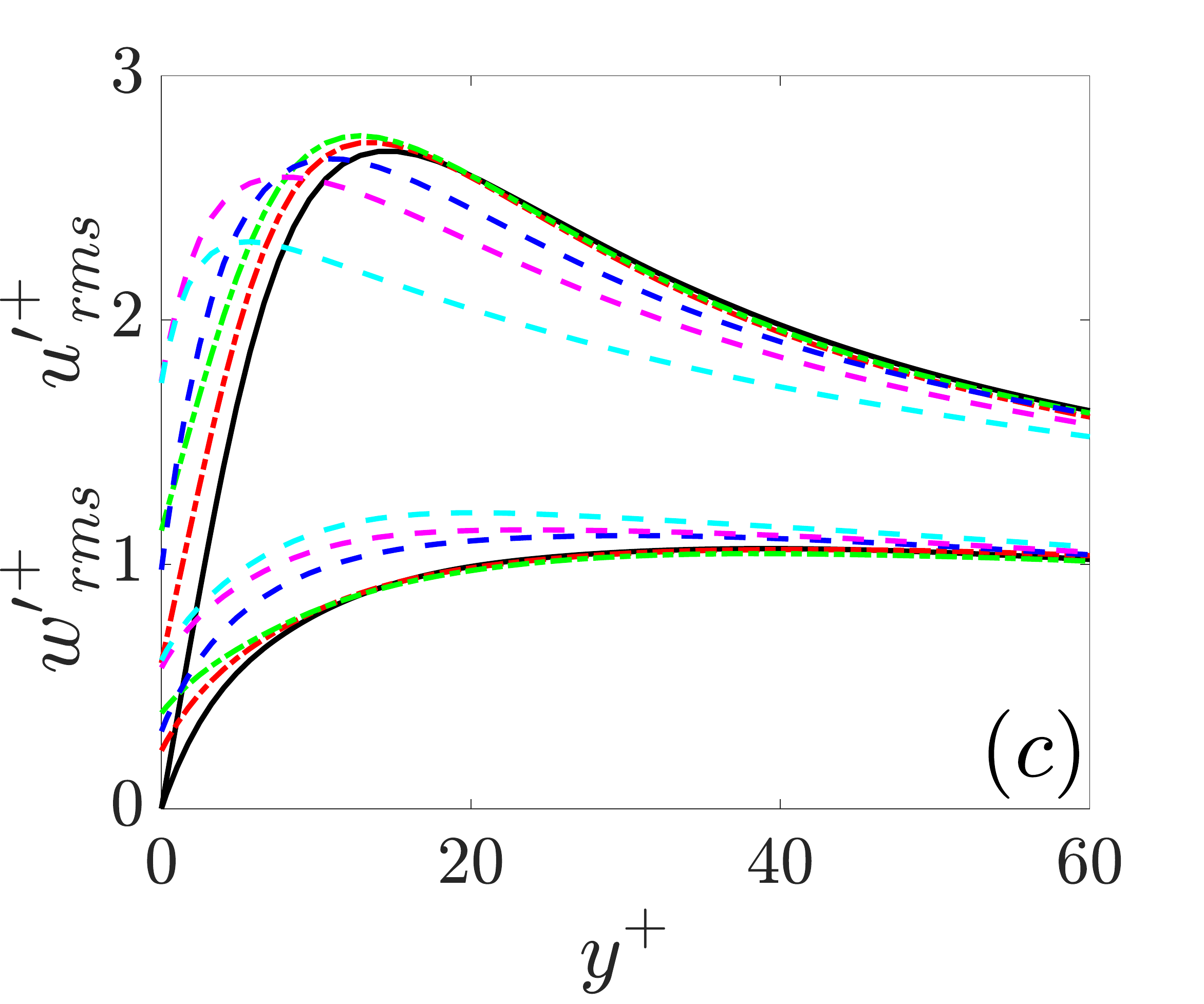}
        {\phantomcaption\label{fig:group2:u_w_rms}}
    \end{subfigure}%
    \begin{subfigure}[tbp]{.4\textwidth}
        \includegraphics[width=1\linewidth]{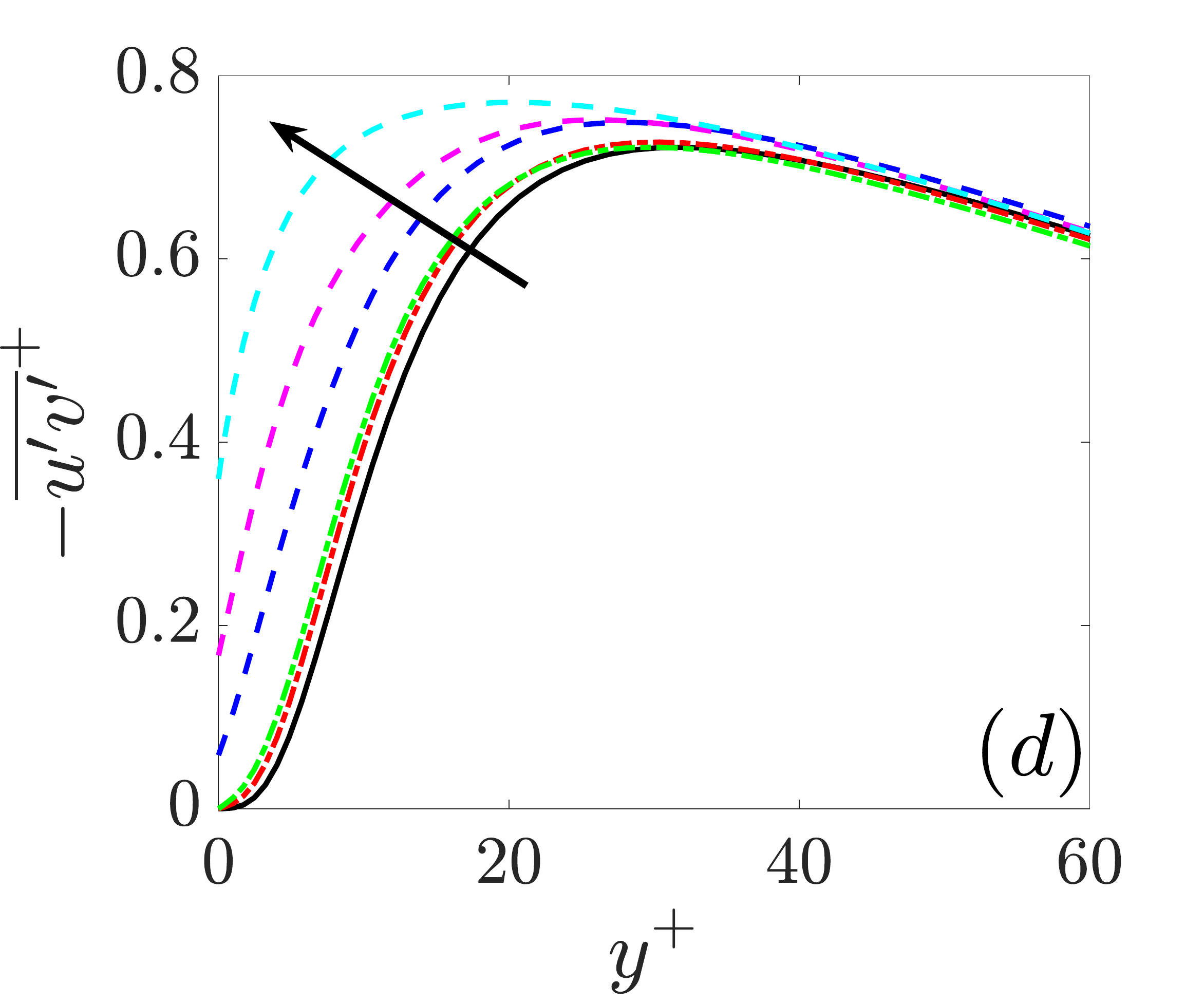}
        {\phantomcaption\label{fig:group2:reynolds_stress}}
    \end{subfigure}
    \vspace*{-5mm}
    \caption{Mean velocity ($a$), r.m.s. velocity fluctuation ($b,c$) and Reynolds shear stress profiles ($d$). \textcolor{red}{$-\;\text{-}\;-$}, L2MX5; \textcolor{green}{$-\;\text{-}\;-$}, L5MX5; \textcolor{blue}{$-\;-$}, L2MZ5; \textcolor{magenta}{$-\;-$}, L5MZ5; \textcolor{cyan}{$-\;-$}, L5MZ10; $-\!-$, smooth-wall data.}
    \label{fig:group2}
    \end{center}
\end{figure}

The general form of the TRM \eqref{eq:TRMv2}, valid for anisotropic surface textures, contains two transpiration factors; one associated with the variations of spanwise shear, ${({m}{\ell})_z}^+$, and another with the variation of streamwise shear, ${({m}{\ell})_x}^+$.
To investigate how the different components in the boundary condition of \eqref{eq:TRMv2} affect the flow, simulations were done where either $m_x = 0$ or $m_z = 0$, thus severing the coupling between their corresponding velocity components and $v$. The simulations are listed in the last 8 rows of \cref{tab:dns}. The results of cases L$2$MX$5$, L$5$MX$5$, L$2$MZ$5$, L$5$MZ$5$ and L$5$MZ$10$ have been gathered in \cref{fig:group2} while the remainder have been omitted for clarity.

%%%%%%%%%%%%%%%%%%%%%%%%%%%%%%%%%%%%%%%%%%%%%%%%%%%%%%%%%%%%%%%%%%%%%%%%%
%%%%%%%%%%%%%%%%%%%%%%%%%%%%%%%%%%%%%%%%%%%%%%%%%%%%%%%%%%%%%%%%%%%%%%%%%
%%%%%%%%%%%%%%%%%%%%%%%%%%%%%%%%%%%%%%%%%%%%%%%%%%%%%%%%%%%%%%%%%%%%%%%%%
%%%%%%%%%%%%%%%%%%%%%%%%%%%%%%%%%%%%%%%%%%%%%%%%%%%%%%%%%%%%%%%%%%%%%%%%%

\subsection{Transpiration due to streamwise shear}
Starting with case L2MX5 (\textcolor{red}{$-\,\text{-}\,-$} in \cref{fig:group2}), the coupling only to streamwise variations of the $u$ velocity shear is kept. This makes it conform to the physical scenario of \cref{fig:TRM:concept}. The result is a weakening of impermeability and the occurrence of ${v^{\prime}}^+$ activity at the boundary (\hyperref[fig:group2:v_rms]{figure \ref*{fig:group2:v_rms}}), but no presence of $-\overline{u^{\prime}v^{\prime}}^+$ (\hyperref[fig:group2:reynolds_stress]{figure \ref*{fig:group2:reynolds_stress}}). A similar situation arises in case L2MX2 (not shown). Compared to the drag of slip-only case L2M0 (${\Delta{U}^+=0.7}$), none to negligible changes in drag occur in cases L2MX2 (${\Delta{U}^+=0.7}$) and L2MX5 (${\Delta{U}^+=0.8}$), respectively. Despite the transpiration length of case L2MX5 being double that of case L2MX2, the results for both are essentially the same and similar to their slip-only counterpart. The virtual-origin effect accounts for the drag change of both cases as shown in \cref{fig:group2:shifted}.

Case L5MX5 (\textcolor{green}{$-\,\text{-}\,-$} in \cref{fig:group2}) exhibits behaviour very similar to that of L2MX5, in that they both result in a slight drag reduction relative to their respective slip-only cases (L2M0 and L5M0 respectively). Going from L5MX5 to L5MX10 (not shown) with double the transpiration length produces no discernible change, which is indicated by ${\Delta{U}^+}$ remaining the same. Transpiration is present at the boundary plane for these cases (\hyperref[fig:group2:v_rms]{figure \ref*{fig:group2:v_rms}}), but no Reynolds shear stress (\hyperref[fig:group2:reynolds_stress]{figure \ref*{fig:group2:reynolds_stress}}).

The behaviour observed in the L$<\!\!\cdot\!\!>$MX$<\!\!\cdot\!\!>$ cases, where transpiration occurs at the boundary but without resulting in the generation of Reynolds shear stress, does not seem to have an analogue in any actual passive surface that the authors are aware of. Considering that the wall-normal mixing of fluid in the near-boundary (or more generally the near-wall) region of canonical turbulence is due to the quasi-streamwise vortices which redistribute momentum along the spanwise and wall-normal directions, explicitly coupling $v$ to just $u$ does not allow such a redistribution to take place at the boundary plane. The minute amounts of ${v^{\prime}}^+$ generated are inactive motions induced by the streamwise streaks (${u^{\prime}}^+$) and thus neutral in generating $-\overline{u^{\prime}v^{\prime}}^+$ as they do not lead to sweep and ejection type events taking place at the boundary plane. Decoupling transpiration from $w$ thus does not permit these structures to move closer to the boundary plane and lead to the generation of $-\overline{u^{\prime}v^{\prime}}^+$ there. The L$<\!\!\cdot\!\!>$MX$<\!\!\cdot\!\!>$ cases have therefore served as control experiments reaffirming the principle role of these near-wall turbulent structures in generating Reynolds shear stress and increasing turbulent skin-friction.

\subsection{Transpiration due to spanwise shear}

%%%%%%%%%%%%%%%%%%%%%%%%%%%%%%%%%%%%%%%%%%%%%%%%%%%%%%%%%%%%%%%%%%%%%%%%%
%%%%%%%%%%%%%%%%%%%%%%%%%%%%%%%%%%%%%%%%%%%%%%%%%%%%%%%%%%%%%%%%%%%%%%%%%
\begin{figure}
    \begin{center}
    \begin{subfigure}[tbp]{.35\textwidth}
        \includegraphics[width=1\linewidth]{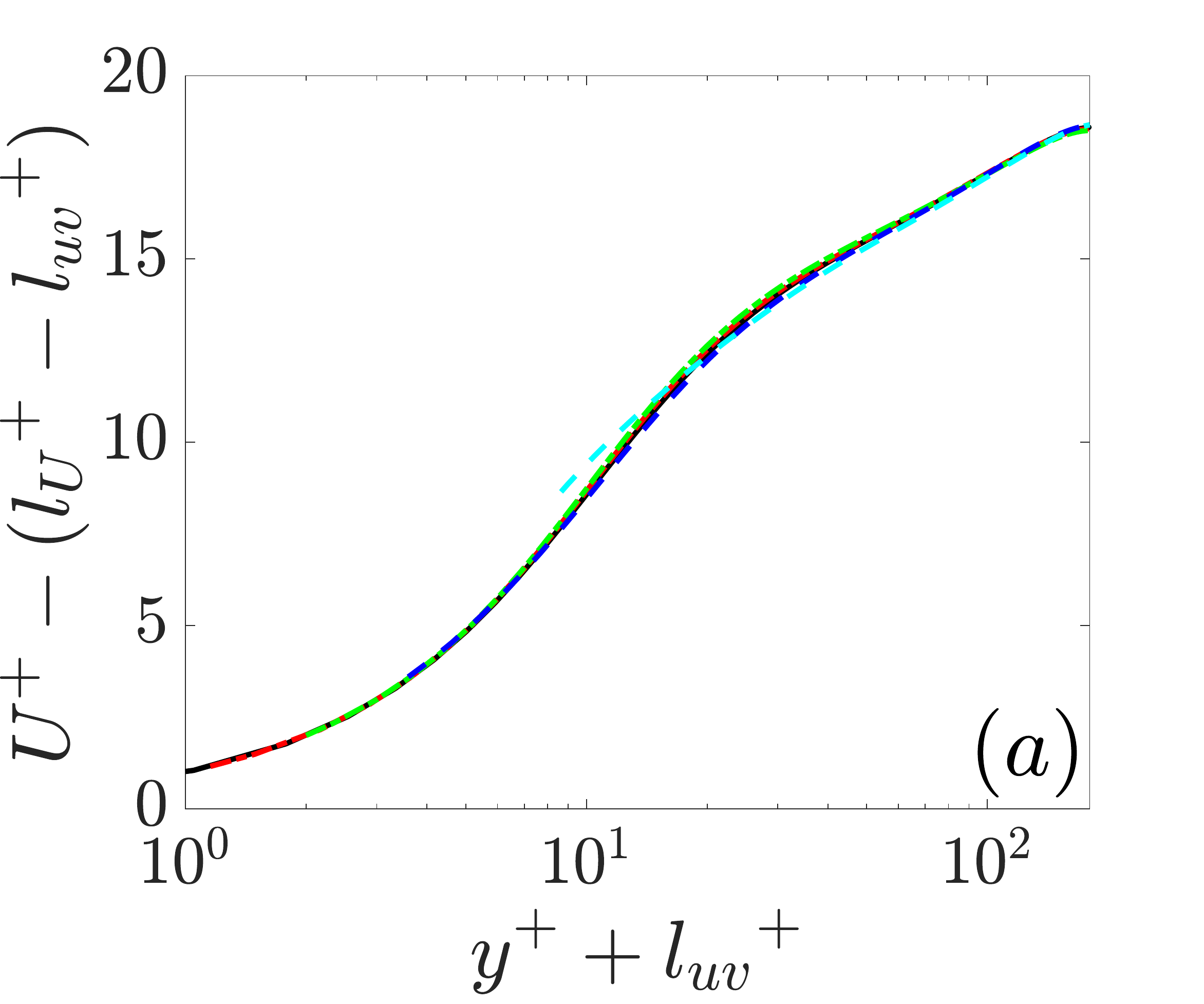}
        {\phantomcaption\label{fig:group2:mean_velocity_shifted}}
        \vspace*{-3mm}
    \end{subfigure}%
    \hspace*{-4mm}
    \begin{subfigure}[tbp]{.35\textwidth}
        \includegraphics[width=1\linewidth]{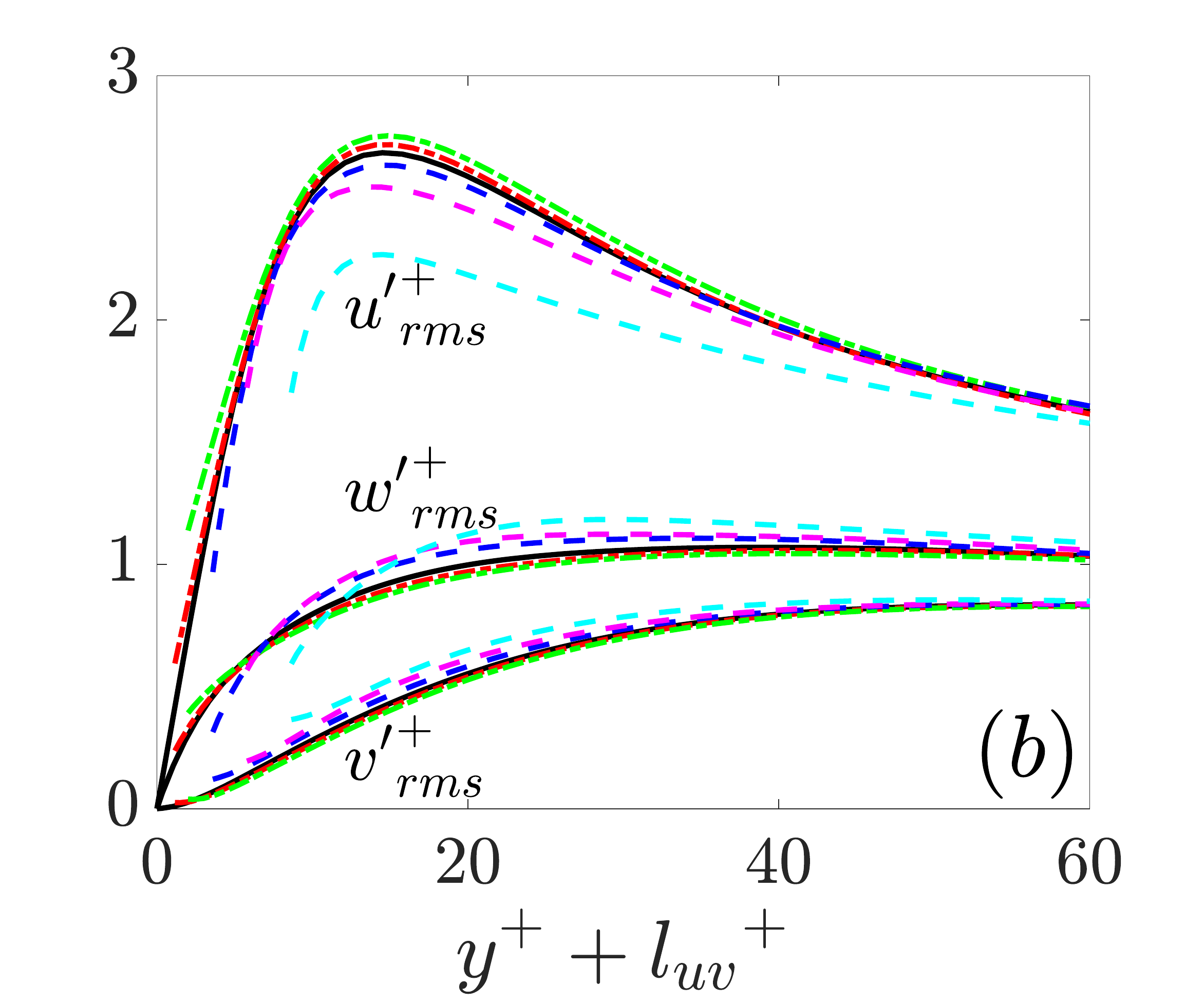}
        {\phantomcaption\label{fig:group2:rms_shifted}}
        \vspace*{-3mm}
    \end{subfigure}%
    \hspace*{-2mm}
    \begin{subfigure}[tbp]{.35\textwidth}
        \includegraphics[width=1\linewidth]{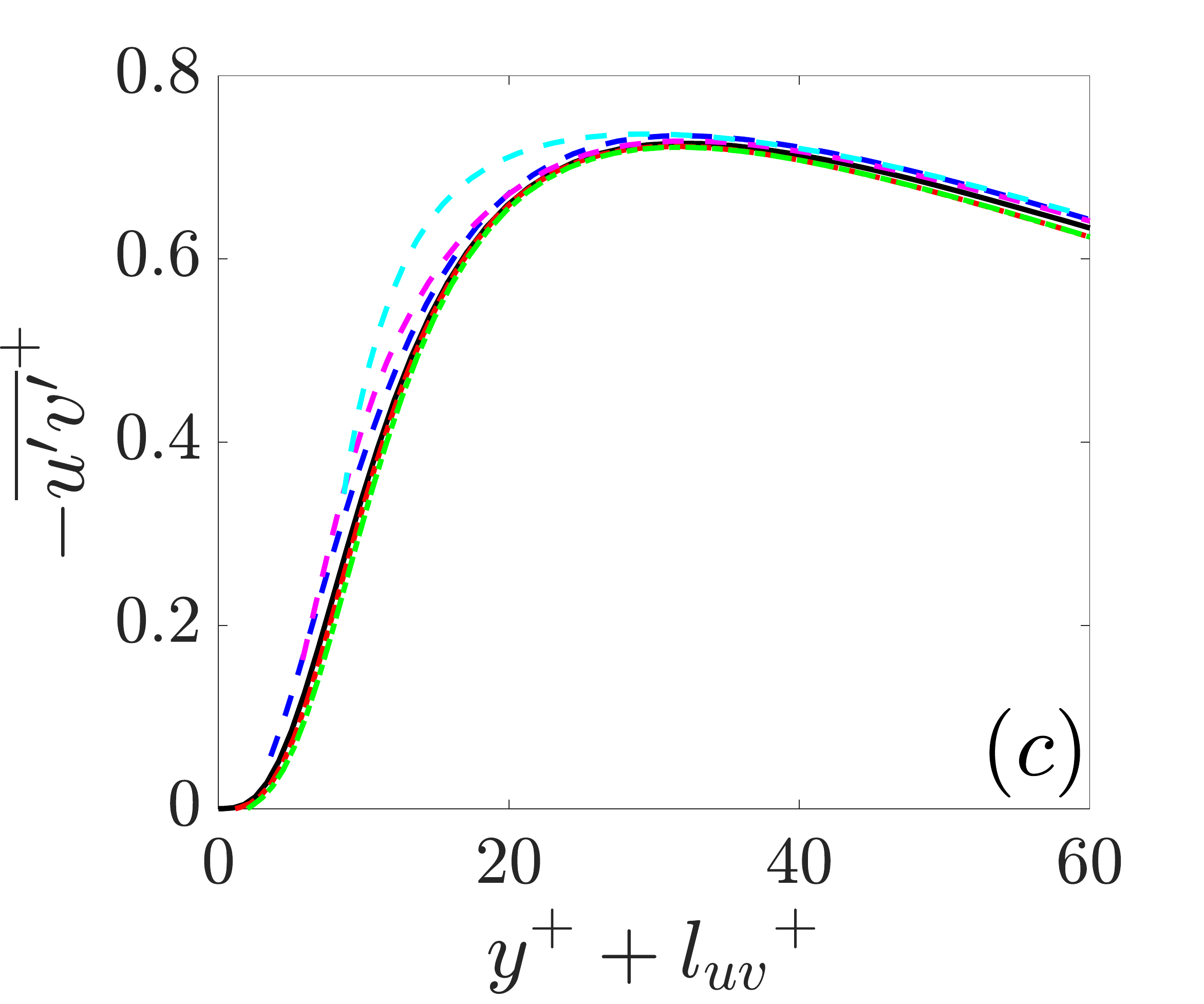}
        {\phantomcaption\label{fig:group2:reynolds_stress_shifted}}
        \vspace*{-3mm}
    \end{subfigure}
    \vspace*{-3mm}
    \caption{Mean velocity ($a$), r.m.s. velocity fluctuation ($b$) and Reynolds shear stress ($c$) profiles with the origin at ${y}^+ = -{\ell_{uv}}^+$ and rescaled with the corresponding ${u_{\tau}}$. \textcolor{red}{$-\,\text{-}\,-$}, L2MX5; \textcolor{green}{$-\,\text{-}\,-$}, L5MX5; \textcolor{blue}{$-\;-$}, L2MZ5; \textcolor{magenta}{$-\;-$}, L5MZ5; \textcolor{cyan}{$-\;-$}, L5MZ10; $-\!-$, smooth-wall data.}
    \label{fig:group2:shifted}
    \end{center}
\end{figure}
%%%%%%%%%%%%%%%%%%%%%%%%%%%%%%%%%%%%%%%%%%%%%%%%%%%%%%%%%%%%%%%%%%%%%%%%%
%%%%%%%%%%%%%%%%%%%%%%%%%%%%%%%%%%%%%%%%%%%%%%%%%%%%%%%%%%%%%%%%%%%%%%%%%

Case L2MZ5 (\textcolor{blue}{$-\;-$} in \cref{fig:group2}) only retains the coupling to the $w$ velocity component. A greater level of transpiration (\hyperref[fig:group2:v_rms]{figure \ref*{fig:group2:v_rms}}) along with the most $-\overline{u^{\prime}v^{\prime}}^+$ activity at the boundary plane is achieved (\hyperref[fig:group2:reynolds_stress]{figure \ref*{fig:group2:reynolds_stress}}) out of all cases with slip lengths of ${\ell_x}^+\!={\ell_z}^+\!=\!2.0$. This results in a noticeable drag increase (${\Delta{U}^+=-1.7}$), particularly with respect to case L2M5 where ${\Delta{U}^+=-1.0}$ and case L2MZ2 (not shown) where the mean flow slip and Reynolds shear stress balance each other and ${\Delta{U}^+=0}$.
The r.m.s. velocity fluctuations of case L2MZ5 also exhibit differences from those of smooth-wall turbulence, with a weakening of ${u^{\prime}}^+$ and strengthening of both ${v^{\prime}}^+$ and ${w^{\prime}}^+$ occurring (\hyperref[fig:group2:u_w_rms]{figure \ref*{fig:group2:u_w_rms}}).
This marks the departure from the smooth-wall turbulence regime, and indeed a suitable collapse with smooth-wall statistics is not obtained for the r.m.s. velocity fluctuations by accounting for the virtual-origin effect (\cref{fig:group2:shifted}), unlike for case L2MZ2 (not shown). For case L5MZ5, the previous transpiration lengths have been kept but the slip lengths have been increased. \Cref{fig:group2:shifted} shows that, similar to case L2MZ5, differences from smooth-wall turbulence are observed for case L5MZ5 but are more pronounced. 

\begin{figure}
   \begin{center}
   \begin{subfigure}[tbp]{.335\textwidth}
     \includegraphics[width=1\linewidth]{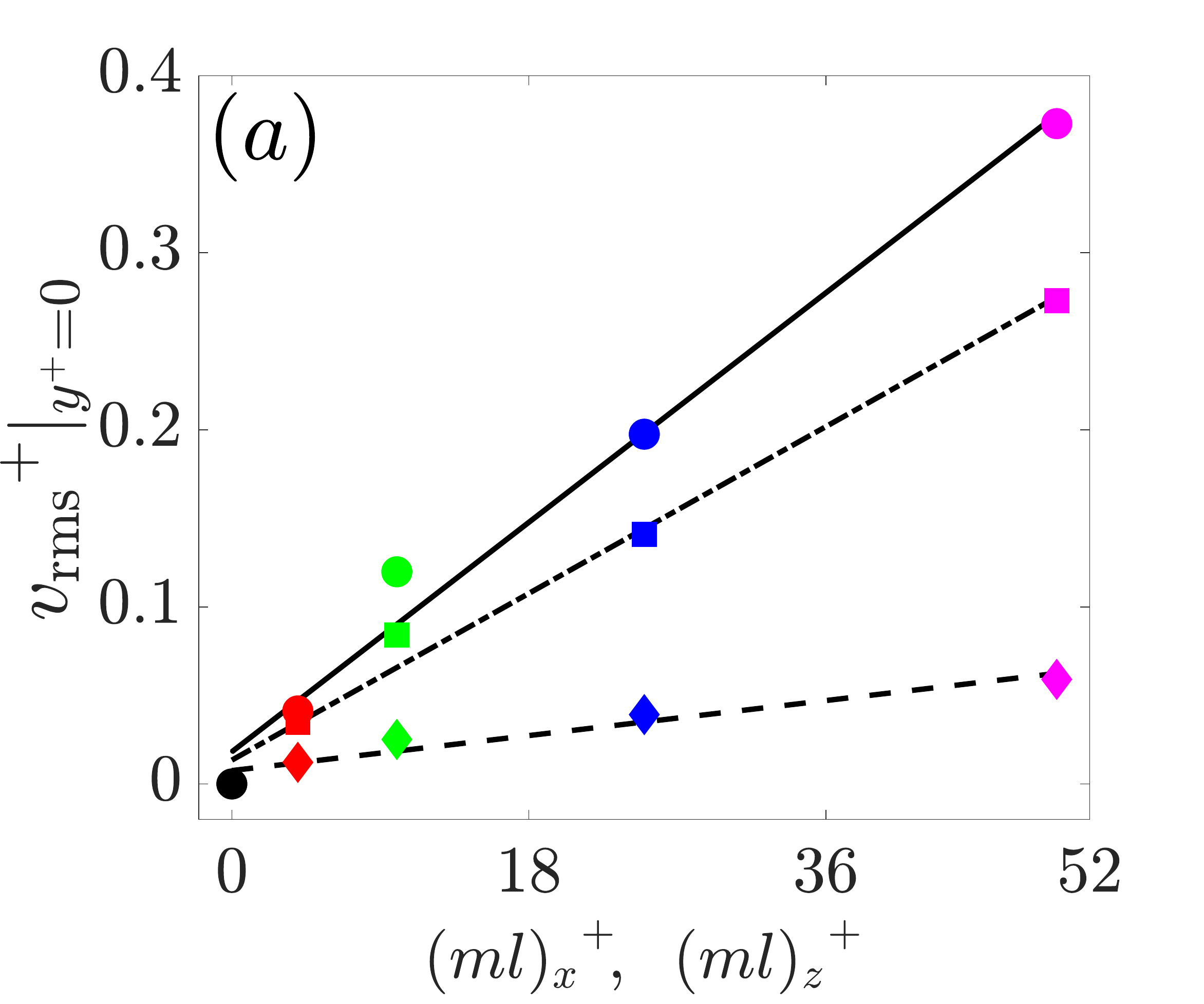}
     {\phantomcaption\label{fig:ml_vs_vrms}}
   \end{subfigure}%
   \begin{subfigure}[tbp]{.335\textwidth}
     \includegraphics[width=1\linewidth]{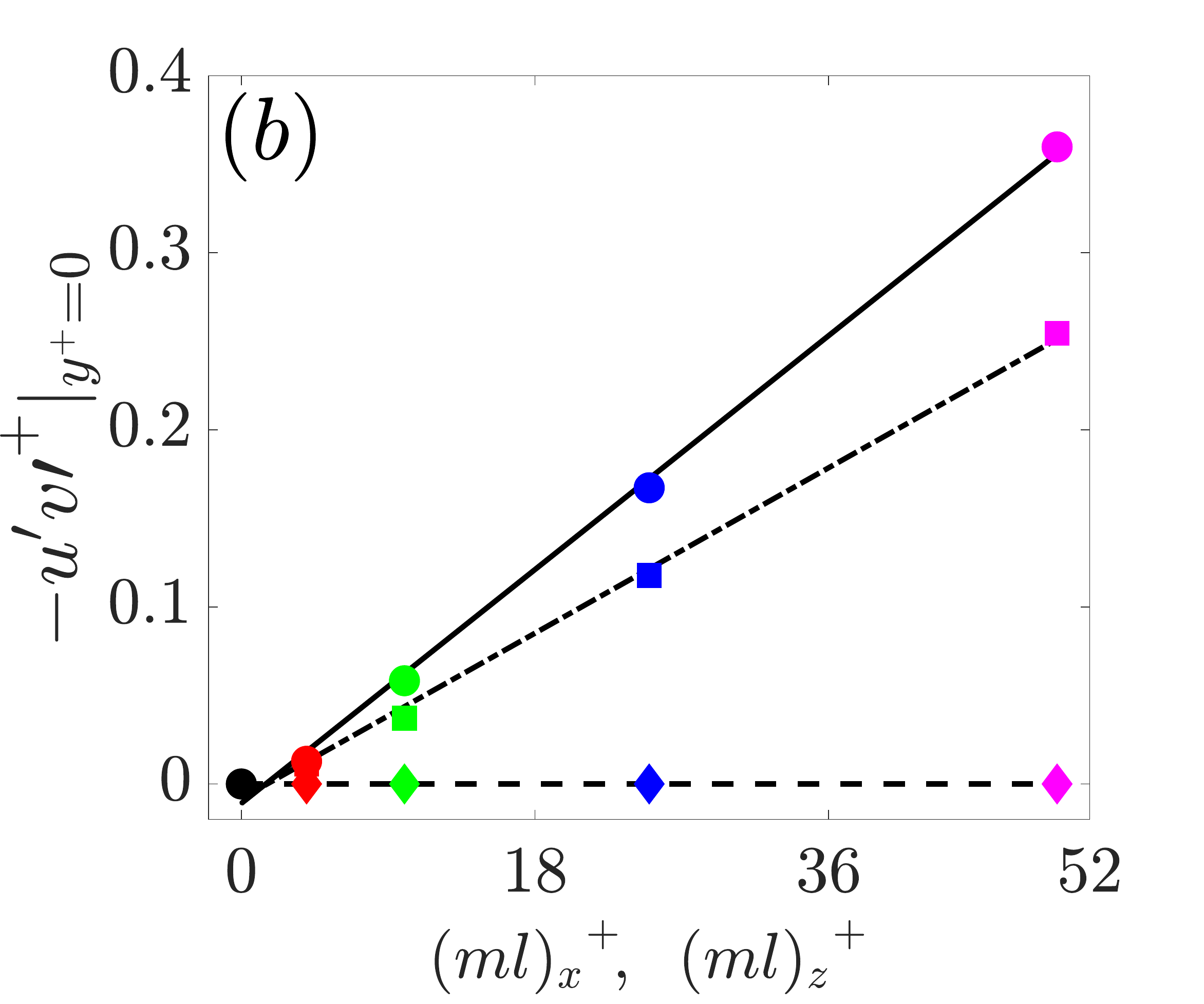}
     {\phantomcaption\label{fig:ml_vs_uv}}
   \end{subfigure}%
   \begin{subfigure}[tbp]{.335\textwidth}
     \includegraphics[width=1\linewidth]{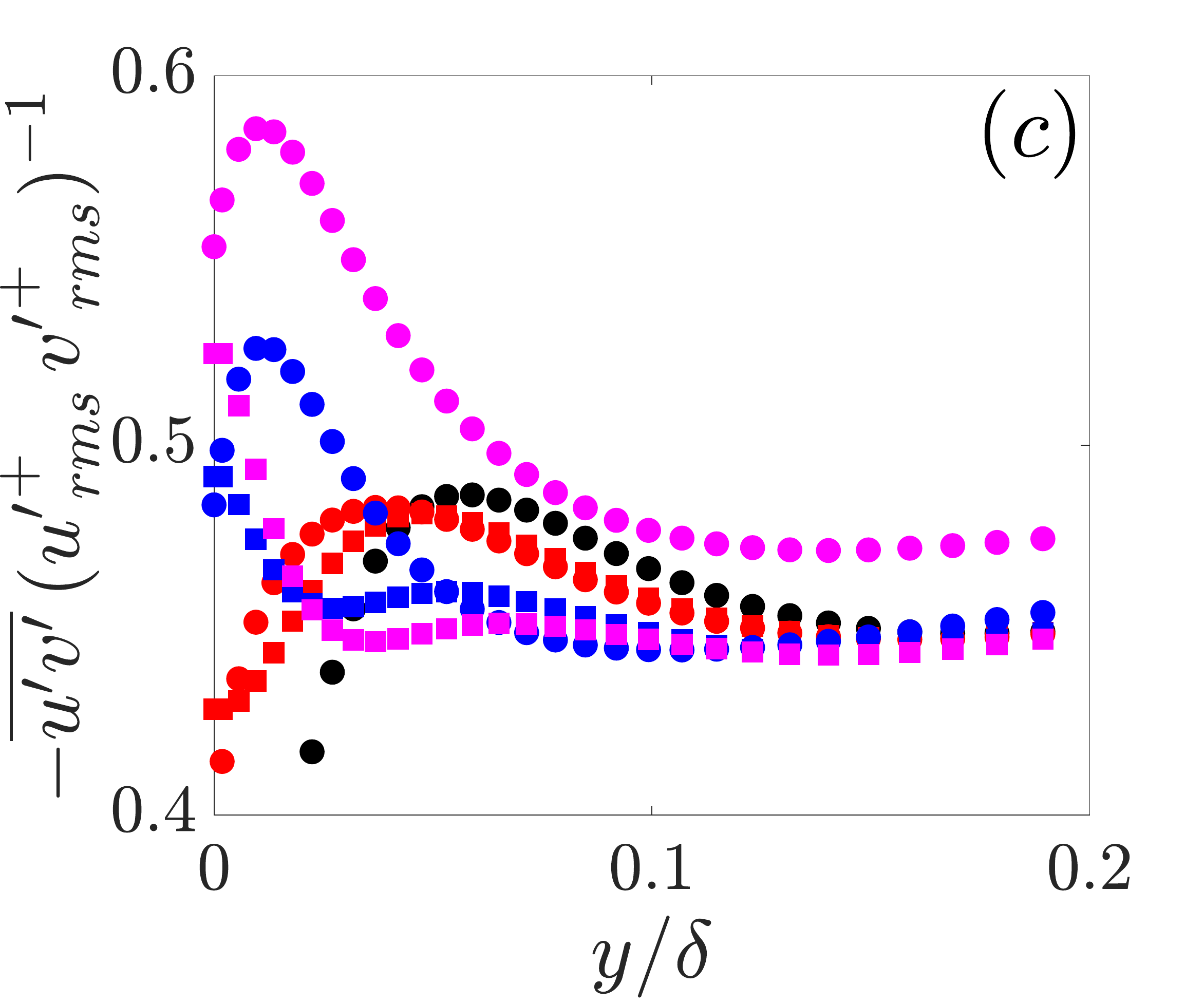}
     {\phantomcaption\label{fig:reystress_vs_uv}}
   \end{subfigure}
   \vspace*{-3mm}
   \caption{Wall-normal velocity fluctuations (\emph{a}), Reynolds shear stress (\emph{b}), correlation coefficient of $u^{\prime}$ and $v^{\prime}$ (\emph{c}) versus ${({m}{\ell})_z}^+$ and ${({m}{\ell})_x}^+$ at the boundary plane (${y^+}=0$). {\Large\textcolor{black}{$\bullet$}}, Smooth-wall data; {\Large\textcolor{red}{$\bullet$}}, L2MZ2; {\Large\textcolor{green}{$\bullet$}}, L2MZ5; {\Large\textcolor{blue}{$\bullet$}}; L5MZ5; {\Large\textcolor{magenta}{$\bullet$}}; L5MZ10; {$\scriptstyle\textcolor{red}{\blacksquare}$}, L2M2; {$\scriptstyle\textcolor{green}{\blacksquare}$}, L2M5; {$\scriptstyle\textcolor{blue}{\blacksquare}$}, L5M5; {$\scriptstyle\textcolor{magenta}{\blacksquare}$}, L5M10; {$\textcolor{red}{\blacklozenge}$}, L2MX2; {$\textcolor{green}{\blacklozenge}$}, L2MX5; {$\textcolor{blue}{\blacklozenge}$}, L5MX5; {$\textcolor{magenta}{\blacklozenge}$}, L5MX10. $-\!-$, linear fit $0.007{({m}{\ell})_z}^+$; {$-\;\text{-}\;-$}, linear fit $0.005{({m}{\ell})}^+$ in ($a$) and ($b$); $-\;-$ linear fit $0.001{({m}{\ell})_x}^+$ in ($a$) and $0{({m}{\ell})_x}^+$ in ($b$). The data for only a few cases are shown in ($c$) for clarity.} 
   \label{fig:ml_vs_vrms_uv}
   \end{center}
\end{figure}

\Cref{fig:ml_vs_vrms} shows that the r.m.s of the wall-normal velocity fluctuations at the boundary plane increase with the the transpiration factors. The intensity of transpiration occurring for cases with spanwise-coupled transpiration only ( {\Large{$\bullet$}} symbols) is considerably more than those with streamwise-coupled transpiration only ({${\blacklozenge}$} symbols). Interestingly, when both transpiration components are active ($\scriptstyle{\blacksquare}$ symbols), the wall transpiration is smaller compared to the L$<\!\!\cdot\!\!>$MZ$<\!\!\cdot\!\!>$ cases. The latter cases exclusively couple transpiration to changes in spanwise shear, leading to a higher frequency of sweep and ejection events at the boundary plane and the greatest degree of Reynolds shear stress generation for a given combination of slip and transpiration lengths. This is confirmed in \cref{fig:ml_vs_uv}, where the Reynolds shear stress at the boundary plane is shown and demonstrates that a redistribution of momentum between the spanwise and wall-normal velocity components is necessary for Reynolds shear-stress generation.

%%%%%%%%%%%%%%%%%%%%%%%%%%%%%%%%%%%%%%%%%%%%%%%%%%%%%%%%%%%%%%%%%%%%%%%%%
%%%%%%%%%%%%%%%%%%%%%%%%%%%%%%%%%%%%%%%%%%%%%%%%%%%%%%%%%%%%%%%%%%%%%%%%%
%%%%%%%%%%%%%%%%%%%%%%%%%%%%%%%%%%%%%%%%%%%%%%%%%%%%%%%%%%%%%%%%%%%%%%%%%
%%%%%%%%%%%%%%%%%%%%%%%%%%%%%%%%%%%%%%%%%%%%%%%%%%%%%%%%%%%%%%%%%%%%%%%%%

\subsection{Changes in the structure of near-wall turbulence}

\label{changesinturbulence}
The TRM has the capacity to modify near-wall turbulence in a similar manner as actual rough surfaces up to a certain extent. In the case of small surface textures, where the modification of turbulence is smooth-wall-like, the TRM serves as a robust model of roughness. In the case of larger surface textures, the TRM is still able to cause the typical characteristics observed in the the presence of roughness. These are now elaborated upon.

The streamwise velocity fluctuations, ${u^{\prime}}^+$, undergo an increase in their near-wall peak for cases L2M2, L2M5, L5M5 (\hyperref[fig:group1:u_w_rms]{figure \ref*{fig:group1:u_w_rms}}) and L2MX5, L5MX5 (\hyperref[fig:group2:u_w_rms]{figure \ref*{fig:group2:u_w_rms}}). \citet{ibrahim2020smoothwalllike} attributed this change to the streamwise streaks ``having more room to decay to zero'', analogous to a thickening of the viscous sub-layer.
In contrast, cases L5M10, L10M10 (\hyperref[fig:group1:u_w_rms]{figure \ref*{fig:group1:u_w_rms}}) and L2MZ5, L5MZ5, L5MZ10 (\hyperref[fig:group2:u_w_rms]{figure \ref*{fig:group2:u_w_rms}}) have lower near-wall peak values compared to the smooth-wall case. These cases demonstrate differences from smooth-wall turbulence, exhibiting a move toward isotropization and hence an analysis presuming the existence of streaks does not remain applicable to them.

\Cref{fig:reystress_vs_uv} shows the correlation coefficient ${-\overline{u^{\prime}v^{\prime}}}/{{u^{\prime}}_{rms}^+}{{v^{\prime}}_{rms}^+}$ as a function of the wall-normal coordinate. The near-wall peak of this coefficient for smooth-wall canonical turbulence marks the position of maximum turbulence production and maximum streamwise fluctuations \citep{kim_moin_moser_1987}. The distribution of the correlation coefficient also serves as an indicator of the structure of turbulence \citep{sabot_comte-bellot_1976}. The peak coefficient for cases L2M2 and L2MZ2 is very close to that of the smooth-wall's, but is displaced further toward the boundary plane. This is consistent with the previously made observation that these cases retain the structure of canonical smooth-wall turbulence with it effectively having undergone a translation closer to the boundary; the very principle underlying the virtual-origin framework.
Cases L5M5, L5MZ5, L5M10 and L5MZ10 unequivocally differ from smooth-wall turbulence. The peak and thus the position of maximum turbulence production is much closer to the boundary plane and considerably enhanced, with the L$<\!\!\cdot\!\!>$MZ$<\!\!\cdot\!\!>$ cases displaying the greatest level of enhancement, all of which is consistent with the observations made thus far.

%%%%%%%%%%%%%%%%%%%%%%%%%%%%%%%%%%%%%%%%%%%%%%%%%%%%%%%%%%%%%%%%%%%%%%%%%
%%%%%%%%%%%%%%%%%%%%%%%%%%%%%%%%%%%%%%%%%%%%%%%%%%%%%%%%%%%%%%%%%%%%%%%%%
\begin{figure}
    \begin{center}
    \hspace*{-1mm}
    \begin{subfigure}[tbp]{.2705\textwidth}
        \includegraphics[width=1\linewidth]{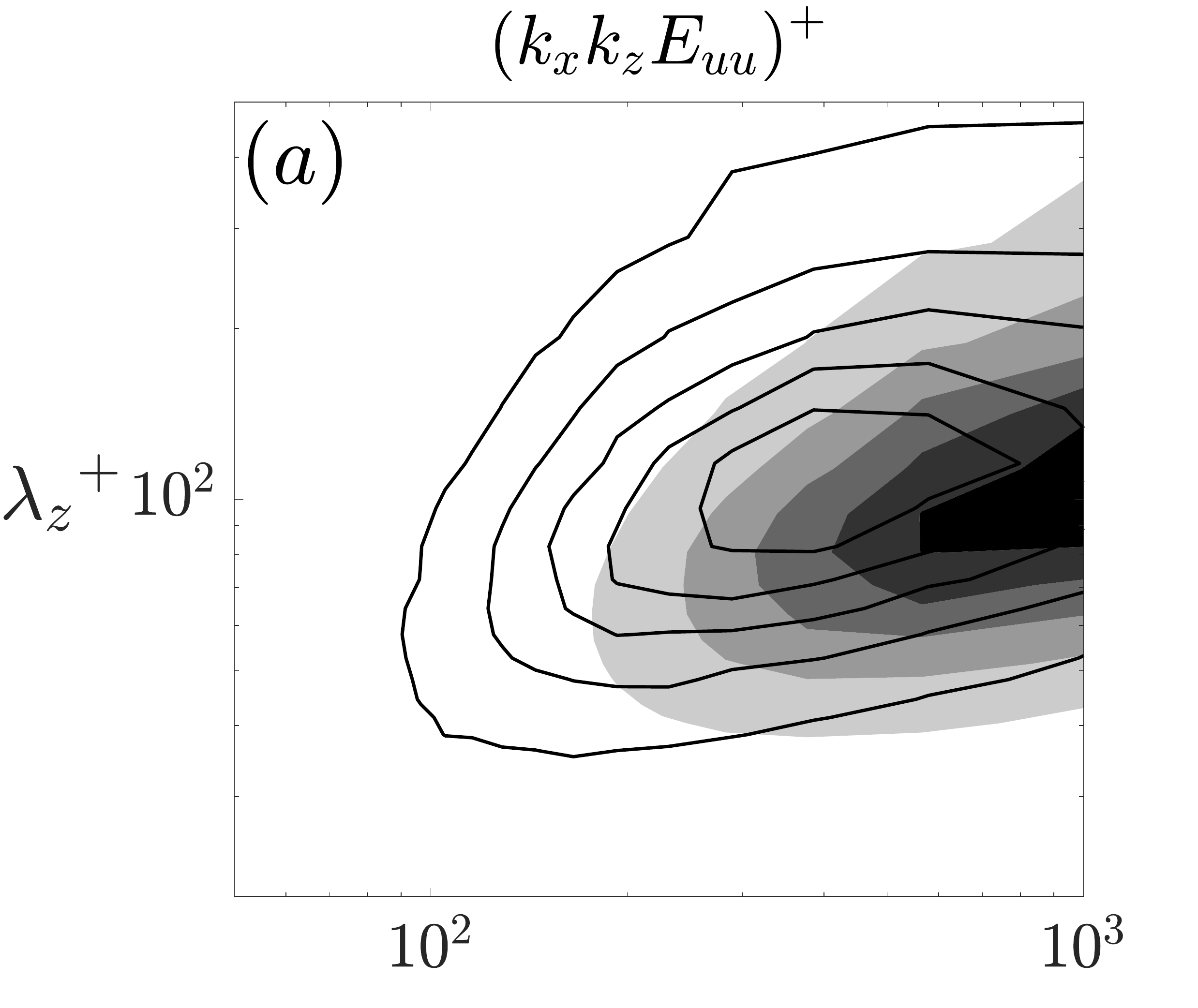}
        {\phantomcaption\label{spectra_L5MZ10:u}}
        \vspace*{-4mm}
    \end{subfigure}%
    \hspace*{-3mm}
    \begin{subfigure}[tbp]{.257\textwidth}
        \includegraphics[width=1\linewidth]{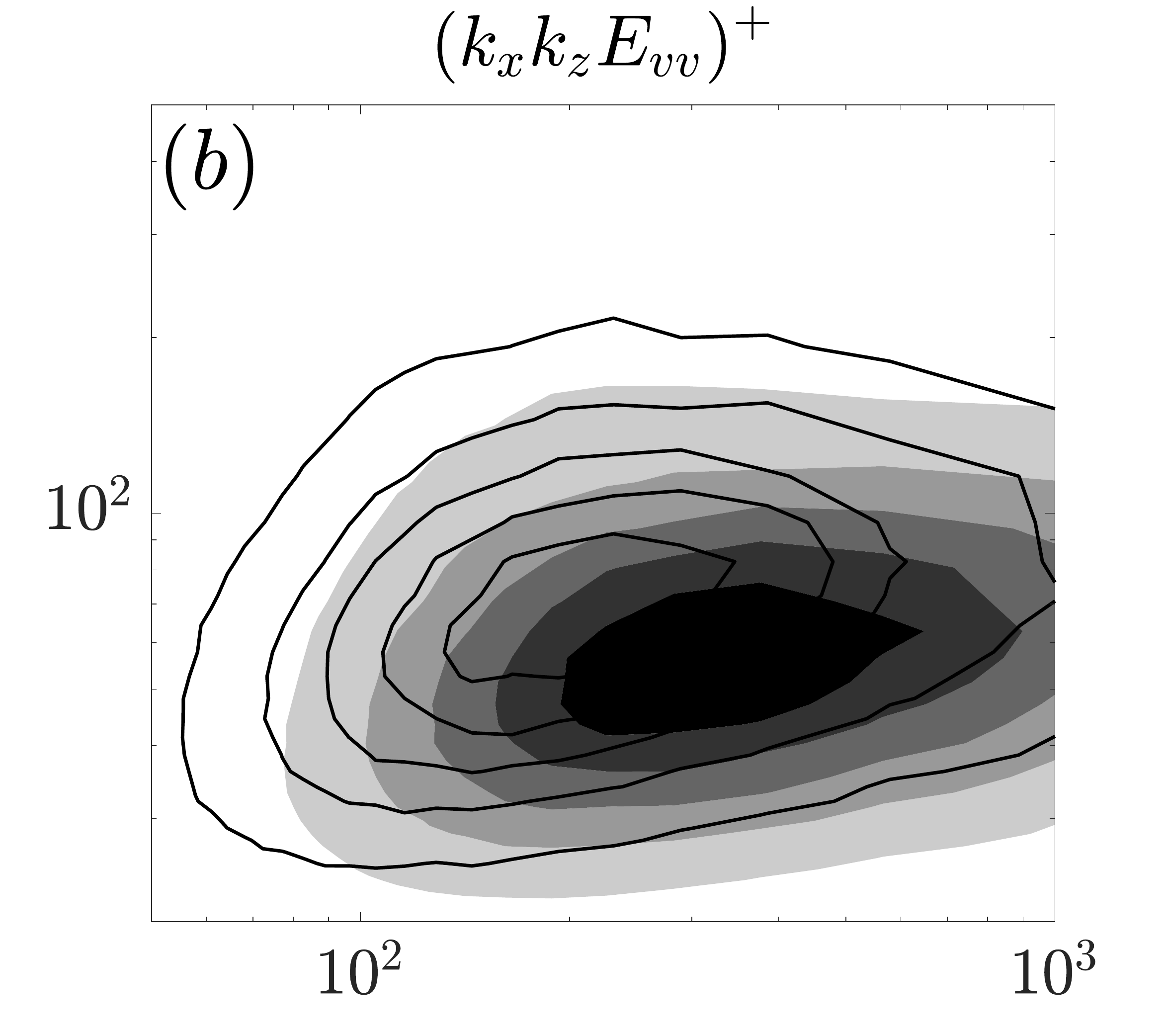}
        {\phantomcaption\label{spectra_L5MZ10:v}}
        \vspace*{-4mm}
    \end{subfigure}%
    \hspace*{-3mm}
    \begin{subfigure}[tbp]{.257\textwidth}
        \includegraphics[width=1\linewidth]{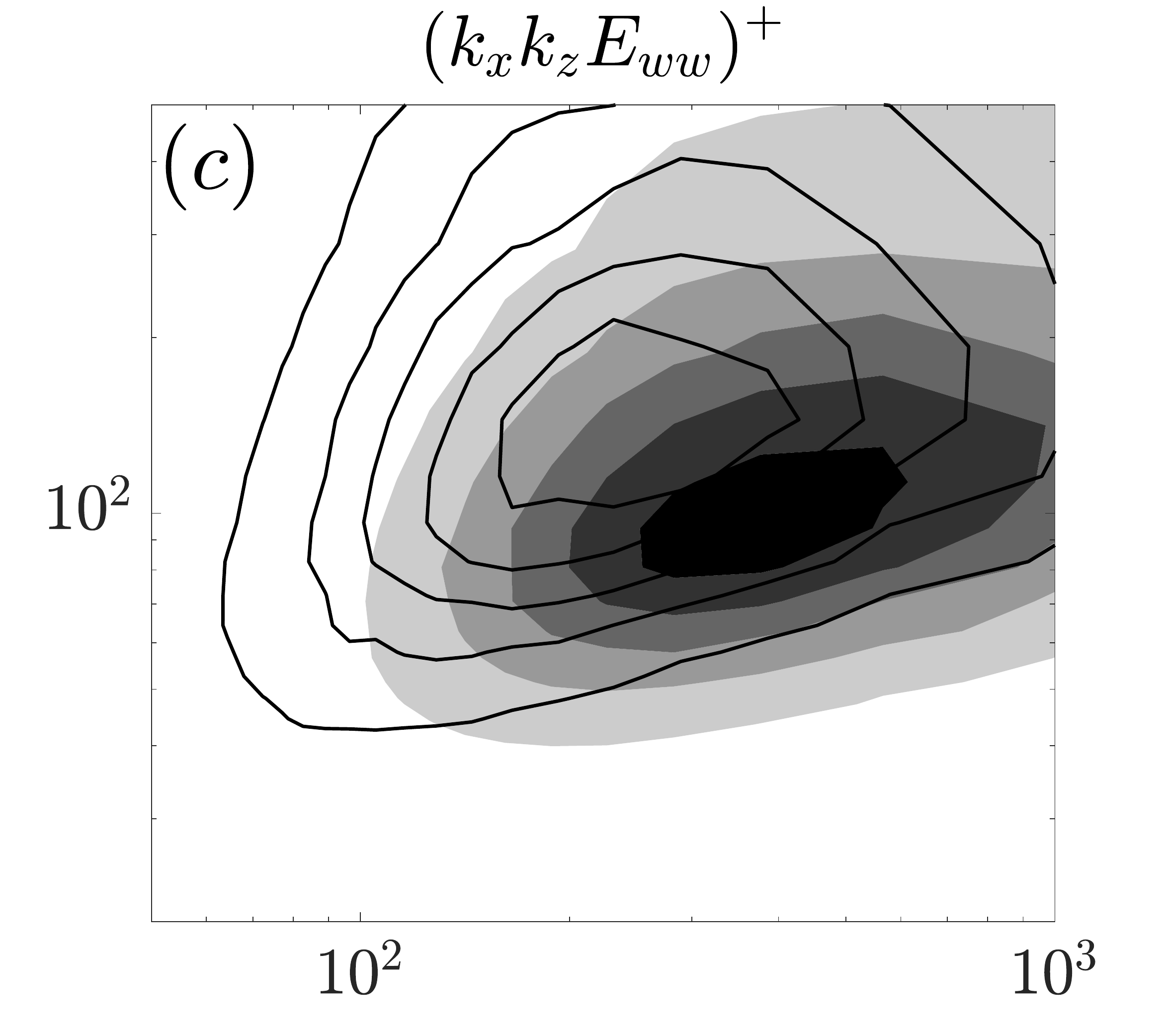}
        {\phantomcaption\label{spectra_L5MZ10:w}}
        \vspace*{-4mm}
    \end{subfigure}%
    \hspace*{-3mm}
    \begin{subfigure}[tbp]{.257\textwidth}
        \includegraphics[width=1\linewidth]{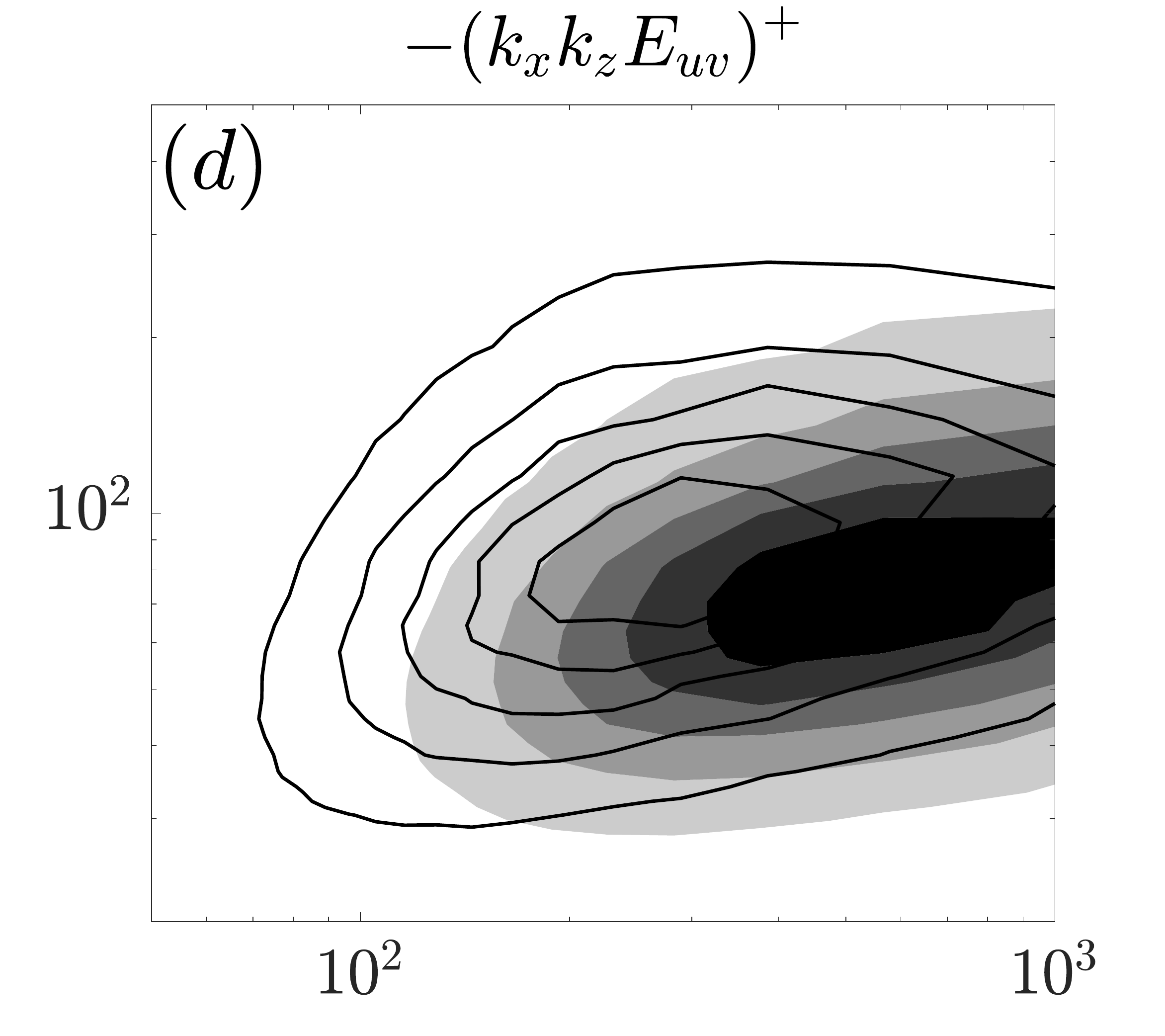}
        {\phantomcaption\label{spectra_L5MZ10:uv}}
        \vspace*{-4mm}
    \end{subfigure}
    \hspace*{-1mm}
    \begin{subfigure}[tbp]{.2705\textwidth}
        \includegraphics[width=1\linewidth]{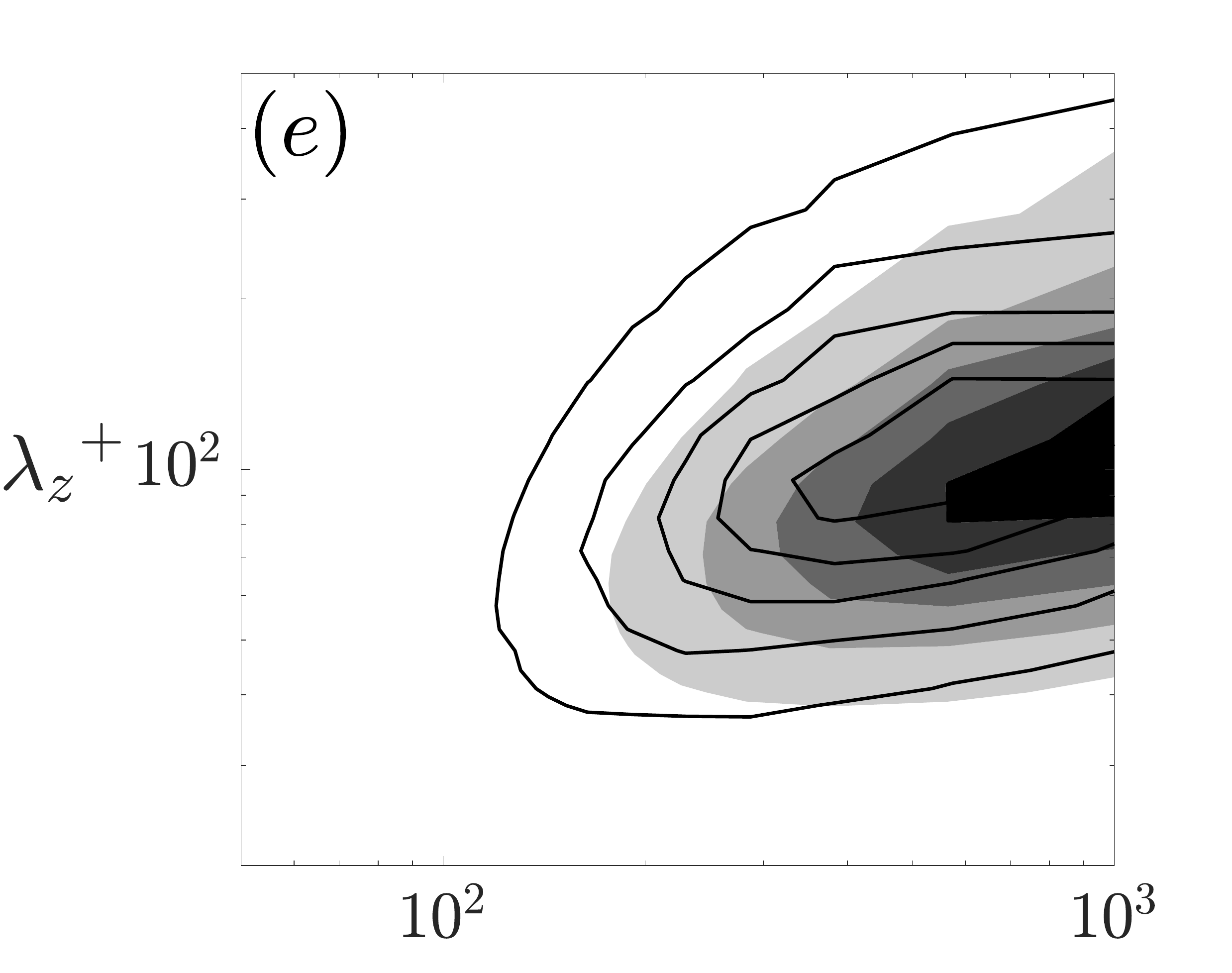}
        {\phantomcaption\label{spectra_L5MZ5:u}}
        \vspace*{-4mm}
    \end{subfigure}%
    \hspace*{-3mm}
    \begin{subfigure}[tbp]{.257\textwidth}
        \includegraphics[width=1\linewidth]{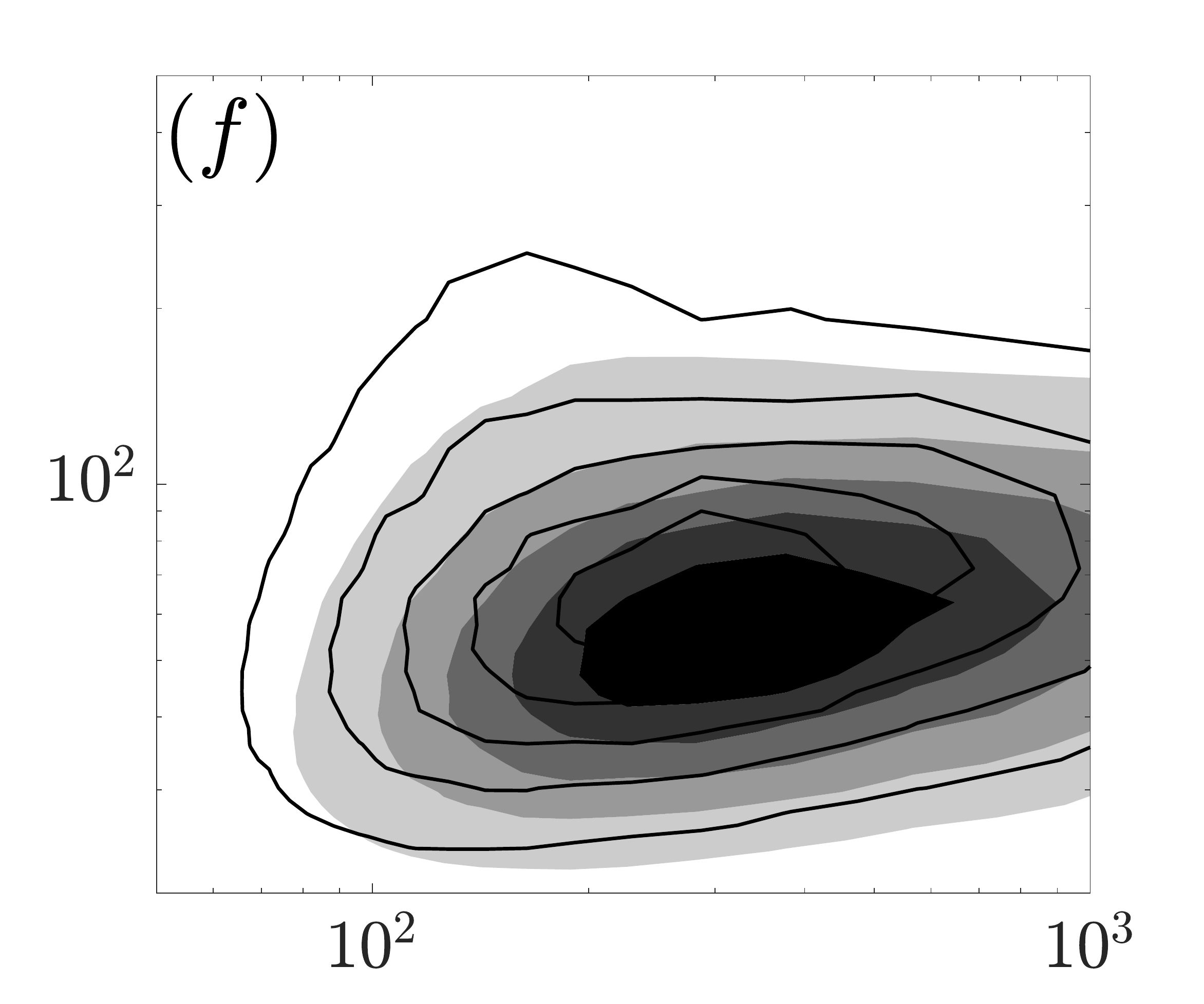}
        {\phantomcaption\label{spectra_L5MZ5:v}}
        \vspace*{-4mm}
    \end{subfigure}%
    \hspace*{-3mm}
    \begin{subfigure}[tbp]{.257\textwidth}
        \includegraphics[width=1\linewidth]{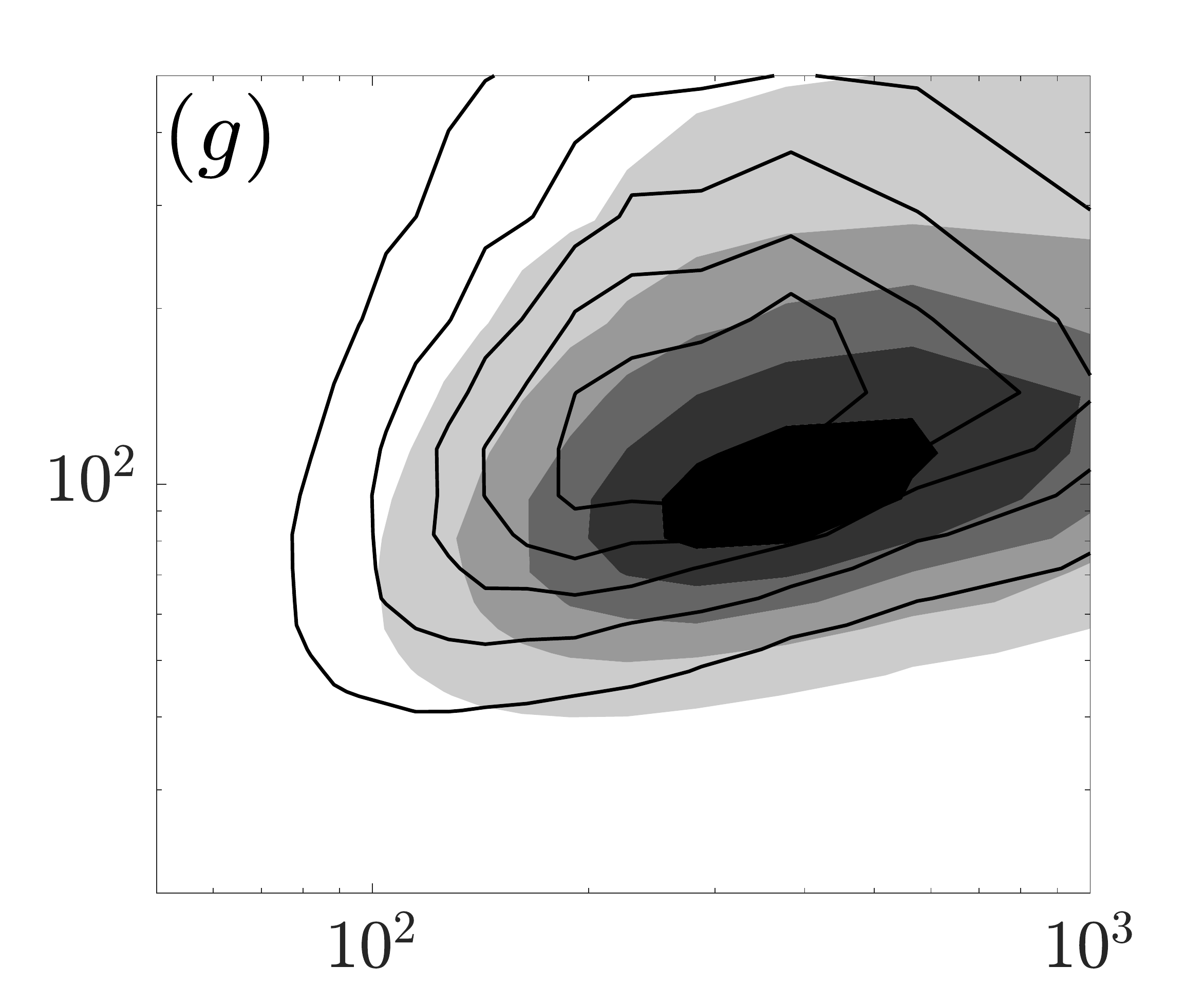}
        {\phantomcaption\label{spectra_L5MZ5:w}}
        \vspace*{-4mm}
    \end{subfigure}%
    \hspace*{-3mm}
    \begin{subfigure}[tbp]{.257\textwidth}
        \includegraphics[width=1\linewidth]{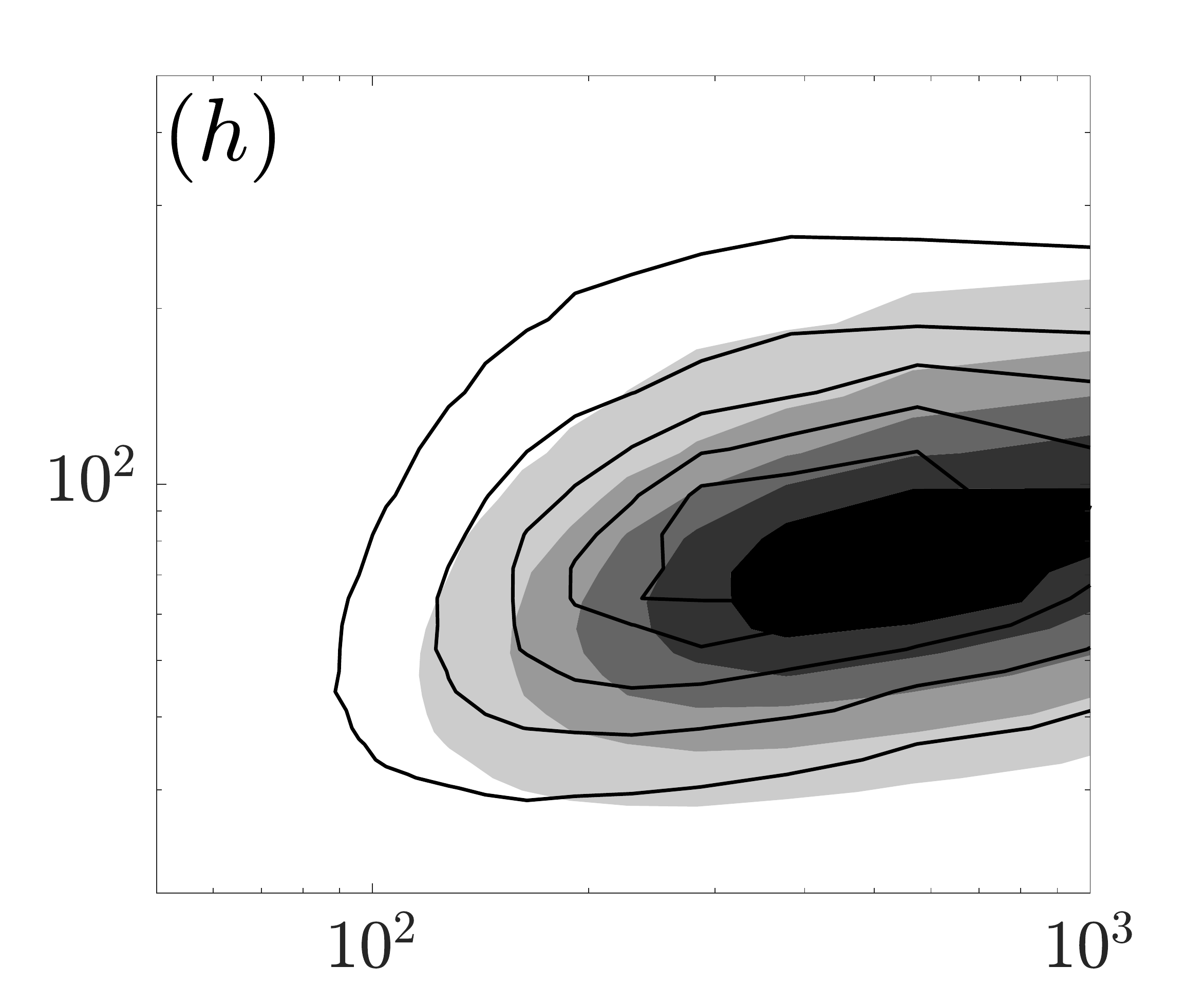}
        {\phantomcaption\label{spectra_L5MZ5:uv}}
        \vspace*{-4mm}
    \end{subfigure}
    \hspace*{-1mm}
    \begin{subfigure}[tbp]{.2705\textwidth}
        \includegraphics[width=1\linewidth]{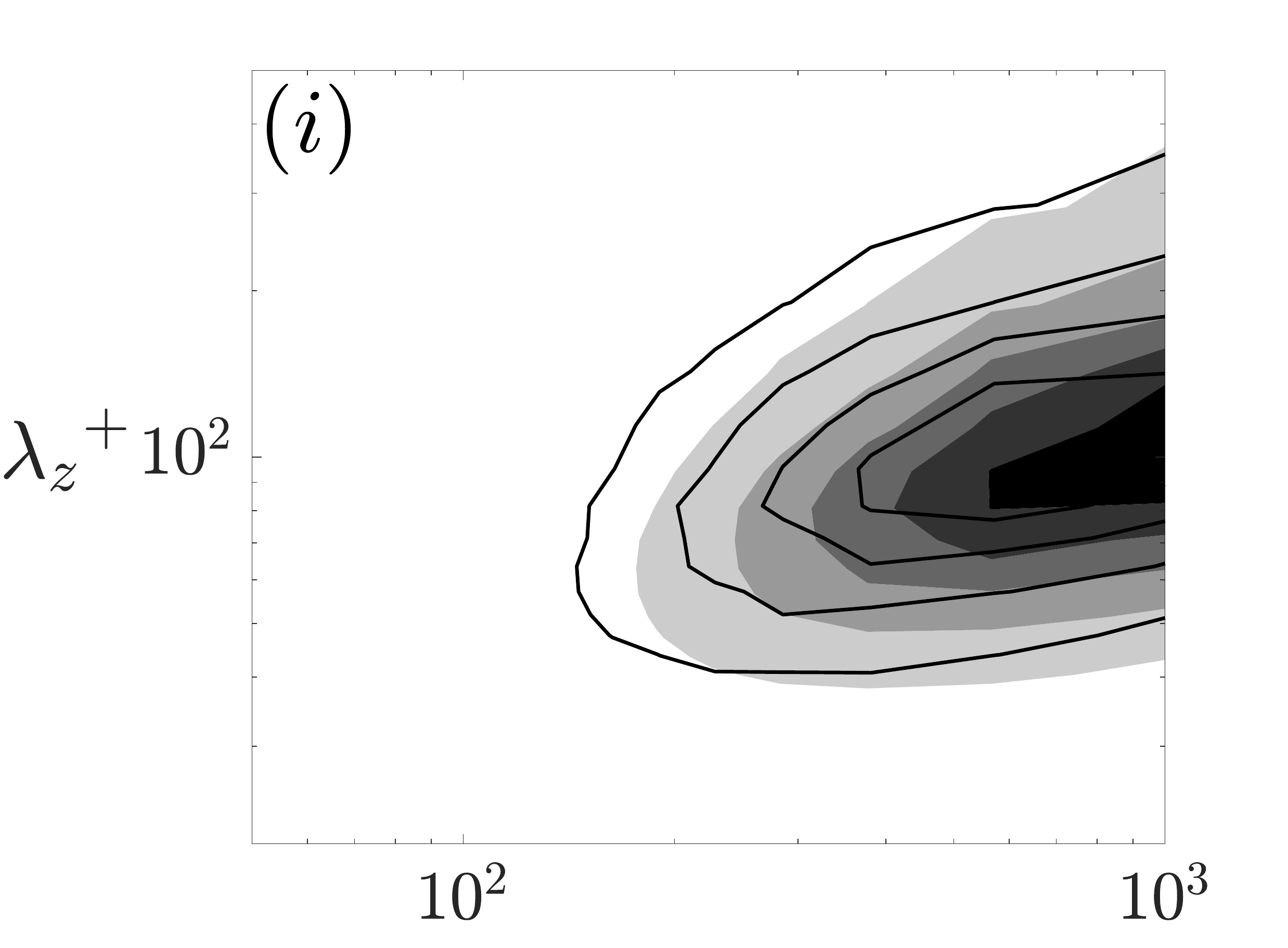}
        {\phantomcaption\label{spectra_L2MZ5:u}}
        \vspace*{-4mm}
    \end{subfigure}%
    \hspace*{-3mm}
    \begin{subfigure}[tbp]{.257\textwidth}
        \includegraphics[width=1\linewidth]{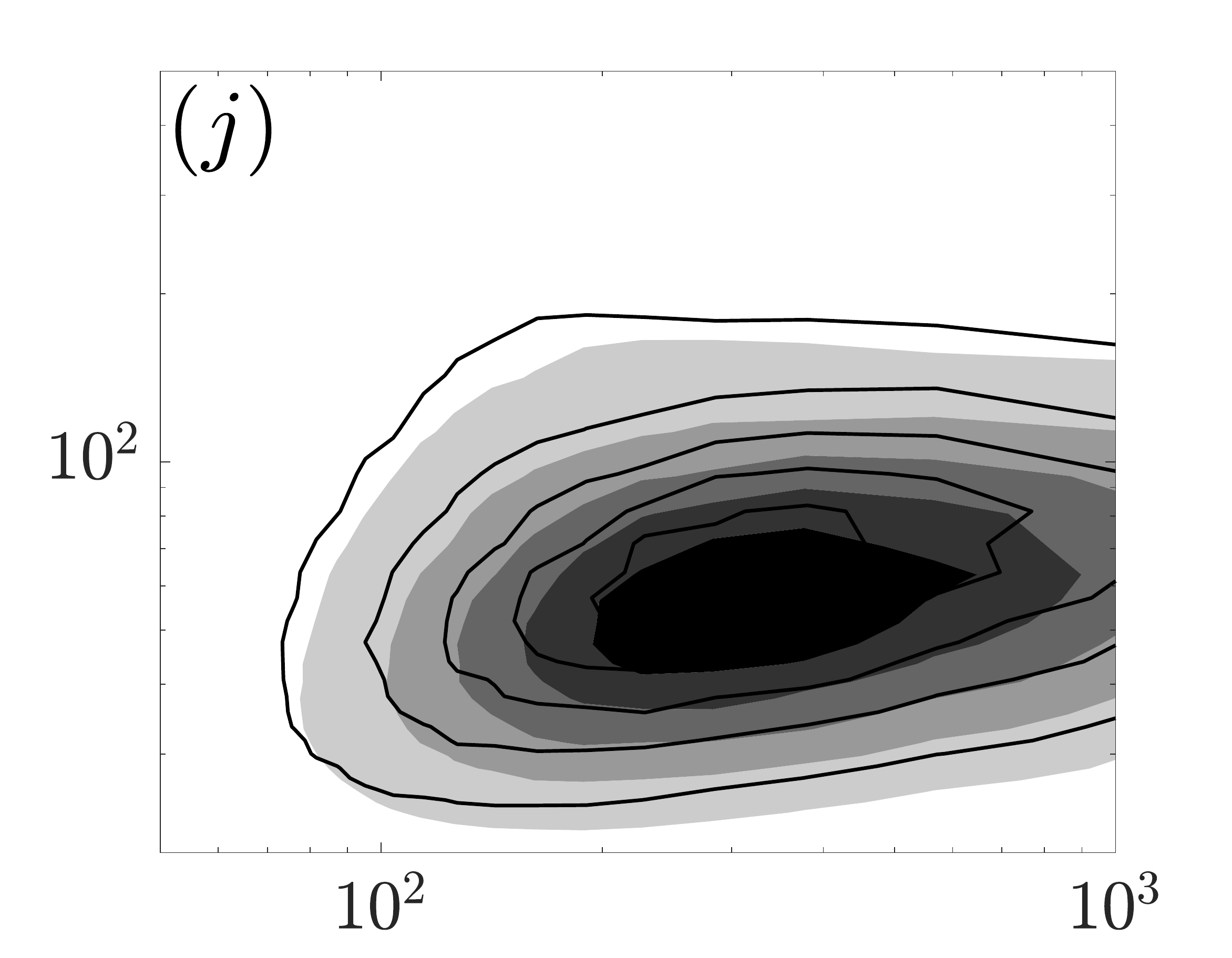}
        {\phantomcaption\label{spectra_L2MZ5:v}}
        \vspace*{-4mm}
    \end{subfigure}%
    \hspace*{-3mm}
    \begin{subfigure}[tbp]{.257\textwidth}
        \includegraphics[width=1\linewidth]{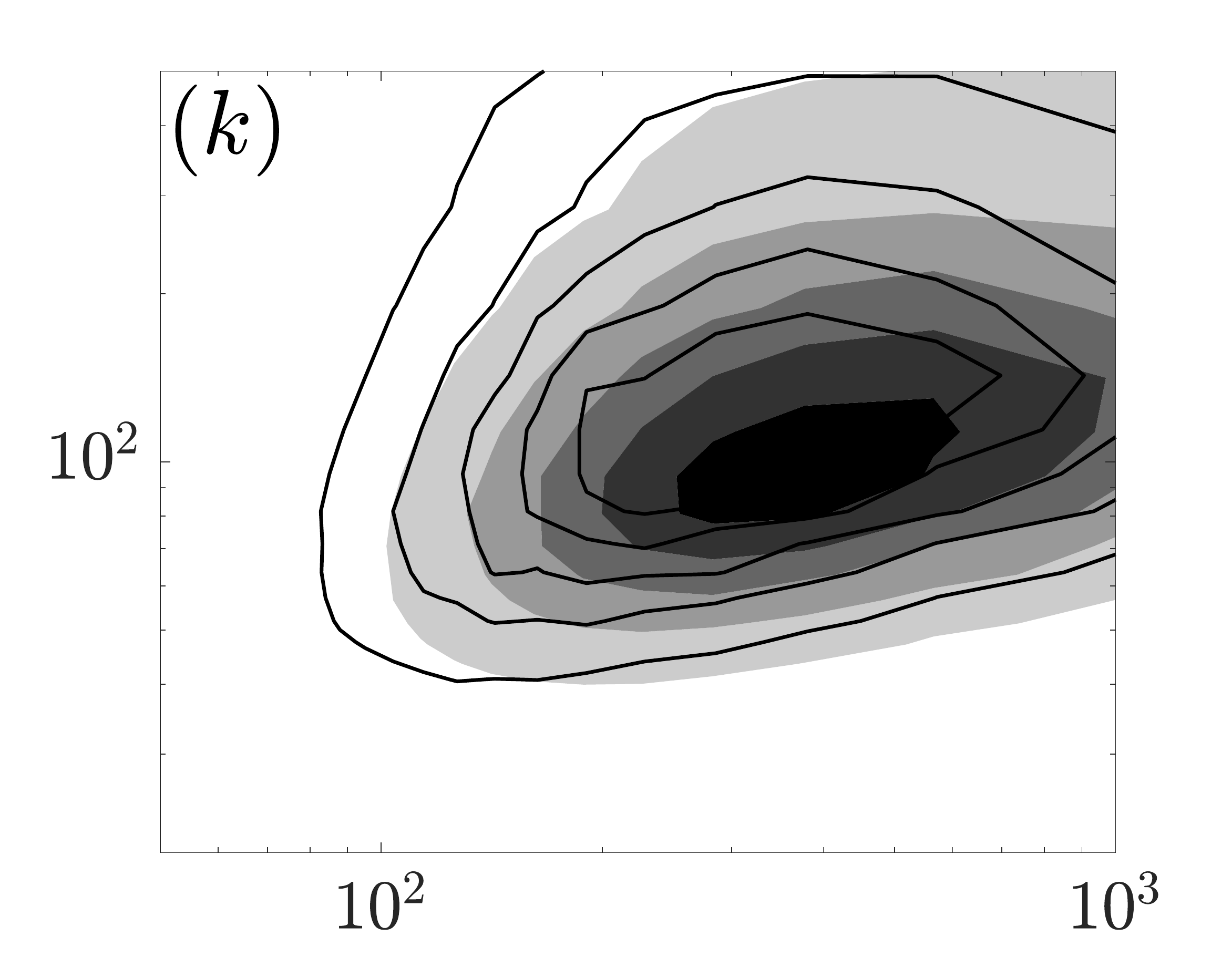}
        {\phantomcaption\label{spectra_L2MZ5:w}}
        \vspace*{-4mm}
    \end{subfigure}%
    \hspace*{-3mm}
    \begin{subfigure}[tbp]{.257\textwidth}
        \includegraphics[width=1\linewidth]{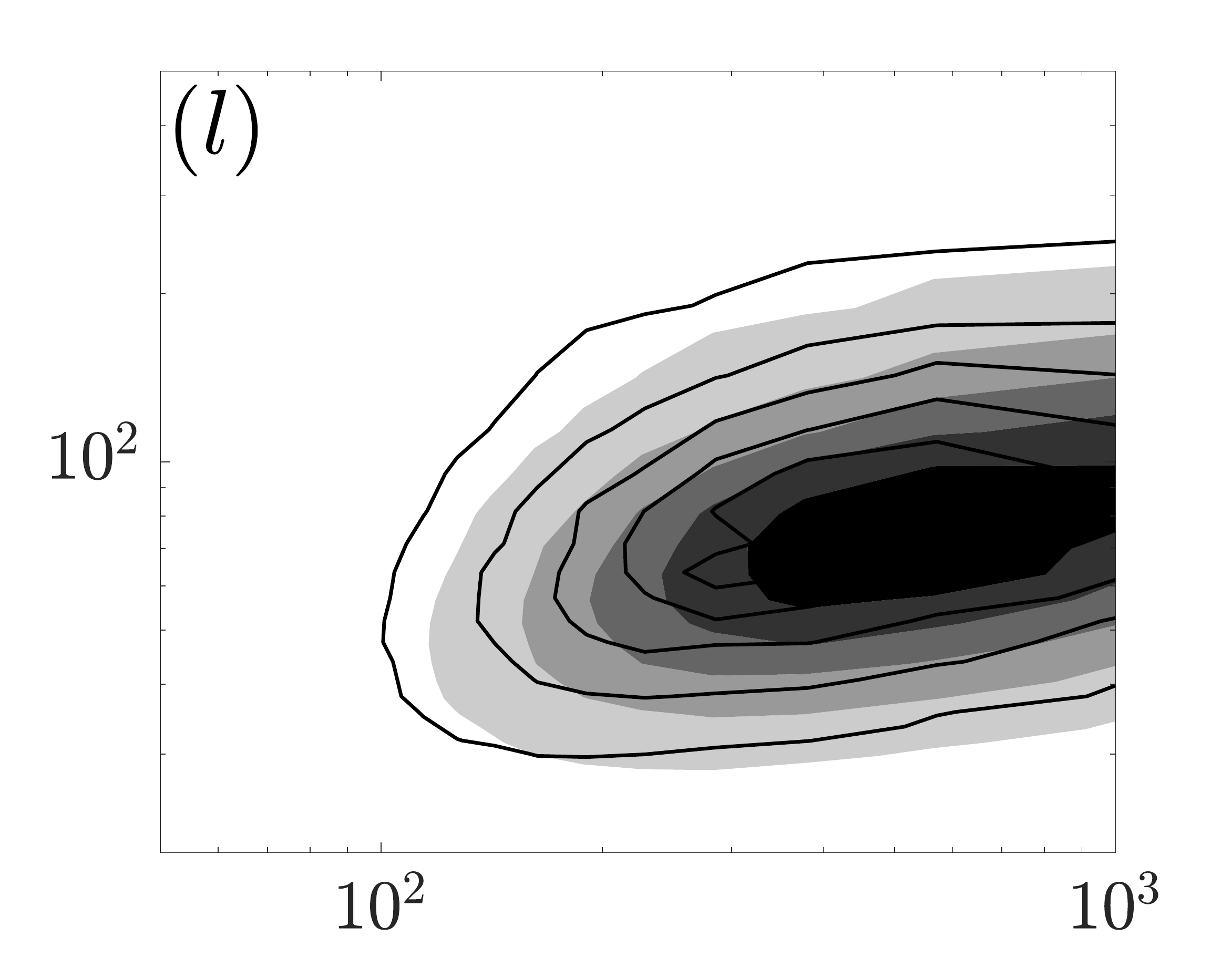}
        {\phantomcaption\label{spectra_L2MZ5:uv}}
        \vspace*{-4mm}
    \end{subfigure}
    \hspace*{-1mm}
    \begin{subfigure}[tbp]{.2705\textwidth}
        \includegraphics[width=1\linewidth]{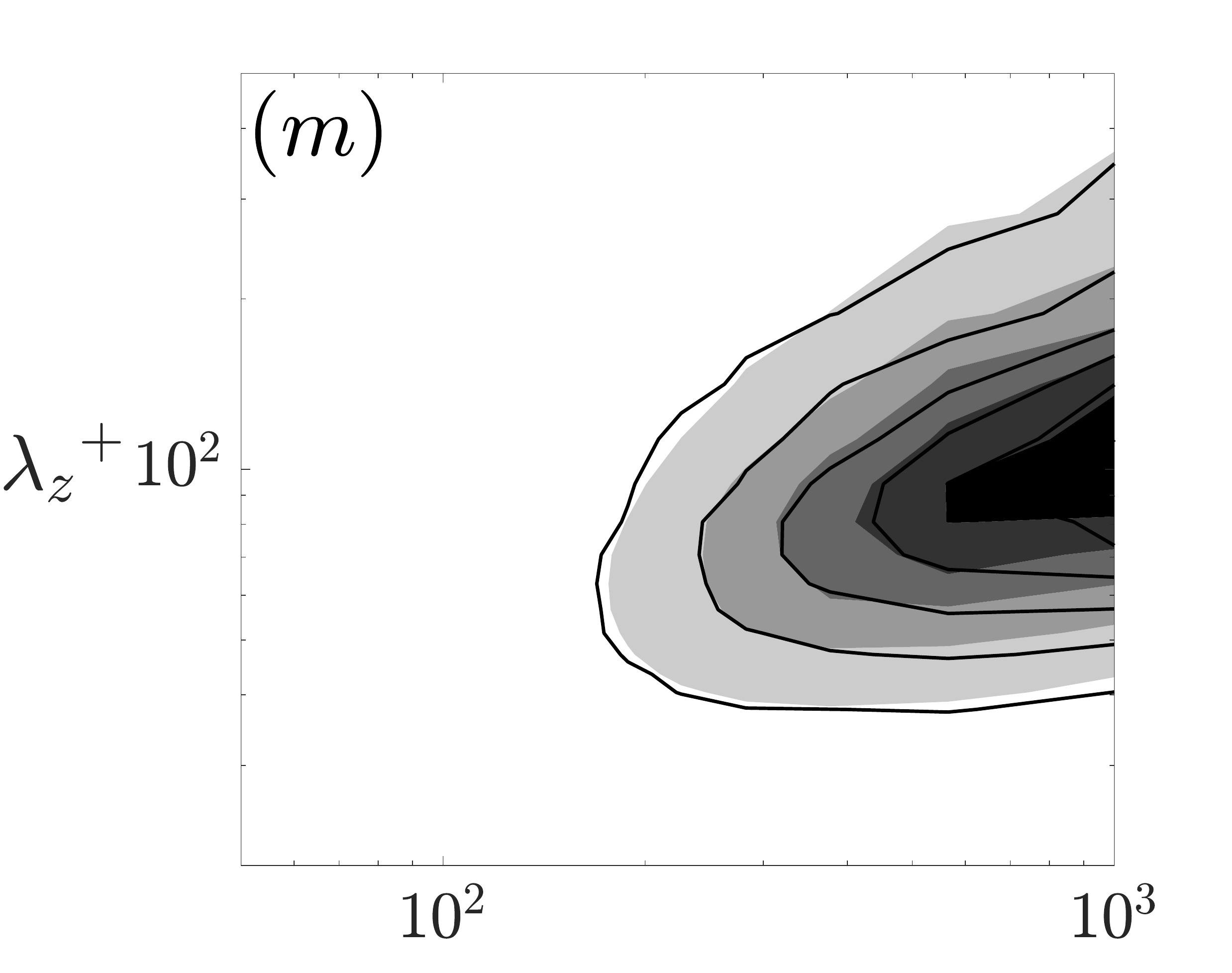}
        {\phantomcaption\label{spectra_L5MX5:u}}
        \vspace*{-4mm}
    \end{subfigure}%
    \hspace*{-3mm}
    \begin{subfigure}[tbp]{.257\textwidth}
        \includegraphics[width=1\linewidth]{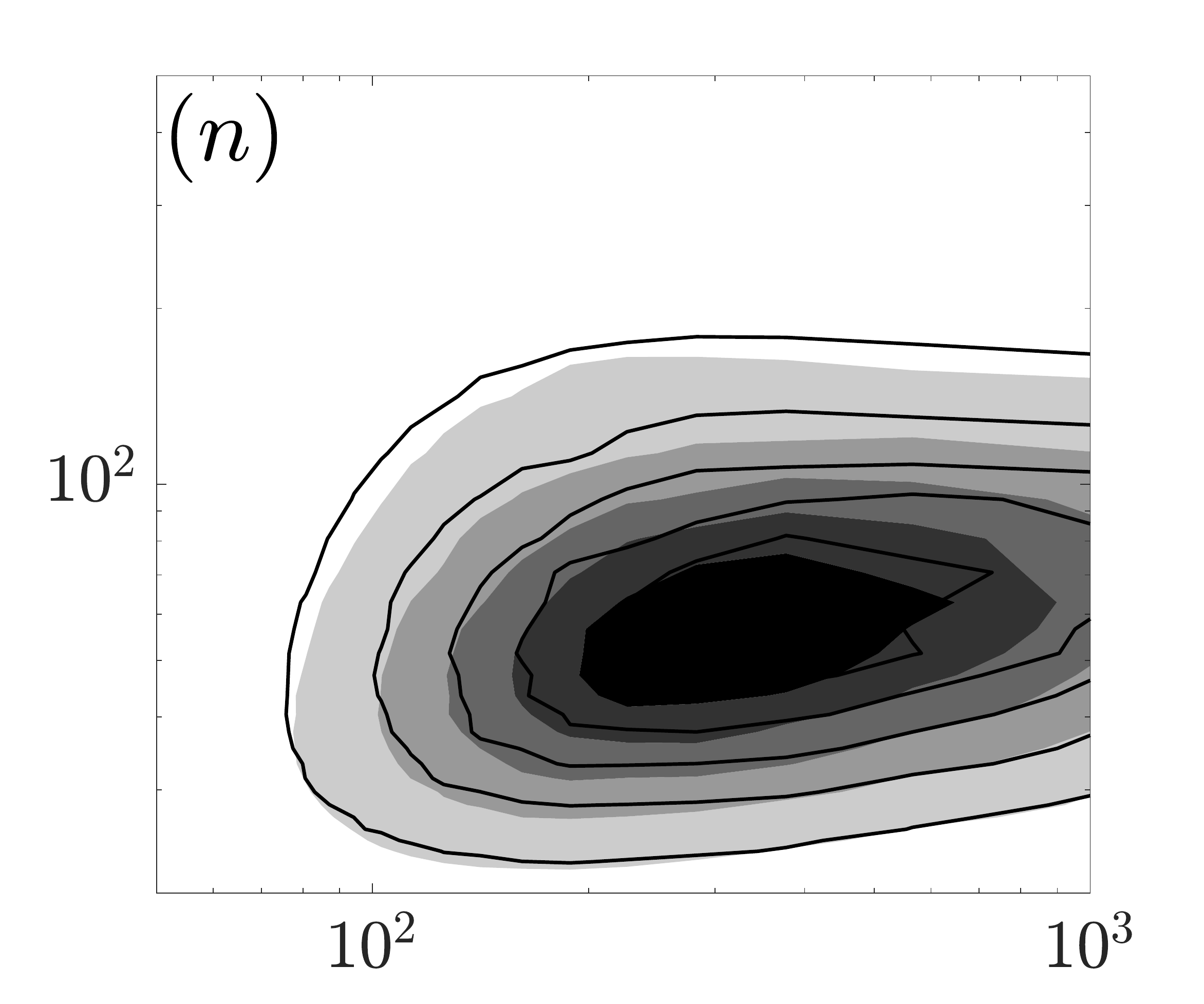}
        {\phantomcaption\label{spectra_L5MX5:v}}
        \vspace*{-4mm}
    \end{subfigure}%
    \hspace*{-3mm}
    \begin{subfigure}[tbp]{.257\textwidth}
        \includegraphics[width=1\linewidth]{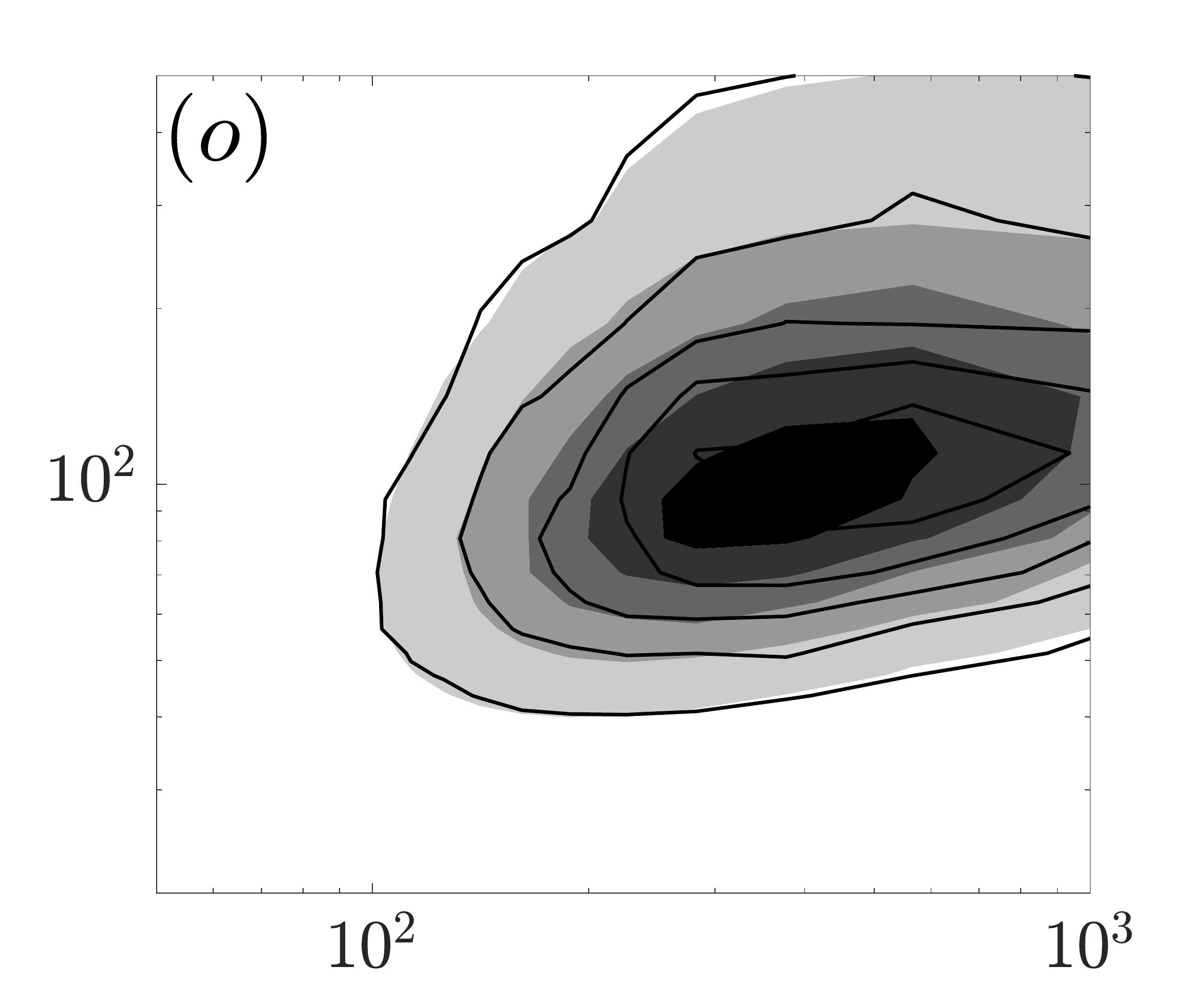}
        {\phantomcaption\label{spectra_L5MX5:w}}
        \vspace*{-4mm}
    \end{subfigure}%
    \hspace*{-3mm}
    \begin{subfigure}[tbp]{.257\textwidth}
        \includegraphics[width=1\linewidth]{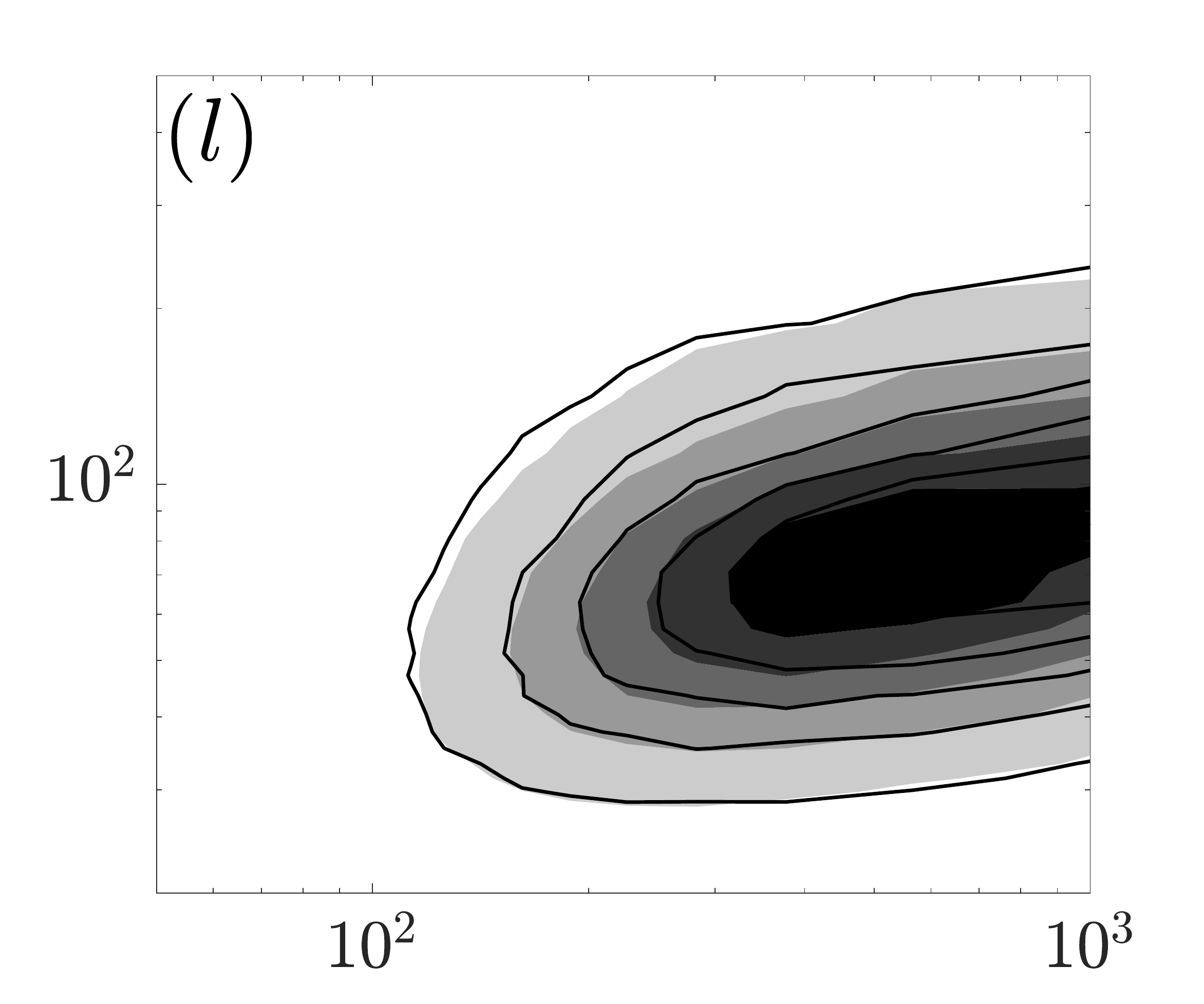}
        {\phantomcaption\label{spectra_L5MX5:uv}}
        \vspace*{-4mm}
    \end{subfigure}
    \hspace*{-1mm}
    \begin{subfigure}[tbp]{.2705\textwidth}
        \includegraphics[width=1\linewidth]{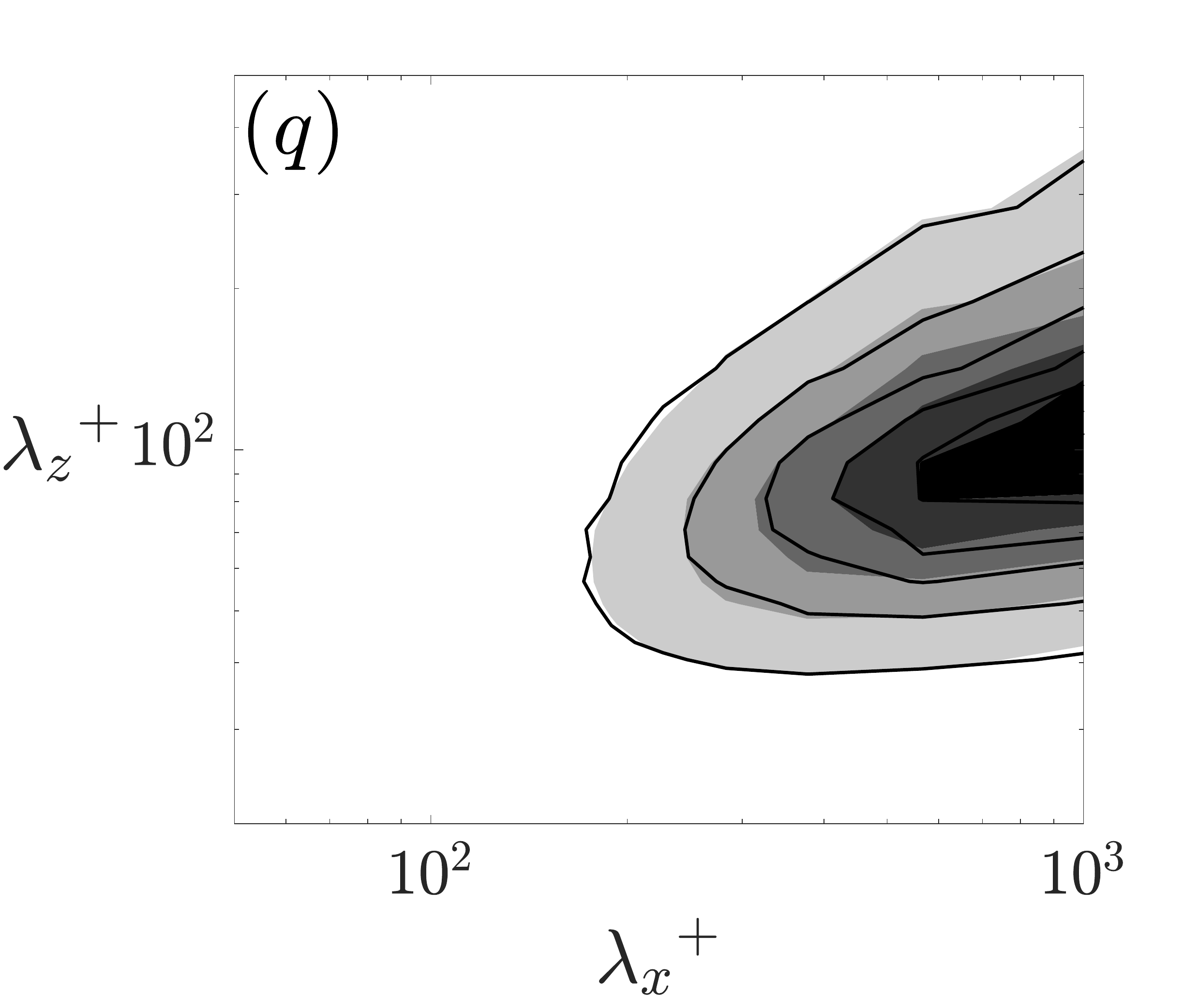}
        {\phantomcaption\label{spectra_L2MX5:u}}
    \end{subfigure}%
    \hspace*{-3mm}
    \begin{subfigure}[tbp]{.257\textwidth}
        \includegraphics[width=1\linewidth]{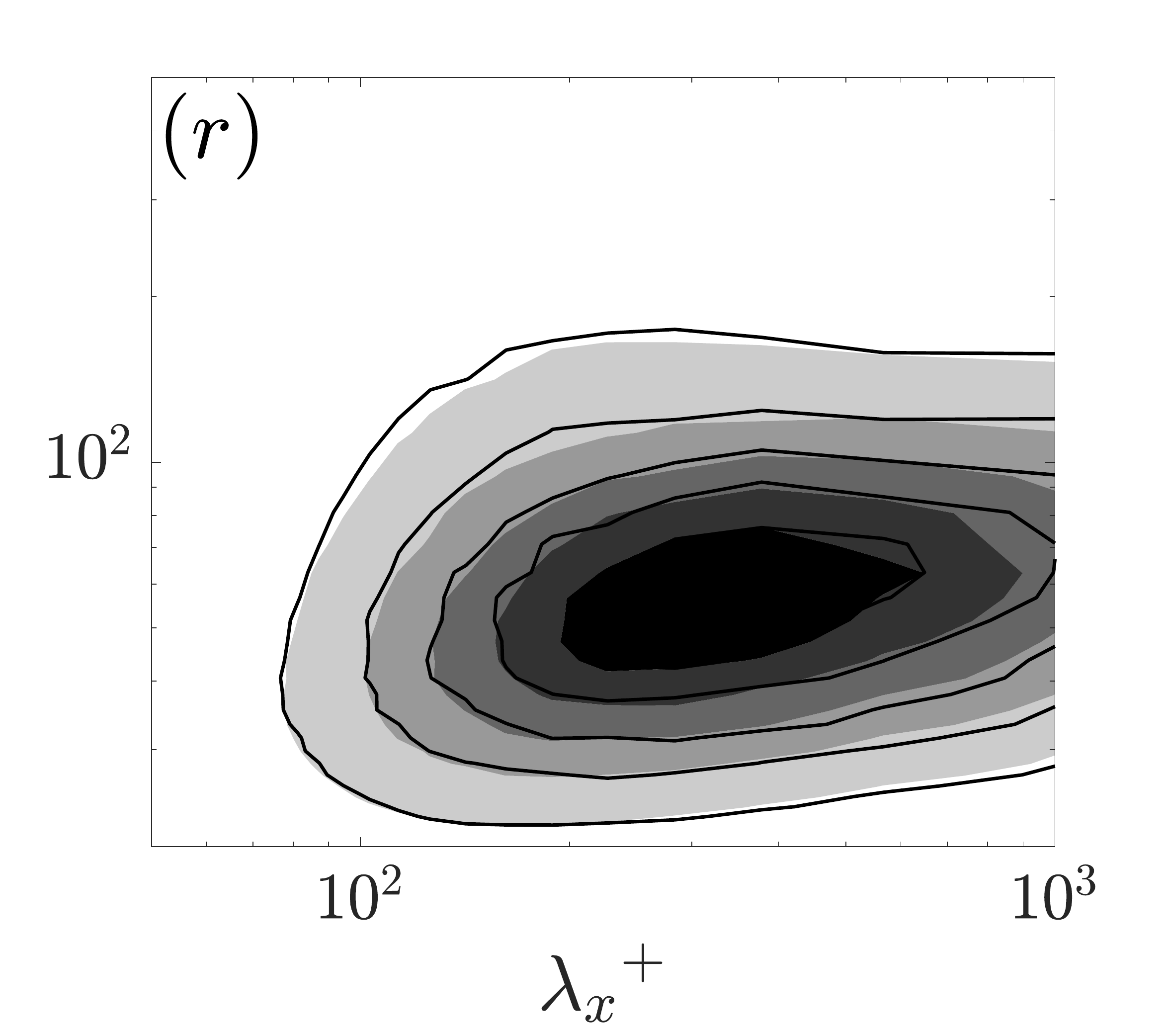}
        {\phantomcaption\label{spectra_L2MX5:v}}
    \end{subfigure}%
    \hspace*{-3mm}
    \begin{subfigure}[tbp]{.257\textwidth}
        \includegraphics[width=1\linewidth]{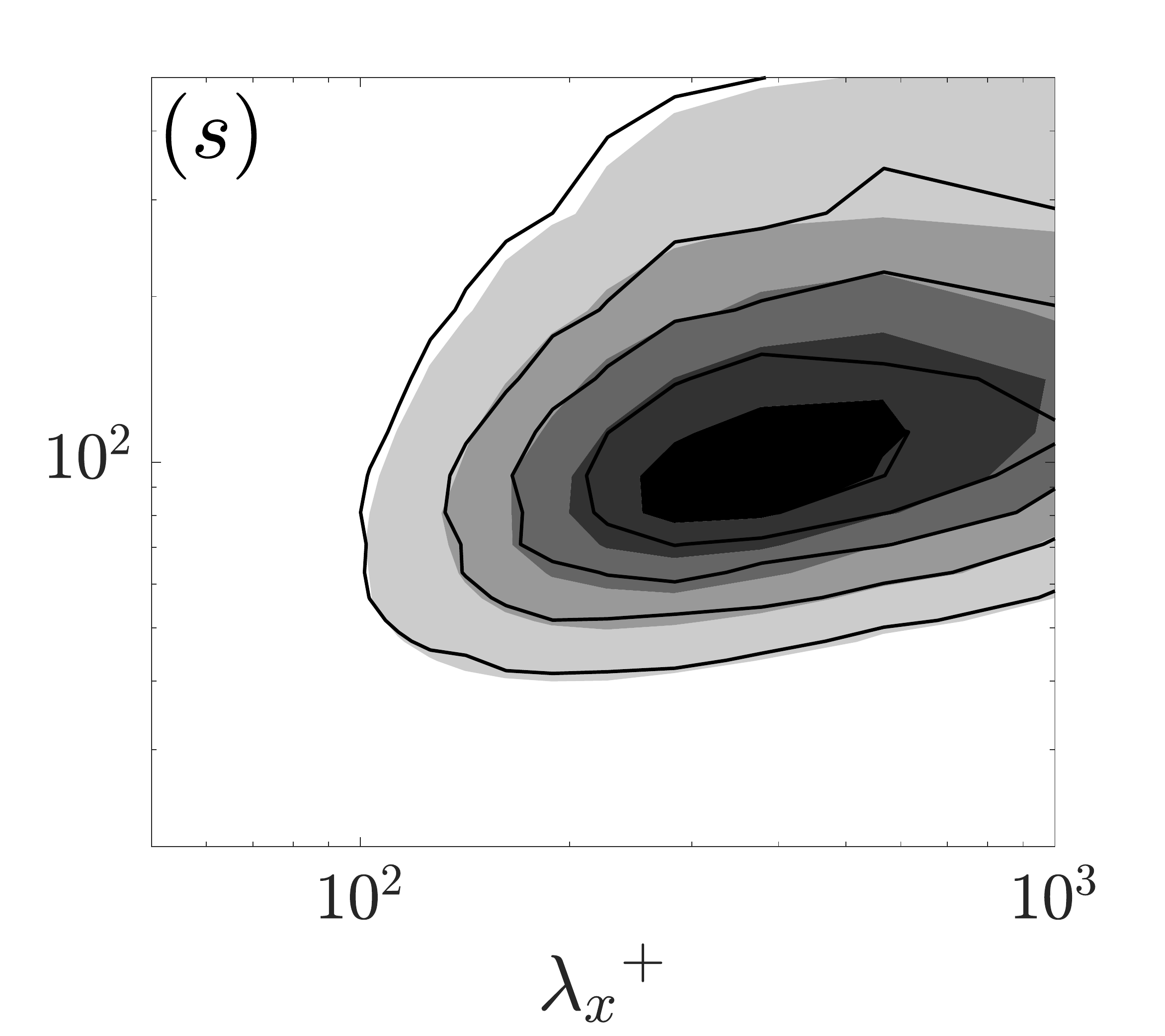}
        {\phantomcaption\label{spectra_L2MX5:w}}
    \end{subfigure}%
    \hspace*{-3mm}
    \begin{subfigure}[tbp]{.257\textwidth}
        \includegraphics[width=1\linewidth]{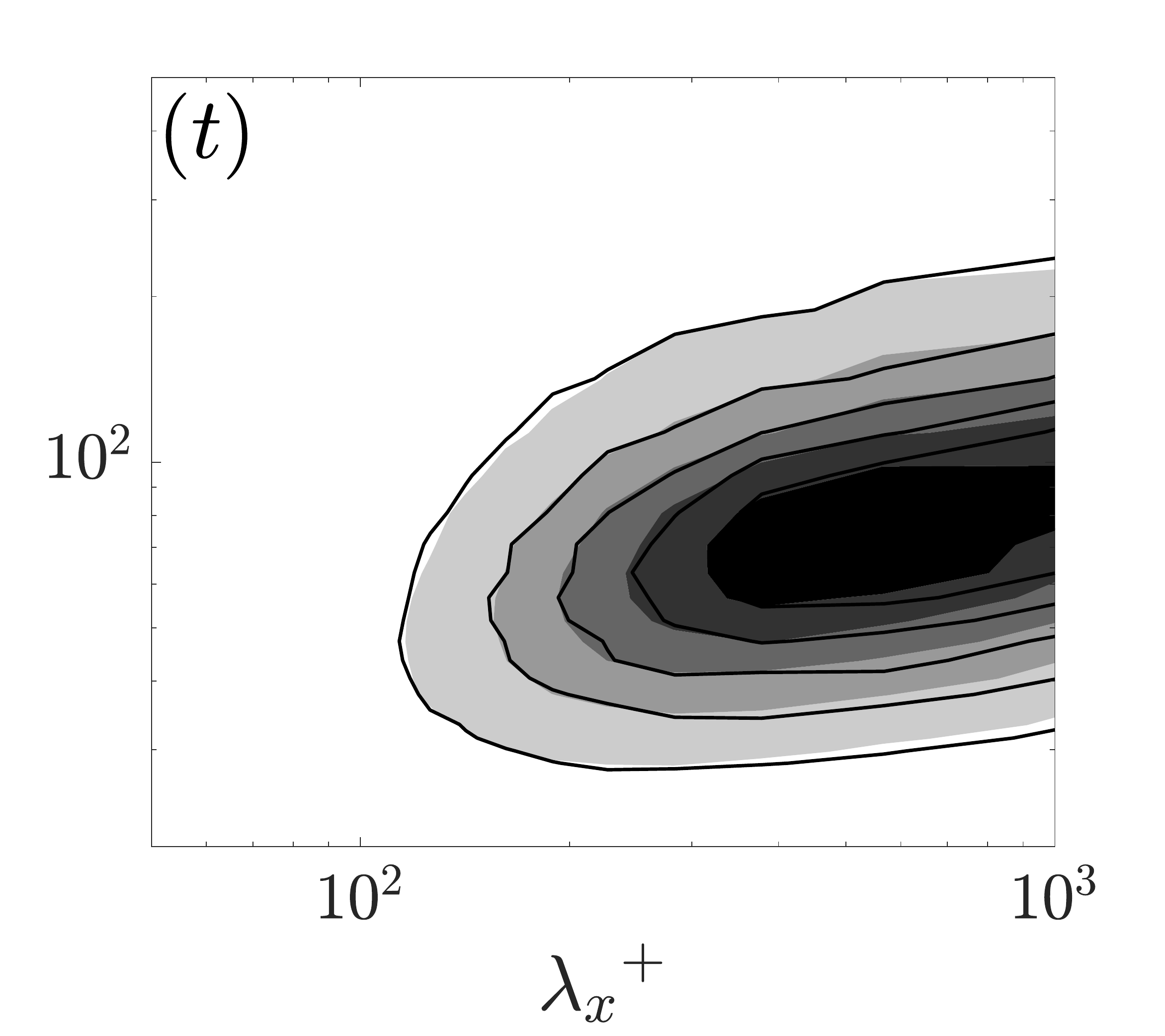}
        {\phantomcaption\label{spectra_L2MX5:uv}}
    \end{subfigure}
    \vspace*{-4mm}
    \caption{Pre-multiplied two dimensional spectral densities of ${u}^2$, ${v}^2$, ${w}^2$ and ${uv}$. L5MZ10 ($a$\textbf{--}$d$); L5MZ5 ($e$\textbf{--}$h$); L2MZ5 ($i$\textbf{--}$l$); L5MX5 ($m$\textbf{--}$p$); L2MX5 ($q$\textbf{--}$t$); shaded regions are the smooth-wall solution at $y^+\!\approx15$ and solid lines are the TRM cases at $y^+\!+{\ell_{uv}}^+\!\approx15$ scaled using $u_{\tau}$ at $y^+\!=-{\ell_{uv}}^+$.}
    \label{fig:spectra2}
    \end{center}
\end{figure}
%%%%%%%%%%%%%%%%%%%%%%%%%%%%%%%%%%%%%%%%%%%%%%%%%%%%%%%%%%%%%%%%%%%%%%%%%
%%%%%%%%%%%%%%%%%%%%%%%%%%%%%%%%%%%%%%%%%%%%%%%%%%%%%%%%%%%%%%%%%%%%%%%%%

Considering the nominal thickness of the viscous sub-layer ($y^+\approx5$) and the position ($y^+\approx20$) and radius ($r^+\approx15$) of the quasi-streamwise vortices \citep{kim_moin_moser_1987}, it stands to reason that a large enough downward displacement of the vortices will cause them to become disrupted. When such a disruption is incurred, the near-wall cycle \citep{jimenez_pinelli_1999} changes and the structure of its turbulence becomes different. This was observed for cases L5M5 and L10M10 in their energy spectra (\cref{fig:spectra1}). 
Similar structural changes can be observed in the energy spectra of cases L2MZ5, L5MZ5 and L5MZ10 (\hyperref[spectra_L5MZ10:u]{figures \ref*{spectra_L5MZ10:u}}-\hyperref[spectra_L5MZ10:u]{\subref*{spectra_L5MZ5:uv}}). In contrast, the previously mentioned smooth-wall-like cases L5MX5 and L2MX5, retain the spectral signature of a smooth-wall. These differences and similarities relative to smooth-wall turbulence are consistent with those observed for the distribution of the correlation coefficient in \cref{fig:reystress_vs_uv}.

In the energy spectra of the wall-normal velocity (second columns of \hyperref[fig:spectra1]{figures \ref*{fig:spectra1}} and \hyperref[fig:spectra2]{\ref*{fig:spectra2}}) the signature commonly associated with the inviscid shear-layer instability emerges for cases L5M5, L10M10 (\hyperref[spectra_L5M5:v]{figures \ref*{spectra_L5M5:v}} and \hyperref[spectra_L10M10:v]{\ref*{spectra_L10M10:v}}) and L5MZ5, L5Z10 (\hyperref[spectra_L5MZ5:v]{figures \ref*{spectra_L5MZ5:v}} and \hyperref[spectra_L5MZ10:v]{\ref*{spectra_L5MZ10:v}}). This indicates that the near-wall turbulence is no longer canonical, and instead exhibits characteristics similar to that of the Kelvin-Helmholtz type instability. These characteristics have been observed in flows over large riblets of certain shapes \citep{garcia-mayoral_2011,endrikat_2021}, large roughness which disrupt the buffer-layer dynamics \citep{Abderrahaman2019} and highly permeable porous structures \citep{jimenez_2001,breugem_boersma_uittenbogaard_2006,manes_poggi_ridolfi_2011,kuwata_suga_2017,gomez_2019}. This instability changes the distribution of turbulent kinetic energy, with  energetic scales over a streamwise range of ${65}\lesssim{{\lambda_{x}}^+}\lesssim{290}$ emerging for spanwise wavelengths of ${{\lambda_{z}}^+}\gtrsim{130}$ (\citealt{garcia-mayoral_2011,endrikat_2021}). This is most evident in the energy spectra of the wall-normal velocity for case L10M10 (\hyperref[spectra_L10M10:v]{figure \ref*{spectra_L10M10:v}}).

In terms of virtual origins; cases L5M5, L5MZ5, L5MZ10 and L10M10 which are not smooth-wall-like have turbulence virtual origins of respectively ${{\ell_{uv}}^+}\approx4.7$, ${{\ell_{uv}}^+}\approx5.8$, ${{\ell_{uv}}^+}\approx8.7$ and ${{\ell_{uv}}^+}\approx7.7$. The first case would place the vortices very close to the domain boundary while the other three cases would make them collide with it. \citet{ibrahim2020smoothwalllike} identified this constraint as being ${{\ell_{uv}}^+}\lesssim5$ for turbulence, or more specifically the profile of the Reynolds shear stress, to remain smooth-wall-like. This is consistent with the preceding explanation about the distance separating the vortices from a regular impenetrable wall. However, case L2MZ5 with ${{\ell_{uv}}^+}\approx3.6$ also exhibits some divergences from smooth-wall-like turbulence, indicative of an additional constraint aside from just the proximity of the vortices to the boundary plane. Based on their results, \citet{ibrahim2020smoothwalllike} established this constraint as being ${\ell_{uv}}^+\lesssim{{{\ell_u}^+}+2}$ for situations where ${{{\ell_u}^+<\ell_{uv}}^+\lesssim5}$. The physical interpretation provided for this by \citet{ibrahim2020smoothwalllike} was that the streamwise streaks become so constricted that they can no longer be sustained. This explains the onset of the divergences seen in case L2MZ5, where ${\ell_{uv}}^+-{{\ell_u}^+}=1.9$.

\section{Characterizing surfaces in terms of slip and transpiration}\label{subsec:surface_characteriziation}

%%%%%%%%%%%%%%%%%%%%%%%%%%%%%%%%%%%%%%%%%%%%%%%%%%%%%%%%%%%%%%%%%%%%%%%%%
%%%%%%%%%%%%%%%%%%%%%%%%%%%%%%%%%%%%%%%%%%%%%%%%%%%%%%%%%%%%%%%%%%%%%%%%%
\begin{figure}
    \begin{center}
    \hspace*{-1mm}
    \begin{subfigure}[tbp]{.34\textwidth}
     \includegraphics[width=1\linewidth]{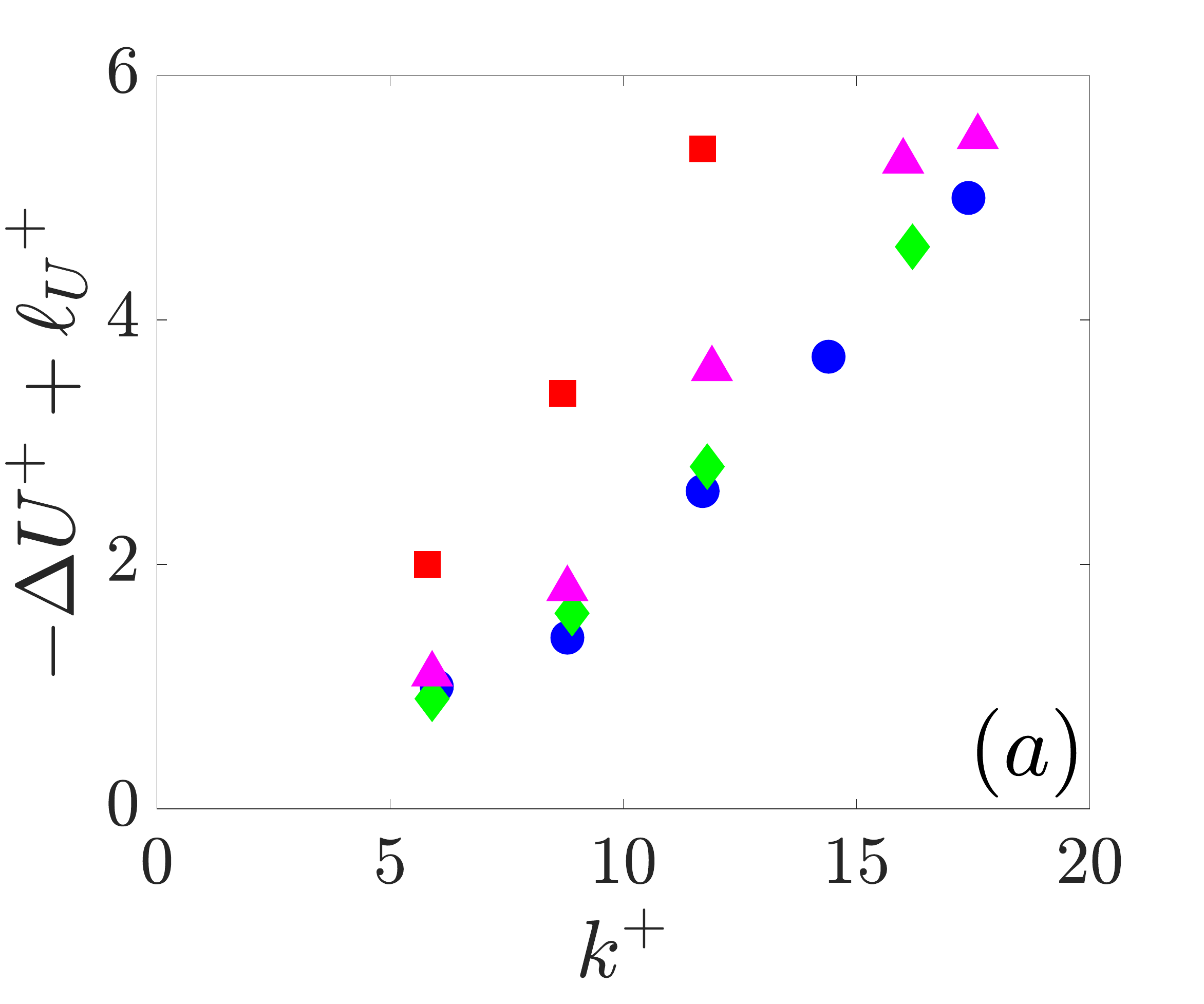}
     {\phantomcaption\label{fig:deltaU_vs_k_abdrhmn}}
    \end{subfigure}%
    \begin{subfigure}[tbp]{.34\textwidth}
     \includegraphics[width=1\linewidth]{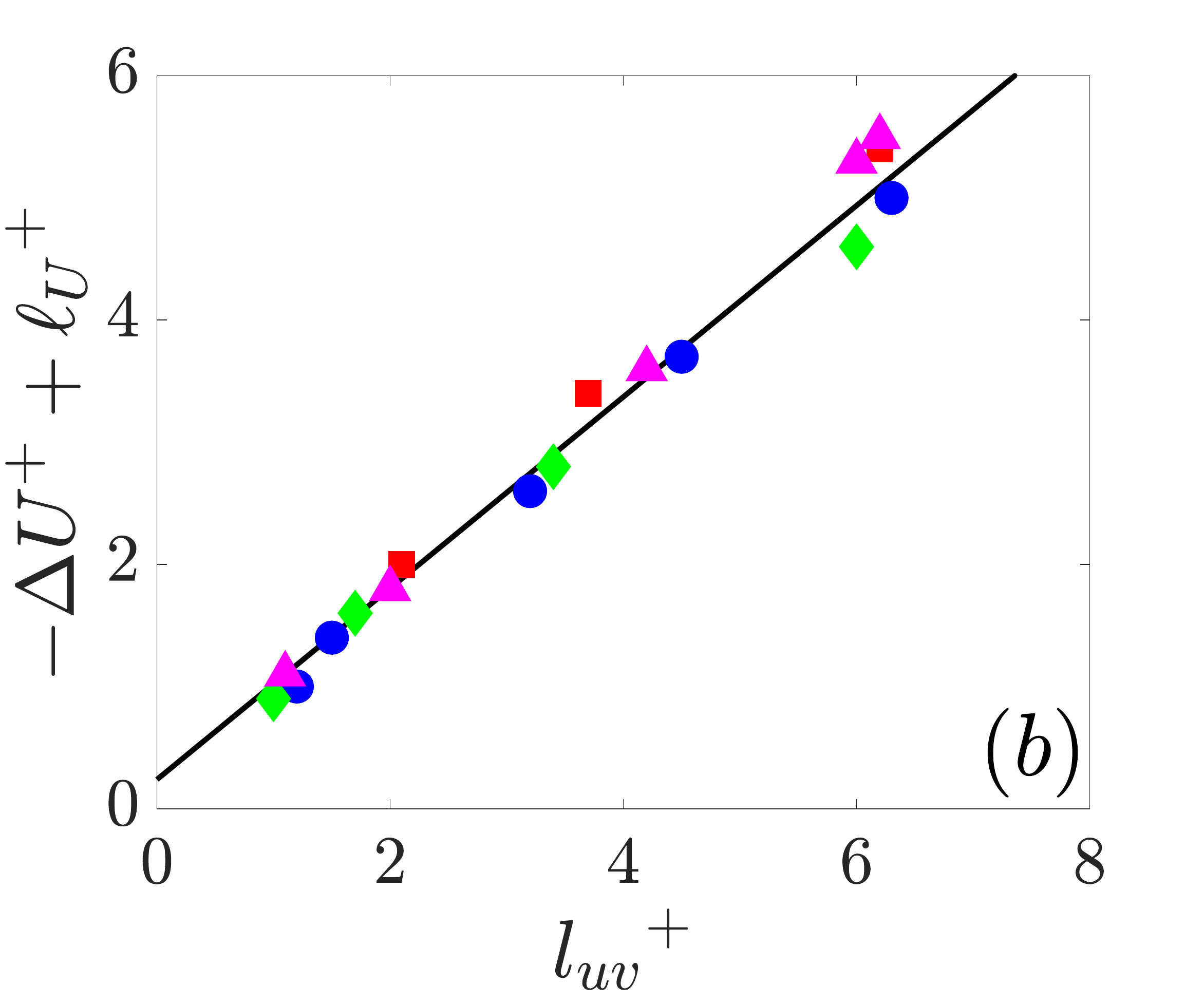}
     {\phantomcaption\label{fig:deltaU_vs_luv_abdrhmn}}
    \end{subfigure}%
    \begin{subfigure}[tbp]{.34\textwidth}
     \includegraphics[width=1\linewidth]{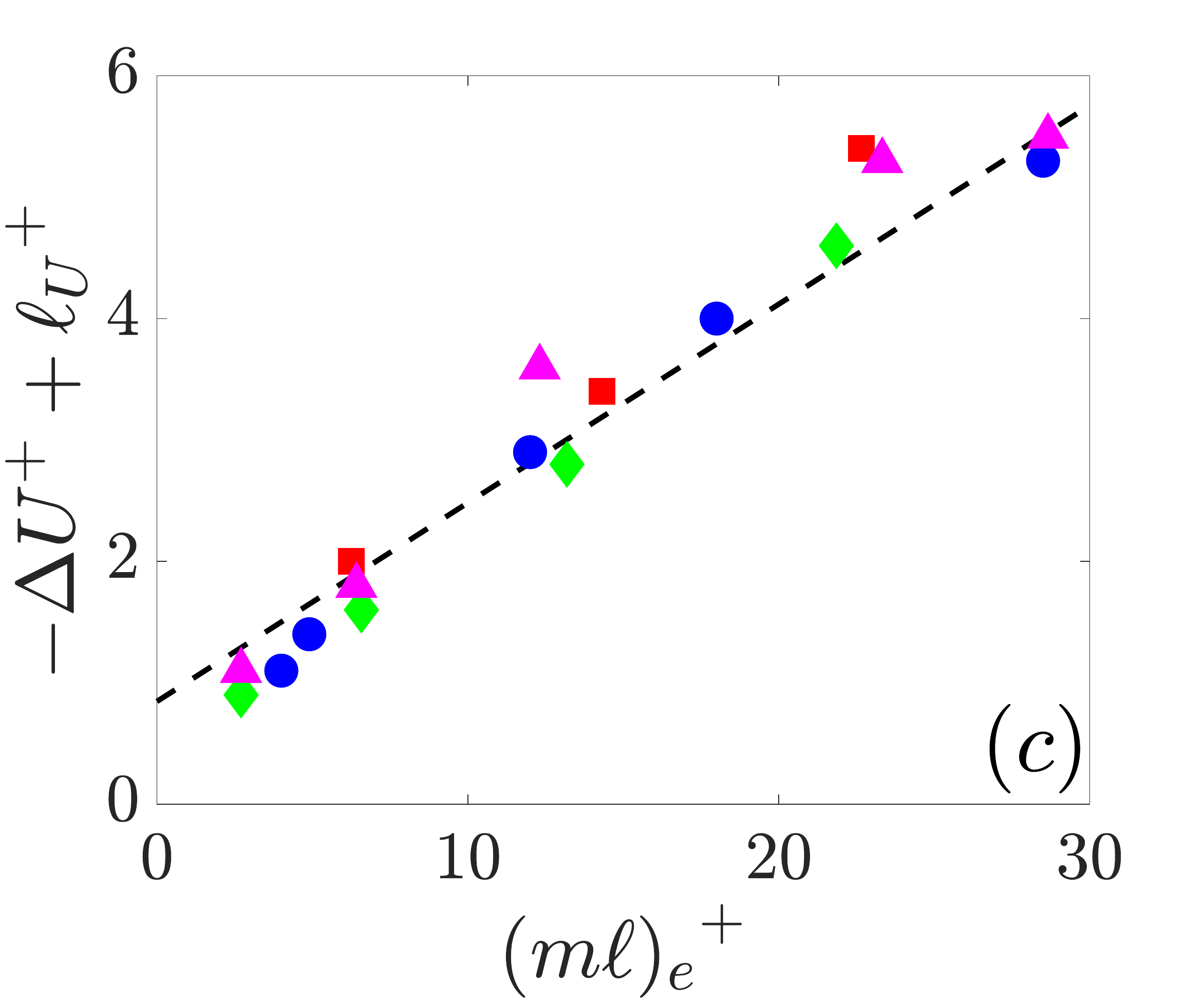}
     {\phantomcaption\label{fig:deltaU_vs_ml_all_surfaces_abdrhmn}}
    \end{subfigure}
    \vspace*{-3mm}
    \caption{${-\Delta{U}^+}\!\!+{{\ell_{U}}^+}$ versus ${k}^+$, ${\ell_{uv}}^+$ and ${({m}{\ell})_e}^+$ in ($a$), ($b$) and ($c$), respectively: {\Large\textcolor{blue}{$\bullet$}}, collocated posts (C); \textcolor{red}{${\blacksquare}$}, collocated posts with alternating heights (CC); \textcolor{green}{${\blacklozenge}$}, spanwise-staggered posts (SZ); \textcolor{magenta}{${\blacktriangle}$}, streamwise-staggered posts (SX). $-\!-$, linear fit $0.78{\ell_{uv}}^+$; $-\;-$ linear fit $0.16{({m}{\ell})_e}^+$. Figures ($a$) and ($b$) reproduced using data in \citet{Abderrahaman2019}. Figure ($c$) shows ${-\Delta{U}^+}\!\!+{{\ell_{U}}^+}$ versus their corresponding values of ${{({m}{\ell})}^+_e}$ as estimated using \eqref{eq:dU_vs_ml_2}.} 
    \label{fig:deltaU_vs_k_luv_abdrhmn}
    \end{center}
\end{figure}
%%%%%%%%%%%%%%%%%%%%%%%%%%%%%%%%%%%%%%%%%%%%%%%%%%%%%%%%%%%%%%%%%%%%%%%%%
%%%%%%%%%%%%%%%%%%%%%%%%%%%%%%%%%%%%%%%%%%%%%%%%%%%%%%%%%%%%%%%%%%%%%%%%%
%%%%%%%%%%%%%%%%%%%%%%%%%%%%%%%%%%%%%%%%%%%%%%%%%%%%%%%%%%%%%%%%%%%%%%%%%
%%%%%%%%%%%%%%%%%%%%%%%%%%%%%%%%%%%%%%%%%%%%%%%%%%%%%%%%%%%%%%%%%%%%%%%%%
\begin{table}
  \centering
  \begin{tabular}{ m{1.8cm}m{0.8cm}m{2.4cm}m{1.0cm}m{1.0cm}m{0.6cm}|m{4.5cm} }
       $\,\mathrm{Case}$ & ${\ell_{U}}^+$ & ${-\Delta{U}^+}\!\!+{{\ell_{U}}^+}$ & ${s_x/k}$ & ${s_z/k}$ & \,\,${k}^+$ & $({-\Delta{U}^+}\!\!+{{\ell_{U}}^+})$ from \eqref{eq:dU_vs_ml} \& \eqref{eq:dU_vs_ml_2}\\
       &&&&&&\\
       $\mathrm{CC}06$ & $1.2$ & $\;\;\;\qquad2.0$ & \,\,$4.0$ & \,\,$4.0$ & $5.8$ & $\qquad\qquad\qquad1.7$\\
       $\mathrm{CC}09$ & $1.6$ & $\;\;\;\qquad3.4$ & \,\,$4.0$ & \,\,$4.0$ & $8.7$ & $\qquad\qquad\qquad3.2$\\
       $\mathrm{CC}12$ & $2.1$ & $\;\;\;\qquad5.4$ & \,\,$4.0$ & \,\,$4.0$ & $11.7$ & $\qquad\qquad\qquad4.5$\\
       $\mathrm{CC}15$ & $2.4$ & $\;\;\;\qquad6.9$ & \,\,$4.0$ & \,\,$4.0$ & $15.4$ & $\qquad\qquad\qquad6.9$\\
       $\mathrm{CC}18$ & $2.4$ & $\;\;\;\qquad7.3$ & \,\,$4.0$ & \,\,$4.0$ & $17.0$ & $\qquad\qquad\qquad7.3$\\
       \vspace*{-4mm}&\vspace*{-4mm}&\vspace*{-4mm}&\vspace*{-4mm}&\vspace*{-4mm}&\vspace*{-4mm}&\vspace*{-4mm}\\
       $\mathrm{SZ}06$ & $0.5$ & $\;\;\;\qquad0.9$ & \,\,$2.0$ & \,\,$4.0$ & $5.9$ & $\qquad\qquad\qquad0.9$\\
       $\mathrm{SZ}09$ & $0.7$ & $\;\;\;\qquad1.6$ & \,\,$2.0$ & \,\,$4.0$ & $8.9$ & $\qquad\qquad\qquad1.6$\\
       $\mathrm{SZ}12$ & $1.3$ & $\;\;\;\qquad2.8$ & \,\,$2.0$ & \,\,$4.0$ & $11.8$ & $\qquad\qquad\qquad2.8$\\
       $\mathrm{SZ}15$ & $1.8$ & $\;\;\;\qquad4.6$ & \,\,$2.0$ & \,\,$4.0$ & $16.2$ & $\qquad\qquad\qquad4.4$\\
       \vspace*{-4mm}&\vspace*{-4mm}&\vspace*{-4mm}&\vspace*{-4mm}&\vspace*{-4mm}&\vspace*{-4mm}&\vspace*{-4mm}\\
       $\mathrm{SX}06$ & $0.5$ & $\;\;\;\qquad1.1$ & \,\,$4.0$ & \,\,$2.0$ & $5.9$ & $\qquad\qquad\qquad0.9$\\
       $\mathrm{SX}09$ & $0.6$ & $\;\;\;\qquad1.8$ & \,\,$4.0$ & \,\,$2.0$ & $8.8$ & $\qquad\qquad\qquad1.5$\\
       $\mathrm{SX}12$ & $1.0$ & $\;\;\;\qquad3.6$ & \,\,$4.0$ & \,\,$2.0$ & $11.9$ & $\qquad\qquad\qquad2.6$\\
       $\mathrm{SX}15$ & $1.2$ & $\;\;\;\qquad5.3$ & \,\,$4.0$ & \,\,$2.0$ & $16.0$ & $\qquad\qquad\qquad4.7$\\
       $\mathrm{SX}18$ & $1.2$ & $\;\;\;\qquad5.5$ & \,\,$4.0$ & \,\,$2.0$ & $17.6$ & $\qquad\qquad\qquad5.7$\\
  \end{tabular}
  \captionof{table}{Slip velocity and roughness function for the collocated posts with alternating heights (CC), spanwise-staggered posts (SZ) and streamwise-staggered posts (SX) cases of \citet{Abderrahaman2019}. The last column shows the estimated ${-\Delta{U}^+}\!\!+{{\ell_{U}}^+}$ using ${{({m}{\ell})}^+_e}$.}
  \label{tab:rough_cases_abdrhmn2}
\end{table}
%%%%%%%%%%%%%%%%%%%%%%%%%%%%%%%%%%%%%%%%%%%%%%%%%%%%%%%%%%%%%%%%%%%%%%%%%
%%%%%%%%%%%%%%%%%%%%%%%%%%%%%%%%%%%%%%%%%%%%%%%%%%%%%%%%%%%%%%%%%%%%%%%%%
A convention in classifying rough surfaces has been to express them in terms of an equivalent sand-grain roughness height, ${k_s}$, \citep{schlichting1937} which originates from the pipe flow measurements of \citet{nikuradse}. The purpose of ${k_s}$ is to serve as a common currency, encapsulating distinct rough surfaces that behave similarly in terms of drag, but it does not provide any utility beyond this. For a given rough surface in the fully rough regime of turbulence, the ratio ${k_s/k}$ asymptotes to a fixed value, providing a measure of simplification in characterizing surfaces for this flow regime. For transitionally rough flows however, ${k_s/k}$ varies with Reynolds number, while different surfaces also transition differently from the hydraulically smooth to the fully rough regime, all of which undermines the utility of ${k_s}$ \citep{jimenez_rough}. 

To assess whether the TRM coefficients may be used for rough surface classification, the transitionally rough DNS data of \citet{Abderrahaman2019} is further used for this purpose. In addition to the collocated posts (C) already discussed in \cref{sec:TRM_vs_rough_DNS} and shown in \cref{fig:RoughGeometry}, \cref{tab:rough_cases_abdrhmn2} reports three more categories of regular rough surfaces investigated by \citet{Abderrahaman2019}. 
The first group (CC) in \cref{tab:rough_cases_abdrhmn2} is composed of collocated posts with alternating heights, while the second (SZ) and third groups (SX) are surfaces with spanwise-staggered posts and streamwise-staggered posts, respectively. It can been observed in \cref{fig:deltaU_vs_k_abdrhmn} that these rough surfaces, for similar values of $k^+$, exhibit different levels of drag. This demonstrates the inadequacy of $k^+$, and by extension ${k_s}^+$, in serving as a suitable parameter for characterising transitionally rough surfaces.

Alternatively, one may classify surfaces using appropriate flow quantities instead \citep{orlandi_leonardi_antonia_2006, Abderrahaman2019}. The findings of \citet{Abderrahaman2019}, which are reproduced in \cref{fig:deltaU_vs_luv_abdrhmn}, demonstrated that the overall effect of the surfaces in \cref{tab:rough_cases_abdrhmn,tab:rough_cases_abdrhmn2} on the turbulent flow are well characterizable using a single flow quantity, namely ${\ell_{uv}}^+$.  

Observing the fact that ${\ell_{uv}}^+$ shows a strong linear scaling with respect to ${({m}{\ell})}^+$, it is reasonable to assume that the various cases of \cref{tab:rough_cases_abdrhmn2} should be characterizable in terms of ${({m}{\ell})}^+$ while not showing the disparity they do when using $k^+$. However, conducting TRM simulations with the aim of reproducing the effects of all these surfaces, as was done in \cref{sec:TRM_vs_rough_DNS} for the collocated posts (C surfaces), would have been a considerable undertaking and fell outside of the scope of this work. Instead, by relying upon a physical intuition of flow over roughness in conjunction with the arguments underlying the TRM boundary conditions, it is examined whether the linear relation between ${-\Delta{U}^+}\!\!+{{\ell_{U}}^+}$ and ${({m}{\ell})}^+$ obtained for the collocated posts can be leveraged for the other surfaces in \cref{tab:rough_cases_abdrhmn2}.

The linear relation between ${-\Delta{U}^+}\!\!+{{\ell_{U}}^+}$ and ${({m}{\ell})}^+$ obtained for the collocated posts and shown in \cref{fig:TranspirationScalingAbdrhmn:sub2} is:
\begin{gather}
    ({-\Delta{U}^+}\!\!+{{\ell_{U}}^+}) \approx {0.185{{({m}{\ell})}^+}+0.36}.\label{eq:dU_vs_ml}
\end{gather}
The geometrical differences of the other surface types (\cref{tab:rough_cases_abdrhmn2}) relative to the collocated posts is mainly in terms of their pitch lengths, such that for a similar height ($k^+$), different levels of drag results. As was argued in \cref{role_of_transpriation_factor}, the pitch of the posts determines the amount of slip velocity which can arise at the crest plane, and the cross-flow slip plays a role in the occurrence of transpiration. Assuming that for a given height, the various surfaces translate to a similar transpiration length (i.e. they permit wall-normal momentum to penetrate to approximately the same depth), then the only effect which must be accounted for is the cross-flow slip, which affects transpiration. Hence, the $m^+$ contribution of the transpiration factor, ${({m}{\ell})}^+$, is assumed to remain fixed, while the ${\ell}^+$ component is adjusted:
% \begin{gather}
%     {-\Delta{U}^+}\!\!+{{\ell_{U}}^+} \approx \alpha \cdot {0.19{{\left ({m}{\ell}\right )}^+}+0.36}\:,\qquad \alpha=\frac{{\ell_X}^+}{{\ell_C}^+}\approx \frac{{\ell_{U,X}}^+}{{\ell_{U,C}}^+}.\label{eq:dU_vs_ml_2}
% \end{gather}
\begin{gather}
    {\left ({m}{\ell}\right )}^+_e \approx \alpha {{\left ({m}{\ell}\right )}^+}\:,\qquad \alpha=\frac{{\ell_X}^+}{{\ell_C}^+}\approx \frac{{\ell_{U,X}}^+}{{\ell_{U,C}}^+}.\label{eq:dU_vs_ml_2}
\end{gather}
Here, ${{\ell_{U}}^+}_X$ is the mean flow origin (slip velocity) of either the C, CC, SX or SZ surfaces of \cref{tab:rough_cases_abdrhmn2}, ${{\ell_{U}}^+}_C$ is the mean flow origin of the collocated posts of \cref{tab:rough_cases_abdrhmn} and ${\left ({m}{\ell}\right )}^+_e$ is the expected transpiration factor.
%In this manner, the expected transpiration factors corresponding to the various CC, SX and SZ surfaces may be approximated.
As a litmus test for the veracity of the approach undertaken here, the drag estimations from using \eqref{eq:dU_vs_ml_2} with \eqref{eq:dU_vs_ml} are gathered in \cref{tab:rough_cases_abdrhmn2} alongside the DNS acquired drag values of \citet{Abderrahaman2019}. The relative closeness of the estimates to the DNS values substantiates the preceding physical argument which lead to \eqref{eq:dU_vs_ml_2}.

\Cref{fig:deltaU_vs_ml_all_surfaces_abdrhmn} plots ${-\Delta{U}^+}\!\!+{{\ell_{U}}^+}$ of the various surfaces against their corresponding values of ${({m}{\ell})}^+_e$. When compared to \cref{fig:deltaU_vs_k_abdrhmn}, it is evident that the transpiration factor serves as a more suitable parameter for characterizing rough surfaces in the transitionally rough regime of turbulence.

The analysis carried out in this section is mainly meant to promote the idea that for transitionally rough surfaces, effective flow quantities such as ${\ell}^+$ and ${m}^+$, which represent the effect of a surface on the flow, may be more suitable for surface characterization rather than surface or surface-derived properties. This could potentially also be exploited for predictive purposes, however drag prediction of rough surfaces remains an outstanding challenge, even for the fully rough flow regime as detailed in a recent review of the topic by \citet{chung_2021}.
\FloatBarrier

\section{Conclusions}
This work has investigated the Transpiration-Resistance Model (TRM), a set of boundary conditions proposed by \citet{Lacis2020} for modeling the hydrodynamic effect of surface micro-textures on turbulent flows. The occurrence of transpiration is an important physical mechanism of drag-altering surfaces, in particular rough surfaces which cause a drag penalty. The TRM expresses transpiration as being due to the variations in shear of the tangential velocities. This permitted investigating three different conditions for transpiration; only due to streamwise variations (L$<\!\!\cdot\!\!>$MX$<\!\!\cdot\!\!>$), only due to spanwise variations (L$<\!\!\cdot\!\!>$MZ$<\!\!\cdot\!\!>$) and due to variations in both (L$<\!\!\cdot\!\!>$M$<\!\!\cdot\cdot\!\!>$). Coupling to streamwise variations only, while leading to relaxed impermeability, caused no drag increase due to an absence of Reynolds shear stress generation at the boundary plane. Conversely, explicit coupling to spanwise shear variations lead to the greatest degree of drag increase due to pronounced turbulent activity taking place at the boundary plane. These simulations reaffirmed that transpiration in wall-bounded canonical turbulence is principally caused by the quasi-streamwise vortices of the near-wall cycle, which to a first-order induce a spanwise flow. 

In terms of surfaces, the effect of small textures would be a displacement of these vortices. This was the effect described by \citet{luchini_manzo_pozzi_1991} for explaining the drag reduction mechanism of riblets. \citet{luchini_1996} conceptualized this effect as a height difference between imaginary smooth-walls perceived by the mean flow and near-wall turbulence with the flow remaining otherwise smooth-wall-like. \citet{ibrahim2020smoothwalllike} expanded this into a virtual origin framework for imposing \emph{a priori} determined virtual origins on the velocities and predicting the resulting drag change. The regime of smooth-wall-like turbulence is also observed for a number of the simulations in this work, with the change in drag being quantified by the difference between the mean flow and turbulence origins expressed by \eqref{eq:virtual-origin}. Certain conditions identified by \citet{ibrahim2020smoothwalllike} which upon violation lead to turbulence ceasing to be smooth-wall-like also remained applicable here. These would be the quasi-streamwise vortices becoming displaced so far downward that they collide with the domain boundary (i.e. ${{\ell_{uv}}^+}\gtrsim5$) and the near-wall streaks becoming too constricted (i.e. ${\ell_{uv}}^+\gtrsim{{{\ell_u}^+}+2}$).

The applicability of the TRM in serving as an effective model for rough surfaces in the transitionally rough turbulence regime was also examined. The geometry-resolving transitionally rough DNS data of \citet{Abderrahaman2019} were used as references for evaluation. The boundary conditions demonstrate the capability of reproducing the behaviour of the rough surfaces examined up to $k^+\approx18$, with the flow over the largest roughness sizes falling outside of the regime of smooth-wall-like turbulence. This seems to indicate that such boundary conditions may also cover the flow conditions in the initial range of this non-smooth-like regime. Clearly, this does not apply to situations where the non-linear interactions between the texture-coherent flow and turbulence become a major component of the near-wall flow dynamics \citep{fairhall_abderrahaman-elena_garcia-mayoral_2019,Abderrahaman2019}.

A brief examination of TRM derived relations obtained for a specific roughness geometry were leveraged to investigate whether characterizing surfaces in terms of effective flow quantities such as slip and transpiration lengths provide a better means of characterizing roughness in the transitionally rough regime. This was predicated upon the previously observed strong linear relation between the turbulence origin, $\ell_{uv}^+$, and the roughness function, $\Delta{U}^+$, by \citet{Abderrahaman2019} for the rough surface geometries they examined. Unlike measures such as sand-grain roughness height, ${k_s}^+$, or effective roughness height, $k^+$, which are inadequate in characterizing surfaces in the transitional regime, the transpiration factor, ${({m}{\ell})}^+$, is shown to better suit this purpose. Indeed, ${({m}{\ell})}^+$ is a measure of the wall-normal flux that can take place across the crest plane.

For potential future investigations, it is necessary to investigate alternative methods for determining the transpiration coefficients of both regular and irregular rough surfaces. Cost-effective approaches, such as minimal-span channels \citep{chung_chan_macdonald_hutchins_ooi_2015} or model-based methods \citep{Chavarin_luhar_Aiaa_2020, ran_zare_jovanovic_2021} may prove useful for such purposes. Indeed, it has been shown that transpiration factor serves the dual purpose of both being an input parameter in the boundary conditions as well as being linearly related to drag. Therefore, once representative values of $(m\ell)^+$ are found, they may be used in CFD applications to investigate more complex geometries. The transpiration lengths obtained in this work by comparing the TRM simulations with geometry-resolving DNS will serve as a reference for developing procedures to determine $m^+$ beyond the Stokes regime.
Finally, a more extensive formulation of the TRM \citep{Lacis2020} derived for modeling the effect of porous media on turbulent flows may be similarly investigated. Many of the flow features observed over rough surfaces are also observed over porous structures, such as the linear regime of drag change due to smooth-wall-like turbulence displacement and the existence of an unstable regime exhibiting a Kelvin-Helmholtz type instability \citep{jimenez_2001,breugem_boersma_uittenbogaard_2006,manes_poggi_ridolfi_2011,kuwata_suga_2017,gomez_2019}. 

\section*{Acknowledgments}
This research was undertaken with the financial support of the Swedish Foundation for Strategic Research (SSF), provided through grant SSF-FFL15-0001. Computational resources were provided by the Swedish National Infrastructure for Computing (SNIC) at the PDC and HPC2N computing centers. The authors also express their gratitude to Dr. Sudhakar Yogaraj at IIT Goa, for his efforts in developing an earlier implementation of the transpiration boundary condition.

\section*{Declaration of Interests}
The authors report no conflicts of interest.

\FloatBarrier
\appendix

\section{Numerical solver validation and grid convergence} \label{app:validation}
%%%%%%%%%%%%%%%%%%%%%%%%%%%%%%%%%%%%%%%%%%%%%%%%%%%%%%%%%%%%%%%%%%%%%%%%%
%%%%%%%%%%%%%%%%%%%%%%%%%%%%%%%%%%%%%%%%%%%%%%%%%%%%%%%%%%%%%%%%%%%%%%%%%
\begin{figure}
    \begin{center}
    \hspace*{-2mm}
    \begin{subfigure}[tbp]{.36\textwidth}
        \includegraphics[width=1\linewidth]{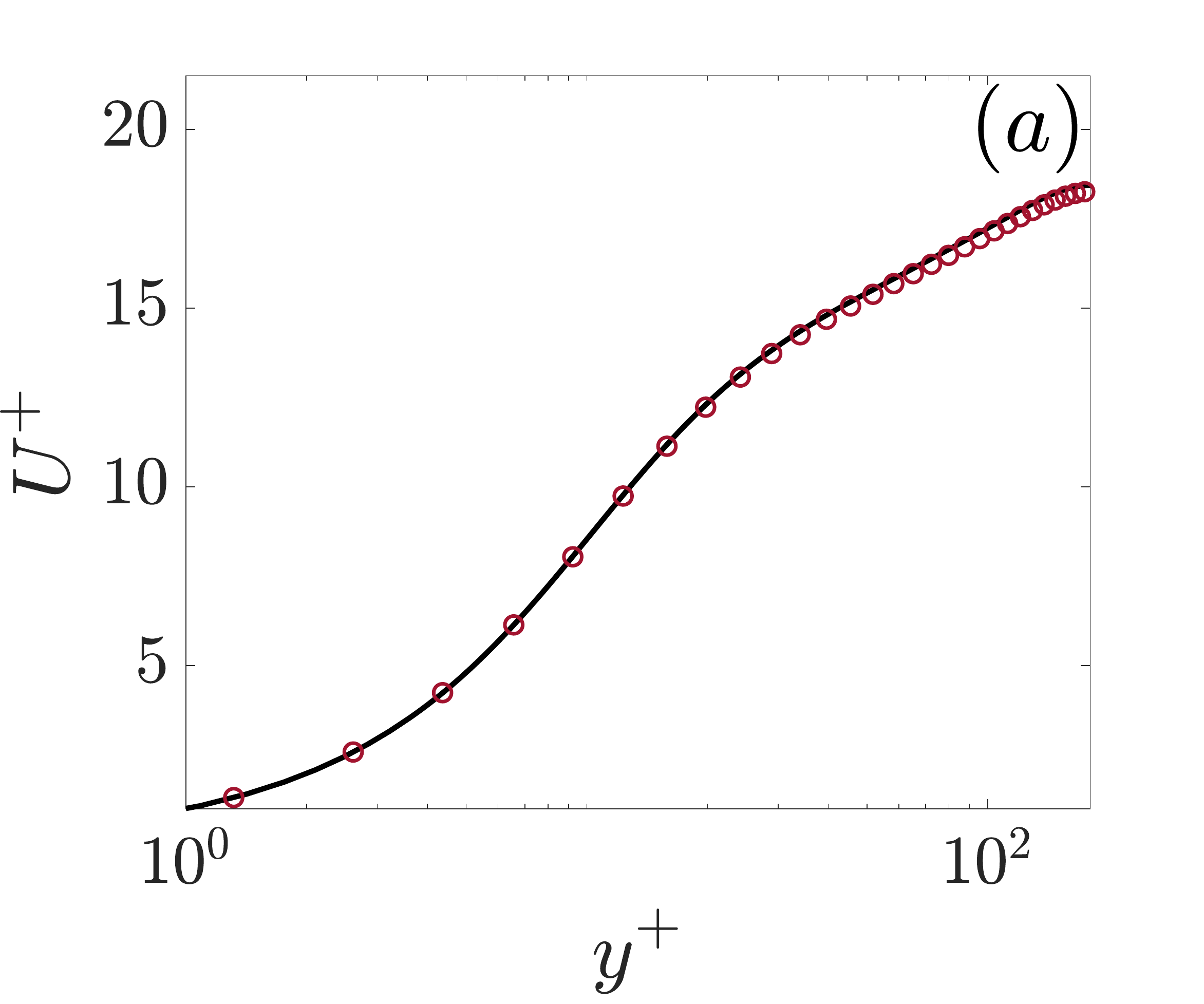}
        {\phantomcaption}
    \end{subfigure}%
    \hspace*{-5mm}
    \begin{subfigure}[tbp]{.36\textwidth}
        \includegraphics[width=1\linewidth]{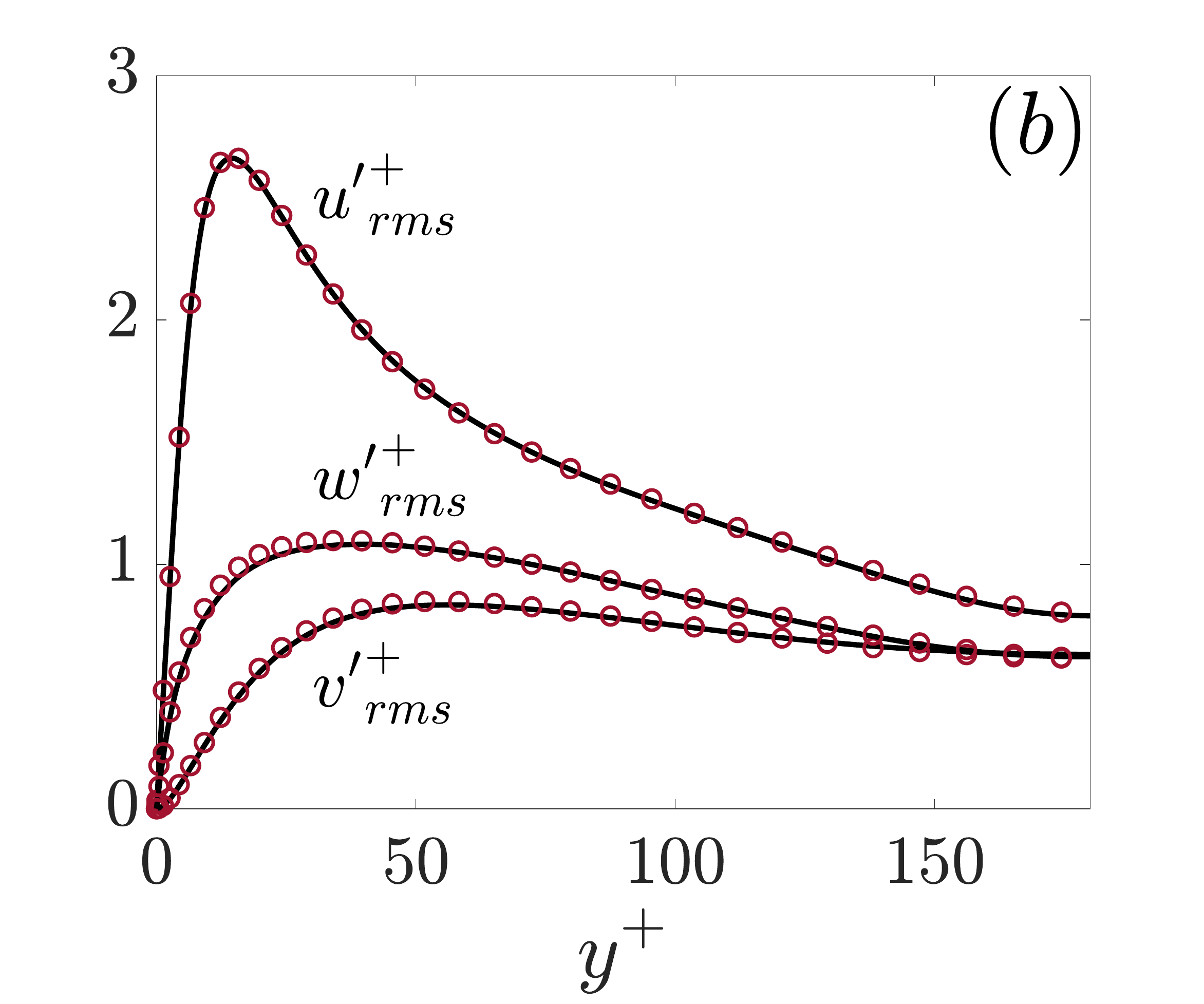}
        {\phantomcaption}
    \end{subfigure}%
    \hspace*{-4mm}
    \begin{subfigure}[tbp]{.36\textwidth}
        \includegraphics[width=1\linewidth]{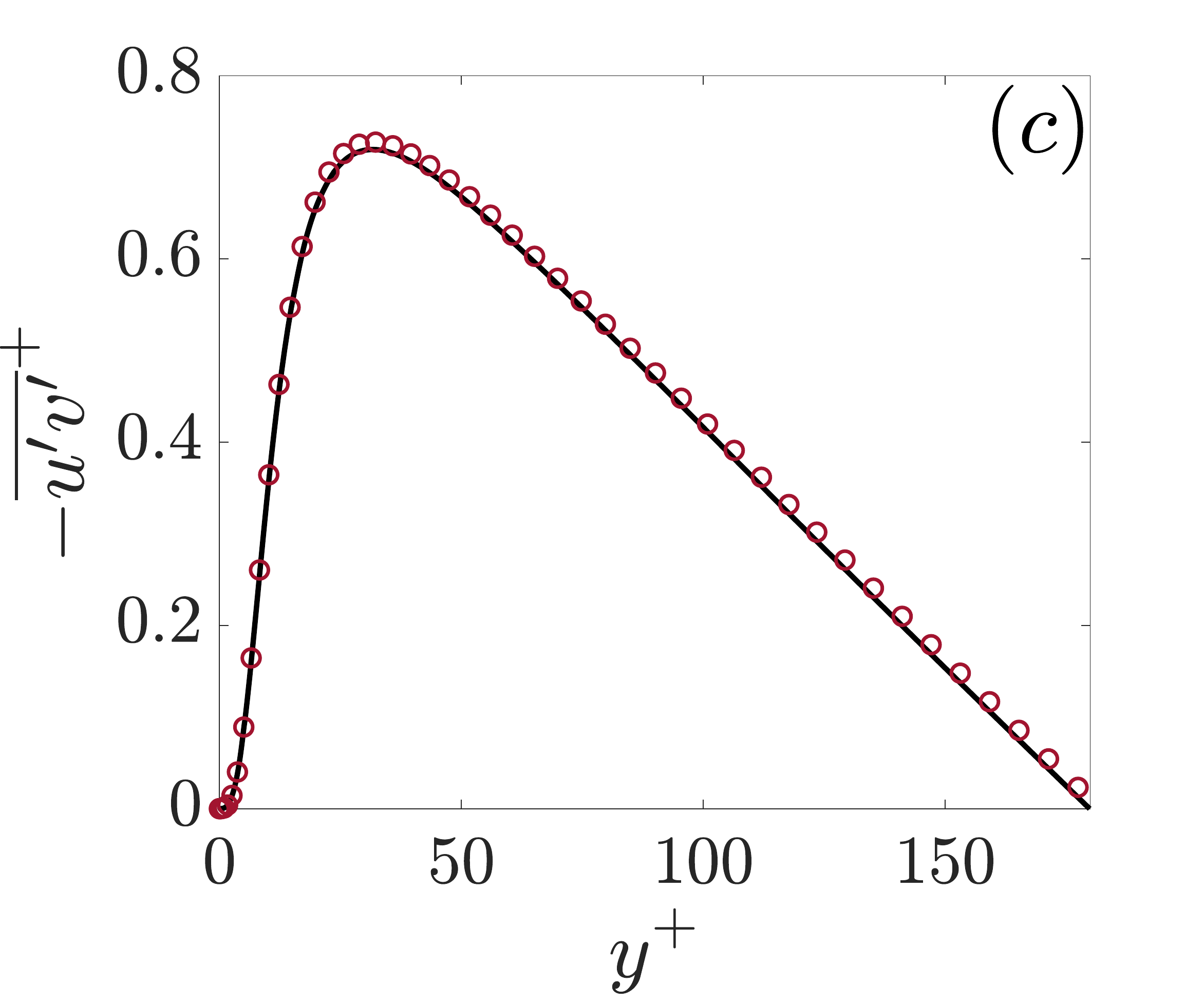}
        {\phantomcaption}
    \end{subfigure}
    \vspace*{-4mm}
    \caption{Mean velocity ($a$), r.m.s. velocity fluctuation ($b$) and Reynolds shear stress profiles ($c$) for a smooth-wall turbulent channel flow. $-\!-$, solver used in this study; {$\scriptstyle\textcolor{app-crimson}{\bigcirc}$}, \citet*{lee_moser_2015}.}
    \label{Validation}
    \end{center}
\end{figure}
%%%%%%%%%%%%%%%%%%%%%%%%%%%%%%%%%%%%%%%%%%%%%%%%%%%%%%%%%%%%%%%%%%%%%%%%%
%%%%%%%%%%%%%%%%%%%%%%%%%%%%%%%%%%%%%%%%%%%%%%%%%%%%%%%%%%%%%%%%%%%%%%%%%
\Cref{Validation} compares the one-point statistics of the solver used in this work against those of \citet{lee_moser_2015}, which were obtained using a spectral solver.
The agreement between the two data-sets is good; with the absolute maximum difference between the mean velocity, r.m.s. velocity fluctuations and Reynolds shear stress not exceeding $1\%$, $2\%$ and $1.2\%$ respectively. Existing discrepancies are due to the lower effective resolution of second-order finite-difference schemes relative to spectral methods at similar grid resolutions as explained by \citet{lee_moser_2015}.

%%%%%%%%%%%%%%%%%%%%%%%%%%%%%%%%%%%%%%%%%%%%%%%%%%%%%%%%%%%%%%%%%%%%%%%%%
%%%%%%%%%%%%%%%%%%%%%%%%%%%%%%%%%%%%%%%%%%%%%%%%%%%%%%%%%%%%%%%%%%%%%%%%%
\begin{figure}
    \begin{center}
    \hspace*{-2mm}
    \begin{subfigure}[tbp]{.36\textwidth}
        \includegraphics[width=1\linewidth]{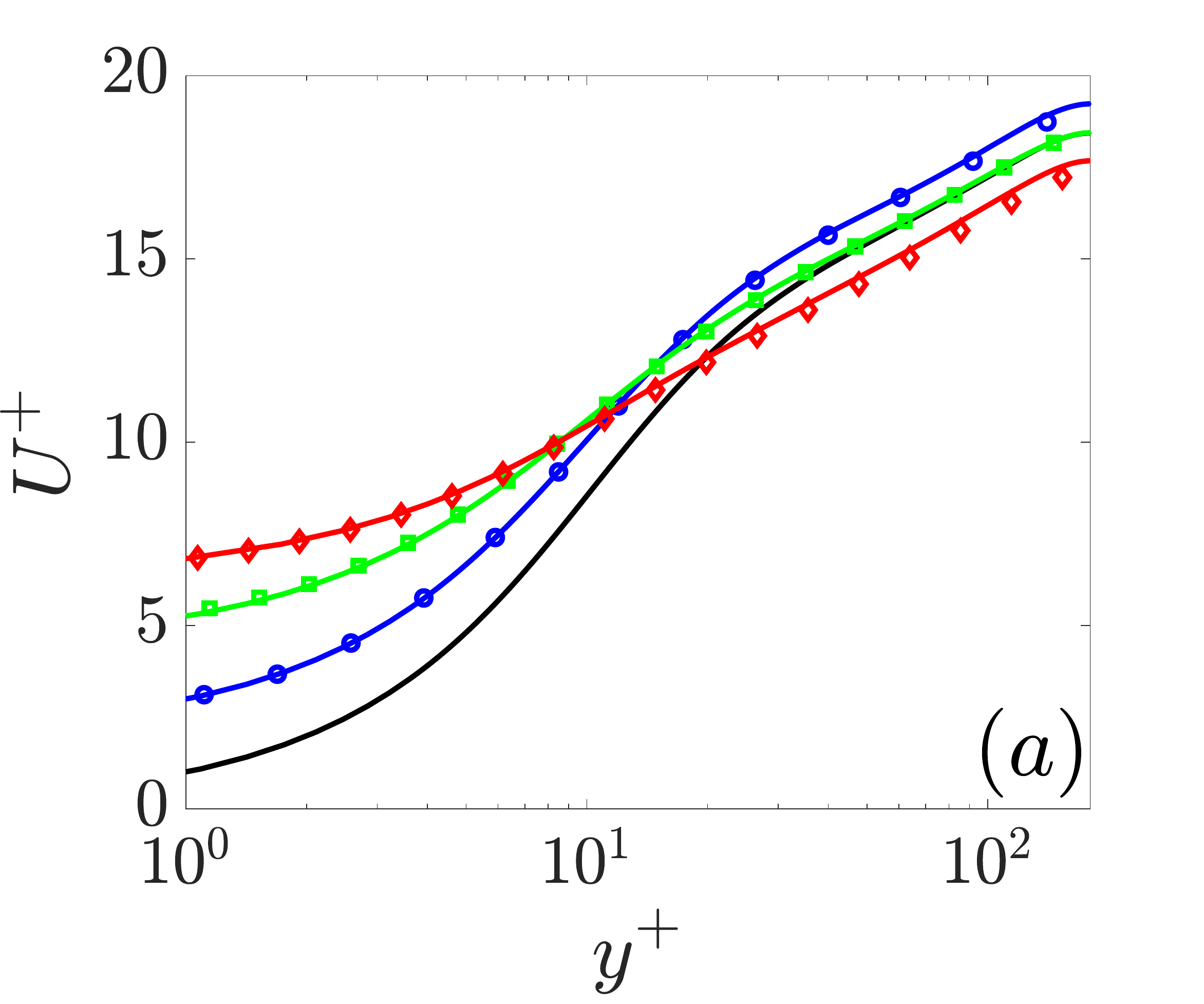}
        {\phantomcaption}
    \end{subfigure}%
    \hspace*{-5mm}
    \begin{subfigure}[tbp]{.36\textwidth}
        \includegraphics[width=1\linewidth]{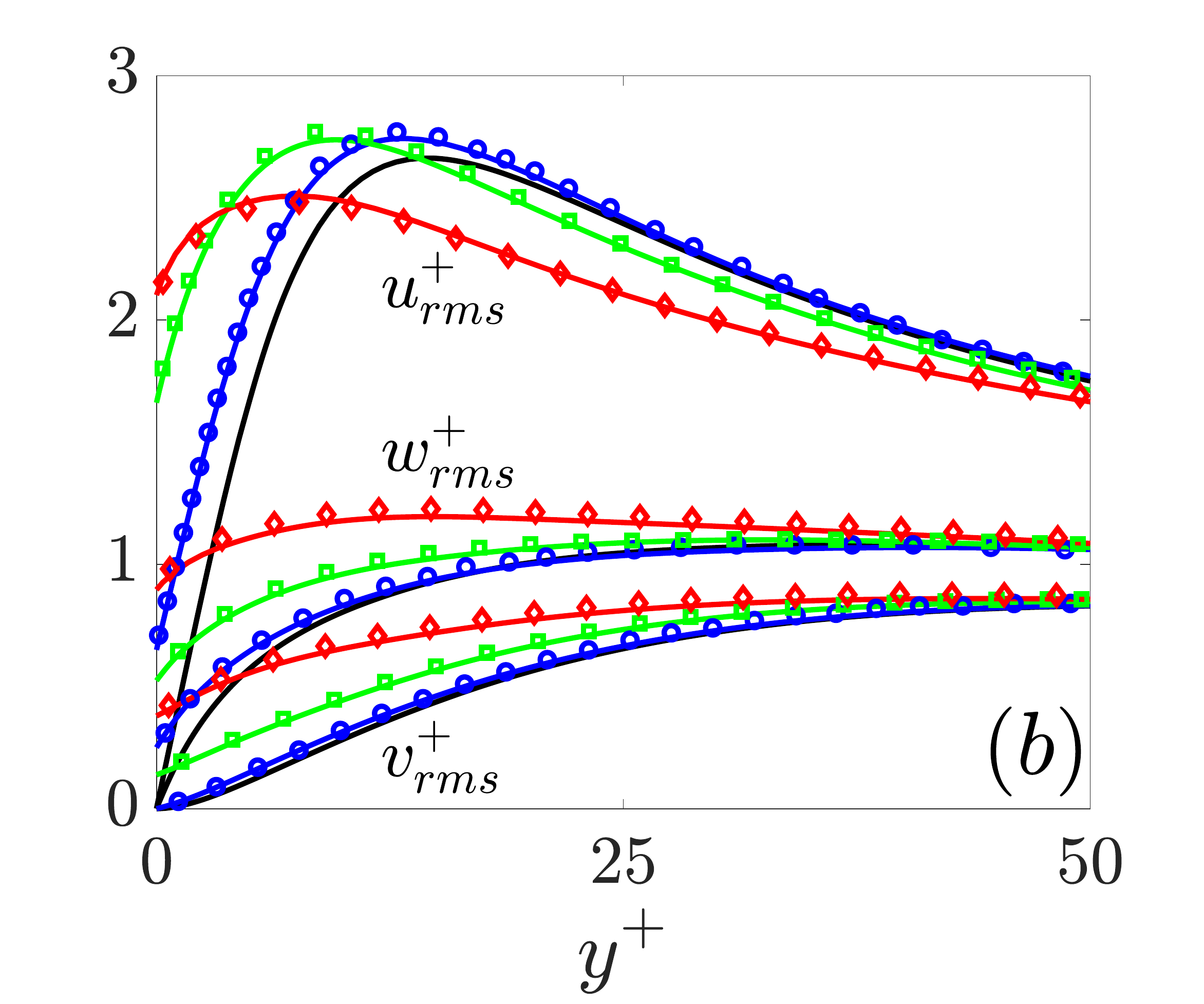}
        {\phantomcaption}
    \end{subfigure}%
    \hspace*{-4mm}
    \begin{subfigure}[tbp]{.36\textwidth}
        \includegraphics[width=1\linewidth]{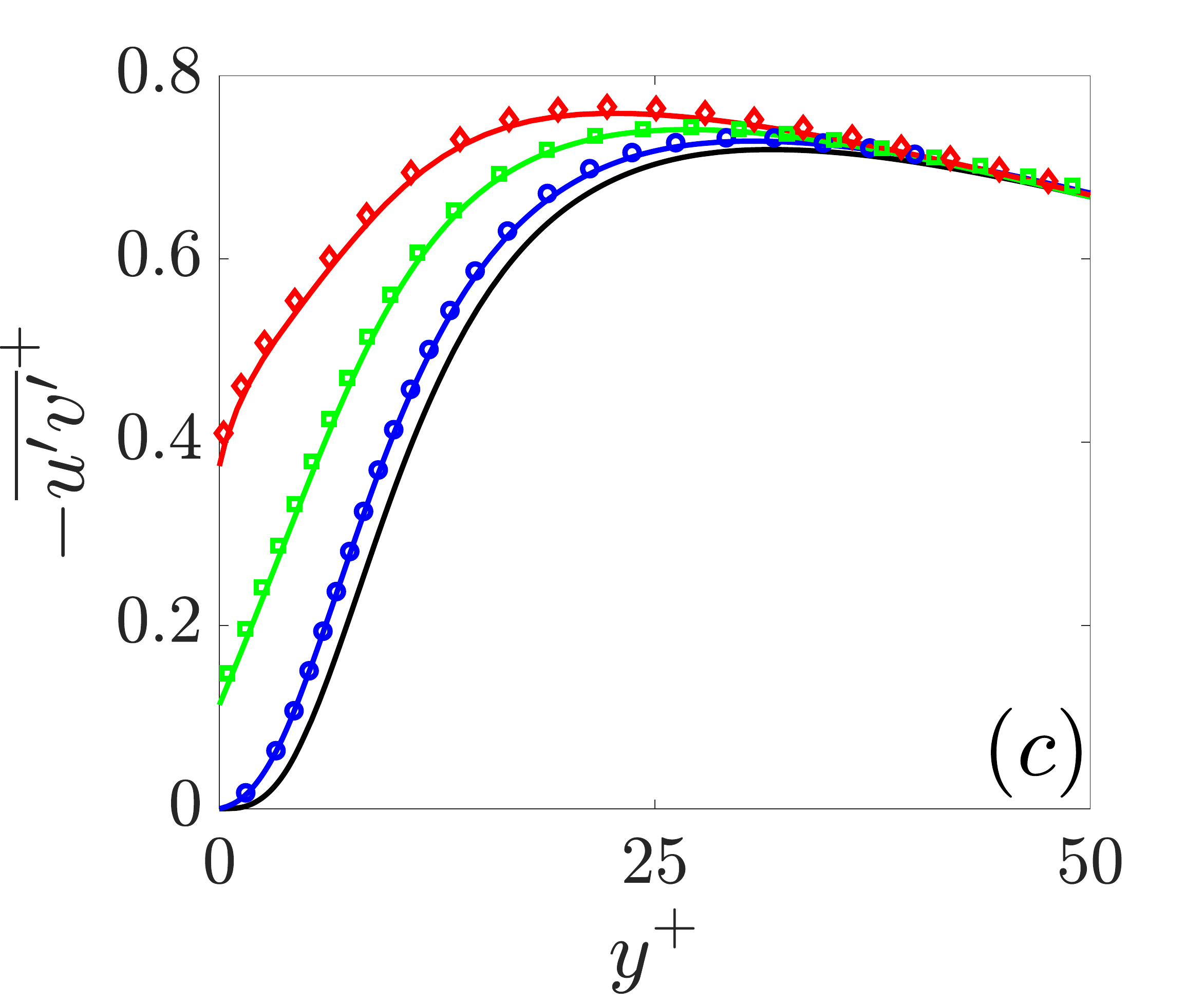}
        {\phantomcaption}
    \end{subfigure}
    \vspace*{-4mm}
    \caption{Comparison of case DHV220, {\small$\textcolor{blue}{\bigcirc}$}, of \citet{GOMEZDESEGURA2020} and cases UWVL2, {$\textcolor{green}{\square}$}, UWVL4, {$\textcolor{red}{\lozenge}$}, of \citet{ibrahim2020smoothwalllike} to cases L2M0, $\textcolor{blue}{-\!-}$, L5M5, $\textcolor{green}{-\!-}$, and L10M10, $\textcolor{red}{-\!-}$. The slip and transpiration lengths used in the boundary conditions are listed in \cref{tab:dns} of \cref{Results_TRM}. The black line represents smooth-wall channel data. Mean velocity $(a)$, r.m.s. velocity fluctuation $(b)$ and Reynolds shear stress $(c)$ profiles.}
    \label{fig:validation_TRM}
    \end{center}
\end{figure}
%%%%%%%%%%%%%%%%%%%%%%%%%%%%%%%%%%%%%%%%%%%%%%%%%%%%%%%%%%%%%%%%%%%%%%%%%
%%%%%%%%%%%%%%%%%%%%%%%%%%%%%%%%%%%%%%%%%%%%%%%%%%%%%%%%%%%%%%%%%%%%%%%%%
For validation of the TRM's Robin boundary conditions; cases L2M0, L5M5 and L10M10 were compared against case DHV220 of \citet{GOMEZDESEGURA2020} along with cases UWVL2 and UWVL4 of \citet{ibrahim2020smoothwalllike}. As mentioned in \cref{sec:TRM}, for isotropic transpiration lengths the wall-normal velocity boundary condition of \eqref{eq:TRMv} becomes similar to \eqref{eq:robin_bdry_cond}. Since matching lengths have been used in the boundary conditions of the selected cases, they are equivalent and thus the results should be the same. Unlike the solver used in this study which uses finite differences discretization in all three grid directions, that used by \citet{GOMEZDESEGURA2020} and \citet{ibrahim2020smoothwalllike} employs spectral discretization in the $x$ and $z$ directions. Nevertheless, as can be seen in \cref{fig:validation_TRM}, the data from the simulations are in good agreement, validating the implementation of the boundary conditions.

%%%%%%%%%%%%%%%%%%%%%%%%%%%%%%%%%%%%%%%%%%%%%%%%%%%%%%%%%%%%%%%%%%%%%%%%%
%%%%%%%%%%%%%%%%%%%%%%%%%%%%%%%%%%%%%%%%%%%%%%%%%%%%%%%%%%%%%%%%%%%%%%%%%
\begin{figure}
    \begin{center}
    \hspace*{-2mm}
    \begin{subfigure}[tbp]{.36\textwidth}
        \includegraphics[width=1\linewidth]{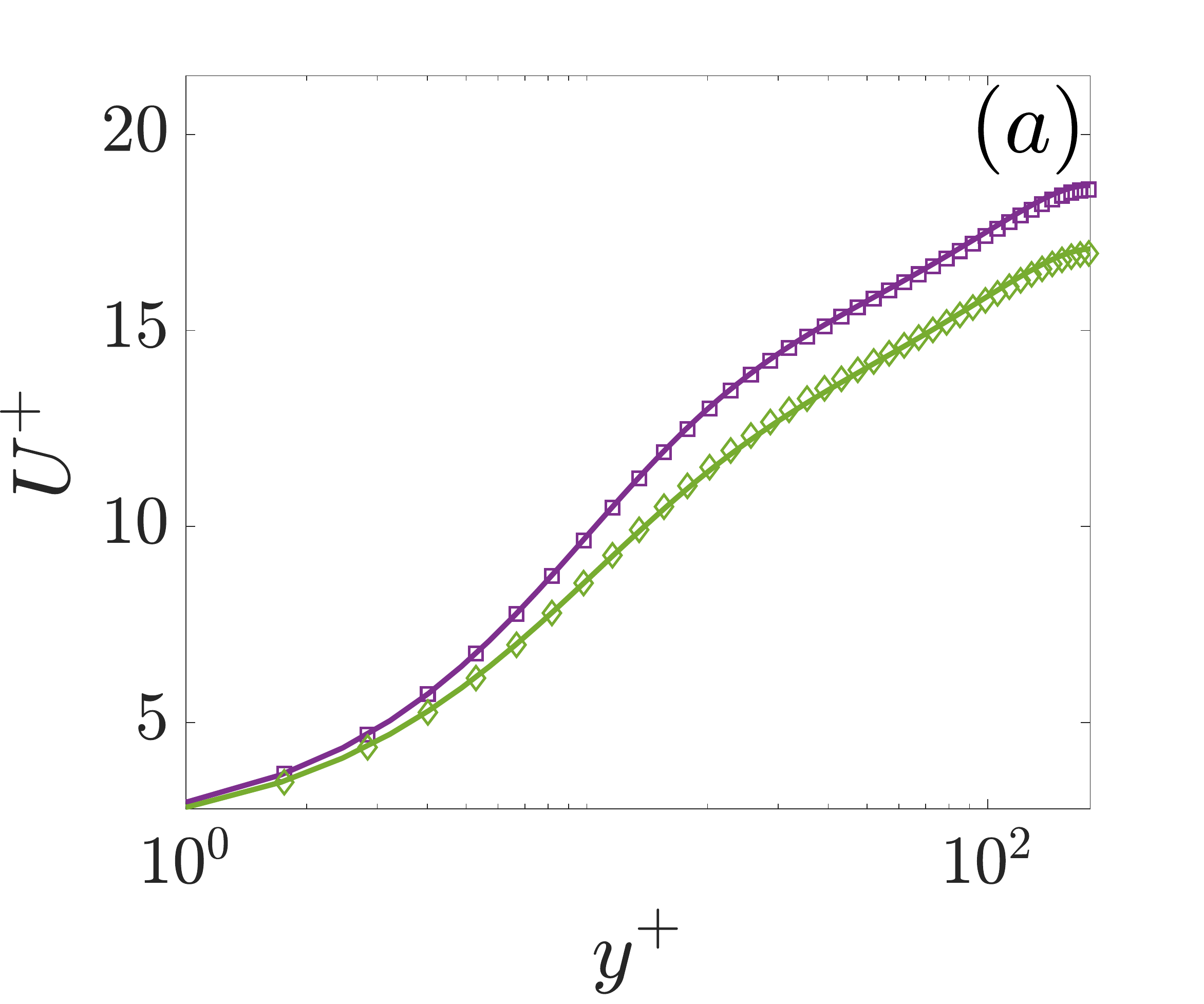}
        {\phantomcaption}
    \end{subfigure}%
    \hspace*{-5mm}
    \begin{subfigure}[tbp]{.36\textwidth}
        \includegraphics[width=1\linewidth]{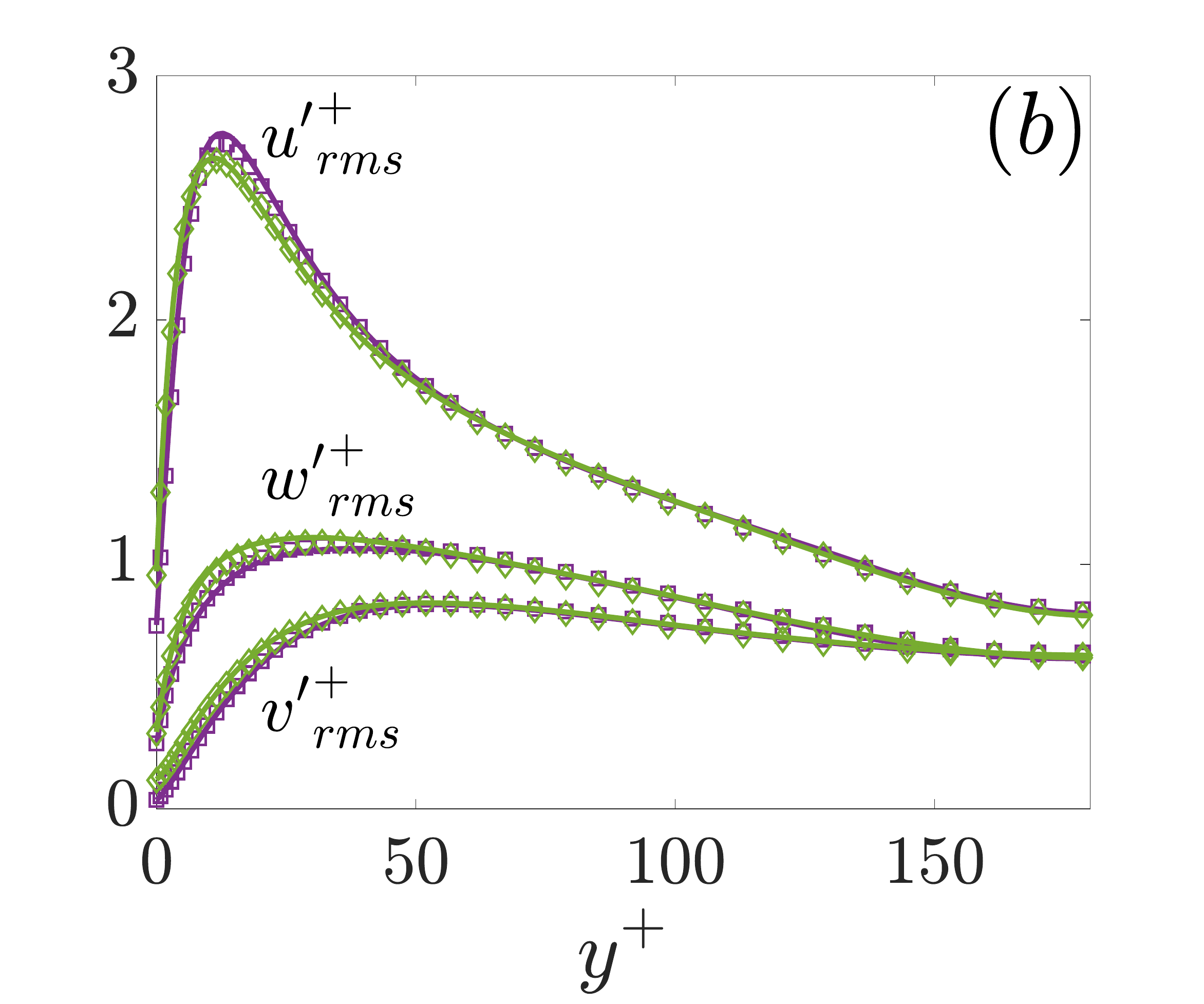}
        {\phantomcaption}
    \end{subfigure}%
    \hspace*{-4mm}
    \begin{subfigure}[tbp]{.36\textwidth}
        \includegraphics[width=1\linewidth]{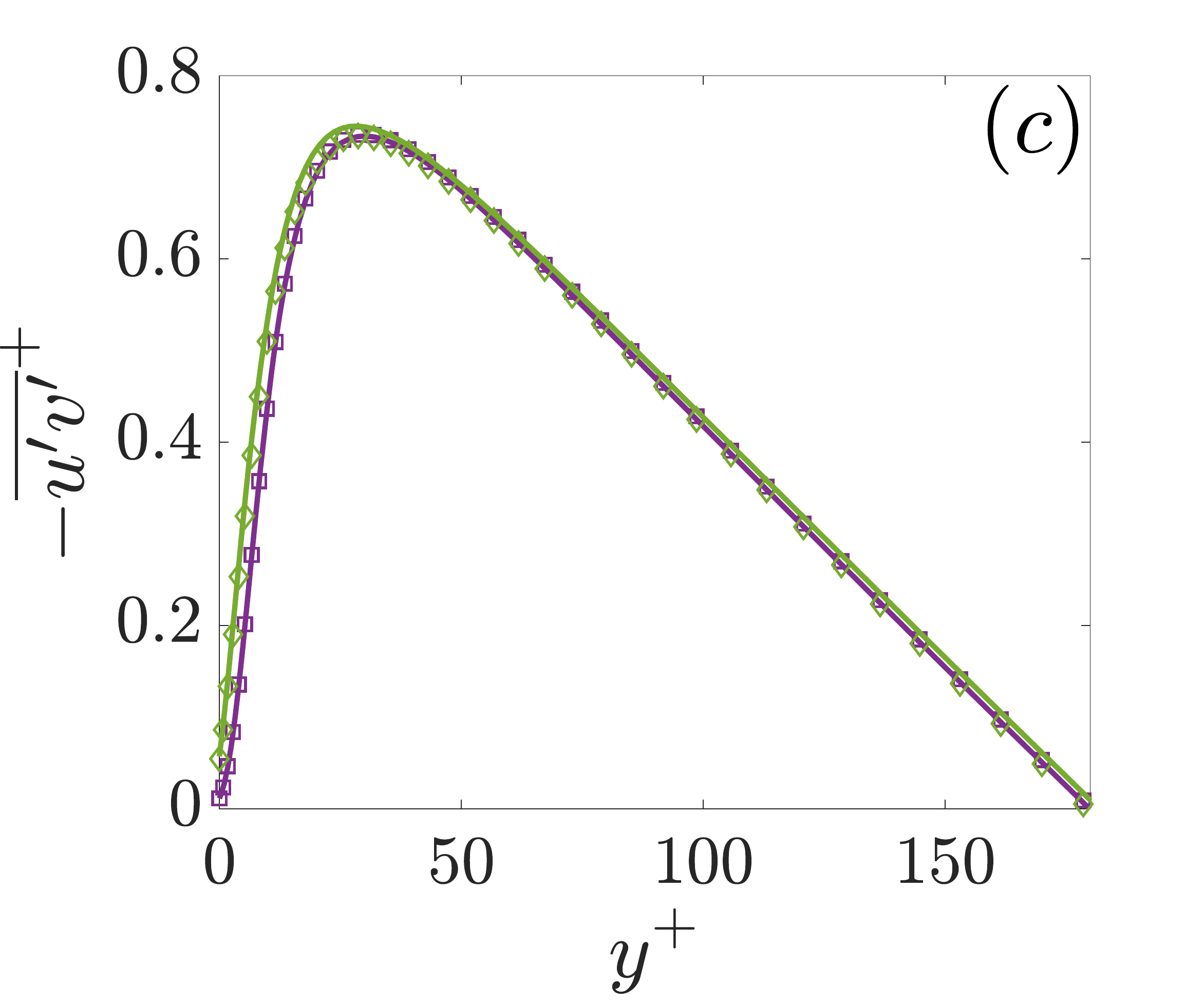}
        {\phantomcaption}
    \end{subfigure}
    \vspace*{-4mm}
    \caption{Mean velocity $(a)$, r.m.s. velocity fluctuation $(b)$ and Reynolds shear stress $(c)$ profiles. \textcolor{app-purple}{${-}$\!${-}$}, L2M2, \textcolor{app-green}{${-}$\!${-}$}, L2MZ5 with $(N_x, N_y, N_z) = (192, 144, 160)$; \textcolor{app-purple}{${-}$\!{${\scriptstyle\square}$}\!${-}$}, L2M2, \textcolor{app-green}{${-}$\!{${\scriptstyle\lozenge}$}\!${-}$}, L2MZ5 with $(N_x, N_y, N_z) = (256, 252, 256)$.}
    \label{Gridconvergence}
    \end{center}
\end{figure}
%%%%%%%%%%%%%%%%%%%%%%%%%%%%%%%%%%%%%%%%%%%%%%%%%%%%%%%%%%%%%%%%%%%%%%%%%
%%%%%%%%%%%%%%%%%%%%%%%%%%%%%%%%%%%%%%%%%%%%%%%%%%%%%%%%%%%%%%%%%%%%%%%%%
\Cref{Gridconvergence} compares two of the TRM simulations, L2M2 and L2MZ5, with higher resolution counterparts where $\Delta{x}^+ = 4.4$, $\Delta{z}^+ = 2.2$, and $0.3\leq\Delta{y}^+\leq2.7$ to assess solution grid independence. The mean velocity profiles for both cases do not exceed a $0.6\%$ difference in absolute maximum. For the r.m.s. velocity fluctuations; L2M2 and L2MZ5 have absolute maximum differences of approximately $0.2\%$ and $0.5\%$ respectively for both ${u^{\prime}}_{rms}^+$ and ${v^{\prime}}_{rms}^+$, while for ${w^{\prime}}_{rms}^+$ it is approximately $1\%$. The Reynolds shear stress for L2M2 has an absolute maximum difference of approximately $0.3\%$, while for L2MZ5 it is $1.4\%$. The results demonstrate satisfactory grid independence, with any additional accuracy obtainable by using a higher resolution grid not justifying the increased simulation cost that would result from doing so.

\section{Cross-analysis with the framework of imposing \emph{a priori} calculated virtual-origins for smooth-wall-like turbulence.} \label{app:comp_to_virtual_origin_framework}

The virtual origin framework has been leveraged in this work to highlight the contributing components to drag modification and identify whether the effect can be solely attributed to that of smooth-wall-like turbulence displacement or alterations of a different nature. It then becomes apt to further compare the results of this work to those of \citet{ibrahim2020smoothwalllike} who proposed the framework.

It should first be emphasized that the objective of \citet{ibrahim2020smoothwalllike} was prediction of the resulting virtual origins of the different velocity components for a given set of slip-length coefficients, ${\ell_x}^+$, ${\ell_y}^+$, ${\ell_z}^+$, used in the Robin boundary conditions of \eqref{eq:robin_bdry_cond}. To do so, they assume that the structure of turbulence remains smooth-wall-like and use the r.m.s. velocity profiles of a smooth-wall solution to \emph{a priori} calculate ${\ell_u}^+$, ${\ell_v}^+$ and ${\ell_w}^+$ using their wall-normal gradients. The calculated origins using smooth-wall data are valid so long as the resulting DNS solutions with the robin boundary conditions fall within the smooth-wall turbulence regime. Within this regime, it also becomes possible to predict the resulting origin for the turbulence, ${y^+=-\ell_{uv}}^+$, using the origins of ${\ell_w}^+$ and ${\ell_v}^+$, which quantify the displacement of the quasi-streamwise vortices.
% \textbf{(a priori to what? I'm not sure I understand what the difference is actually: For tangential components, do you mean that $\ell^+_u, \ell^+_w$ were inputs in their analysis? while TRM inputs are $\ell^+_x, \ell^+_z$? where virtual origins obtained by shift rms profile may be different than the imposed slip lengths).}

% \textbf{I guess we addressing referee comments with the text below. However, I'm not sure what message we are trying to convey. Is it related to the "fundamental assumption" mentioned above? Do you want to emphasize that it is perhaps not a very good one?}
There is also a subtle difference between the virtual origins of the velocities, ${\ell_u}^+$, ${\ell_v}^+$ and ${\ell_w}^+$, and that of the turbulence, ${\ell_{uv}}^+$. The latter defines the imaginary smooth-wall that the near-wall cycle (essentially the quasi-streamwise vortices) perceive while the former represent the imaginary walls where the velocity components are expected to decay to zero from their values at the boundary plane ($y^+=0$). Hence, while shifting the origin to ${\ell_{uv}}^+$ recovers smooth-wall-like turbulence profiles, there would still be some discrepancy in ${u^{\prime}}^+$, ${v^{\prime}}^+$, ${w^{\prime}}^+$ and $-\overline{u^{\prime}v^{\prime}}^+$ for the first few layers of $y^+$. This is evident in the results of \citet{ibrahim2020smoothwalllike} and in the rough-wall simulations of \citet{Abderrahaman2019} which remained smooth-wall-like, with the latter authors observing that ``Very near the rough surface, different variables would extrapolate to zero at different heights, as is particularly evident for ${u^{\prime}}^+$ and ${w^{\prime}}^+$ which experience virtual origins shallower than ${\ell_{uv}}^+$'' (see figure 13 of the respective paper). This was also pointed out by \citet{ibrahim2020smoothwalllike}, where they state that ``The offsets ${\ell_v}^+$ and ${\ell_w}^+$ are merely prescribed, \emph{a priori} values to quantify the offset in $v$ and $w$ caused by the surface, but turbulence would react to their combined effect, perceiving a single origin if it is to remain smooth-wall-like.'' Additionally, the weakening of permeability to any degree will subject the otherwise viscous-dominated near-wall region to greater perturbations and degrade its laminar-like quality. Hence, \emph{a priori} determined virtual origins derived from smooth-wall statistics will not be entirely representative of the actual resulting virtual origins of ${u^{\prime}}^+$, ${v^{\prime}}^+$ and ${w^{\prime}}^+$ from DNS where length coefficients have been imposed using boundary conditions for the velocity components, even if the overall turbulence statistics remain smooth-wall-like.

The preceding explanations are the reason why, for cases with matching boundary condition coefficients, the \emph{a posteriori} calculated virtual origins in \cref{tab:dns} differ from the \emph{a priori} ones of \citet{ibrahim2020smoothwalllike}. As well be examined further below, this discrepancy does not extend to the origin of the turbulence for matching cases which retain smooth-wall-like turbulence.

For the cases in \cref{tab:dns} involving a transpiration boundary condition, the virtual origin of the turbulence, ${\ell_{uv}}^+$, generally falls in-between those of the spanwise and wall-normal velocities, ${\ell_w}^+$ and ${\ell_v}^+$. This is consistent with the observations made by \citet{gomez_2018}, \citet{GOMEZDESEGURA2020} and \citet{ibrahim2020smoothwalllike}. The underlying physical mechanism implied by this was explained in \cref{sec:virtual}.

For drag reducing cases in which the imposed transpiration length was always smaller than the spanwise slip length (${\ell_y}^+<{\ell_z}^+$), but the predicted origin for $v$ was deeper than $w$ (${\ell_v}^+>{\ell_w}^+$), \citet{ibrahim2020smoothwalllike} observed that ${\ell_{uv}}^+={\ell_{w}}^+$. None of the cases in \cref{tab:dns} have transpiration lengths which are smaller than the slip lengths. The drag neutral case of L2M2 has ${\ell_v}^+\approx2.4$ and ${\ell_w}^+\approx1.7$, but this does not result in ${\ell_{uv}}^+={\ell_{w}}^+$. Interestingly, case DHV222 of \citet{gomez_2018} is similar to case L2M2 and the virtual origins reported for it are ${\ell_v}^+\approx2.3$ and ${\ell_w}^+\approx2.0$ \citep[$1.7$ when curvature is accounted for, as later reported by][]{GOMEZDESEGURA2020}.
% This is indicative that the condition of ${\ell_{uv}}^+={\ell_{w}}^+$ when ${\ell_v}^+>{\ell_w}^+$ is not generally applicable to all cases that fall within the smooth-wall-like turbulence regime, at the very least it does not seem to apply to \emph{a posteriori} calculated origins.

%%%%%%%%%%%%%%%%%%%%%%%%%%%%%%%%%%%%%%%%%%%%%%%%%%%%%%%%%%%%%%%%%%%%%%%%%
%%%%%%%%%%%%%%%%%%%%%%%%%%%%%%%%%%%%%%%%%%%%%%%%%%%%%%%%%%%%%%%%%%%%%%%%%
\begin{figure}
    \begin{center}
    \hspace*{-2mm}
    \begin{subfigure}[tbp]{.36\textwidth}
        \includegraphics[width=1\linewidth]{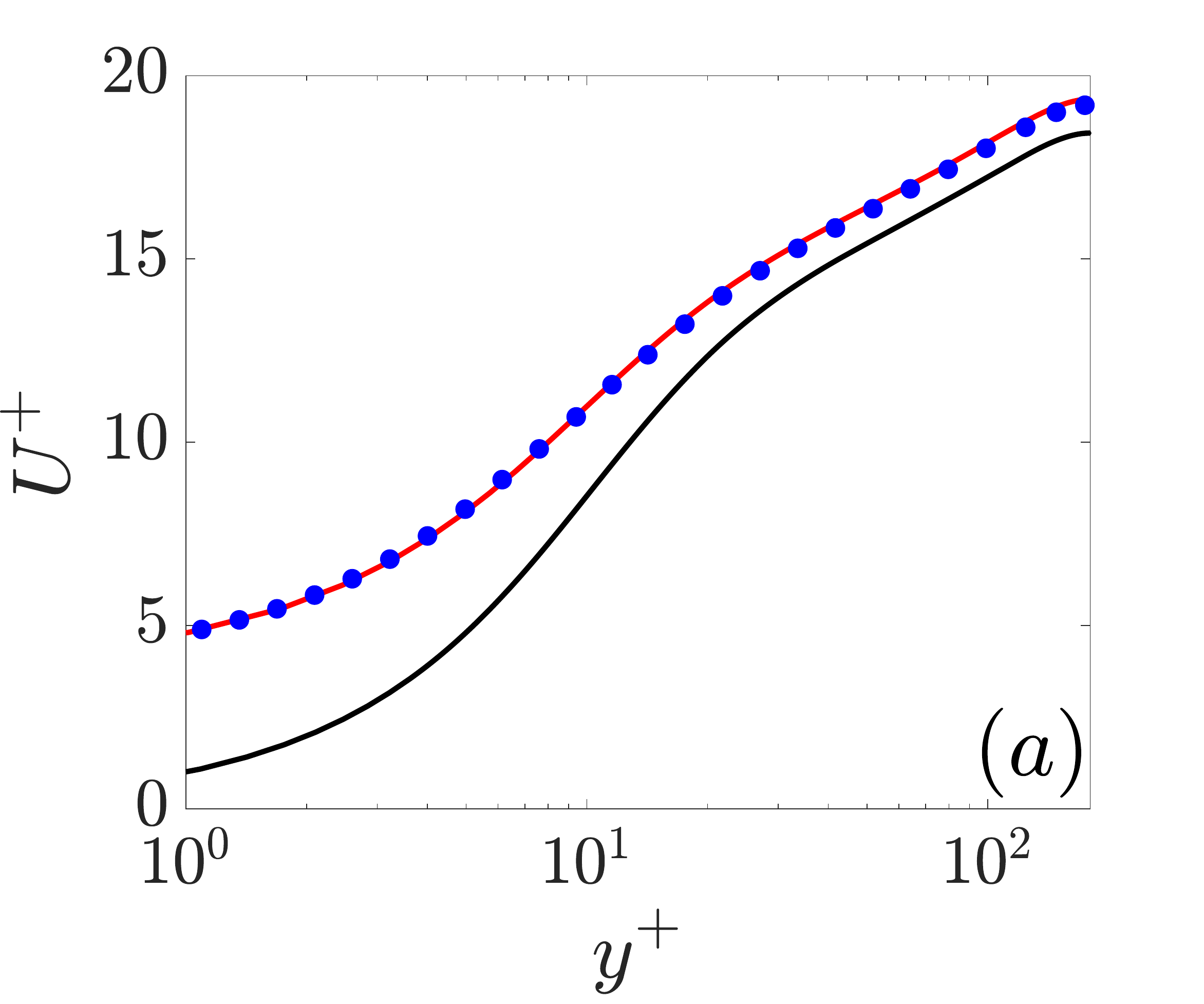}
        {\phantomcaption}
    \end{subfigure}%
    \hspace*{-5mm}
    \begin{subfigure}[tbp]{.36\textwidth}
        \includegraphics[width=1\linewidth]{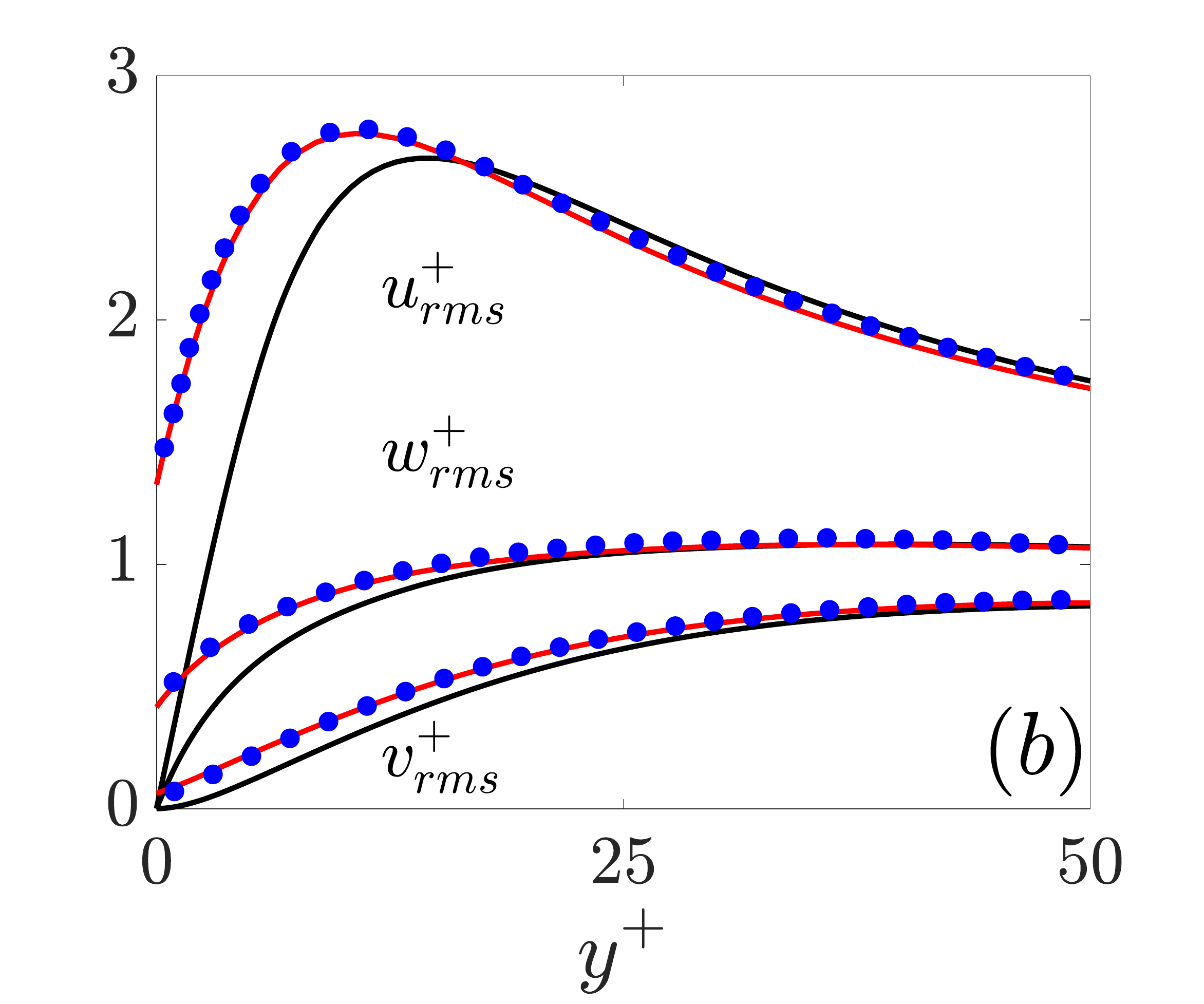}
        {\phantomcaption}
    \end{subfigure}%
    \hspace*{-4mm}
    \begin{subfigure}[tbp]{.36\textwidth}
        \includegraphics[width=1\linewidth]{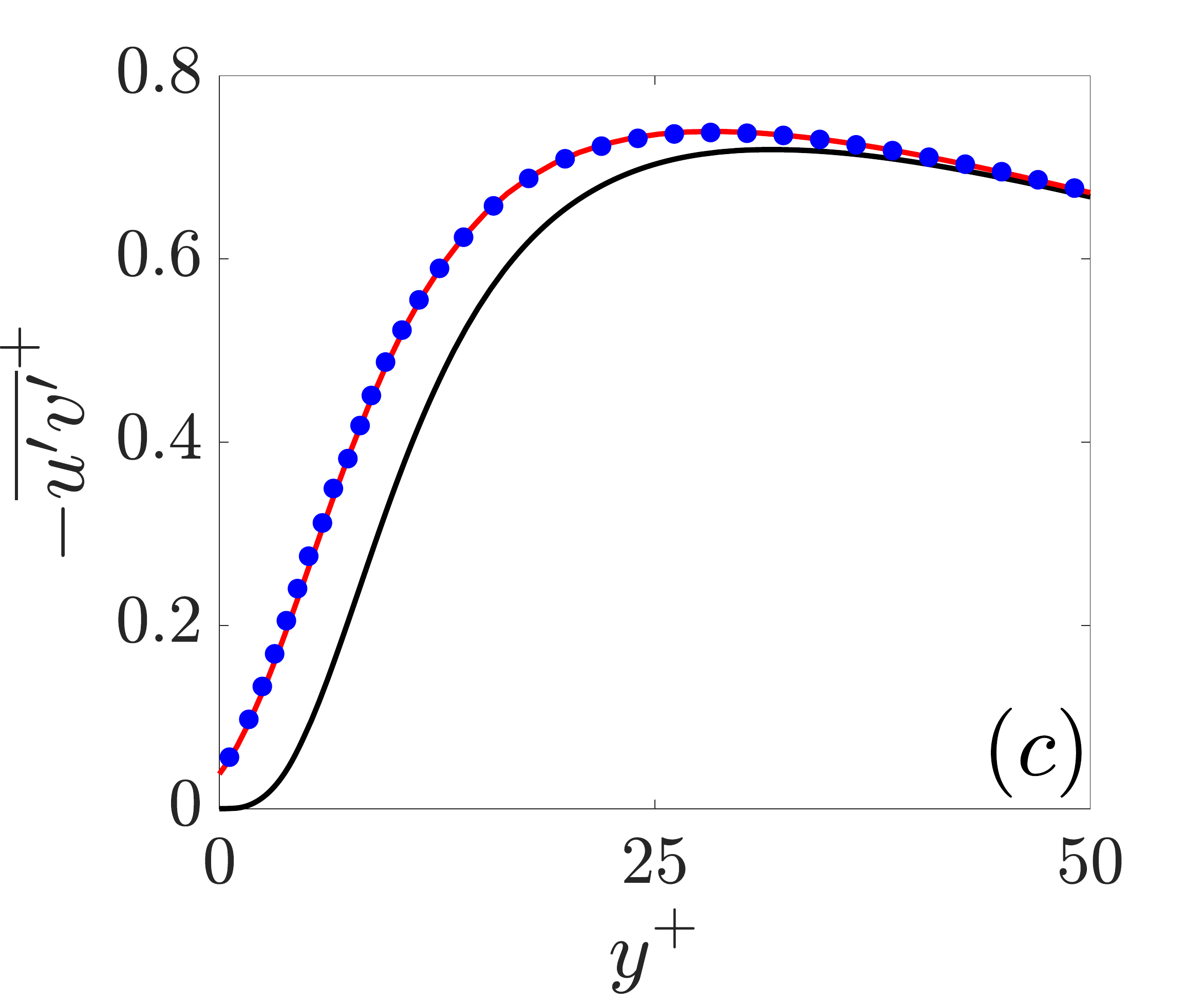}
        {\phantomcaption}
    \end{subfigure}
    \vspace*{-4mm}
    \caption{Comparison of case UWV4, {$\textcolor{blue}{\bullet}$}, of \citet{ibrahim2020smoothwalllike} to case L4M2.5, $\textcolor{red}{-\!-}$. The slip and transpiration lengths used were ${\ell_x}^+={\ell_z}^+=4.0$ and ${m_x}^+={m_z}^+=2.5$. The black line represents smooth-wall data. Mean velocity $(a)$, r.m.s. velocity fluctuations $(b)$ and Reynolds shear stress $(c)$.}
    \label{fig:validation_UWV4}
    \end{center}
\end{figure}
%%%%%%%%%%%%%%%%%%%%%%%%%%%%%%%%%%%%%%%%%%%%%%%%%%%%%%%%%%%%%%%%%%%%%%%%%
%%%%%%%%%%%%%%%%%%%%%%%%%%%%%%%%%%%%%%%%%%%%%%%%%%%%%%%%%%%%%%%%%%%%%%%%%
To better assess this observed difference, case UWV4 of \citet{ibrahim2020smoothwalllike} was replicated. The case is designated L4M2.5 and its raw results were in agreement with that of UWV4 (\cref{fig:validation_UWV4}). The \emph{a posteriori} estimated virtual origins for L4M2.5 however were ${\ell_u}^+\approx3.2$, ${\ell_w}^+\approx3.0$ and ${\ell_v}^+\approx2.8$, unlike the \emph{a priori} obtained origins for UWV4 which were ${\ell_u}^+\approx3.6$, ${\ell_w}^+\approx2.9$ and ${\ell_v}^+\approx3.9$. The \emph{a posteriori} calculated mean flow and turbulence origins for both UWV4 and L4M2.5 are in agreement \citep[${\ell_U}^+\approx3.9$, ${\ell_{uv}}^+\approx3.0$ for L4M2.5 and ${\ell_U}^+\approx3.8$, ${\ell_{uv}}^+\approx2.9$ for UWV4 of][]{ibrahim2020smoothwalllike}, which is to be expected since the simulation results were in agreement.

% Based on the analysis made thus far, the predictive conditions proposed by \citet{ibrahim2020smoothwalllike} should be applied when the aim is to predict virtual origins for the smooth-wall-like regime of turbulence due to the imposition of velocity boundary conditions, and not generally to DNS data resulting from such boundary conditions.   

% It should be mentioned however that the proposed expression by \citet{ibrahim2020smoothwalllike} for predicting ${\ell_{uv}}^+$ using the \emph{a priori} calculated ${\ell_w}^+$ and ${\ell_v}^+$ (see equation 4.5 of the respective paper) remains valid when used with the \emph{a posteriori} determined ${\ell_w}^+$ and ${\ell_v}^+$ of case L4M2.5, and also with those of case L2M2.

Beyond their drag reducing cases, the virtual origin framework remained inapplicable to the cases of \citet{ibrahim2020smoothwalllike} which demonstrated drag increase as most of them fell outside of the regime smooth-wall-like turbulence. The drag increasing cases of this work also fall outside of the smooth-wall-like regime. Hence, there are no discrepancies where such cases are concerned.

\FloatBarrier
\bibliographystyle{jfm}
\bibliography{main}

\begin{thebibliography}{52}
\expandafter\ifx\csname natexlab\endcsname\relax\def\natexlab#1{#1}\fi
\def\au#1{#1} \def\ed#1{#1} \def\yr#1{#1}\def\at#1{#1}\def\jt#1{\textit{#1}}
  \def\bt#1{#1}\def\bvol#1{\textbf{#1}} \def\vol#1{#1} \def\pg#1{#1}
  \def\publ#1{#1}\def\arxiv#1{#1}\def\org#1{#1}\def\st#1{\textit{#1}}

\bibitem[Abderrahaman-Elena {\em et~al.\/}(2019)Abderrahaman-Elena, Fairhall \&
  Garc{\'{\i}}a-Mayoral]{Abderrahaman2019}
{\sc \au{Abderrahaman-Elena, N.}, \au{Fairhall, C.T.} \&
  \au{Garc{\'{\i}}a-Mayoral, R.}} \yr{2019}  \at{Modulation of near-wall
  turbulence in the transitionally rough regime}.  \jt{Journal of Fluid
  Mechanics}  \bvol{865},  \pg{1042–1071}.

\bibitem[Bechert \& Bartenwerfer(1989)]{bechert_bartenwerfer_1989}
{\sc \au{Bechert, D.W.} \& \au{Bartenwerfer, M.}} \yr{1989}  \at{The viscous
  flow on surfaces with longitudinal ribs}.  \jt{Journal of Fluid Mechanics}
  \bvol{206},  \pg{105–129}.

\bibitem[Bose \& Park(2018)]{bose_2018}
{\sc \au{Bose, S.~T.} \& \au{Park, G.~I.}} \yr{2018}  \at{Wall-modeled
  large-eddy simulation for complex turbulent flows}.  \jt{Annual Review of
  Fluid Mechanics}  \bvol{50}~(1),  \pg{535--561}.

\bibitem[Bottaro(2019)]{bottaro_2019}
{\sc \au{Bottaro, A.}} \yr{2019}  \at{Flow over natural or engineered surfaces:
  an adjoint homogenization perspective}.  \jt{Journal of Fluid Mechanics}
  \bvol{877},  \pg{P1}.

\bibitem[Breugem {\em et~al.\/}(2006)Breugem, Boersma \&
  Uittenbogaard]{breugem_boersma_uittenbogaard_2006}
{\sc \au{Breugem, W.P.}, \au{Boersma, B.J.} \& \au{Uittenbogaard, R.E.}}
  \yr{2006}  \at{The influence of wall permeability on turbulent channel flow}.
   \jt{Journal of Fluid Mechanics}  \bvol{562},  \pg{35–72}.

\bibitem[Busse \& Sandham(2012)]{busse_2012}
{\sc \au{Busse, A.} \& \au{Sandham, N.D.}} \yr{2012}  \at{Influence of an
  anisotropic slip-length boundary condition on turbulent channel flow}.
  \jt{Physics of Fluids}  \bvol{24}~(5),  \pg{55--111}.

\bibitem[Chavarin \& Luhar(2020)]{Chavarin_luhar_Aiaa_2020}
{\sc \au{Chavarin, Andrew} \& \au{Luhar, Mitul}} \yr{2020}  \at{Resolvent
  analysis for turbulent channel flow with riblets}.  \jt{AIAA Journal}
  \bvol{58}~(2),  \pg{589--599}.

\bibitem[Chorin(1968)]{Chorin1968}
{\sc \au{Chorin, A.~J.}} \yr{1968}  \at{Numerical solution of the navier-stokes
  equations}.  \jt{Mathematics of Computation}  \bvol{22}~(104),
  \pg{745--762}.

\bibitem[Chung {\em et~al.\/}(2015)Chung, Chan, MacDonald, Hutchins \&
  Ooi]{chung_chan_macdonald_hutchins_ooi_2015}
{\sc \au{Chung, D.}, \au{Chan, L.}, \au{MacDonald, M.}, \au{Hutchins, N.} \&
  \au{Ooi, A.}} \yr{2015}  \at{A fast direct numerical simulation method for
  characterising hydraulic roughness}.  \jt{Journal of Fluid Mechanics}
  \bvol{773},  \pg{418–431}.

\bibitem[Chung {\em et~al.\/}(2021)Chung, Hutchins, Schultz \&
  Flack]{chung_2021}
{\sc \au{Chung, D.}, \au{Hutchins, N.}, \au{Schultz, M.P.} \& \au{Flack, K.A.}}
  \yr{2021}  \at{Predicting the drag of rough surfaces}.  \jt{Annual Review of
  Fluid Mechanics}  \bvol{53}~(1),  \pg{439--471}.

\bibitem[Clauser(1956)]{CLAUSER19561}
{\sc \au{Clauser, F.H.}} \yr{1956} {\em The Turbulent Boundary Layer\/},
  \st{Advances in Applied Mechanics},  \vol{vol.~4},  \pg{pp. 1 -- 51}.
  \publ{Elsevier}.

\bibitem[Costa(2018)]{costa2018}
{\sc \au{Costa, P.}} \yr{2018}  \at{A fft-based finite-difference solver for
  massively-parallel direct numerical simulations of turbulent flows}.
  \jt{Computers \& Mathematics with Applications}  \bvol{76}~(8),  \pg{1853 --
  1862}.

\bibitem[Endrikat {\em et~al.\/}(2021)Endrikat, Modesti, Garc{\'{\i}}a-Mayoral,
  Hutchins \& Chung]{endrikat_2021}
{\sc \au{Endrikat, S.}, \au{Modesti, D.}, \au{Garc{\'{\i}}a-Mayoral, R.},
  \au{Hutchins, N.} \& \au{Chung, D.}} \yr{2021}  \at{Influence of riblet
  shapes on the occurrence of kelvin–helmholtz rollers}.  \jt{Journal of
  Fluid Mechanics}  \bvol{913},  \pg{A37}.

\bibitem[Fairhall {\em et~al.\/}(2019)Fairhall, Abderrahaman-Elena \&
  Garc{\'{\i}}a-Mayoral]{fairhall_abderrahaman-elena_garcia-mayoral_2019}
{\sc \au{Fairhall, C.T.}, \au{Abderrahaman-Elena, N.} \&
  \au{Garc{\'{\i}}a-Mayoral, R.}} \yr{2019}  \at{The effect of slip and surface
  texture on turbulence over superhydrophobic surfaces}.  \jt{Journal of Fluid
  Mechanics}  \bvol{861},  \pg{88–118}.

\bibitem[Forooghi {\em et~al.\/}(2018{\natexlab{{\em a\/}}})Forooghi, Stroh,
  Schlatter \& Frohnapfel]{Forooghi_rough_2018}
{\sc \au{Forooghi, .}, \au{Stroh, A.}, \au{Schlatter, P.} \& \au{Frohnapfel,
  B.}} \yr{2018{\natexlab{{\em a\/}}}}  \at{Direct numerical simulation of flow
  over dissimilar, randomly distributed roughness elements: A systematic study
  on the effect of surface morphology on turbulence}.  \jt{Phys. Rev. Fluids}
  \bvol{3},  \pg{044605}.

\bibitem[Forooghi {\em et~al.\/}(2018{\natexlab{{\em b\/}}})Forooghi,
  Frohnapfel, Magagnato \& Busse]{Forooghi_2018}
{\sc \au{Forooghi, Pourya}, \au{Frohnapfel, Bettina}, \au{Magagnato, Franco} \&
  \au{Busse, Angela}} \yr{2018{\natexlab{{\em b\/}}}}  \at{A modified
  parametric forcing approach for modelling of roughness}.  \jt{International
  Journal of Heat and Fluid Flow}  \bvol{71},  \pg{200--209}.

\bibitem[Fukagata {\em et~al.\/}(2006)Fukagata, Kasagi \&
  Koumoutsakos]{Fukagata_2006}
{\sc \au{Fukagata, K.}, \au{Kasagi, N.} \& \au{Koumoutsakos, P.}} \yr{2006}
  \at{A theoretical prediction of friction drag reduction in turbulent flow by
  superhydrophobic surfaces}.  \jt{Physics of Fluids}  \bvol{18}~(5),
  \pg{051703}.

\bibitem[Garc{\'{\i}}a-Mayoral {\em et~al.\/}(2019)Garc{\'{\i}}a-Mayoral,
  {G{\'{o}}mez-de-Segura} \& Fairhall]{Garcia_Mayoral_2019}
{\sc \au{Garc{\'{\i}}a-Mayoral, R.}, \au{{G{\'{o}}mez-de-Segura}, G.} \&
  \au{Fairhall, C.T.}} \yr{2019}  \at{The control of near-wall turbulence
  through surface texturing}.  \jt{Fluid Dynamics Research}  \bvol{51}~(1),
  \pg{011410}.

\bibitem[Garc{\'{\i}}a-Mayoral \& Jiménez(2011)]{garcia-mayoral_2011}
{\sc \au{Garc{\'{\i}}a-Mayoral, R.} \& \au{Jiménez, J.}} \yr{2011}
  \at{Hydrodynamic stability and breakdown of the viscous regime over riblets}.
   \jt{Journal of Fluid Mechanics}  \bvol{678},  \pg{317–347}.

\bibitem[{G{\'{o}}mez-de-Segura} {\em et~al.\/}(2018){G{\'{o}}mez-de-Segura},
  Fairhall, MacDonald, Chung \& Garc{\'{\i}}a-Mayoral]{gomez_2018}
{\sc \au{{G{\'{o}}mez-de-Segura}, G.}, \au{Fairhall, C.~T.}, \au{MacDonald,
  M.}, \au{Chung, D.} \& \au{Garc{\'{\i}}a-Mayoral, R.}} \yr{2018}
  \at{Manipulation of near-wall turbulence by surface slip and permeability}.
  \jt{Journal of Physics: Conference Series}  \bvol{1001},  \pg{012011}.

\bibitem[{G{\'{o}}mez-de-Segura} \& Garc{\'{\i}}a-Mayoral(2019)]{gomez_2019}
{\sc \au{{G{\'{o}}mez-de-Segura}, G.} \& \au{Garc{\'{\i}}a-Mayoral, R.}}
  \yr{2019}  \at{Turbulent drag reduction by anisotropic permeable substrates
  – analysis and direct numerical simulations}.  \jt{Journal of Fluid
  Mechanics}  \bvol{875},  \pg{124–172}.

\bibitem[{G{\'{o}}mez-de-Segura} \&
  Garc{\'{\i}}a-Mayoral(2020)]{GOMEZDESEGURA2020}
{\sc \au{{G{\'{o}}mez-de-Segura}, G.} \& \au{Garc{\'{\i}}a-Mayoral, R.}}
  \yr{2020}  \at{Imposing virtual origins on the velocity components in direct
  numerical simulations}.  \jt{International Journal of Heat and Fluid Flow}
  \bvol{86},  \pg{108675}.

\bibitem[Ibrahim {\em et~al.\/}(2021)Ibrahim, {G{\'{o}}mez-de-Segura}, Chung \&
  Garc{\'{\i}}a-Mayoral]{ibrahim2020smoothwalllike}
{\sc \au{Ibrahim, J.I.}, \au{{G{\'{o}}mez-de-Segura}, G.}, \au{Chung, D.} \&
  \au{Garc{\'{\i}}a-Mayoral, R.}} \yr{2021} The smooth-wall-like behaviour of
  turbulence over drag-altering surfaces: a unifying virtual-origin framework.

\bibitem[Jiménez(2004)]{jimenez_rough}
{\sc \au{Jiménez, J.}} \yr{2004}  \at{Turbulent flows over rough walls}.
  \jt{Annual Review of Fluid Mechanics}  \bvol{36}~(1),  \pg{173--196},
  \arxiv{arXiv: https://doi.org/10.1146/annurev.fluid.36.050802.122103}.

\bibitem[Jiménez \& Pinelli(1999)]{jimenez_pinelli_1999}
{\sc \au{Jiménez, J.} \& \au{Pinelli, A.}} \yr{1999}  \at{The autonomous cycle
  of near-wall turbulence}.  \jt{Journal of Fluid Mechanics}  \bvol{389},
  \pg{335–359}.

\bibitem[Jiménez {\em et~al.\/}(2001)Jiménez, Uhlmann, Pinelli \&
  Kawahara]{jimenez_2001}
{\sc \au{Jiménez, J.}, \au{Uhlmann, M.}, \au{Pinelli, A.} \& \au{Kawahara,
  G.}} \yr{2001}  \at{Turbulent shear flow over active and passive porous
  surfaces}.  \jt{Journal of Fluid Mechanics}  \bvol{442},  \pg{89–117}.

\bibitem[Kim \& Moin(1985)]{kim1985}
{\sc \au{Kim, J.} \& \au{Moin, P.}} \yr{1985}  \at{Application of a
  fractional-step method to incompressible navier-stokes equations}.
  \jt{Journal of Computational Physics}  \bvol{59}~(2),  \pg{308--323}.

\bibitem[Kim {\em et~al.\/}(1987)Kim, Moin \& Moser]{kim_moin_moser_1987}
{\sc \au{Kim, J.}, \au{Moin, P.} \& \au{Moser, R.}} \yr{1987}  \at{Turbulence
  statistics in fully developed channel flow at low reynolds number}.
  \jt{Journal of Fluid Mechanics}  \bvol{177},  \pg{133–166}.

\bibitem[Kline {\em et~al.\/}(1967)Kline, Reynolds, Schraub \&
  Runstadler]{kline_1967}
{\sc \au{Kline, S.J.}, \au{Reynolds, W.C.}, \au{Schraub, F.A.} \&
  \au{Runstadler, P.W.}} \yr{1967}  \at{The structure of turbulent boundary
  layers}.  \jt{Journal of Fluid Mechanics}  \bvol{30}~(4),  \pg{741–773}.

\bibitem[Kuwata \& Suga(2017)]{kuwata_suga_2017}
{\sc \au{Kuwata, Y.} \& \au{Suga, K.}} \yr{2017}  \at{Direct numerical
  simulation of turbulence over anisotropic porous media}.  \jt{Journal of
  Fluid Mechanics}  \bvol{831},  \pg{41–71}.

\bibitem[Lacis {\em et~al.\/}(2020)Lacis, Sudhakar, Pasche \&
  Bagheri]{Lacis2020}
{\sc \au{Lacis, U.}, \au{Sudhakar, Y.}, \au{Pasche, S.} \& \au{Bagheri, S.}}
  \yr{2020}  \at{Transfer of mass and momentum at rough and porous surfaces}.
  \jt{Journal of Fluid Mechanics}  \bvol{884},  \pg{A21}.

\bibitem[Lee \& Moser(2015)]{lee_moser_2015}
{\sc \au{Lee, M.} \& \au{Moser, R.D.}} \yr{2015}  \at{Direct numerical
  simulation of turbulent channel flow up to $\mathit{Re}_{{\it\tau}}\approx
  5200$}.  \jt{Journal of Fluid Mechanics}  \bvol{774},  \pg{395–415}.

\bibitem[Leonardi {\em et~al.\/}(2003)Leonardi, Orlandi, Smalley, Djenidi \&
  Antonia]{leonardi_orlandi_smalley_djenidi_antonia_2003}
{\sc \au{Leonardi, S.}, \au{Orlandi, P.}, \au{Smalley, R.~J.}, \au{Djenidi, L.}
  \& \au{Antonia, R.~A.}} \yr{2003}  \at{Direct numerical simulations of
  turbulent channel flow with transverse square bars on one wall}.  \jt{Journal
  of Fluid Mechanics}  \bvol{491},  \pg{229–238}.

\bibitem[Li \& Laizet(2010)]{2DECOMPFFTA}
{\sc \au{Li, N.} \& \au{Laizet, S.}} \yr{2010} {2DECOMP}\&{FFT} - a highly
  scalable 2{D} decomposition library and {FFT} interface.  \bt{In {\em Cray
  User Group 2010: Simulation comes of age\/}}.

\bibitem[Luchini(1996)]{luchini_1996}
{\sc \au{Luchini, P}} \yr{1996} Reducing the turbulent skin friction.  \bt{In
  {\em Computational methods in applied sciences' 96 (Paris, 9-13 September
  1996)\/}},  \pg{pp. 465--470}.  \publ{John Wiley \& Sons Ltd}.

\bibitem[Luchini(2015)]{luchini_2015}
{\sc \au{Luchini, P.}} \yr{2015} The relevance of longitudinal and transverse
  protrusion heights for drag reduction by a superhydrophobic surface.  \bt{In
  {\em Proc. European Drag Reduction and Flow Control Meeting—EDRFMC 2015;
  March 23--26\/}},  \pg{pp. 81--82}.

\bibitem[Luchini {\em et~al.\/}(1991)Luchini, Manzo \&
  Pozzi]{luchini_manzo_pozzi_1991}
{\sc \au{Luchini, P.}, \au{Manzo, F.} \& \au{Pozzi, A.}} \yr{1991}
  \at{Resistance of a grooved surface to parallel flow and cross-flow}.
  \jt{Journal of Fluid Mechanics}  \bvol{228},  \pg{87–109}.

\bibitem[Manes {\em et~al.\/}(2011)Manes, Poggi \&
  Ridolfi]{manes_poggi_ridolfi_2011}
{\sc \au{Manes, C.}, \au{Poggi, D.} \& \au{Ridolfi, L.}} \yr{2011}
  \at{Turbulent boundary layers over permeable walls: scaling and near-wall
  structure}.  \jt{Journal of Fluid Mechanics}  \bvol{687},  \pg{141–170}.

\bibitem[Min \& Kim(2004)]{Min_2004}
{\sc \au{Min, T.} \& \au{Kim, J.}} \yr{2004}  \at{Effects of hydrophobic
  surface on skin-friction drag}.  \jt{Physics of Fluids}  \bvol{16}~(7),
  \pg{L55--L58}.

\bibitem[Nikuradse(1933)]{nikuradse}
{\sc \au{Nikuradse, J.}} \yr{1933}  \at{Strömungsgesetze in rauhen rohren}.
  \jt{V.D.I. Forschungsheft}  \bvol{361}.

\bibitem[Orlandi \& Leonardi(2006)]{orlandi_2_3}
{\sc \au{Orlandi, P.} \& \au{Leonardi, S.}} \yr{2006}  \at{{D}{N}{S} of
  turbulent channel flows with two- and three-dimensional roughness}.
  \jt{Journal of Turbulence}  \bvol{7},  \pg{N73}.

\bibitem[Orlandi {\em et~al.\/}(2006)Orlandi, Leonardi \&
  Antonia]{orlandi_leonardi_antonia_2006}
{\sc \au{Orlandi, P.}, \au{Leonardi, S.} \& \au{Antonia, R.~A.}} \yr{2006}
  \at{Turbulent channel flow with either transverse or longitudinal roughness
  elements on one wall}.  \jt{Journal of Fluid Mechanics}  \bvol{561},
  \pg{279–305}.

\bibitem[Panton(1999)]{Panton}
{\sc \au{Panton, R.}} \yr{1999} Self-sustaining mechanisms of wall turbulence -
  a review.  \bt{In {\em 37th Aerospace Sciences Meeting and Exhibit\/}}.

\bibitem[Ran {\em et~al.\/}(2021)Ran, Zare \&
  Jovanović]{ran_zare_jovanovic_2021}
{\sc \au{Ran, W.}, \au{Zare, A.} \& \au{Jovanović, M.~R.}} \yr{2021}
  \at{Model-based design of riblets for turbulent drag reduction}.  \jt{Journal
  of Fluid Mechanics}  \bvol{906},  \pg{A7}.

\bibitem[Sabot \& Comte-Bellot(1976)]{sabot_comte-bellot_1976}
{\sc \au{Sabot, J.} \& \au{Comte-Bellot, G.}} \yr{1976}  \at{Intermittency of
  coherent structures in the core region of fully developed turbulent pipe
  flow}.  \jt{Journal of Fluid Mechanics}  \bvol{74}~(4),  \pg{767–796}.

\bibitem[Schlichting(1937)]{schlichting1937}
{\sc \au{Schlichting, H.}} \yr{1937}  \bt{Experimental investigation of the
  problem of surface roughness}. {\em Tech. Rep.\/} 823.  \org{National
  Advisory Committee for Aeronautics}.

\bibitem[Schlichting(1979)]{Schlichting}
{\sc \au{Schlichting, H.}} \yr{1979} {\em Boundary Layer Theory\/}.
  \publ{McGraw-Hill}.

\bibitem[Spalart \& McLean(2011)]{Spalart_2011}
{\sc \au{Spalart, P.R.} \& \au{McLean, J.D.}} \yr{2011}  \at{Drag reduction:
  enticing turbulence, and then an industry}.  \jt{Philosophical Transactions
  of the Royal Society A: Mathematical, Physical and Engineering Sciences}
  \bvol{369}~(1940),  \pg{1556--1569}.

\bibitem[Spalart {\em et~al.\/}(1991)Spalart, Moser \& Rogers]{SPALART1991}
{\sc \au{Spalart, P.~R.}, \au{Moser, R.~D.} \& \au{Rogers, M.~M.}} \yr{1991}
  \at{Spectral methods for the navier-stokes equations with one infinite and
  two periodic directions}.  \jt{Journal of Computational Physics}
  \bvol{96}~(2),  \pg{297--324}.

\bibitem[Sudhakar {\em et~al.\/}(2021)Sudhakar, L{\=a}cis, Pasche \&
  Bagheri]{sudhakar2020}
{\sc \au{Sudhakar, Y.}, \au{L{\=a}cis, U.}, \au{Pasche, S.} \& \au{Bagheri,
  S.}} \yr{2021}  \at{Higher-order homogenized boundary conditions for flows
  over rough and porous surfaces}.  \jt{Transport in Porous Media}  \pg{pp.
  1--42}.

\bibitem[Wesseling(2009)]{Wesseling2009}
{\sc \au{Wesseling, P.}} \yr{2009} {\em Principles of computational fluid
  dynamics\/}, ,  \vol{vol.~29}.  \publ{Springer Science \& Business Media}.

\bibitem[Zampogna {\em et~al.\/}(2019)Zampogna, Magnaudet \&
  Bottaro]{zampogna_2019}
{\sc \au{Zampogna, G.~A.}, \au{Magnaudet, J.} \& \au{Bottaro, A.}} \yr{2019}
  \at{Generalized slip condition over rough surfaces}.  \jt{Journal of Fluid
  Mechanics}  \bvol{858},  \pg{407–436}.

\end{thebibliography}

\end{document}